\documentclass{article}

\usepackage{arxiv}

\usepackage[utf8]{inputenc} 
\usepackage[T1]{fontenc}    
\usepackage{hyperref}       
\usepackage{url}            
\usepackage{booktabs}       
\usepackage{amsfonts}       
\usepackage{nicefrac}       
\usepackage{microtype}      
\usepackage{lipsum}
\usepackage{graphicx}
\graphicspath{ {./images/} }
\usepackage{epstopdf}
\ifpdf
  \DeclareGraphicsExtensions{.eps,.pdf,.png,.jpg}
\else
  \DeclareGraphicsExtensions{.eps}
\fi

\usepackage{algorithm}
\usepackage{algpseudocode}
\usepackage{threeparttable}

\usepackage{caption}
\usepackage{subcaption}
\usepackage{mathtools}
\usepackage{amssymb}
\usepackage{multirow}
\usepackage{siunitx}
\usepackage{xcolor}

\usepackage[shortlabels]{enumitem}
\setlist[enumerate]{leftmargin=.5in}
\setlist[itemize]{leftmargin=.5in}

\newenvironment{codefont}{\fontfamily{lmtt}\selectfont}{\par}
\DeclareTextFontCommand{\codetext}{\codefont}

\newtheorem{remark}{Remark}

\title{Efficient Bayesian calibration of many-parameter system models}

\author{
 Promit Chakroborty \\
  Department of Civil and Environmental Engineering\\
  Vanderbilt University\\
  Nashville, TN 37212 \\
  \texttt{promit.chakroborty@vanderbilt.edu} \\
   \And
 Sankaran Mahadevan \\
  Department of Civil and Environmental Engineering\\
  Vanderbilt University\\
  Nashville, TN 37212 \\
  \texttt{sankaran.mahadevan@vanderbilt.edu} \\
}

\begin{document}
\maketitle
\begin{abstract}

Computer models of complex engineering systems rely on proper tuning of their model parameters to ensure accurate predictions of the system behavior. The challenge of effectively calibrating many-parameter models is the difficulty of sampling in high-dimensional spaces and the computational expense of generating a large number of samples to characterize the calibrated parameter distributions. The method of active subspaces has been shown to be effective at constructing low-dimensional latent spaces for Bayesian inverse problems when the misfit function (i.e., negative log-likelihood) is treated as the function of interest. On the other hand, works that implement surrogate modeling for inference often focus on approximating the predictive model itself. In this work, an integrated dimension reduction and surrogate modeling framework for efficient and robust model calibration based on the Kennedy O’Hagan framework is proposed, with the following key components. First, an active subspace of the misfit function is identified. Then, a surrogate model for the misfit is constructed in this low-dimensional latent space. Care is taken to ensure that the assumed probabilistic structure of the misfit surrogate is compatible with the structure imposed on the misfit by the observation noise and computer model discrepancy. Further, a generalized likelihood function is defined that can account for the misfit surrogate uncertainty along with the other usual sources of uncertainty, e.g., experimental noise, model inadequacy, etc. This general formulation is shown to be valid for surrogates of any deterministic bijective function of the original likelihood, not just the misfit. Finally, a strategy for incorporating the uncertainty in identifying the active subspace is included. The behavior of the assembled calibration framework is demonstrated through two examples - an analytical polynomial with $ 10 $ to $ 100 $ parameters, and an $ 11 $-dimensional vehicle side-impact test problem. Overall, the proposed method is shown to be efficient at reducing the dimension of the parameter space and sampling from the posterior, even using simple MCMC algorithms like vanilla Metropolis-Hastings. Simultaneously, it comprehensively accounts for all the sources of uncertainty arising from the modeling workflow (based on the Kennedy \& O'Hagan formulation) as well as the numerical tools leveraged in the framework itself.


\end{abstract}


\section{Introduction}
\label{section:introduction}

Computational models enable many design and analysis tasks for engineering systems by simulating their behavior under arbitrary operating conditions that may be impossible or infeasible to recreate experimentally. Faithfully predicting the key quantities of interest that characterize the system's response is paramount for these models, so they are often complex in structure and expensive to create and evaluate. 
In turn, they are usually built to be applicable in a variety of contexts or to a class of systems with similar behavior to maximize their utility. The exact value of the ``model parameters'' for a given system is frequently unknown a priori; to accurately analyze a particular problem using a model, its parameters must first be calibrated based on experimental observations of the problem. The topic of model calibration has been extensively studied~\cite{kennedy2001bayesian, sung2024review, mullins2016bayesian}, and has applications across the spectrum of science and engineering fields, from biology and healthcare~\cite{rocha2022bayesian, sung2020calibration} to hydrology and environmental sciences~\cite{gupta2006model, Goh01112013, cheng2021hierarchical}, from nuclear physics~\cite{higdon2015bayesian, kejzlar2020statistical} to materials sciences~\cite{generale2022bayesian}, and beyond~\cite{thelen2023comprehensive, liyanage2022efficient, kenett2022digital}. Since physical observations of complex engineering systems are usually sparse and noisy, and there are an assortment of errors and uncertainties present in any computational modeling workflow, calibrating model parameters requires inference under uncertainty. Bayes' Theorem forms the basis of a robust and powerful framework that is applied in such cases to obtain distributions of parameter values that best match the model to the observed behavior of a system of interest~\cite{gamerman2006markov}. 
Therefore, \textit{Bayesian model calibration} has received significant interest in the literature~\cite{kennedy2001bayesian, sung2024review, mullins2016bayesian}.

The posterior parameter densities obtained as solutions of the Bayesian model calibration problem are proportional to the product of an assumed prior density based on a priori knowledge and a likelihood function that compares the model output predictions for a given set of parameters against experimental data on the system. For practical problems, the posteriors are almost never available in closed form. Instead, they are sampled using Monte Carlo-based simulation techniques, and these sample sets are subsequently used to construct the ``empirical'' posterior density function, or compute statistics of the posterior (such as the mean, mode, or percentiles). Procedures that can efficiently sample complex-shaped distributions are thus a key enabler of Bayesian calibration. \textit{Markov chain Monte Carlo} (MCMC) methods are one such class of algorithms. 
Popular for their ability to sample from density functions known only up to a normalization constant and of arbitrary shape, a variety of algorithms designed for specialized sampling tasks~\cite{haario2006dram, VrugtterBraak, goodman2010ensemble, falcioni1999biased, ching2007transitional, chakroborty2026intrepid} have been developed from the foundational \textit{Metropolis-Hastings} algorithm~\cite{metropolis1953equation, hastings1970}, and several of them have been used for Bayesian inference~\cite{qian2003monte, ching2007transitional, generale2022bayesian, lewis2017bayesian}. Other works have focused on reducing the number of samples required to effectively represent the posterior distribution, both with~\cite{straub2015bayesian} and without~\cite{neal2021robust} MCMC-based samplers.

Accurately representing the calibrated distribution may require tens of thousands of samples, depending on the complexity of its shape and the dimensionality of the parameter vector. 
Since each sample draw involves running an expensive computer model (for the likelihood), this direct sampling approach can quickly become infeasible. Analogous difficulties arise across most engineering uncertainty quantification tasks, and a common solution is to replace the expensive computer model with a comparatively cheaper \textit{surrogate model} (also called an emulator), thus vastly reducing the overall computational cost of the analysis. Numerous methods for building such approximations have been developed in the literature, from reduced-order physics models~\cite{lucia2004reduced} to purely data-driven regression techniques such as Gaussian Process Regression~\cite{RasmussenWilliamsGPR, AKMCS}, Polynomial Chaos Expansions~\cite{xiu2002wiener, sharma2024physics}, and Neural Networks~\cite{haykin2009neural}, as well as multifidelity techniques that combine multiple models with varying costs and predictive accuracies~\cite{peherstorfer2018survey, chakroborty2023general}. For Bayesian inference specifically, surrogates are primarily used to emulate the computer model, indirectly reducing the cost of evaluating the likelihood function~\cite{jorgensen2022efficient, kapusuzoglu2023multi, rocha2022bayesian, myren2021comparison}. Few works focus on the likelihood itself, although Wagner et al.~\cite{wagner2021bayesian} use Polynomial Chaos Expansion to approximate the likelihood and provide analytical expressions for the posterior density. A group of likelihood-free calibration methods called approximate Bayesian computation~\cite{beaumont2002approximate} also leverage emulators in some cases to improve efficiency~\cite{jorgensen2022efficient}. This paper, however, focuses on the more traditional ways of sampling the posterior by directly formulating and computing the likelihood at candidate points.

Another problem arises due to the \textit{curse of dimensionality} when several parameters must be inferred simultaneously. Not only does the number of samples required to effectively explore the parameter space increase rapidly with dimension, but most sampling algorithms struggle with degeneracy in high-dimensional spaces~\cite{Bengtsson_2008, ObstaclestoHighDimensionalParticleFiltering, AgapiouIS}. The simplest strategy to achieve \textit{dimension reduction} involves identifying only those parameters that the model is highly sensitive to and fixing the rest at nominal values (e.g., their mean), thus reducing the effective dimensionality of the parameter space while keeping the information loss low~\cite{sobol2001global, tarantola2007estimating}. More sophisticated solutions aim to find and retain a small number of directions or curves in the parameter space along which the model changes the most. These latent spaces can be constructed in broadly two ways - through the principle of sufficient dimension reduction~\cite{li1991sliced, cook2005sufficient, bura2001estimating, hsing2009rkhs, yeh2008nonlinear} (also known as inverse regression methods) or using the theory of \textit{active subspaces}~\cite{ConstantineActiveSubspacesBook2015, zahm2020gradient, bridges2019activemanifold, romor2022kernel} - and have been applied to the task of Bayesian calibration to great success~\cite{cui2014likelihood, constantine2016accelerating, lewis2017bayesian}.

Studies that explicitly couple dimension reduction and surrogate modeling for the task of model calibration are largely limited to special cases; for example, problems where the computer model is a field quantity or is approximated as one, such that low-dimensional representations arise naturally out of truncated Karhunen-Loève expansions~\cite{marzouk2009dimensionality} or sparse polynomial chaos surrogates~\cite{elsheikh2014efficient}. A more general integrated dimension reduction and surrogate modeling framework has been proposed for forward prediction problems by Guo et al.~\cite{guo2023investigation} and extended to include active learning~\cite{guo2024active}. In the present work, we follow a similarly broad strategy as~\cite{guo2023investigation} for coupling dimension reduction and surrogate modeling to efficiently solve the model calibration problem as defined by Kennedy \& O'Hagan~\cite{kennedy2001bayesian}, making minimal assumptions about the types of parameters that can be calibrated while ensuring theoretical compatibility of the chosen tools and accounting for the additional uncertainties introduced through the selected approximations. Authors like Cui et al.~\cite{cui2014likelihood} and Constantine et al.~\cite{constantine2016accelerating} suggest that dimension reduction for Bayesian inference should optimize for the likelihood function (or a simple transformation thereof). Following their suggestion, our algorithm constructs a low-dimensional subspace and an emulator for the negative log-likelihood, which is a departure from the more common practice of approximating the predictive model with a surrogate.

Section~\ref{section:dimension_reduction} introduces the theory of active subspaces and details the dimension reduction methodology used. Section~\ref{section:surrogate_and_generalized_likelihood} discusses the surrogate construction and its impact on the likelihood function. The calibrated posteriors are formally defined in Section~\ref{section:calibration_posterior} along with certain practical considerations around the informativeness vs. uncertainty content of the inferred parameter distributions in engineering applications. Finally, the framework is applied to two numerical examples in Section~\ref{section:numerical_examples}. But first, the key contributions of our manuscript are detailed below.

\subsection{Key components and contributions of the proposed methodology}
\label{section:contributions}

Both surrogate modeling and dimension reduction are tools that rely on multiple implicit and nuanced assumptions. These may not be compatible in some problems and can lead to nonsensical calibration results if coupled ad hoc within a Bayesian inference framework. More concerning is the realization that a cursory glance may not always reveal broken assumptions and the lack of meaningfulness in the results produced, posing a significant risk to the trustworthiness and robustness of heuristic calibration algorithms applied in practice. Hence, this paper focuses on making deliberate choices regarding the preferred dimension reduction and surrogate modeling tools, ensuring theoretical compatibility while maximizing computational efficiency and ease of implementation. Our work has the following key components:
\begin{enumerate}
    \item A low-dimensional latent space is identified that uses the minimum number of variables to capture most of the variation in the negative log-likelihood function (comparing the model prediction to the observations).
    \item A surrogate that maps the latent-space vector to the outputs of the negative log-likelihood function is assembled. The probabilistic form of this surrogate is first carefully derived based on the modeling assumptions as per the Kennedy \& O'Hagan framework~\cite{kennedy2001bayesian}.
    \item A ``generalized likelihood'' is defined that can incorporate the prediction error of the negative log-likelihood emulator along with the other uncertainties already present. This generalized likelihood is shown to be valid for surrogates of any deterministic bijective function of the likelihood.
    \item A simple bootstrapping routine as per~\cite{ConstantineActiveSubspacesBook2015} is implemented to quantify the uncertainty in constructing the latent space. This uncertainty is considered in generating the training data for the surrogate, and again when reconstructing the calibrated samples from the latent space back to the original space.
    \item In addition to sampling the full calibrated posterior, our algorithm also generates samples from a ``conditional active posterior'' at no additional cost, which serves as a valuable compromise between the uncertainty and informativeness of the parameter vector. The conditional active posterior allows for significantly more confident inference of the parameters compared to the full posterior, without meaningfully affecting the uncertainty in the model output predictions.
    \item Both the full posterior and the conditional active posterior are available through cheap-to-evaluate density functions, allowing for direct functional analyses and optimization-based point parameter estimation in addition to Monte Carlo statistical estimation.
\end{enumerate}

The novel contributions of this paper are discussed next.
\begin{enumerate}
    \item To the best of the author's knowledge, this work is the first to develop a flexible mathematical framework wherein the assumptions of both dimension reduction and surrogate modeling are compatible in the context of model parameter calibration. The focus on calibration allows the proposed method to target the likelihood function directly (instead of the QoI function as in previous literature) for dimension reduction and emulation, which improves the performance of these tools for inference problems. On the other hand, by minimizing the assumptions on the calibration quantities and on the type and behavior of the QoI function, the proposed algorithm remains applicable across a broad spectrum of calibration problems (see Section~\ref{section:notation}), unlike more specialized algorithms in the literature that have incorporated both dimension reduction and surrogate modeling in the past.
    \item The ``generalized likelihood'' formulated in this paper is original, and provides a way for the prediction uncertainty of surrogates of quantities other than the QoI function to be included in Bayes' theorem.
    \item The proposed framework allows for a comprehensive inclusion of all sources of uncertainty arising due to the various mappings (active subspace, surrogate model). In particular, the estimation uncertainty of the latent space is accounted for in a unique way in the proposed methodology, which is a component of the uncertainty that is often unaccounted for in existing inference algorithms.
    \item The present work also presents a rigorous treatment of posterior uncertainty analysis, moving beyond heuristics and providing a sound theoretical basis for balancing informativeness vs. uncertainty of the calibration parameters through the definition and use of the ``conditional active posterior''.
\end{enumerate}

\section{Notation and problem setup}
\label{section:notation}

Let $ \Upsilon \left( \mathbf{X} \right) $ denote a physical process or engineering system, whose outcome depends on some operating conditions specified through ``external inputs'' $ \mathbf{X} = \begin{bmatrix} x_1 & \dots & x_{d_\text{in}} \end{bmatrix}^T \in \mathbb{R}^{d_\text{in}} $. A computer model of this system, denoted $ f \left( \mathbf{X}, \boldsymbol{\Theta}_f \right) \in \mathbb{R}^{d_\text{out}} $, simulates its behavior for each $ \mathbf{X} $ by simultaneously predicting multiple output quantities of interest (QoI). In addition to $ \mathbf{X} $, this QoI function also depends on some parameter inputs $ \boldsymbol{\Theta}_f = \begin{bmatrix} \theta_1 & \dots & \theta_{d_f} \end{bmatrix}^T \in \mathbb{R}^{d_f} $, which capture a variety of additional attributes and conditions that affect the model's behavior, e.g., features of the model's mathematical construction, or physical process properties not encoded in the external inputs. To optimally represent the system under study, the specific, fixed values of these parameters resulting in the ``best fit'' of the model must be inferred from the experimentally observed behavior of the true process $ \Upsilon $. However, no model is perfect, and usually even the best-fit $ f \left( \mathbf{X}, \boldsymbol{\Theta}_f \right) \neq \Upsilon \left( \mathbf{X} \right) $ due to limitations in the modeling workflow. While there is no universally correct or optimal way to model this mismatch, the seminal work by Kennedy and O’Hagan~\cite{kennedy2001bayesian} proposed the following sufficiently general formulation:
\begin{equation}
    \label{eqn:process_to_model}
    \Upsilon \left( \mathbf{X} \right) = \kappa f \left( \mathbf{X}, \boldsymbol{\Theta}_f \right) + \eta \left( \mathbf{X} \right)
\end{equation}
where $ \kappa $ is an unknown but constant regression parameter and $ \eta \left( \mathbf{X} \right) \in \mathbb{R}^{d_\text{out}} $ represents a model inadequacy function. Many works refer to $ \eta \left( \mathbf{X} \right) $ as a model discrepancy term.

On the other hand, experimental observations of $ \Upsilon $ also include errors and uncertainties unrelated to the modeling procedure. It is common to model these uncertainties using an additive term~\cite{kennedy2001bayesian} as
\begin{equation}
    \label{eqn:observation_to_process}
    \mathbf{Y} \left( \mathbf{X} \right) = \Upsilon \left( \mathbf{X} \right) + \varepsilon_{\mathbf{Y}} \left( \mathbf{X} \right)
\end{equation}
where $ \varepsilon_{\mathbf{Y}} \left( \mathbf{X} \right) \in \mathbb{R}^{d_\text{out}} $ denotes the stochastic experimental noise, and $ \mathbf{Y} \left( \mathbf{X} \right) = \begin{bmatrix} y_1 \left( \mathbf{X} \right) & \dots & y_{d_\text{out}} \left( \mathbf{X} \right) \end{bmatrix}^T \in \mathbb{R}^{d_\text{out}} $ is the random function corresponding to the act of experimentally observing the QoIs that characterize the behavior of $ \Upsilon \left( \mathbf{X} \right) $. (These are the same QoIs predicted by $ f \left( \mathbf{X}, \boldsymbol{\Theta}_f \right) $.) Thus, the relationship between the model predictions and the output observations can be written as
\begin{equation}
    \label{eqn:observation_to_model}
    \mathbf{Y} \left( \mathbf{X} \right) = \kappa f \left( \mathbf{X}, \boldsymbol{\Theta}_f \right) + \eta \left( \mathbf{X} \right) + \varepsilon_{\mathbf{Y}} \left( \mathbf{X} \right)
\end{equation}
Since several of these terms exist to represent uncertainties in the modeling and experimental setup, $ \mathbf{Y} \left( \mathbf{X} \right) $, $ \varepsilon_{\mathbf{Y}} \left( \mathbf{X} \right) $, $ f \left( \mathbf{X}, \boldsymbol{\Theta}_f \right) $, and $ \eta \left( \mathbf{X} \right) $ are all random fields in the most general case. However, a wide variety of simplifying assumptions have been used in the literature, e.g., modeling $ f \left( \mathbf{X}, \boldsymbol{\Theta}_f \right) $ as a deterministic function of $ \mathbf{X} $ and $ \boldsymbol{\Theta}_f $, setting $ \kappa = 1 $ or treating $ \eta \left( \mathbf{X} \right) $ as independent of $ \mathbf{X} $.

It is clear from Eq.~\eqref{eqn:observation_to_model} that the model parameters $ \boldsymbol{\Theta}_f $ cannot be inferred without knowing the specifics of the model discrepancy $ \eta \left( \mathbf{X} \right) $, the regression constant $ \kappa $, and the experimental noise $ \varepsilon_{\mathbf{Y}} \left( \mathbf{X} \right) $. Additionally, the existence of uncertainties at every step of the modeling-to-observation workflow implies that there is no single optimal value of any of these quantities, but rather a set of values with varying probability of accurately capturing the observed behavior of $ \Upsilon $. Inferring these quantities using the Bayesian calibration framework is hence the natural conclusion. To do so, appropriate probabilistic structures must be chosen for $ f \left( \mathbf{X}, \boldsymbol{\Theta}_f \right) $, $ \eta \left( \mathbf{X} \right) $, and $ \varepsilon_{\mathbf{Y}} \left( \mathbf{X} \right) $ to construct a robust and informative likelihood that can effectively incorporate the experimental observations. The Kennedy and O'Hagan approach~\cite{kennedy2001bayesian} models $ f \left( \mathbf{X}, \boldsymbol{\Theta}_f \right) $ and $ \eta \left( \mathbf{X} \right) $ as Gaussian processes and $ \varepsilon_{\mathbf{Y}} $ as a zero-mean Gaussian distribution (independent of $ \mathbf{X} $); the hyperparameters corresponding to the mean and covariance functions of the aforementioned quantities are inferred simultaneously with $ \boldsymbol{\Theta}_f $. This Gaussian structure is in keeping with engineering practice, where it is customary to treat the experimental noise as normally distributed with zero mean if there is no systematic bias in the measurement. Other works have built on~\cite{kennedy2001bayesian}, proposing different distribution structures for the model discrepancy~\cite{ling2014selection} or alternative procedures for calibrating it~\cite{maupin2020model}, while retaining the basic form of Eq.~\eqref{eqn:observation_to_model}.

We adopt the simultaneous calibration paradigm and the assumptions on $ f \left( \mathbf{X}, \boldsymbol{\Theta}_f \right) $, $ \eta \left( \mathbf{X} \right) $, and $ \varepsilon_{\mathbf{Y}} $ from~\cite{kennedy2001bayesian} in the present work. To formulate our integrated dimension reduction and Bayesian calibration framework, the parameter vector $ \boldsymbol{\Theta} = \begin{bmatrix} \theta_1 & \dots & \theta_{d_p} \end{bmatrix}^T \in \mathbb{R}^{d_\text{p}} $ (also called the calibration vector) is constructed as the collection of $ \boldsymbol{\Theta}_f $, $ \kappa $, and the hyperparameters defining the mean and covariance functions of $ f \left( \mathbf{X}, \boldsymbol{\Theta}_f \right) $, $ \eta \left( \mathbf{X} \right) $, and $ \varepsilon_{\mathbf{Y}} $. Consequently, we can write
\begin{equation}
    \label{eqn:general_model_distribution_assumptions_KOH_framework}
    \begin{gathered}
        f \left( \mathbf{X}, \boldsymbol{\Theta}_f \right) \sim \mathcal{N} \left( \boldsymbol{\mu}_f \left( \mathbf{X}, \boldsymbol{\Theta} \right), \boldsymbol{\Sigma}_f \left( \mathbf{X}, \boldsymbol{\Theta} \right) \right) \; \text{, } \; \eta \left( \mathbf{X} \right) \sim \mathcal{N} \left( \boldsymbol{\mu}_\eta \left( \mathbf{X}, \boldsymbol{\Theta} \right), \boldsymbol{\Sigma}_\eta \left( \mathbf{X}, \boldsymbol{\Theta} \right) \right) \; \text{, and } \; \varepsilon_{\mathbf{Y}} \sim \mathcal{N} \left( \boldsymbol{0}, \boldsymbol{\Sigma}_\varepsilon \left( \boldsymbol{\Theta} \right) \right) \\
        \Rightarrow \mathbf{Y} \left( \mathbf{X} \right) \sim \mathcal{N} \left( \kappa \boldsymbol{\mu}_f \left( \mathbf{X}, \boldsymbol{\Theta} \right) + \boldsymbol{\mu}_\eta \left( \mathbf{X}, \boldsymbol{\Theta} \right) , \boldsymbol{\Sigma}_{\mathbf{Y}} \left( \boldsymbol{\Theta} \right) \right) \qquad \text{(from Eq.~\eqref{eqn:observation_to_model})} \\
        \text{where } \boldsymbol{\Sigma}_{\mathbf{Y}} \left( \boldsymbol{\Theta} \right) = \boldsymbol{\Sigma}_f \left( \mathbf{X}, \boldsymbol{\Theta} \right) + \boldsymbol{\Sigma}_\eta \left( \mathbf{X}, \boldsymbol{\Theta} \right) + \boldsymbol{\Sigma}_\varepsilon \left( \boldsymbol{\Theta} \right)
    \end{gathered}
\end{equation}

If the parameters are assumed to follow a prior distribution with joint density function $ \pi \left( \boldsymbol{\Theta} \right) $, then the posterior density $ p \left( \boldsymbol{\Theta} \right) $ according to Bayes' theorem is given by
\begin{equation}
    \label{eqn:Bayes_theorem}
    p \left( \boldsymbol{\Theta} \right) \propto \mathcal{L} \left( \boldsymbol{\Theta} \right) \pi \left( \boldsymbol{\Theta} \right)
\end{equation}
Here, the likelihood function $ \mathcal{L} \left( \boldsymbol{\Theta} \right) $ is written in terms of the misfit function $ \mathcal{M} \left( \boldsymbol{\Theta} \right) $, which measures the ``goodness of fit'' of the model by comparing the predicted QoIs against experimentally observed values at a finite number of input conditions. Let $ \mathfrak{D} $ be the set of experimental data having $ D $ tuples of observed input-output pairs, where $ \left( \mathbf{X}^*, \mathbf{Y}^* \right) \in \mathfrak{D} $ denotes one such tuple. Then, Eq.~\eqref{eqn:general_model_distribution_assumptions_KOH_framework} implies that
\begin{multline}
\label{eqn:misfit_definition}
    \mathcal{M} \left( \boldsymbol{\Theta} \right) = \\
    \sum_{\forall \left( \mathbf{X}^*, \mathbf{Y}^* \right) \in \mathfrak{D}} \frac{1}{2} \left[ \left\{ \mathbf{Y}^* - \kappa \boldsymbol{\mu}_f \left( \mathbf{X}^*, \boldsymbol{\Theta} \right) - \boldsymbol{\mu}_\eta \left( \mathbf{X}^*, \boldsymbol{\Theta} \right) \right\}^T \left\{ \boldsymbol{\Sigma}_{\mathbf{Y}} \left( \boldsymbol{\Theta} \right) \right\}^{-1} \left\{ \mathbf{Y}^* - \kappa \boldsymbol{\mu}_f \left( \mathbf{X}^*, \boldsymbol{\Theta} \right) - \boldsymbol{\mu}_\eta \left( \mathbf{X}^*, \boldsymbol{\Theta} \right) \right\} \right]
\end{multline}
\begin{equation}
    \label{eqn:misfit_to_likelihood}
    \text{and } \mathcal{L} \left( \boldsymbol{\Theta} \right) = \exp{\left( - \mathcal{M} \left( \boldsymbol{\Theta} \right) \right)}
\end{equation}

The calibration procedure detailed in this work infers the values of all the components of $ \boldsymbol{\Theta} $ simultaneously, which includes not just the model parameters but also the hyperparameters defining the Gaussian distributions characterizing each of the terms in the modeling-to-observation workflow. No distinction is made in any of the key components of our framework (viz., the dimension reduction procedure or the surrogate model construction) between the model parameters, some of which may have engineering interpretations, and the other inferred quantities, which may not. Specifically, our method's dimension reduction (see Section~\ref{section:likelihood_informed_subspace}) focuses purely on mathematical convenience and the computational benefits of sampling from and constructing a surrogate in low-dimensional domains, and not on the explainability of the resultant latent parameter vector. As a result, an analyst should not attempt to ascribe any physical meaning to the low-dimensional representation of $ \boldsymbol{\Theta} $; at most, information about the sensitivities of the parameters can be extracted. This is an extension of the known issue of identifiability that arise when a calibration problem includes both model parameters and model discrepancy~\cite{maupin2020model, white2023discrepancy}, whereby the presence of additional parameters alters the interpretation of the model parameters in the calibration problem.

To simplify the notation and improve the clarity of the discussion in the rest of this paper, we make the following simplifying assumptions:
\begin{enumerate}
    \item $ \boldsymbol{\Sigma}_{\mathbf{Y}} \left( \boldsymbol{\Theta} \right) \equiv \boldsymbol{\Sigma}_{\mathbf{Y}} $ is a constant matrix (i.e., not dependent on the parameter vector $ \boldsymbol{\Theta} $) and is prescribed a priori instead of being calibrated along with $ \boldsymbol{\Theta} $.
    \item The QoI function $ f \left( \mathbf{X}, \boldsymbol{\Theta}_f \right) $ is a deterministic function (not a random field), and thus only contributes the model parameters $ \boldsymbol{\Theta}_f $ to the calibration vector $ \boldsymbol{\Theta} $. We use the shorthand $ f_{\boldsymbol{\Theta}}^* = f \left( \mathbf{X}^*, \boldsymbol{\Theta} \right) $ in the subsequent sections.
    \item The QoI function is assumed to be perfectly adequate at capturing the true process $ \Upsilon $, and the model discrepancy term $ \eta \left( \mathbf{X} \right) $ is omitted henceforth. We emphasize that this is done purely for clarity of notation and ease of discussion; the derivations in this paper can be expanded in a straightforward manner to include an explicit model discrepancy term.
    \item The components of the QoI function are assumed to be uncorrelated without loss of generality, since this can always be achieved using affine transformations. Additionally, the components of the experimental noise $ \varepsilon_{\mathbf{Y}} \left( \mathbf{X} \right) $ are assumed to be independent. Therefore, $ \boldsymbol{\Sigma}_{\mathbf{Y}} = \operatorname{diag} \left( \begin{bmatrix} \sigma_{\mathbf{Y}, 1}^2 & \dots & \sigma_{\mathbf{Y}, d_\text{out}}^2 \end{bmatrix} \right) $.
\end{enumerate}
In the rest of this manuscript, we will build upon the above notation and assumptions to facilitate our construction of the integrated efficient Bayesian calibration framework. (We will also see in Section~\ref{section:generalized_likelihood} how the above definition of the likelihood is expanded to accept surrogate predictions of the misfit. This broader definition is termed the ``\textit{generalized likelihood}''.) However, these assumptions are only used for notational simplicity; none of them are necessary for the mathematical derivations of the subsequent sections, and analogous results can be derived in a straightforward manner for cases that do not include these assumptions. Moreover, even the Gaussianity assumptions borrowed from Kennedy \& O'Hagan~\cite{kennedy2001bayesian} can be replaced with other probabilistic structures. Doing so will change the definition of the misfit function but not the general steps and derivations that make up the proposed method.

\section{Dimension reduction for Bayesian calibration}
\label{section:dimension_reduction}

In this section, we introduce the theory of active subspaces, our chosen dimension reduction tool, and discuss its implementation details for our proposed framework. However, other dimension reduction tools can be used instead if preferred; the framework's general construction remains valid.

\subsection{Active subspaces (AS)}
\label{section:active_subsapces}

The method of active subspaces, introduced by Constantine~\cite{ConstantineActiveSubspacesBook2015}, aims to identify a small number of directions in the domain of a function of interest along which most of its variability is captured. The directions that maximize the expected value of the squared directional derivative - also known as the derivative-based global sensitivity measure (DGSM) - are used to form a vector space, which serves as the low-dimensional ``Active Subspace'' of the function.

Let $ h \left( \mathbf{u} \right) $ be a scalar-valued, square-integrable, and differentiable function, with the inputs $ \mathbf{u} \in \mathbb{R}^n $ distributed according to the probability density function $ \rho_\mathbf{u} \left( \mathbf{u} \right) $ such that, without loss of generality, $ \mathbb{E}_\rho \left[ \mathbf{u} \right] = \boldsymbol{0} $. (Here, $ \mathbb{E}_\rho \left[ \cdot \right] $ denotes the expected value operation with respect to the density $ \rho $.) To identify the directions along which the DGSM of $ h \left( \mathbf{u} \right) $, i.e., the variance of the gradient of $ h \left( \mathbf{u} \right) $, are maximized, we first compute the ``uncentered gradient covariance matrix'' as
\begin{equation}
    \label{eqn:uncentered_gradient_covariance_matrix}
    \mathbf{C}_h = \mathbb{E}_\rho \left[ \left( \nabla h \left( \mathbf{u} \right) \right) \left( \nabla h \left( \mathbf{u} \right) \right)^T \right] = \int \left( \nabla h \left( \mathbf{u} \right) \right) \left( \nabla h \left( \mathbf{u} \right) \right)^T \rho_\mathbf{u} \left( \mathbf{u} \right) d\mathbf{u}
\end{equation}
From here, a simple application of the theory of Principal Component Analysis reveals that the leading eigenvectors of $ \mathbb{E}_\rho \left[ \left( \nabla h \left( \mathbf{u} \right) \right) \left( \nabla h \left( \mathbf{u} \right) \right)^T \right] $ are the directions we want. Rigorous proofs are provided in~\cite{ConstantineActiveSubspacesBook2015,constantine2014active}, including the following result:
\begin{remark}
\label{remark:directional_gradient_variance}
    Let $ \left( \lambda_i, \mathbf{w}_i \right) $, $ i = 1, \dots, n $, be the $ i $-th eigenpair of $ \mathbb{E}_\rho \left[ \left( \nabla h \left( \mathbf{u} \right) \right) \left( \nabla h \left( \mathbf{u} \right) \right)^T \right] $ in decreasing order. Then,
    \begin{equation}
        \label{eqn:directional_gradient_variance}
        \mathbb{E}_\rho \left[ \left( \left( \nabla h \left( \mathbf{u} \right) \right)^T \mathbf{w}_i \right)^2 \right] = \lambda_i
    \end{equation}
\end{remark}
Clearly, the function varies less along the later eigenvectors, by virtue of corresponding to the smaller eigenvalues and hence having smaller values of the squared directional gradient. Hence, to capture the majority of the variability of the function, it is enough to limit ourselves to the first $ m < n $ eigenvectors. If $ \mathbf{W}_m = \begin{bmatrix} \mathbf{w}_1 & \dots & \mathbf{w}_m \end{bmatrix} $ denotes the matrix whose columns are the first $ m $ eigenvectors of $ \mathbb{E}_\rho \left[ \left( \nabla h \left( \mathbf{u} \right) \right) \left( \nabla h \left( \mathbf{u} \right) \right)^T \right] $, then $ \mathbf{W}_m^T \mathbf{u} $ can be used as the low-dimensional representation of the input to the function, and the column-space of $ \mathbf{W}_m $ is called the active subspace of $ h \left( \mathbf{u} \right) $. Choosing the number of directions $ m $ depends to some extent on the task for which the active subspace is used; hence, we discuss it further in Section~\ref{section:likelihood_informed_subspace}.


\subsection{The likelihood-informed subspace (LIS)}
\label{section:likelihood_informed_subspace}

For Bayesian inference, we want to leverage active subspace-based dimension reduction to obtain an efficient, low-dimensional representation of the likelihood, which depends on the QoI function. The first instinct may be to construct the active subspace for $ f \left( \mathbf{X}, \boldsymbol{\Theta} \right) $ directly. However, there are a number of reasons why constructing the ``likelihood-informed subspace'', i.e., the AS for the misfit function $ \mathcal{M} \left( \boldsymbol{\Theta} \right) $, is more beneficial:
\begin{enumerate}
    \item The QoI function is often vector-valued. While Section~\ref{section:active_subsapces} only describes the method of active subspaces for scalar-valued functions, extensions to vector-valued functions have been developed recently~\cite{zahm2020gradient}. However, the relative contributions of the individual components are decided heuristically in these extensions. By contrast, the misfit function is always scalar-valued, with the components combined according to the covariance matrix of the observation noise. This allows for a much more straightforward and mathematically robust application of active subspace theory.
    \item The misfit - and by extension the likelihood - is usually a non-linear function of $ f \left( \mathbf{X}, \boldsymbol{\Theta} \right) $. 
    Hence, the optimal directions for capturing most of the variability in the QoIs are not, in general, the same as those for the likelihood or misfit. If the objective is model calibration, it may be better to form the active subspace for the misfit directly rather than the QoIs.
\end{enumerate}
The above insights are supported by prior works~\cite {cui2014likelihood, constantine2016accelerating} that also treat the misfit as the function of interest when leveraging dimension reduction for Bayesian inference.

Consistent with Section~\ref{section:active_subsapces}, to learn a low-dimensional representation for the misfit function, we need to compute the eigendecomposition of its uncentered gradient covariance matrix $ \mathbf{C}_\mathcal{M} $. In practice, this is estimated by evaluating the misfit gradient at a finite number of samples drawn from the distribution of the system parameters, which in our case is the prior distribution $ \pi \left( \boldsymbol{\Theta} \right) $. Instead of estimating $ \mathbf{C}_\mathcal{M} $ directly, we can instead form the gradient data matrix $ \hat{\mathbf{B}}_\mathcal{M} $:
\begin{equation}
    \label{eqn:gradient_data_matrix}
    \hat{\mathbf{B}}_\mathcal{M} = \frac{1}{\sqrt{M}} \begin{bmatrix} \nabla \mathcal{M} \left( \boldsymbol{\Theta}_1 \right) & \dots & \nabla \mathcal{M} \left( \boldsymbol{\Theta}_M \right) \end{bmatrix}
\end{equation}
whose $ k $-th column is the gradient computed at the sample $ \boldsymbol{\Theta}_k \in \mathfrak{A} $, with $ \mathfrak{A} = \left\{ \boldsymbol{\Theta}_k \sim \pi \left( \boldsymbol{\Theta} \right) : k = 1, \dots, M \right\} $ being the training data set for the active subspace. The Singular Value Decomposition (SVD) of $ \hat{\mathbf{B}}_\mathcal{M} $ supplies the estimated eigenpairs $ \left( \hat{\lambda}_j , \hat{\mathbf{w}}_j \right) $, $ j = 1, \dots, d_p $, as the diagonal elements of the matrix of estimated eigenvalues $ \hat{\Lambda} $ and the columns of the matrix of estimated eigenvectors $ \hat{\mathbf{W}} $, respectively, where
\begin{equation}
    \label{eqn:SVD_of_gradient_data_matrix}
    \hat{\mathbf{B}}_\mathcal{M} = \hat{\mathbf{W}} \sqrt{\hat{\Lambda}} \hat{V}^T
\end{equation}
Here, $ \hat{\cdot} $ implies a finite-sample approximation, and $ \hat{V} $ is a unitary matrix obtained from the SVD procedure. It is worth noting that the gradient of the misfit, which is necessary for the procedure described in this section, can be obtained from gradients of the QoI function for negligible additional cost. (It is assumed that evaluating $ f \left( \mathbf{X}, \boldsymbol{\Theta} \right) $ is much more expensive than elementary algebra operations. Indeed, for practical engineering problems, where $ f \left( \mathbf{X}, \boldsymbol{\Theta} \right) $ is usually a numerical simulation program, model evaluations or gradient evaluations are usually several orders of magnitude more expensive than analytical computations.) 
\begin{equation}
    \label{eqn:misfit_gradient}
    \nabla \mathcal{M} \left( \boldsymbol{\Theta} \right) = - \left[ \left( \nabla_{\boldsymbol{\Theta}} f \left( \mathbf{X}^*, \boldsymbol{\Theta} \right) \right)^T \boldsymbol{\Sigma}_{\mathbf{Y}}^{-1} \left( \mathbf{Y} - f_{\boldsymbol{\Theta}}^* \right) \right]
\end{equation}
This implies that the same set of QoI evaluations (including gradient computations) can be used to construct the active subspace for both the QoIs and the misfit function. Hence, it is relatively cheap and straightforward to learn both subspaces and use the one with better performance. For example, while the LIS may perform better purely for calibration, the active subspace of the QoI function may be beneficial when other downstream UQ tasks need to be performed after calibration, and the intention is to use the same low-dimensional representation for all analyses. In this manuscript, we restrict ourselves to only Bayesian calibration tasks and build our framework on the assumption that the likelihood-informed subspace is used. Finally, to decide how many gradient evaluations $ M $ are enough to estimate the eigendecomposition with acceptable accuracy, some heuristic guidance is provided in~\cite{ConstantineActiveSubspacesBook2015}. To estimate the first $ m $ eigenpairs with high confidence,
\begin{equation}
    \label{eqn:gradient_sample_size_heuristic_AS}
    M = \alpha m \ln{d_p}
\end{equation}
where $ \alpha \in \left[ 2, 10 \right] $ is an oversampling factor, with the interval recommended based on epistemic experience of the authors in~\cite{ConstantineActiveSubspacesBook2015}.

Learning the LIS is, in effect, a process of rotation and down-selection in the system's parameter space. To obtain the low-dimensional parameter vector, first, the matrix of eigenvectors $ \mathbf{W} $ is used to construct the rotated parameter vector $ \boldsymbol{\Xi} = \begin{bmatrix} \xi_1 & \dots & \xi_{d_p} \end{bmatrix}^T = \mathbf{W}^T \boldsymbol{\Theta} $. It is then partitioned into vectors $ \boldsymbol{\Xi}_{d_l} = \begin{bmatrix} \xi_1 & \dots & \xi_{d_l} \end{bmatrix}^T $ and $ \boldsymbol{\Xi}_\perp = \begin{bmatrix} \xi_{d_l+1} & \dots & \xi_{d_p} \end{bmatrix}^T $ (with the integer $ 1 \leq d_l \leq d_p $), with the former being the desired low-dimensional parameter vector and the latter being the vector of ignored variables (or ``inactive'' variables, to use the terminology of~\cite{ConstantineActiveSubspacesBook2015,constantine2016accelerating,constantine2014active}). They can equivalently be computed as
\begin{gather}
    \boldsymbol{\Xi}_{d_l} = \mathbf{W}_{d_l}^T \boldsymbol{\Theta} \;\; \text{ where } \;\; \mathbf{W}_{d_l} = \begin{bmatrix} \mathbf{w}_1 & \dots & \mathbf{w}_{d_l} \end{bmatrix} \label{eqn:latent_space_projection} \\
    \boldsymbol{\Xi}_\perp = \mathbf{W}_\perp^T \boldsymbol{\Theta} \;\; \text{ where } \;\; \mathbf{W}_\perp = \begin{bmatrix} \mathbf{w}_{d_l+1} & \dots & \mathbf{w}_{d_p} \end{bmatrix} \label{eqn:inactive_variables_projection} \\
    \text{and } \mathbf{W} = \begin{bmatrix} \mathbf{W}_{d_l} & \mathbf{W}_\perp \end{bmatrix} \label{eqn:full_projection_matrix_composition}
\end{gather}
In this manuscript, we use the term \textit{projection} to denote the process of transforming an instance of the parameter vector $ \boldsymbol{\Theta} $ into its corresponding low-dimensional representation $ \boldsymbol{\Xi}_{d_l} $ as $ \boldsymbol{\Xi}_{d_l} = \mathbf{W}_{d_l}^T \boldsymbol{\Theta} $. This process involves some information loss, since the dimensionality of the vector is reduced by discarding the inactive variables $ \boldsymbol{\Xi}_\perp $. 
On the other hand, we use \textit{reconstruction} to refer to the process of representing a LIS parameter vector sample $ \boldsymbol{\Xi}_{d_l} $ in the original parameter space as $ \boldsymbol{\Theta} = \mathbf{W}_{d_l} \boldsymbol{\Xi}_{d_l} $, which is equivalent to the rotation $ \boldsymbol{\Theta} = \mathbf{W} \boldsymbol{\Xi} $ assuming the inactive variables $ \boldsymbol{\Xi}_\perp = \boldsymbol{0} $. Note that in the above equations, we have used the exact eigenvector matrix $ \mathbf{W} $ to represent the exact mathematical transformations and definitions. In practice, the estimated eigenvectors from $ \hat{\mathbf{W}} $ are used, introducing some new uncertainties that must also be quantified; this is addressed in Section~\ref{section:projection_uncertainty_quantification}.

Two useful bounds, proved in~\cite{constantine2014active} and~\cite{constantine2016accelerating}, govern the selection of the number $ l $ of retained variables. In the interest of clarity, they are presented here in plain language and not through mathematical expressions.
\begin{enumerate}
    \item The error in the estimated active subspace, i.e., $ e \left( {\mathbf{W}_{d_l}} \right) = \lVert \mathbf{W}_{d_l} \mathbf{W}_{d_l}^T - \hat{\mathbf{W}}_{d_l} \hat{\mathbf{W}}_{d_l}^T \rVert $, is bounded above by $ \gamma_1 \frac{\lambda_1}{\lambda_{d_l} - \lambda_{d_l+1}} $, where $ \gamma_1 $ is a proportionality constant. (The reader is referred to~\cite{constantine2014active} for more details.)
    \item The Hellinger distance between the exact posterior distribution (defined through $ \mathcal{M} \left( \boldsymbol{\Theta} \right) $) and the approximated posterior distribution (defined through $ \mathcal{M} \left( \mathbf{W}_{d_l} \boldsymbol{\Xi}_{d_l} \right) $) is bounded above by $ \gamma_2 \left[ \left( e \left( {\mathbf{W}_{d_l}} \right) \sqrt{\sum_{j=1}^l \lambda_j} \right) + \sqrt{\sum_{j=l+1}^{d_p} \lambda_j} \right] $, where $ \gamma_2 $ is a proportionality constant. (The reader is referred to~\cite{constantine2016accelerating} for more details.)
\end{enumerate}
In effect, a good active subspace minimizes both errors simultaneously, for which $ d_l $ should be picked such that it corresponds to a large spectral gap $ \left( \lambda_{d_l} - \lambda_{d_l+1} \right) $ and a small contribution from the inactive variables as quantified by $ \sum_{j=d_l+1}^{d_p} \lambda_j $. At present, there is no rigorous guidance on how much relative weight to assign to these two factors or how to optimize for them. Therefore, in this work, we simply select $ d_l $ according to the largest observed spectral gap once $ \sum_{j=d_l+1}^{d_p} \lambda_j < \tau $ for some pre-defined threshold $ \tau $. Future work may explore more rigorous selection criteria to determine the appropriate number of active subspace directions $ d_l $.

\subsubsection{Quantifying the projection uncertainty}
\label{section:projection_uncertainty_quantification}

Since a finite number of samples are used to estimate the eigenvectors in Eq.~\eqref{eqn:SVD_of_gradient_data_matrix}, the obtained directions for the low-dimensional representation are inexact. This uncertainty must be accounted for in the calibration procedure. A simple way to quantify the estimation uncertainty in the eigenvectors is to use the \textit{Bootstrap algorithm}, and a procedure for estimating the estimation error in the active subspace $ e \left( {\mathbf{W}_{d_l}} \right) $ using Bootstrapping is provided in~\cite{ConstantineActiveSubspacesBook2015}. The same procedure also produces replicates of $ \hat{\mathbf{\Lambda}} $ and $ \hat{\mathbf{W}} $, which are of primary interest to us here. The steps involved are as follows:
\begin{enumerate}
    \item Populate the active subspace training data set $ \mathfrak{A} $ 
    and form the nominal gradient data matrix $ \hat{\mathbf{B}}_\mathcal{M} $ as per Eq.~\eqref{eqn:gradient_data_matrix}. Subsequently, compute the nominal estimated eigenvalue matrix $ \hat{\Lambda} $ and the nominal estimated eigenvector matrix $ \hat{\mathbf{W}} $ as per Eq.~\eqref{eqn:SVD_of_gradient_data_matrix}.
    \item Decide the desired number of bootstrap samples $ N $, in other words, the number of replicates.
    \item Form $ N $ index sets $ J_i = \left\{ j_{i1}, \dots, j_{iM} \right\} $, $ i = 1, \dots, N $, s.t. $ j_{ik} $ is drawn randomly from the set of integers $ \left\{ 1, \dots, M \right\} $.
    \item Thus, the $ i $-th replicate of the gradient data matrix is formed as
        \begin{equation}
            \hat{\mathbf{B}}_\mathcal{M}^{(i)} = \frac{1}{\sqrt{M}} \begin{bmatrix} \nabla \mathcal{M} \left( \boldsymbol{\Theta}_{j_{i1}} \right) & \dots & \nabla \mathcal{M} \left( \boldsymbol{\Theta}_{j_{iM}} \right) \end{bmatrix}
        \end{equation}
    \item The singular value decomposition of $ \hat{\mathbf{B}}_\mathcal{M}^{(i)} $ produces the $ i $-th replicate of the estimated eigenvalue matrix as $ \hat{\Lambda}^{(i)} $ and estimated eigenvector matrix as $ \hat{\mathbf{W}}^{(i)} $. These replicates can be conceptualized as samples drawn from the underlying random matrices $ \hat{\Lambda} \left( \omega_N \right) $ and $ \hat{\mathbf{W}} \left( \omega_N \right) $, respectively, where $ \omega_N $ is a latent random variable that encodes the uncertainty of the estimation using $ N $ samples.
    \item Using $ \hat{\Lambda} $ to select the number of active subspace directions $ d_l $, $ N $ replicates of the LIS projection matrix can be formed as $ \hat{\mathbf{W}}_{d_l}^{(i)} $, $ i = 1, \dots, N $.
\end{enumerate}
Note that here we do not use $ \hat{\Lambda}^{(i)} $ to estimate replicates of the number of LIS directions $ d_l $. This is because, in our experience, the SVD of the gradient data matrix is stable enough for the value of $ d_l $ to not change across the different bootstrap samples. On the other hand, it is problematic to have multiple LIS constructions with different dimensionalities. Since we construct only one surrogate for the misfit, and do so in the LIS (see Section~\ref{section:training_misfit_surrogate}), there is no way to handle the projection and reconstruction steps across the various replicates if they do not all have the same dimensions. Hence, we restrict our estimation of the LIS dimension $ d_l $ to only using the nominal eigenvalue matrix $ \hat{\Lambda} $.

The replicates $ \hat{\mathbf{W}}_{d_l}^{(i)} $ as above will be used to account for uncertainty in the active subspace estimate across all projection and reconstruction steps in the proposed framework. The details will be discussed in their respective sections.

\section{Computing the generalized likelihood}
\label{section:surrogate_and_generalized_likelihood}

Once the likelihood-informed subspace has been discovered, we would like to create a fast-running surrogate in this subspace to approximate the computationally expensive likelihood function during posterior sampling. Appropriately leveraging the dimension reduction necessitates constructing the surrogate to map the low-dimensional parameter vector $ \Xi_{d_l} $ to the likelihood prediction. Since the LIS by construction minimizes the information loss of the low-dimensional representation with respect to the misfit function (the directions with the largest variance fraction of the misfit gradient are retained), it is most sensible to approximate $ \mathcal{M} \left( \boldsymbol{\Theta} \right) $ with the surrogate. In this section, we discuss the nuances of constructing a surrogate to predict the misfit function as defined in Eq.~\eqref {eqn:misfit_definition}. We further explain its impact on the likelihood and present a broader definition of the likelihood - the ``generalized likelihood'' - to account for this effect.

\subsection{Deriving the appropriate form for the misfit surrogate}
\label{section:misfit_surrogate_form}

A surrogate that is robust and trustworthy for uncertainty quantification tasks provides not just a prediction of the function of interest but also an estimate of the uncertainty in this prediction, making the prediction a random field rather than a deterministic function. Consequently, it is important to be mindful of any pre-existing assumptions about the probabilistic behavior of the quantity of interest and ensure that the surrogate is compatible with them.

In the case of the misfit function, the Gaussianity assumptions in Section~\ref{section:notation} imply that $ \mathcal{M} \left( \boldsymbol{\Theta} \right) $ is a Gamma random field. This can be shown as follows. Let the set $ \mathfrak{D} $ of experimental data have $ D $ sets of observations $ \left( \mathbf{X}^*, \mathbf{Y}^* \right) $ as per Section~\ref{section:notation}. Now, $ \mathbf{Y}^* - f \left( \mathbf{X}^*, \boldsymbol{\Theta} \right) = \varepsilon_\mathbf{Y} \sim \mathcal{N} \left( \boldsymbol{0}, \boldsymbol{\Sigma}_{\mathbf{Y}} \right) $, with $ \boldsymbol{\Sigma}_{\mathbf{Y}} = \operatorname{diag} \left( \begin{bmatrix} \sigma_{\mathbf{Y}, 1}^2 & \dots & \sigma_{\mathbf{Y}, d_\text{out}}^2 \end{bmatrix} \right) $ being a diagonal matrix, since the components of the output are uncorrelated. ($ \sigma_{\mathbf{Y}, j}^2 $ is thus the variance of the $ j $-th component of the observation noise.) From here, we can rewrite the misfit function of Eq.~\eqref{eqn:misfit_definition} as
\begin{equation}
    \label{eqn:misfit_simplified}
    \mathcal{M} \left( \boldsymbol{\Theta} \right) = \frac{1}{2} \sum_{\mathfrak{D}} \sum_{j=1}^{d_\text{out}} \left( \frac{ y_j - f_{\boldsymbol{\Theta}, j}^* }{\sigma_{\mathbf{Y}, j}} \right)^2
\end{equation}
where $ y_j $ and $ f_{\boldsymbol{\Theta}, j}^* $ are the $ j $-th component of the observation and the QoI function, respectively. It is clear to see that each individual term in the double-sum is the square of a standard Normal random variable. From here, an elementary application of the properties of standard Normal distributions, Chi-squared distributions, and Gamma distributions leads to the following result, with $ \Gamma \left( \alpha, \theta \right) $ denoting a Gamma distribution with shape parameter $ \alpha $ and scale parameter $ \theta $.
\begin{equation}
    \label{eqn:distribution_of_misfit}
    \mathcal{M} \left( \boldsymbol{\Theta} \right) \sim \Gamma \left( \frac{Dd_\text{out}}{2}, 1 \right)
\end{equation}
To allow for some level of modeling inaccuracy and imprecision, we build our misfit surrogate $ \mathcal{G}_\mathcal{M} \left( \boldsymbol{\Xi}_{d_l} \right) $ on the low-dimensional parameter vector as a Gamma random field of the form
\begin{equation}
    \label{eqn:misfit_surrogate_form}
    \mathcal{G}_\mathcal{M} \left( \boldsymbol{\Xi}_{d_l} \right) \sim \Gamma \left(\alpha \left( \boldsymbol{\Xi}_{d_l} \right) + \frac{Dd_\text{out}}{2}, c \left( \boldsymbol{\Xi}_{d_l} \right) \right)
\end{equation}
where $ \alpha \left( \boldsymbol{\Xi}_{d_l} \right) $ and $ c \left( \boldsymbol{\Xi}_{d_l} \right) $ are deterministic hyperparameter functions that allow for the shape and scale of $ \mathcal{G}_\mathcal{M} \left( \boldsymbol{\Xi}_{d_l} \right) $ to change with $ \boldsymbol{\Xi}_{d_l} $, and obey the following relations 
\begin{gather}
    c \left( \boldsymbol{\Xi}_{d_l} \right) = \frac{\operatorname{\mathbb{V}ar} \left[ \mathcal{G}_\mathcal{M} \left( \boldsymbol{\Xi}_{d_l} \right) | \boldsymbol{\Xi}_{d_l} \right]}{\mathbb{E} \left[ \mathcal{G}_\mathcal{M} \left( \boldsymbol{\Xi}_{d_l} \right) | \boldsymbol{\Xi}_{d_l} \right]} \label{eqn:misfit_surrogate_scale_parameter_relationship} \\
    \alpha \left( \boldsymbol{\Xi}_{d_l} \right) = \frac{\left( \mathbb{E} \left[ \mathcal{G}_\mathcal{M} \left( \boldsymbol{\Xi}_{d_l} \right) | \boldsymbol{\Xi}_{d_l} \right] \right)^2}{\operatorname{\mathbb{V}ar} \left[ \mathcal{G}_\mathcal{M} \left( \boldsymbol{\Xi}_{d_l} \right) | \boldsymbol{\Xi}_{d_l} \right]} - \frac{Dd_\text{out}}{2}\label{eqn:misfit_surrogate_shape_parameter_relationship}
\end{gather}

To learn the above hyperparameter functions of $ \mathcal{G}_\mathcal{M} \left( \boldsymbol{\Xi}_{d_l} \right) $, we fit regression curves to $ \mathbb{E} \left[ \mathcal{G}_\mathcal{M} \left( \boldsymbol{\Xi}_{d_l} \right) | \boldsymbol{\Xi}_{d_l} \right] $ and $ \operatorname{\mathbb{V}ar} \left[ \mathcal{G}_\mathcal{M} \left( \boldsymbol{\Xi}_{d_l} \right) | \boldsymbol{\Xi}_{d_l} \right] $. Section~\ref{section:training_misfit_surrogate} discusses the steps involved in generating the training data and fitting the regression curves.

\subsection{Training the misfit surrogate}
\label{section:training_misfit_surrogate}

There are two sources of uncertainty that must be accounted for by the misfit surrogate:
\begin{enumerate}
    \item The reconstruction error introduced as a result of the low-dimensional approximation, i.e., discarding the inactive variables, and
    \item The projection error of the surrogate training data, which is necessarily generated in the original space but is used by the surrogate in the latent space.
\end{enumerate}
We train a separate regression curve for each case, viz., the reconstruction surrogate $ \mathcal{G}_\text{rec} \left( \boldsymbol{\Xi}_{d_l} \right) $ and the projection surrogate $ \mathcal{G}_\text{proj} \left( \boldsymbol{\Xi}_{d_l} \right) $, respectively. These are then combined to predict $ \mathbb{E} \left[ \mathcal{G}_\mathcal{M} \left( \boldsymbol{\Xi}_{d_l} \right) | \boldsymbol{\Xi}_{d_l} \right] $ and $ \operatorname{\mathbb{V}ar} \left[ \mathcal{G}_\mathcal{M} \left( \boldsymbol{\Xi}_{d_l} \right) | \boldsymbol{\Xi}_{d_l} \right] $ at any LIS parameter $ \boldsymbol{\Xi}_{d_l} $, which finally allows for the construction of $ \mathcal{G}_\mathcal{M} \left( \boldsymbol{\Xi}_{d_l} \right) $.

\subsubsection{The reconstruction surrogate}
\label{section:reconstruction_surrogate}

Numerous distinct points in the original parameter space will have the same low-dimensional representation $ \boldsymbol{\Xi}_{d_l} $ (corresponding to different realizations of $ \boldsymbol{\Xi}_\perp $), resulting in multiple values of $ \mathcal{M} \left( \boldsymbol{\Theta} \right) $ for the same LIS parameter $ \boldsymbol{\Xi}_{d_l} $. Of course, it is infeasible to collect replicates of $ \mathcal{M} \left( \boldsymbol{\Theta} \right) $ at each $ \boldsymbol{\Xi}_{d_l} $, as this would correspond to multiple QoI function evaluations for each LIS training input. Instead, we undertake the following steps:
\begin{enumerate}
    \item Construct the surrogate training data set $ \mathfrak{T} = \left\{ \boldsymbol{\Theta}_k \sim \pi \left( \boldsymbol{\Theta} \right) : k = 1, \dots, n_\mathfrak{T} \right\} $.
    \item Evaluate $ \mathcal{M} \left( \boldsymbol{\Theta} \right) $ at each point in $ \mathfrak{T} $ to get $ \mathfrak{T}_\mathcal{M} = \left\{ \mathcal{M} \left( \boldsymbol{\Theta}_k \right) : \boldsymbol{\Theta}_k \in \mathfrak{T} \text{ and } k = 1, \dots, n_\mathfrak{T} \right\} $.
    \item Form the LIS projection matrix $ \hat{\mathbf{W}}_{d_l} $ from $ \hat{\mathbf{W}} $ and $ \hat{\mathbf{\Lambda}} $ (Eq.~\eqref{eqn:SVD_of_gradient_data_matrix}), and use it to project the points in $ \mathfrak{T} $ to their respective low-dimensional representations. Hence form $ \mathfrak{T}_{d_l} = \left\{ \hat{\mathbf{W}}_{d_l}^T \boldsymbol{\Theta}_k : \boldsymbol{\Theta}_k \in \mathfrak{T} \text{ and } k = 1, \dots, n_\mathfrak{T} \right\} $.
    \item Fit $ \mathcal{G}_\text{rec} \left( \boldsymbol{\Xi}_{d_l} \right) : \mathfrak{T}_{d_l} \mapsto \mathfrak{T}_\mathcal{M} $ as a random field regression curve that allows for non-zero prediction variance at the training points. Any regression method is acceptable, as long as it predicts both a mean value $ \mu_\text{rec} \left( \boldsymbol{\Xi}_{d_l} \right) $ and a prediction variance $ \left( \sigma_\text{rec} \left( \boldsymbol{\Xi}_{d_l} \right) \right)^2 $ at each $ \boldsymbol{\Xi}_{d_l} $. 
    \item At any LIS parameter value $ \boldsymbol{\Xi}_{d_l} $, the above random field can be used to obtain the mean $ \mu_\text{rec} \left( \boldsymbol{\Xi}_{d_l} \right) $ and variance $ \left( \sigma_\text{rec} \left( \boldsymbol{\Xi}_{d_l} \right) \right)^2 $ in the nominal misfit prediction. (Note that $ \mathcal{G}_\text{rec} \left( \boldsymbol{\Xi}_{d_l} \right) $ is a scalar-valued random field since the misfit function is scalar-valued.)
\end{enumerate}

By generating the training data through random sampling of the full prior $ \pi \left( \boldsymbol{\Theta} \right) $ (thus varying both $ \boldsymbol{\Xi}_{d_l} $ and $ \boldsymbol{\Xi}_\perp $) and allowing for the reconstruction surrogate to contain non-zero variance at the training points, we approximate the uncertainty resulting from the dimension reduction without producing replicates of $ \mathcal{M} \left( \boldsymbol{\Theta} \right) $ for each $ \boldsymbol{\Xi}_{d_l} $.

\subsubsection{The projection surrogate}
\label{section:projection_surrogate}

The uncertainty stemming from the finite-sample estimation of the uncentered gradient covariance $ \hat{C}_\mathcal{M} $ (equivalently the formation of the gradient data matrix $ \hat{\mathbf{B}}_\mathcal{M} $) causes imprecise estimation of the projection matrix $ \mathbf{W}_{d_l} $, which must then be propagated into the misfit surrogate. Section~\ref{section:projection_uncertainty_quantification} discusses how to quantify this estimation uncertainty by generating replicates $ \hat{\mathbf{W}}_{d_l}^{(i)} $, $ i = 1, \dots, N $ of the projection matrix. In this section, we provide the following steps to construct the projection surrogate $ \mathcal{G}_\text{proj} \left( \boldsymbol{\Xi}_{d_l} \right) $, which utilizes these replicates to inject the projection uncertainty into the misfit surrogate $ \mathcal{G}_\mathcal{M} \left( \boldsymbol{\Xi}_{d_l} \right) $.
\begin{enumerate}
    \item Form the LIS projection matrix $ \hat{\mathbf{W}}_{d_l} $ from $ \hat{\mathbf{W}} $ and $ \hat{\mathbf{\Lambda}} $ (Eq.~\eqref{eqn:SVD_of_gradient_data_matrix}).
    \item Generate the replicates of the projection matrix $ \hat{\mathbf{W}}_{d_l}^{(i)} $, $ i = 1, \dots, N $ as in Section~\ref{section:projection_uncertainty_quantification}.
    \item Build $ \mathfrak{T} $, $ \mathfrak{T}_\mathcal{M} $, and $ \mathfrak{T}_{d_l} $ as in Section~\ref{section:reconstruction_surrogate}.
    \item Construct replicates of the LIS surrogate training data set \\ $ \mathfrak{T}_{d_l}^{(i)} = \left\{ \left( \hat{\mathbf{W}}_{d_l}^{(i)} \right)^T \boldsymbol{\Theta}_k : \boldsymbol{\Theta}_k \in \mathfrak{T} \text{ and } k = 1, \dots, n_\mathfrak{T} \right\} $.
    \item At this point, the projection uncertainty is represented by LIS parameter training set replicates $ \mathfrak{T}_{d_l}^{(i)} $ that all map to the same set of outputs $ \mathfrak{T}_\mathcal{M} $. Unfortunately, there is no simple way to build a surrogate from uncertain inputs to known outputs. Instead, we need to use $ \mathfrak{T}_{d_l}^{(i)} $ to construct replicates of $ \mathfrak{T}_\mathcal{M} $ that can all map to the same nominal parameter set $ \mathfrak{T}_{d_l} $. We use a simple first-order Taylor approximation as follows:
    \begin{enumerate}
        \item Evaluate $ \mathfrak{T}_{\nabla \mathcal{M}} = \left\{ \nabla \mathcal{M} \left( \boldsymbol{\Theta}_k \right) : \boldsymbol{\Theta}_k \in \mathfrak{T} \text{ and } k = 1, \dots, n_\mathfrak{T} \right\} $.
        \item Compute QoI prediction replicate sets $ \mathfrak{T}_\mathcal{M}^{(i)} = \left\{ \mathsf{M}_{\boldsymbol{\Theta}_k}^{(i)} : \boldsymbol{\Theta}_k \in \mathfrak{T} \text{ and } k = 1, \dots, n_\mathfrak{T} \right\} $, where
        \begin{equation}
        \label{eqn:projection_misfit_replicate_definition}
            \mathsf{M}_{\boldsymbol{\Theta}_k}^{(i)} = \mathcal{M} \left( \boldsymbol{\Theta}_k \right) + \left[ \nabla \mathcal{M} \left( \boldsymbol{\Theta}_k \right) \cdot \left( \hat{\mathbf{W}}_{d_l}^T \boldsymbol{\Theta}_k - \left( \hat{\mathbf{W}}_{d_l}^{(i)} \right)^T \boldsymbol{\Theta}_k \right) \right]
        \end{equation}
    \end{enumerate}
    \item Quantifying the effect of the LIS estimation uncertainty on the misfit surrogate requires only an estimation of the additional variance in the misfit prediction introduced by the inexact projection. To achieve this, we compute the variance of the replicates corresponding to each training point from the sets $ \mathfrak{T}_\mathcal{M}^{(i)} $, and fit $ \mathcal{G}_\text{proj} \left( \boldsymbol{\Xi}_{d_l} \right) $ as a regression curve to the variance values.
    \begin{enumerate}
        \item Define the set $ \mathfrak{T}_{V} = \left\{ V_{\boldsymbol{\Theta}_k} : \boldsymbol{\Theta}_k \in \mathfrak{T} \text{ and } k = 1, \dots, n_\mathfrak{T} \right\} $, where
        \begin{equation}
            \label{eqn:replicates_variance}
            V_{\boldsymbol{\Theta}_k} = \frac{1}{N-1} \sum_{i=1}^N \left[ \mathsf{M}_{\boldsymbol{\Theta}_k}^{(i)} - \left( \frac{1}{N} \sum_{i=1}^N \mathsf{M}_{\boldsymbol{\Theta}_k}^{(i)} \right) \right]^2
        \end{equation}
        \item Fit $ \mathcal{G}_\text{proj} \left( \boldsymbol{\Xi}_{d_l} \right) : \mathfrak{T}_{d_l} \mapsto \mathfrak{T}_V $; any sufficiently expressive regression or interpolation curve will suffice. 
    \end{enumerate}
    \item At any LIS parameter value $ \boldsymbol{\Xi}_{d_l} $, the additional variance $ v_\text{proj} \left( \boldsymbol{\Xi}_{d_l} \right) $ stemming from the projection uncertainty is predicted by the above surrogate $ \mathcal{G}_\text{proj} \left( \boldsymbol{\Xi}_{d_l} \right) $. 
\end{enumerate}

We note that the above procedure, by requiring gradient evaluations of the misfit function, has a natural synergy with the method of active subspaces, which also requires them. Ideally, the training data set for the active subspace $ \mathfrak{A} $ should be the same as the surrogate training data set $ \mathfrak{T} $. However, since $ \mathfrak{A} $ requires the parameter values to be directly sampled from the prior $ \pi \left( \boldsymbol{\Theta} \right) $, it is possible that the surrogate will not be acceptably trained if $ \mathfrak{T} = \mathfrak{A} $. This is more likely to happen if $ \pi \left( \boldsymbol{\Theta} \right) $ is a distribution with narrow modes and has a very different shape from the posterior. In such cases, the calibration procedure will require likelihood approximations for parameter values far from the prior's mode, thus necessitating that the misfit surrogate extrapolate heavily despite being trained only near the modes. One solution to this issue is to start with $ \mathfrak{T} = \mathfrak{A} $ to maximize the use of existing gradient evaluations, and then enrich $ \mathfrak{T} $ through active learning to ensure robust predictive accuracy of the misfit surrogate across the whole domain at the minimum possible additional cost. Developing an appropriate active learning scheme is beyond the scope of this paper and left for future work.

\subsubsection{Assembling the misfit surrogate}
\label{section:misfit_surrogate_assembly}

Assembling the final misfit surrogate from the reconstruction and projection surrogates is straightforward. First, we construct the regression curves for the mean $ \mathbb{E} \left[ \mathcal{G}_\mathcal{M} \left( \boldsymbol{\Xi}_{d_l} \right) | \boldsymbol{\Xi}_{d_l} \right] $ and variance $ \operatorname{\mathbb{V}ar} \left[ \mathcal{G}_\mathcal{M} \left( \boldsymbol{\Xi}_{d_l} \right) | \boldsymbol{\Xi}_{d_l} \right] $ of the misfit surrogate $ \mathcal{G}_\mathcal{M} \left( \boldsymbol{\Xi}_{d_l} \right) $ as
\begin{gather}
    \mathbb{E} \left[ \mathcal{G}_\mathcal{M} \left( \boldsymbol{\Xi}_{d_l} \right) | \boldsymbol{\Xi}_{d_l} \right] = \mu_\text{rec} \left( \boldsymbol{\Xi}_{d_l} \right) \label{eqn:misfit_surrogate_gamma_field_mean_function} \\
    \operatorname{\mathbb{V}ar} \left[ \mathcal{G}_\mathcal{M} \left( \boldsymbol{\Xi}_{d_l} \right) | \boldsymbol{\Xi}_{d_l} \right] = \left( \sigma_\text{rec} \left( \boldsymbol{\Xi}_{d_l} \right) \right)^2 + v_\text{proj} \left( \boldsymbol{\Xi}_{d_l} \right) \label{eqn:misfit_surrogate_gamma_field_variance_function}
\end{gather}
where $ \mu_\text{rec} \left( \boldsymbol{\Xi}_{d_l} \right) $ and $ \left( \sigma_\text{rec} \left( \boldsymbol{\Xi}_{d_l} \right) \right)^2 $ are the mean and variance functions of the reconstruction surrogate $ \mathcal{G}_\text{rec} $ (as defined in section~\ref{section:reconstruction_surrogate}), while $ v_\text{proj} \left( \boldsymbol{\Xi}_{d_l} \right) $ is the prediction of the projection surrogate $ \mathcal{G}_\text{proj} $ (as defined in section~\ref{section:projection_surrogate}). Then, the hyperparameter functions $ \alpha \left( \boldsymbol{\Xi}_{d_l} \right) $ and $ c \left( \boldsymbol{\Xi}_{d_l} \right) $ of the misfit surrogate $ \mathcal{G}_\mathcal{M} \left( \boldsymbol{\Xi}_{d_l} \right) $ can be computed from Eqs.~\eqref{eqn:misfit_surrogate_scale_parameter_relationship} and~\eqref{eqn:misfit_surrogate_shape_parameter_relationship}.

\subsection{Formulating the generalized likelihood and using the misfit surrogate}
\label{section:generalized_likelihood}

Existing Bayesian calibration frameworks that incorporate surrogate modeling nearly always restrict their formulation to cases where the QoI function is approximated. For example, the well-known Kennedy-O'Hagan framework~\cite{kennedy2001bayesian} leverages Gaussianity assumptions on the observation noise and the QoI surrogate, allowing the usual likelihood (Eq.~\eqref{eqn:misfit_to_likelihood}) to be used with a minor modification to the definition of the misfit (Eq.~\eqref{eqn:misfit_definition}) that accounts for the surrogate's predictive uncertainty. This is different from our case, where the surrogate instead predicts the misfit directly. Here, the likelihood must be revisited from first principles, and a more general form must be derived that can handle stochastic approximations of the misfit function itself. In this section, we go one step further and define the ``generalized likelihood'' that is applicable regardless of what function is approximated by the surrogate, as long as it is related to the original ``model-vs.-observation'' likelihood deterministically. Then, we derive its specific form resulting from the assumptions of our proposed calibration framework.

In essence, the likelihood of a set of parameter values is the probability that the system can have the observed behavior for that set of parameter values. Computing this probability requires a way to predict the system's behavior for the chosen parameter values, as well as assumptions about the probabilistic behavior of the prediction and the noise associated with the observation procedure. Building on the notation in Section~\ref{section:notation}, where the QoI function is the model used to predict the system's behavior, we can write this probability as
\begin{equation}
    \label{eqn:generalized_likelihood}
    \mathcal{L} \left( \boldsymbol{\Theta}, \delta \right) = \int_{f_{\boldsymbol{\Theta}}^*} \rho \left( \mathbf{Y}^* | f_{\boldsymbol{\Theta}}^* \right) \cdot \rho \left( f_{\boldsymbol{\Theta}}^* | \boldsymbol{\Theta}, \delta \right) d f_{\boldsymbol{\Theta}}^*
\end{equation}
where $ \delta $ is a term that accounts for the surrogate model error, $ \rho \left( \cdot \right) $ denotes a probability density function, and all other symbols are as defined earlier. The key insight here is that, in general, the prediction of the system's behavior may be a stochastic function of the parameter vector (and a surrogate error parameter that is also inferred), while the observed output, in turn, differs from the system prediction by some noise. Since the only known quantities are the pre-selected parameter vector $ \boldsymbol{\Theta} $ and the observed input-output pair $ \left( \mathbf{X}^*, \mathbf{Y}^* \right) $, all possible values of the QoI function $ f_{\boldsymbol{\Theta}}^* $ must be accounted for, and hence it is marginalized over.

When the QoI function (which is treated as deterministic in this paper as per Section~\ref{section:notation}) is evaluated directly (i.e., the original physics model is used), there is no surrogate error and $ \rho \left( f_{\boldsymbol{\Theta}}^* | \boldsymbol{\Theta}, \delta \right) \equiv 1 $. Then, assuming Gaussian noise in the output (again per Section~\ref{section:notation}), Eq.~\eqref{eqn:generalized_likelihood} simplifies into Eqs.~\eqref{eqn:misfit_definition} and~\eqref{eqn:misfit_to_likelihood}, demonstrating the model-vs.-observation likelihood to be a special case of the generalized likelihood. If a surrogate is used as an intermediate predictor, the specific modeling assumptions dictate the forms of the two density functions in Eq.~\eqref{eqn:generalized_likelihood}, and computing the one-dimensional integral produces the generalized likelihood of the chosen system parameters. For example, under Gaussian observation noise and for a Gaussian process surrogate predicting the QoI, Eq.~\eqref{eqn:generalized_likelihood} results in the familiar structure presented in the original Kennedy-O'Hagan paper~\cite{kennedy2001bayesian}, as shown in Appendix~\ref{appendix:deriving_KOH_from_generalized_likelihood}.

Our proposed framework has two key modeling assumptions: (a) the noise in the observation $ \mathbf{Y} $ is Gaussian, and (b) the surrogate predicts the misfit function as defined in Eq.~\eqref{eqn:misfit_definition}, whose form is itself a result of the previous assumption. From these, we can develop expressions for the two densities in Eq.\eqref{eqn:generalized_likelihood} which will allow us to evaluate the generalized likelihood. First, assumption (a) leads to
\begin{equation}
    \label{eqn:proposed_framework_probability_density_y_given_f}
    \rho \left( \mathbf{Y}^* | f_{\boldsymbol{\Theta}}^* \right) \propto \exp{\left[ - \mathcal{M} \left( \boldsymbol{\Theta} \right) \right]}
\end{equation}
where $ \mathcal{M} \left( \boldsymbol{\Theta} \right) $ is defined as per Eq.~\eqref{eqn:misfit_definition}. Next, to satisfy assumption (b), we first observe that $ \rho \left( f_{\boldsymbol{\Theta}}^* | \boldsymbol{\Theta}, \delta \right) = \rho \left( \mathcal{M} \left( \boldsymbol{\Theta} \right) | \boldsymbol{\Theta}, \delta \right) $, since Eq.~\eqref{eqn:misfit_definition} implies that $ \mathcal{M} \left( \boldsymbol{\Theta} \right) $ is a deterministic function of $ f_{\boldsymbol{\Theta}}^* $. Now, $ \mathcal{M} \left( \boldsymbol{\Theta} \right) $ is too expensive to evaluate directly, and so the proposed framework draws upon the misfit surrogate as formulated in Section~\ref{section:misfit_surrogate_form}. Adding a surrogate error term, which is also modeled as a Gamma distribution for mathematical simplicity, we can write
\begin{align}
    \mathcal{M} \left( \boldsymbol{\Theta} \right) &= \mathcal{G}_\mathcal{M} \left( \boldsymbol{\Xi}_{d_l} \right) + G_\delta &&\text{ where } \;\; G_\delta \sim \Gamma \left( \delta, c \left( \boldsymbol{\Xi}_{d_l} \right) \right) \label{eqn:surrogate_approximation_of_misfit} \\
    \Rightarrow \mathcal{M} \left( \boldsymbol{\Theta} \right) &= \mathcal{G}_{\mathcal{M}, \delta} \left( \boldsymbol{\Xi}_{d_l}, \delta \right) &&\text{ where } \;\; \mathcal{G}_{\mathcal{M}, \delta} \left( \boldsymbol{\Xi}_{d_l}, \delta \right) \sim \Gamma \left( \delta + \alpha \left( \boldsymbol{\Xi}_{d_l} \right) + \frac{Dd_\text{out}}{2}, c \left( \boldsymbol{\Xi}_{d_l} \right) \right) \label{eqn:approximated_misfit_distribution_with_discrepancy}
\end{align}
This allows us to assert the following, since $ \boldsymbol{\Xi}_{d_l} $ a deterministic transformation of $ \boldsymbol{\Theta} $.
\begin{equation}
    \label{eqn:proposed_framework_probability_density_f_given_theta}
    \rho \left( f_{\boldsymbol{\Theta}}^* | \boldsymbol{\Theta}, \delta \right) = \rho_{\mathcal{G}_{\mathcal{M}, \delta}} \left( \boldsymbol{\Xi}_{d_l}, \delta \right)
\end{equation}
where $ \rho_{\mathcal{G}_{\mathcal{M}, \delta}} \left( \boldsymbol{\Xi}_{d_l}, \delta \right) $ is the probability density of the Gamma random variable $ \mathcal{G}_{\mathcal{M}, \delta} \left( \boldsymbol{\Xi}_{d_l}, \delta \right) $ with the specified values of $ \boldsymbol{\Xi}_{d_l} $ and $ \delta $. (Strictly speaking, the transformation of $ \boldsymbol{\Xi}_{d_l} $ from $ \boldsymbol{\Theta} $ is stochastic due to the finite sample approximation of the LIS. However, as discussed in the previous sections, for practical purposes we simply use the nominal projection matrix $ \mathbf{W}_{d_l} $, and build the misfit surrogate to account for the projection uncertainty. This allows for a single deterministic transformation from $ \boldsymbol{\Theta} $ to $ \boldsymbol{\Xi}_{d_l} $.) Combining Eq.~\eqref{eqn:proposed_framework_probability_density_y_given_f},~\eqref{eqn:proposed_framework_probability_density_f_given_theta}, and~\eqref{eqn:approximated_misfit_distribution_with_discrepancy}, the generalized likelihood in Eq.~\eqref{eqn:generalized_likelihood} becomes
\begin{equation}
    \label{eqn:proposed_framework_likelihood}
    \mathcal{L} \left( \boldsymbol{\Theta}, \delta \right) \approx \mathcal{L} \left( \boldsymbol{\Xi}_{d_l}, \delta \right) = \mathbb{E}_{\rho_{\mathcal{G}_{\mathcal{M}, \delta}}} \left[ \exp{\left( - \mathcal{G}_{\mathcal{M}, \delta} \left( \boldsymbol{\Xi}_{d_l}, \delta \right) \right)} \right]
\end{equation}
This expectation can easily be computed numerically; when values of $ \boldsymbol{\Xi}_{d_l} $ and $ \delta $ are specified, all the parameters of $ \mathcal{G}_{\mathcal{M}, \delta} \left( \boldsymbol{\Xi}_{d_l}, \delta \right) $ are known, and it can be sampled in a straightforward manner.

To conclude this section, we remark that the above steps can be followed to derive specialized forms of the generalized likelihood from Eq.~\eqref{eqn:generalized_likelihood} for all instances of a broad class of modeling frameworks. Any function of the parameters (or inputs, in cases of input inference) can be approximated by a surrogate and used to compute the generalized likelihood, as long as the probability of the observed output given the predicted quantity can be computed deterministically. (For example, in our case, the probability of $ \mathbf{Y} $ given a misfit function value prediction is evaluated by simply exponentiating the negative of the predicted misfit.)

\section{Calibration using the generalized likelihood}
\label{section:calibration_posterior}

So far, we have constructed a cheap and accurate low-dimensional surrogate for the misfit function (i.e., negative log-likelihood function), which is the most complex and computationally expensive component of the posterior distribution. The final step of our framework is to utilize this surrogate to make the calibrated system parameter posteriors cheap to evaluate and easy to sample from. To this end, this section defines (a) an equivalent calibration problem that lies solely in the latent space, which allows for robust and efficient sampling from the posterior of the low-dimensional parameter vector, and (b) a simple transformation between the posterior distribution of the low-dimensional parameter vector and the full posterior distribution of the system parameters, allowing for the generated samples and statistics of the low-dimensional posterior to be transformed into the full original system parameter space. We also discuss how to use these results in practice.

\subsection{Constructing the calibrated posterior distribution}
\label{section:posterior_definition}

We begin by considering the transformation from $ \boldsymbol{\Theta} $ to $ \boldsymbol{\Xi} $, which involves the uncertain rotation matrix $ \hat{\mathbf{W}} \left( \omega_N \right) $. Given the prior $ \pi \left( \boldsymbol{\Theta} \right) $ on the system parameters,
\begin{align}
    \pi \left( \boldsymbol{\Xi} | \omega_N \right) &= \pi \left( \left( \hat{\mathbf{W}} \left( \omega_N \right) \right)^T \boldsymbol{\Theta} \right) \label{eqn:rotated_parameter_prior_given_rotation_matrix} \\
    \Rightarrow \pi \left( \boldsymbol{\Xi} \right) &= \int \pi \left( \left( \hat{\mathbf{W}} \left( \omega_N \right) \right)^T \boldsymbol{\Theta} \right) \rho \left( \omega_N \right) d \omega_N , \label{eqn:rotated_parameter_prior_marginalized_over_rotation_matrix}
\end{align}
where the prior of $ \boldsymbol{\Xi} $ is marginalized over all possible rotation matrices arising from the eigendecomposition of the uncentered gradient covariance matrix (see Section~\ref{section:likelihood_informed_subspace}), with $ \rho \left( \omega_N \right) $ denoting the probability density of these rotation matrices. (Note that the Jacobian of the rotation from $ \boldsymbol{\Theta} $ to $ \boldsymbol{\Xi} $ is $ \hat{\mathbf{W}} \left( \omega_N \right) $, which has unit determinant.) When the system parameters have a standard Gaussian prior (i.e., $ \pi \left( \boldsymbol{\Theta} \right) = \phi_{d_p} \left( \boldsymbol{\Theta} \right) $, where $ \phi_a (\cdot) $ denotes the $ a $-dimensional standard normal density function), this becomes
\begin{equation}
    \pi \left( \boldsymbol{\Xi} | \omega_N \right) = \phi_{d_p} \left( \boldsymbol{\Theta} \right) \Rightarrow \pi \left( \boldsymbol{\Xi} \right) = \phi_{d_p} \left( \boldsymbol{\Xi} \right) = \phi_{d_p} \left( \boldsymbol{\Theta} \right) , \label{eqn:marginal_rotated_parameter_prior_standard_normal}
\end{equation}
since the standard Normal distribution is radially symmetric and rotations preserve distance from the origin. Decomposing this prior in terms of the active and inactive variables gives
\begin{equation}
    \label{eqn:decomposition_of_prior}
    \pi \left( \boldsymbol{\Xi} \right) = \pi \left( \boldsymbol{\Xi}_{d_l} \right) \cdot \pi \left( \boldsymbol{\Xi}_{\perp} | \boldsymbol{\Xi}_{d_l} \right)
\end{equation}
Observe that when $ \pi \left( \boldsymbol{\Theta} \right) = \phi_{d_p} \left( \boldsymbol{\Theta} \right) $, Eq.~\eqref{eqn:marginal_rotated_parameter_prior_standard_normal} implies that $ \pi \left( \boldsymbol{\Xi}_{d_l} \right) = \phi_{d_l} \left( \boldsymbol{\Xi}_{d_l} \right) $ and $ \pi \left( \boldsymbol{\Xi}_\perp \right) = \phi_{(d_p - d_l)} \left( \boldsymbol{\Xi}_\perp \right) $. Next, Eq.~\eqref{eqn:decomposition_of_prior} permits a formulation of the calibration problem purely in the latent space by simply substituting the misfit surrogate-based low-dimensional generalized likelihood approximation (Eq.~\eqref{eqn:proposed_framework_likelihood}) into Bayes' Theorem (Eq.~\eqref{eqn:Bayes_theorem}) to get
\begin{align}
    p \left( \boldsymbol{\Xi} \right) &= \mathcal{L} \left( \boldsymbol{\Xi} \right) \pi \left( \boldsymbol{\Xi}_{d_l} \right) \pi \left( \boldsymbol{\Xi}_{\perp} | \boldsymbol{\Xi}_{d_l} \right) \\
    \Rightarrow p \left( \boldsymbol{\Xi} \right) &\approx \mathcal{L} \left( \boldsymbol{\Xi}_{d_l}, \delta \right) \pi \left( \boldsymbol{\Xi}_{d_l} \right) \pi \left( \boldsymbol{\Xi}_{\perp} | \boldsymbol{\Xi}_{d_l} \right), \label{eqn:decomposition_of_posterior}
\end{align}
which allows us to define
\begin{equation}
    \label{eqn:posterior_of_low-dimensional_parameters}
    p \left( \boldsymbol{\Xi}_{d_l} \right) = \mathcal{L} \left( \boldsymbol{\Xi}_{d_l}, \delta \right) \pi \left( \boldsymbol{\Xi}_{d_l} \right)
\end{equation}
As a consequence of the dimension reduction utilized in the generalized likelihood formulation, the posterior of the rotated parameter vector $ p \left( \boldsymbol{\Xi} \right) $ is the composition of the posterior of the low-dimensional parameter vector $ p \left( \boldsymbol{\Xi}_{d_l} \right) $ and the \textit{conditional prior of the inactive variables} $ \pi \left( \boldsymbol{\Xi}_{\perp} | \boldsymbol{\Xi}_{d_l} \right) $, i.e.,
\begin{equation}
    \label{eqn:parameter_posterior_composition}
    p \left( \boldsymbol{\Xi} \right) = p \left( \boldsymbol{\Xi}_{d_l} \right) \cdot \pi \left( \boldsymbol{\Xi}_{\perp} | \boldsymbol{\Xi}_{d_l} \right)
\end{equation}
The implications of the above equation will be discussed further in Section~\ref{section:calibrated_samples_uncertainty_vs_informativeness}. From here, obtaining the posterior distribution for the original system parameters involves another rotation using $ \hat{\mathbf{W}} \left( \omega_N \right) $ as
\begin{align}
    p \left( \boldsymbol{\Theta} | \omega_N \right) &= p \left( \hat{\mathbf{W}} \left( \omega_N \right) \boldsymbol{\Xi} \right) \label{eqn:system_parameter_posterior_given_rotation_matrix_basic_definition} \\
    \Rightarrow p \left( \boldsymbol{\Theta} | \omega_N \right) &= \mathcal{L} \left( \left( \hat{\mathbf{W}}_{d_l} \left( \omega_N \right) \right)^T \boldsymbol{\Theta}, \delta \right) \pi \left( \left( \hat{\mathbf{W}}_{d_l} \left( \omega_N \right) \right)^T \boldsymbol{\Theta} \right) \pi \left( \left( \hat{\mathbf{W}}_\perp \left( \omega_N \right) \right)^T \boldsymbol{\Theta} | \left( \hat{\mathbf{W}}_{d_l} \left( \omega_N \right) \right)^T \boldsymbol{\Theta} \right) \label{eqn:system_parameter_posterior_given_rotation_matrix} \\
    \Rightarrow p \left( \boldsymbol{\Theta} \right) &= \int p \left( \boldsymbol{\Theta} | \omega_N \right) \rho \left( \omega_N \right) d \omega_N , \label{eqn:system_parameter_posterior_marginalized_over_rotation_matrix}
\end{align}

This posterior can then be used to compute any statistic of the system parameters as
\begin{equation}
    \label{eqn:general_system_parameter_statisitcs}
    \mathbb{E}_p \left[ g \left( \boldsymbol{\Theta} \right) \right] = \int g \left( \boldsymbol{\Theta} \right) p \left( \boldsymbol{\Theta} \right) d \boldsymbol{\Theta} = \iint g \left( \boldsymbol{\Theta} \right) p \left( \boldsymbol{\Theta} | \omega_N \right) \rho \left( \omega_N \right) d \omega_N d \boldsymbol{\Theta}
\end{equation}
The benefit of the double-integral form will become clear in Section~\ref{section:posterior_in_practice}. Past works have underscored the necessity of incorporating the uncertainty stemming from the dimension reduction component using asymptotic analysis~\cite{kim2020post}, highlighting how inferred posteriors can be overconfident when the dimension reduction uncertainty is ignored.

\subsection{Using the calibrated posterior distribution in practice}
\label{section:posterior_in_practice}

It is nearly impossible for $ p \left( \boldsymbol{\Theta} \right) $ to be obtained in closed form. However, since $ \mathcal{L} \left( \boldsymbol{\Xi}_{d_l}, \delta \right) $ is cheap to evaluate, and samples of $ \hat{\mathbf{W}} \left( \omega_N \right) $ are available as in Section~\ref{section:projection_uncertainty_quantification}, it is possible to estimate $ p \left( \boldsymbol{\Theta} \right) $ as
\begin{equation}
    \label{eqn:estimate_of_system_parameter_posterior}
    \hat{p} \left( \boldsymbol{\Theta} \right) = \frac{1}{N} \sum_{i=1}^N \mathcal{L} \left( \left( \hat{\mathbf{W}}_{d_l}^{(i)} \right)^T \boldsymbol{\Theta}, \delta \right) \pi \left( \left( \hat{\mathbf{W}}_{d_l}^{(i)} \right)^T \boldsymbol{\Theta} \right) \pi \left( \left( \hat{\mathbf{W}}_\perp^{(i)} \right)^T \boldsymbol{\Theta} | \left( \hat{\mathbf{W}}_{d_l}^{(i)} \right)^T \boldsymbol{\Theta} \right)
\end{equation}
Eq.~\eqref{eqn:estimate_of_system_parameter_posterior} allows for certain geometric properties of the posterior to be calculated numerically without the use of samples. In particular, the mode of the posterior (i.e., the maximum a posteriori estimate) of the calibrated parameters, can be evaluated via numerical optimization of the estimated density function above.

In cases where samples are desired from the posterior, however, it is inadvisable to use the posterior defined in Eq.~\eqref{eqn:estimate_of_system_parameter_posterior}, which is marginalized over realizations of the projection matrix, as it is defined on the full high-dimensional system parameter vector. As described in Section~\ref{section:introduction}, most sampling algorithms struggle with degeneracy in high-dimensional spaces; this is the primary reason for employing dimension reduction in the first place. Instead, following the procedure in~\cite{constantine2016accelerating}, we first generate samples from the much lower dimensional density $ p \left( \boldsymbol{\Xi}_{d_l} \right) $, using any robust sampling algorithm; in this manuscript, we use Markov chain Monte Carlo (MCMC) sampling. Let a set of $ n $ calibrated low-dimensional parameter vector samples be denoted as
\begin{equation}
    \label{eqn:low_dimensional_parameter_sample_set}
    \mathfrak{C}_{\boldsymbol{\Xi}_{d_l}} = \left\{ \boldsymbol{\Xi}_{{d_l}, k} \sim p \left( \boldsymbol{\Xi}_{d_l} \right) : k = 1, \dots, n \right\}
\end{equation}
Simultaneously, generate a set of $ n $ samples from the conditional prior of the inactive variables given the low-dimensional parameter vector as
\begin{equation}
\label{eqn:inactive_variable_parameter_sample_set}
    \mathfrak{C}_{\boldsymbol{\Xi}_\perp} = \left\{ \boldsymbol{\Xi}_{\perp, k} \sim \pi \left( \boldsymbol{\Xi}_\perp | \boldsymbol{\Xi}_{d_l} \right) : k = 1, \dots, n \right\}
\end{equation}
(When the priors are standard normal, these sample sets can be generated independently.) Now, a nominal set of samples from the posterior $ p \left( \boldsymbol{\Theta} \right) $ can be constructed as
\begin{equation}
    \label{eqn:nominal_set_of_posterior_samples}
    \mathfrak{C}_{\boldsymbol{\Theta}} = \left\{ \boldsymbol{\Theta}_k = \hat{\mathbf{W}} \boldsymbol{\Xi}_k = \hat{\mathbf{W}} \begin{bmatrix} \boldsymbol{\Xi}_{{d_l}, k}^T & \boldsymbol{\Xi}_{\perp, k}^T \end{bmatrix}^T : \boldsymbol{\Xi}_{{d_l}, k} \in \mathfrak{C}_{\boldsymbol{\Xi}_{d_l}} , \boldsymbol{\Xi}_{\perp, k} \in \mathfrak{C}_{\boldsymbol{\Xi}_\perp} ,  k = 1, \dots, n \right\}
\end{equation}
To account for the projection uncertainty, replicate sets of the calibrated posterior samples are computed using the bootstrapped projection matrices from Section~\ref{section:projection_uncertainty_quantification}
\begin{equation}
    \label{eqn:bootstrap_set_of_posterior_samples}
    \mathfrak{C}_{\boldsymbol{\Theta}}^{(i)} = \left\{ \boldsymbol{\Theta}_k^{(i)} = \hat{\mathbf{W}}^{(i)} \boldsymbol{\Xi}_k = \hat{\mathbf{W}}^{(i)} \begin{bmatrix} \boldsymbol{\Xi}_{{d_l}, k}^T & \boldsymbol{\Xi}_{\perp, k}^T \end{bmatrix}^T : \boldsymbol{\Xi}_{{d_l}, k} \in \mathfrak{C}_{\boldsymbol{\Xi}_{d_l}} , \boldsymbol{\Xi}_{\perp, k} \in \mathfrak{C}_{\boldsymbol{\Xi}_\perp} ,  k = 1, \dots, n \right\}
\end{equation}
It is computationally challenging to evaluate the marginal posterior density values $ \hat{p} \left( \boldsymbol{\Theta} \right) $ from these (or any) sample sets, since the quantities are high-dimensional. However, statistics of the posterior can still be easily computed using Monte Carlo estimates of the double-integral form from Eq.~\eqref{eqn:general_system_parameter_statisitcs} as
\begin{equation}
    \label{eqn:general_system_parameter_statisitcs_monte_carlo_estimator}
    \hat{\mathbb{E}}_p \left[ g \left( \boldsymbol{\Theta} \right) \right] = \sum_{i=1}^N \frac{1}{N} \left[ \sum_{k=1}^n \frac{ g \left( \boldsymbol{\Theta}_k^{(i)} \right) }{n} \right] \;\; \text{ , where } \boldsymbol{\Theta}_k^{(i)} \in \mathfrak{C}_{\boldsymbol{\Theta}}^{(i)}
\end{equation}

\subsection{Balancing inferred uncertainty with parameter informativeness}
\label{section:calibrated_samples_uncertainty_vs_informativeness}

The expressions in Section~\ref{section:posterior_definition} reveal an interesting paradox. The inactive directions in the parameter space, by virtue of not affecting the misfit function much, are discarded in the generalized likelihood as defined in Eq.~\eqref{eqn:proposed_framework_likelihood}. Thus, the prior uncertainty along these directions is propagated unchanged to the calibrated posteriors of the system parameters, per Eq.~\eqref{eqn:system_parameter_posterior_marginalized_over_rotation_matrix}. Consequently, the aspects of the system parameter vector that do not meaningfully affect the system observable still contribute significant uncertainty to the calibrated results. Further, the more inactive components there are, i.e., the fewer directions in the parameter space that the system observable is sensitive to, the more uncertain the parameter calibration is. While this is a mathematical fact and makes intuitive sense, it is highly undesirable from a practical standpoint. In terms of utility, we not only want to infer the system parameters with high confidence, but also reduce the consideration provided to those aspects of the parameter set that do not affect the system observable. Hence, instead of the full parameter vector, it may often be beneficial to work with the ``\textit{conditional active parameter vector}'', i.e.,
\begin{equation}
    \label{eqn:conditional_active_parameter_vector}
    \boldsymbol{\Theta}_{\boldsymbol{\mu}_\perp} = \mathbf{W} \boldsymbol{\Xi}_{\boldsymbol{\mu}_\perp} = \mathbf{W} \begin{bmatrix} \boldsymbol{\Xi}_{d_l}^T & \boldsymbol{\mu}_\perp^T \end{bmatrix}^T,
\end{equation}
where the inactive variables $ \boldsymbol{\Xi}_\perp $ are fixed at their corresponding prior mean values $ \boldsymbol{\mu}_\perp $. This is especially simplified in the case of standard Gaussian priors, since $ \boldsymbol{\mu}_\perp = \boldsymbol{0} $, resulting in
\begin{equation}
    \label{eqn:conditional_active_parameter_vector_normal_prior}
    \boldsymbol{\Theta}_{\boldsymbol{\mu}_\perp} = \mathbf{W} \begin{bmatrix} \boldsymbol{\Xi}_{d_l}^T & \boldsymbol{0}^T \end{bmatrix}^T = \mathbf{W}_{d_l} \boldsymbol{\Xi}_{d_l}
\end{equation}

Correspondingly, we can limit our attention to the ``\textit{conditional active posterior distribution}''
\begin{equation}
\label{eqn:conditional_active_posterior_general_definition}
    p \left( \boldsymbol{\Theta}_{\boldsymbol{\mu}_\perp} \right) = \int \mathcal{L} \left( \left( \hat{\mathbf{W}}_{d_l} \left( \omega_N \right) \right)^T \boldsymbol{\Theta}_{\boldsymbol{\mu}_\perp}, \delta \right) \pi \left( \left( \hat{\mathbf{W}}_{d_l} \left( \omega_N \right) \right)^T \boldsymbol{\Theta}_{\boldsymbol{\mu}_\perp} \right) \pi \left( \boldsymbol{\mu}_\perp | \left( \hat{\mathbf{W}}_{d_l} \left( \omega_N \right) \right)^T \boldsymbol{\Theta}_{\boldsymbol{\mu}_\perp} \right) \rho \left( \omega_N \right) d \omega_N , 
\end{equation}
which can be estimated using samples of the projection matrix as
\begin{equation}
    \label{eqn:estimate_of_active_posterior}
    \hat{p} \left( \boldsymbol{\Theta}_{\boldsymbol{\mu}_\perp} \right) = \frac{1}{N} \sum_{i=1}^N \mathcal{L} \left( \left( \hat{\mathbf{W}}_{d_l}^{(i)} \right)^T \boldsymbol{\Theta}_{\boldsymbol{\mu}_\perp}, \delta \right) \pi \left( \left( \hat{\mathbf{W}}_{d_l}^{(i)} \right)^T \boldsymbol{\Theta}_{\boldsymbol{\mu}_\perp} \right) \pi \left( \boldsymbol{\mu}_\perp | \left( \hat{\mathbf{W}}_{d_l}^{(i)} \right)^T \boldsymbol{\Theta}_{\boldsymbol{\mu}_\perp} \right) ,
\end{equation}
or be constructed as an empirical distribution from the sample set(s)
\begin{gather}
    \mathfrak{C}_{\boldsymbol{\Theta}_{\boldsymbol{\mu}_\perp}} = \left\{ \boldsymbol{\Theta}_{{\boldsymbol{\mu}_\perp}, k} = \hat{\mathbf{W}} \boldsymbol{\Xi}_{{\boldsymbol{\mu}_\perp}, k} = \hat{\mathbf{W}} \begin{bmatrix} \boldsymbol{\Xi}_{{d_l}, k}^T & \boldsymbol{\mu}_\perp^T \end{bmatrix}^T : \boldsymbol{\Xi}_{{d_l}, k} \in \mathfrak{C}_{\boldsymbol{\Xi}_{d_l}} ,  k = 1, \dots, n \right\} \label{eqn:nominal_set_of_active_posterior_samples} \\
    \mathfrak{C}_{\boldsymbol{\Theta}_{\boldsymbol{\mu}_\perp}}^{(i)} = \left\{ \boldsymbol{\Theta}_{{\boldsymbol{\mu}_\perp}, k}^{(i)} = \hat{\mathbf{W}}^{(i)} \boldsymbol{\Xi}_{{\boldsymbol{\mu}_\perp}, k} = \hat{\mathbf{W}}^{(i)} \begin{bmatrix} \boldsymbol{\Xi}_{{d_l}, k}^T & \boldsymbol{\mu}_\perp^T \end{bmatrix}^T : \boldsymbol{\Xi}_{{d_l}, k} \in \mathfrak{C}_{\boldsymbol{\Xi}_{d_l}} ,  k = 1, \dots, n \right\} \label{eqn:bootstrap_set_of_active_posterior_samples}
\end{gather}
where $ \mathfrak{C}_{\boldsymbol{\Xi}_{d_l}} $ is as defined in Eq.~\eqref{eqn:low_dimensional_parameter_sample_set}. Statistics of this posterior can be computed using Eq.~\eqref{eqn:general_system_parameter_statisitcs_monte_carlo_estimator} by substituting $ \boldsymbol{\Theta}_{{\boldsymbol{\mu}_\perp}, k}^{(i)} $ for $ \boldsymbol{\Theta}_{k}^{(i)} $. We also note here that samples from the conditional active posterior are obtained essentially for free from our proposed methodology. Both the full posterior and the conditional active posterior are obtained from the same set of MCMC samples in the latent space.

The conditional active parameter vector will always be maximally inferred from the observed output data, since it only includes those aspects of the parameter set that the system observable is sensitive to. There is also minimal risk of improperly inferring the system parameters, since changes along the inactive variables do not meaningfully change the system output. However, it is important to note that the output is not entirely unaffected by the inactive variables. While the QoI's sensitivity to the inactive variables is low by design, it may not be exactly zero. Hence, the conditional active posterior distribution underestimates the uncertainty in the inferred system parameters by some amount, with the magnitude of the error depending on the strength of the active subspace, i.e., the information loss incurred by the low-dimensional representation of the parameter vector. Perhaps a more holistic way to balance a parameter's uncertainty with its sensitivity is to modify the prior distribution to depend on these sensitivities, so that the prior uncertainty along directions that do not meaningfully inform the QoIs is not entirely removed but is taken to be low relative to the active directions. However, identifying and incorporating such priors is far from a trivial task. In the current workflow, it is the dimension reduction procedure that identifies the relative sensitivities of the different directions. However, constructing the likelihood-informed subspace requires the prior distributions to already be selected. Reconciling these issues and identifying appropriately proportioned priors along the active and inactive directions is thus left for future work.

\section{Numerical Examples}
\label{section:numerical_examples}

In this section, we study the behavior of the proposed calibration framework by applying it to two test problems with known behavior and prescribed input-output observations: first, a constructed polynomial function and then a numerical example considering side-impact crashworthiness of a vehicle. The goal of our proposed methodology is to enable the calibration of large parameter sets while comprehensively incorporating every source of uncertainty introduced by our analysis framework in addition to those already present in the computer model construction and experimental data collection. Hence, our results focus solely on the accuracy of misfit prediction under dimension reduction and surrogate approximation, as well as the quality of the calibration achieved. These observations are supported by comparisons between the probability distribution of the QoIs under the prior and posterior parameter densities, which highlight the uncertainty reduction and the increased concentration of the posterior samples around the observed output. 
We believe this to be sufficient numerical verification for our method, since (a) the generalized likelihood formulation used to account for the misfit prediction uncertainty of the surrogate is exact, with the mathematical derivation provided in Section~\ref{section:generalized_likelihood}, (b) Constantine et al.~\cite{constantine2016accelerating} have validated the use of active subspaces for the misfit function, and (c) the use of surrogates to replace expensive functions in engineering uncertainty quantification tasks is established practice. To the best of our knowledge, there is no existing model calibration procedure that accounts for the full suite of uncertainties accounted for within our framework. Since the solutions to Bayesian inference problems depend inextricably on the uncertainties captured by the likelihood formulation, it would be an unequal comparison to juxtapose our results against those obtained from another method that includes a different set of uncertainties than our own. Therefore, we omit any such potentially misleading comparisons here.

Some implementation decisions for the proposed algorithm are common across the subsequent sections. First, both the reconstruction surrogate and the projection surrogate that make up the misfit surrogate as per Section~\ref{section:training_misfit_surrogate} use Gaussian Process Regression (GPR)~\cite{RasmussenWilliamsGPR} with the covariance kernel composed of a Matérn 5/2 kernel along with an additive white noise kernel, and are trained using $ 500 $ samples drawn uniformly at random from the set $ \left[ -3, 3 \right]^{d_p} $. These training samples are independent of those used for identifying the active subspace (the specifics for which vary across the examples and are discussed in their respective sections.) Second, the basic Metropolis-Hastings~\cite{metropolis1953equation} algorithm with a standard normal proposal distribution is used to generate the posterior samples in the latent space, with $ 20 $ independent chains run in parallel for $ 25,000 $ iterations each. The implementation of the GPR surrogate modeling and MCMC sampling are deliberately kept as simple as possible, to keep the focus on the behavior of the proposed framework in its simplest case. Far more sophisticated sampling and surrogate modeling tools can be used, which will significantly improve the efficiency of the algorithm.

\subsection{Example 1: A Polynomial Test Function}
\label{section:example_1_polynomial}

\begin{figure}[t!bhp]
\centering
\begin{subfigure}{.40\textwidth}
  \centering
  \includegraphics[width=\textwidth]{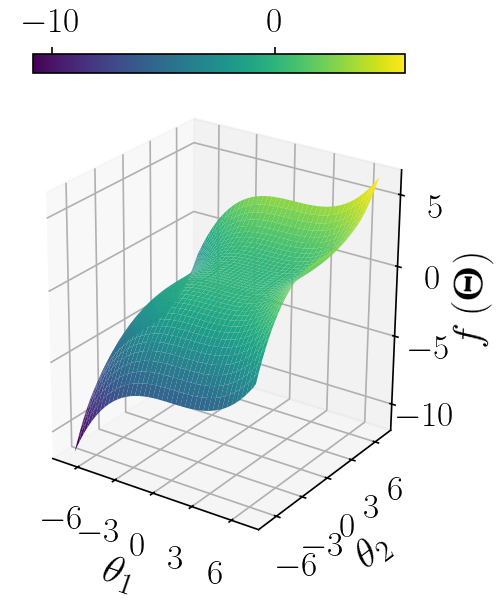}
  \caption{}
  \label{fig:example_1_ploynomial_3d_plot}
\end{subfigure}%
\begin{subfigure}{.40\textwidth}
  \centering
  \includegraphics[width=\textwidth]{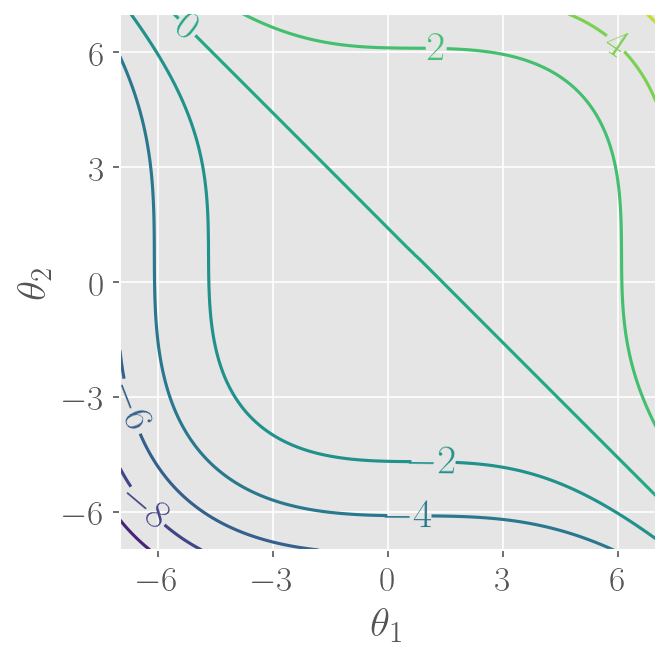}
  \caption{}
  \label{fig:example_1_polynomial_contour_plot}
\end{subfigure}
\caption{Depiction of the polynomial function (Eq.~\eqref{eqn:example_1_function}) from example 1 for $ d_p = 2 $. Figure (a) shows a 3-D plot, while (b) shows a contour plot.}
\label{fig:example_1_polynomial_visualization}
\end{figure}

Consider the following polynomial function (visualized for $ d_p = 2 $ in Figure~\ref{fig:example_1_polynomial_visualization}):
\begin{multline}
    \label{eqn:example_1_function}
    f \left( \boldsymbol{\Theta} \right) = \frac{1}{10} \left[ \sum_{j=1}^{d_p} \frac{j}{20} \left( \theta_j - \frac{1}{\sqrt{d_p}} \right) \right]^{\nicefrac{7}{5}} + \frac{1}{10} \left[ \sum_{j=1}^{d_p} \frac{\left( d_p - j \right)^{0.9}}{20} \left( \theta_j - \frac{1}{\sqrt{d_p}} \right) \right]^{\nicefrac{7}{5}} + \\ \frac{\gamma}{10} \sum_{j=1}^{d_p} \left[ \frac{1}{2} \left( \theta_j - \frac{1}{\sqrt{d_p}} \right) \right]^3
\end{multline}
with each $ \theta_j $ having a standard Normal prior distribution. 
The first two terms involve linear combinations of the parameters, while the third term breaks this linearity. By construction, therefore, the above polynomial can be well approximated in a two-dimensional latent space (i.e., $ d_l = 2 $) for small values of $ \gamma $, with the approximation growing poorer as $ \gamma $ is increased. Additionally, it has an inflection point at $ \theta_j = \nicefrac{1}{\sqrt{d_p}} $, $ j = 1, \dots, d_p $, which is also the only point in the domain where $ f \left( \boldsymbol{\Theta} \right) = 0 $. We use this polynomial to test the performance and robustness of our Bayesian calibration framework against the dimensionality of the parameter space and the strength of the active subspace (i.e., how well the latent space description of the function approximates the true function) using three cases, as listed in Table~\ref{tab:example_1_test_cases}. The underlying ``true'' value of the parameter vector is taken to be $ \theta_j = \nicefrac{1}{\sqrt{d_p}} $, $ j = 1, \dots, d_p $ such that $ f_{\boldsymbol{\Theta}}^* = 0 $, and the observation uncertainty is assumed Gaussian with standard deviation $ \boldsymbol{\Sigma}_{\mathbf{Y}} = 0.1 $.

\begin{table}[t!bhp]
    \centering
    \begin{tabular}{c|c|c|c}
        Name & Characteristics & $ \gamma $ & $ d_p $ \\
        \hline \hline
        Case 1 & Moderate dimensional with strong active subspace & $ 0.05 $ & $ 10 $ \\
        Case 2 & Moderate dimensional with weaker active subspace & $ 0.12 $ & $ 10 $ \\
        Case 3 & High dimensional with strong active subspace & $ 0.6 $ & $ 100 $ \\
        \hline
    \end{tabular}
    \caption{Cases for the polynomial test function with corresponding values of $ \gamma $ and $ d_p $.}
    \label{tab:example_1_test_cases}
\end{table}

\begin{figure}[t!bhp]
\centering
\begin{subfigure}{.30\textwidth}
  \centering
  \includegraphics[width=\textwidth]{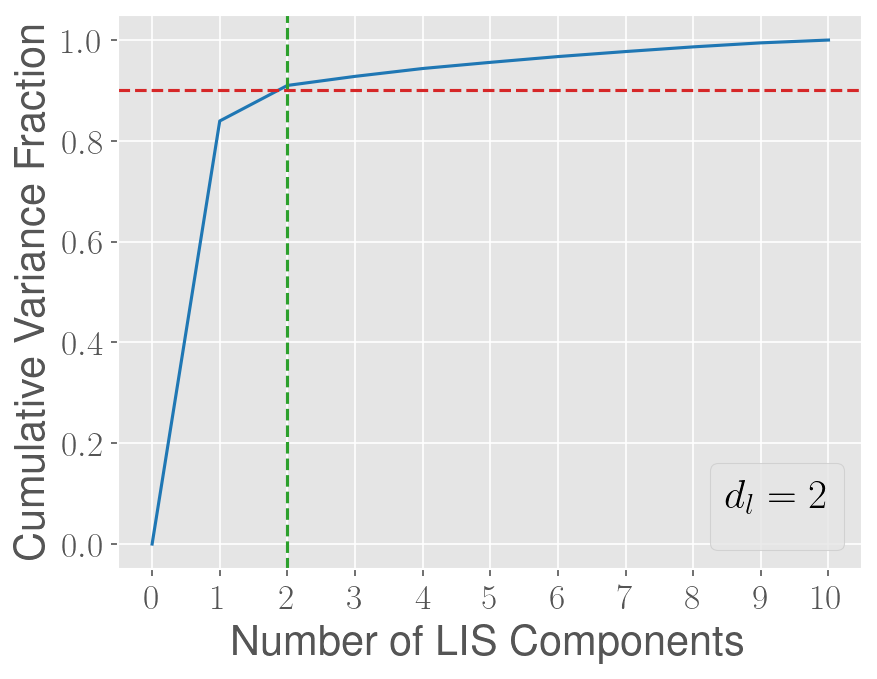}
  \caption{}
  \label{fig:example_1_case_1_cumulative_variance_gradient}
\end{subfigure}%
\begin{subfigure}{.30\textwidth}
  \centering
  \includegraphics[width=\textwidth]{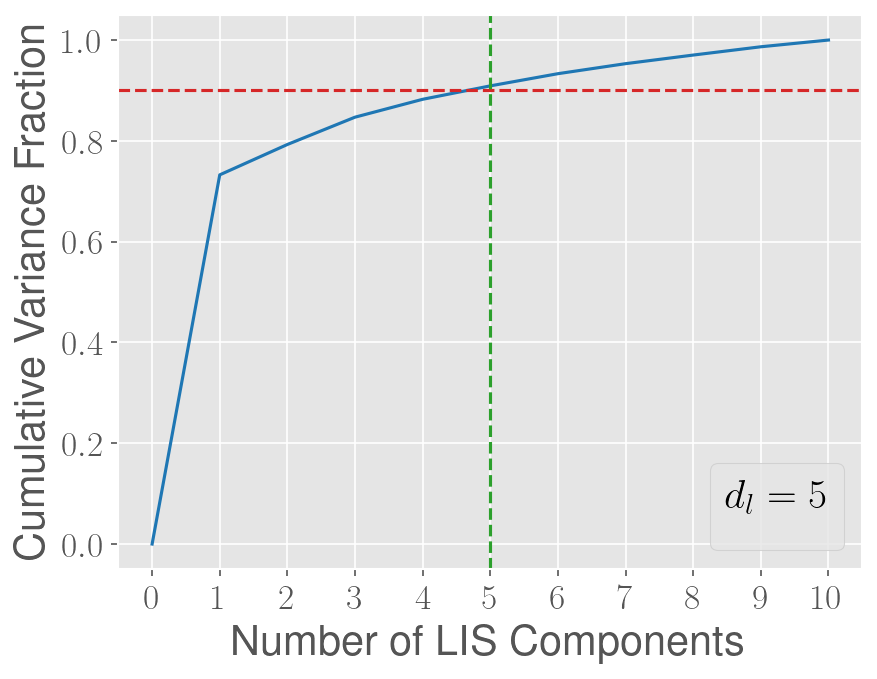}
  \caption{}
  \label{fig:example_1_case_2_cumulative_variance_gradient}
\end{subfigure}%
\begin{subfigure}{.30\textwidth}
  \centering
  \includegraphics[width=\textwidth]{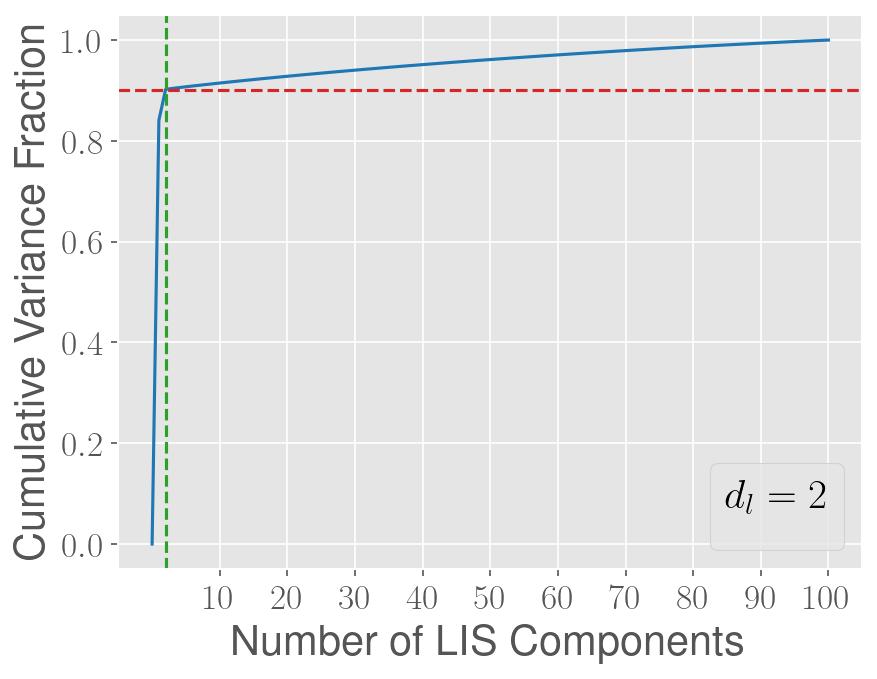}
  \caption{}
  \label{fig:example_1_case_3_cumulative_variance_gradient}
\end{subfigure}
\begin{subfigure}{.30\textwidth}
  \centering
  \includegraphics[width=\textwidth]{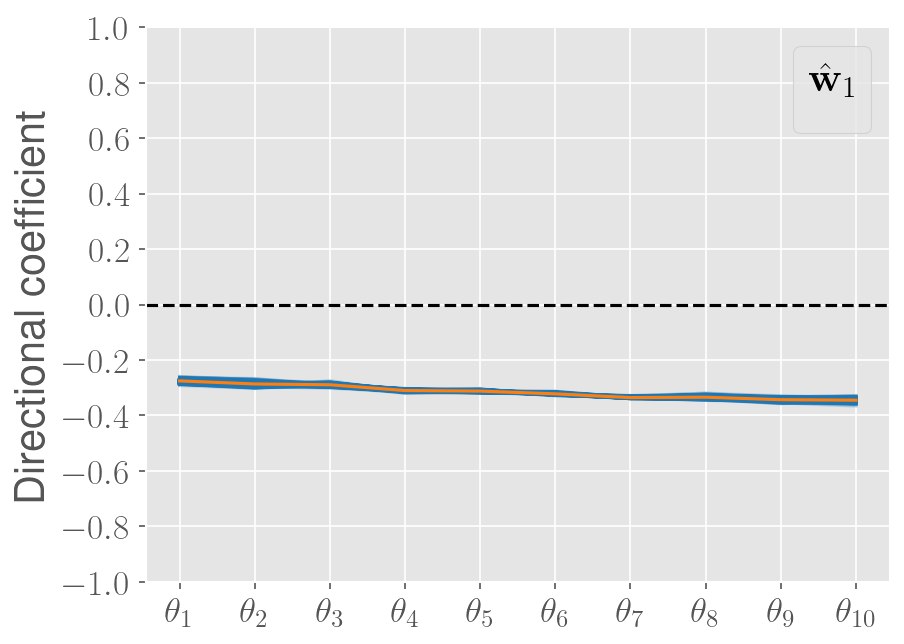}
  \caption{}
  \label{fig:example_1_case_1_projection_vector_1}
\end{subfigure}%
\begin{subfigure}{.30\textwidth}
  \centering
  \includegraphics[width=\linewidth]{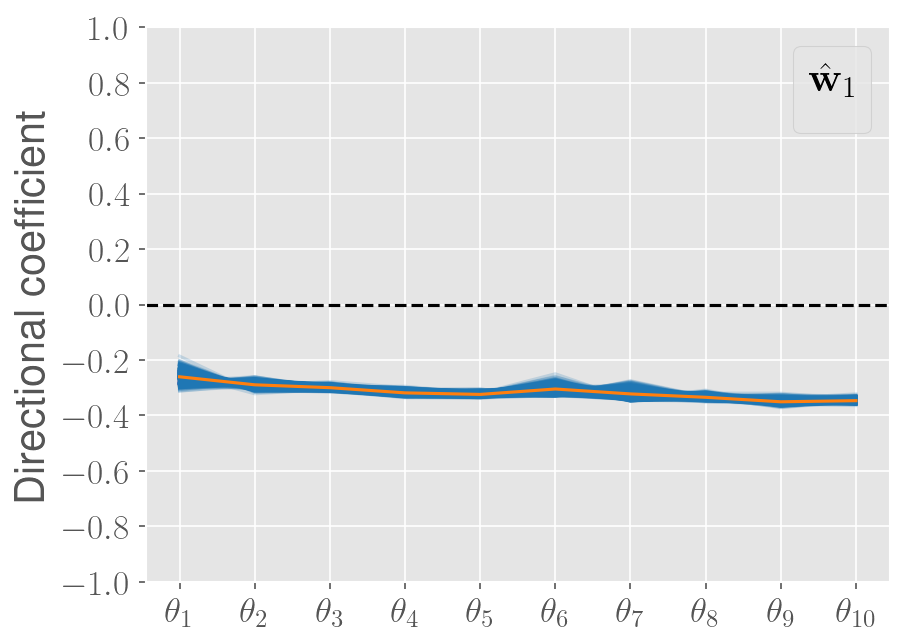}
  \caption{}
  \label{fig:example_1_case_2_projection_vector_1}
\end{subfigure}%
\begin{subfigure}{.30\textwidth}
  \centering
  \includegraphics[width=\linewidth]{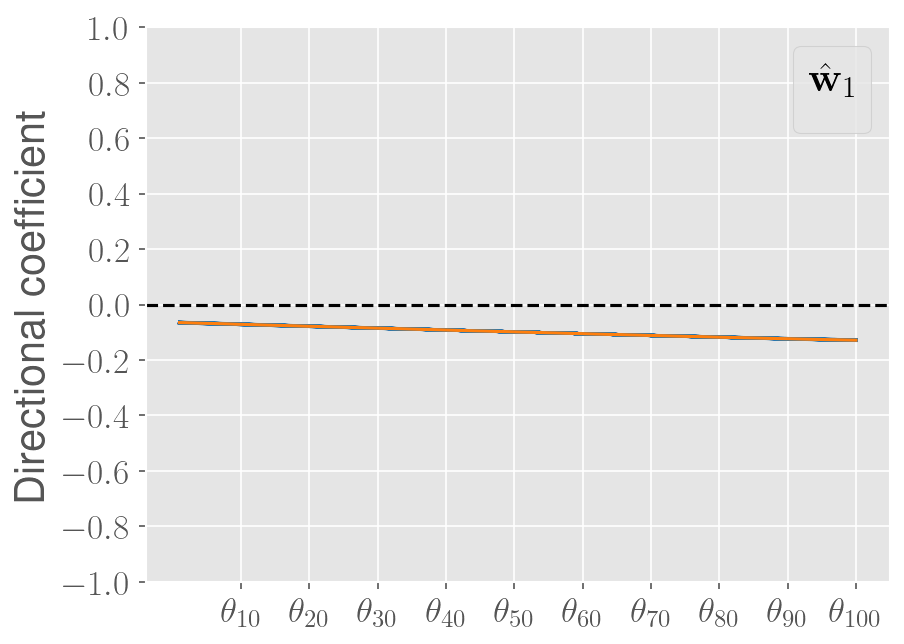}
  \caption{}
  \label{fig:example_1_case_3_projection_vector_1}
\end{subfigure}
\begin{subfigure}{.30\textwidth}
  \centering
  \includegraphics[width=\textwidth]{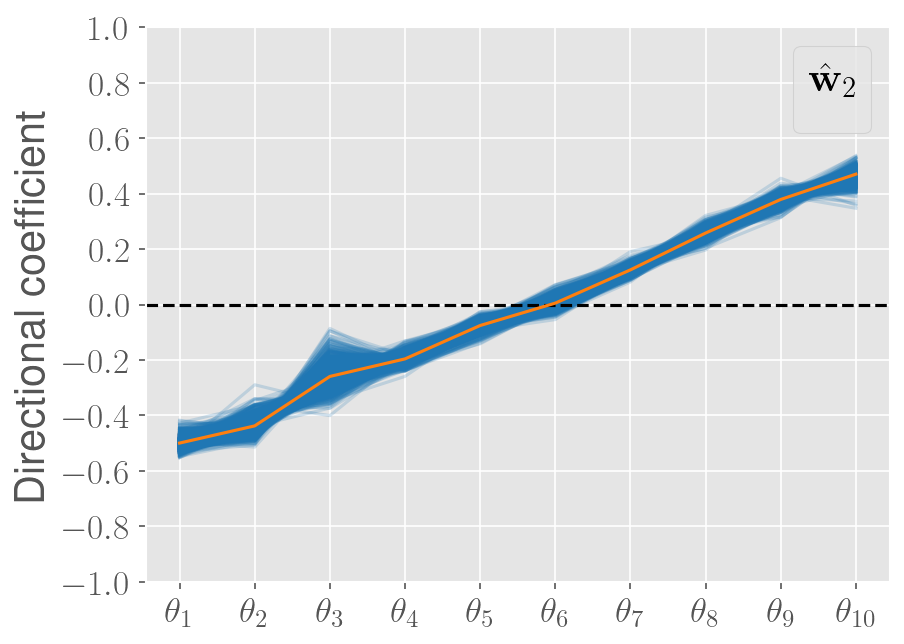}
  \caption{}
  \label{fig:example_1_case_1_projection_vector_2}
\end{subfigure}%
\begin{subfigure}{.30\textwidth}
  \centering
  \includegraphics[width=\linewidth]{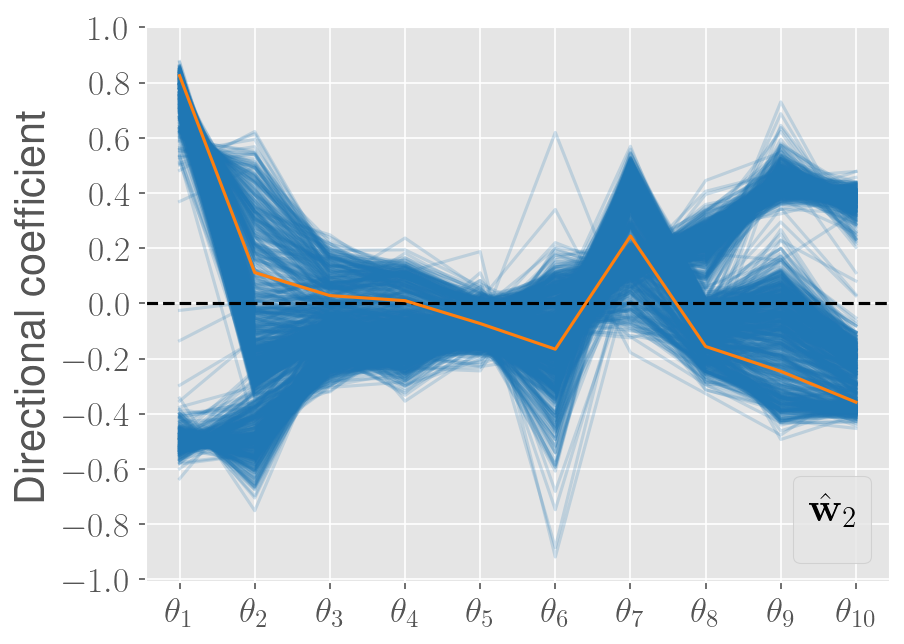}
  \caption{}
  \label{fig:example_1_case_2_projection_vector_2}
\end{subfigure}%
\begin{subfigure}{.30\textwidth}
  \centering
  \includegraphics[width=\linewidth]{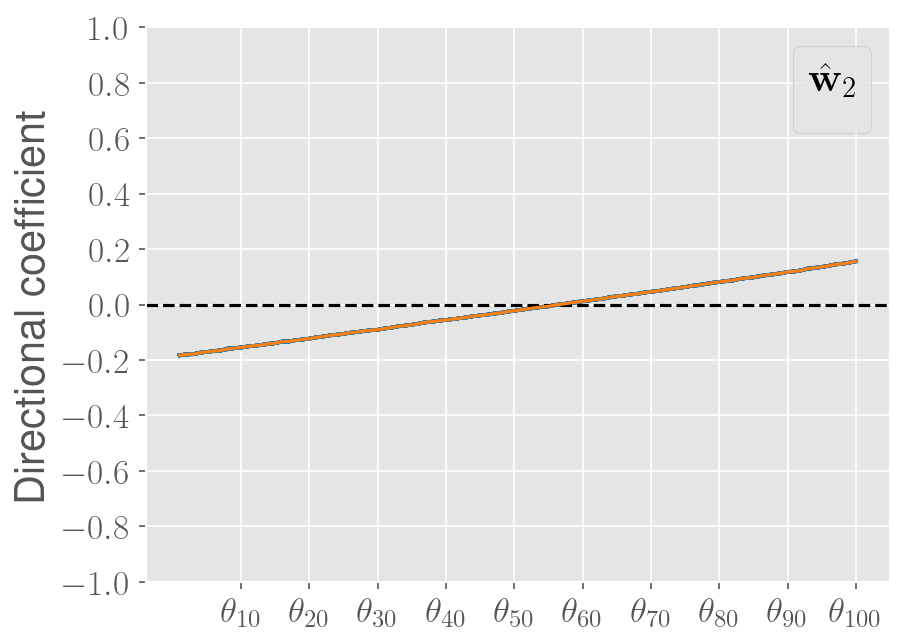}
  \caption{}
  \label{fig:example_1_case_3_projection_vector_2}
\end{subfigure}
\caption{Active subspace results for the polynomial function in example 1. Figures (a), (b), and (c) plot the cumulative gradient variance explained as a function of the number of eigenvectors included for Cases 1, 2, and 3, respectively. In these figures, the red dashed line represents the $ 90 \% $ variance fraction threshold, which is used to select $ d_l $, while the green dashed line marks the chosen value of $ d_l $. Figures (d), (e), and (f) visualize $ \hat{\mathbf{w}}_1 $ for Cases 1, 2, and 3, with the nominal vector in orange and $ 1000 $ bootstrapped replicates in blue. Figures (g), (h), and (i) visualize $ \hat{\mathbf{w}}_2 $ for Cases 1, 2, and 3, respectively, following the same scheme.}
\label{fig:active_subspace_results_for_example_1}
\end{figure}

\begin{figure}[t!bhp]
\centering
\begin{subfigure}{.30\textwidth}
  \centering
  \includegraphics[width=\textwidth]{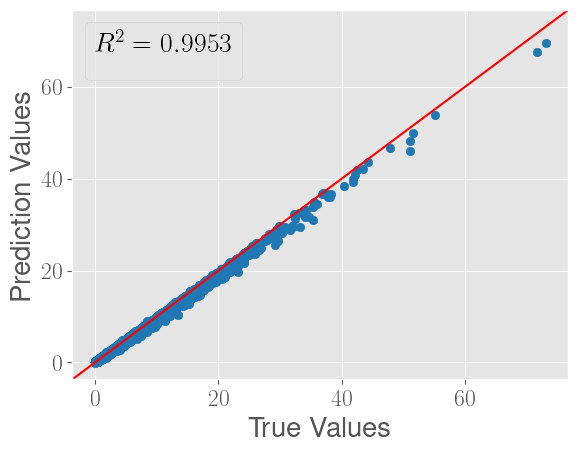}
  \caption{}
  \label{fig:example_1_case_1_surrogate_quality}
\end{subfigure}%
\begin{subfigure}{.30\textwidth}
  \centering
  \includegraphics[width=\linewidth]{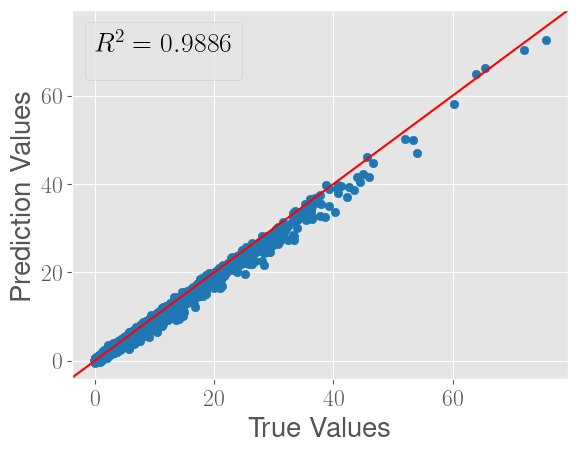}
  \caption{}
  \label{fig:example_1_case_2_surrogate_quality}
\end{subfigure}%
\begin{subfigure}{.30\textwidth}
  \centering
  \includegraphics[width=\linewidth]{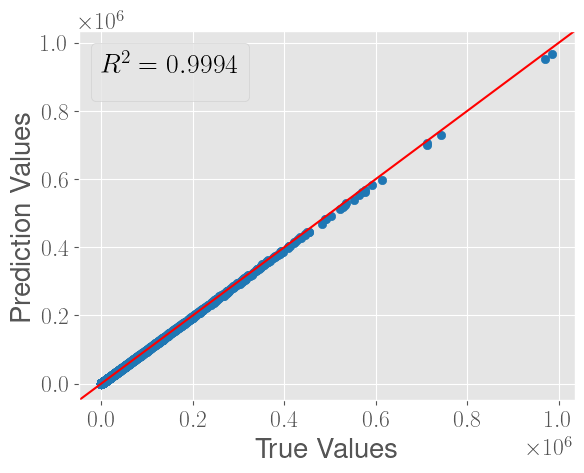}
  \caption{}
  \label{fig:example_1_case_3_surrogate_quality}
\end{subfigure}
\caption{Misfit surrogate quality results for example 1. Figure (a) plots the predicted misfit values vs. the true misfit values for Case 1, (b) for Case 2, and (c) for Case 3. Each figure includes the $ R^2 $ value, and the $ x=y $ line (i.e., $ \SI{45}{\degree} $ line) highlighted in red.}
\label{fig:example_1_surrogate_quality}
\end{figure}

\begin{figure}[t!bhp]
\centering
\begin{subfigure}{.48\textwidth}
  \centering
  \includegraphics[width=\textwidth]{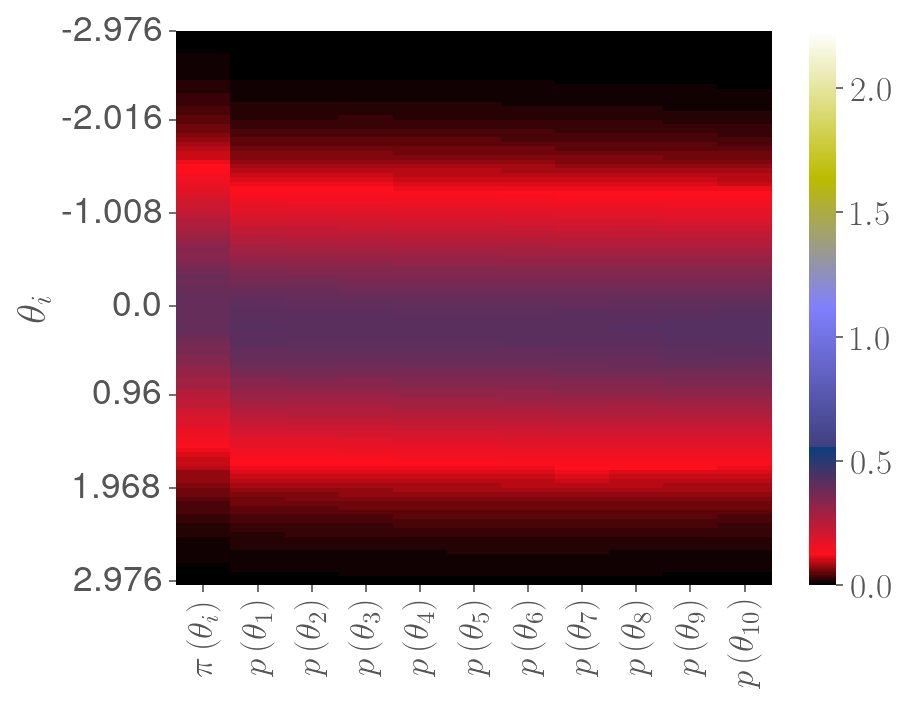}
  \caption{}
  \label{fig:example_1_case_1_full_posterior_heatmap}
\end{subfigure}%
\begin{subfigure}{.48\textwidth}
  \centering
  \includegraphics[width=\textwidth]{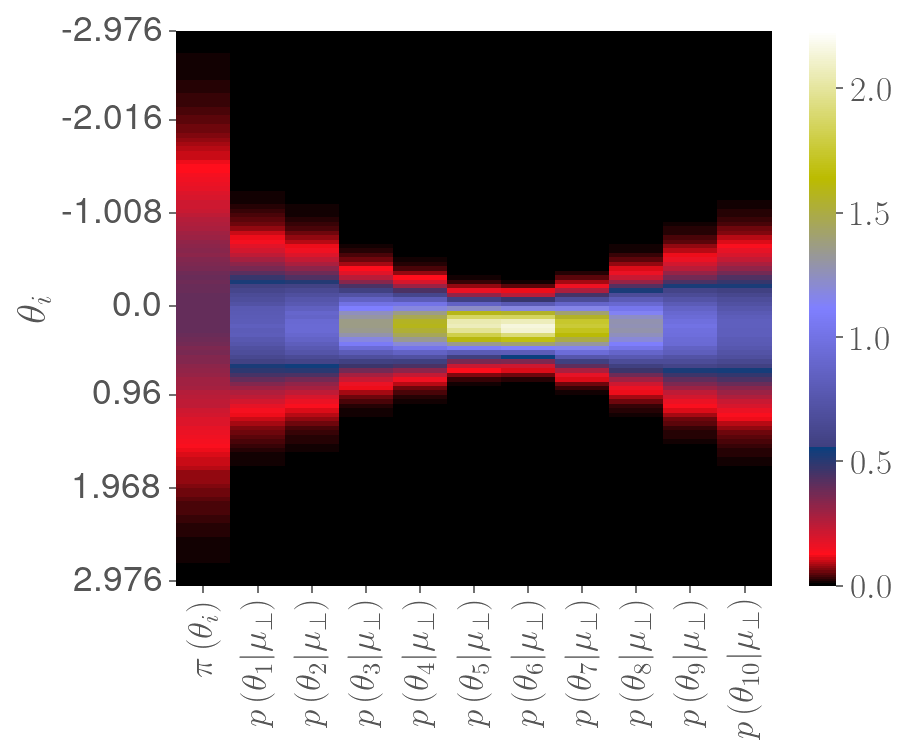}
  \caption{}
  \label{fig:example_1_case_1_conditional_posterior_heatmap}
\end{subfigure}
\begin{subfigure}{.48\textwidth}
  \centering
  \includegraphics[width=\textwidth]{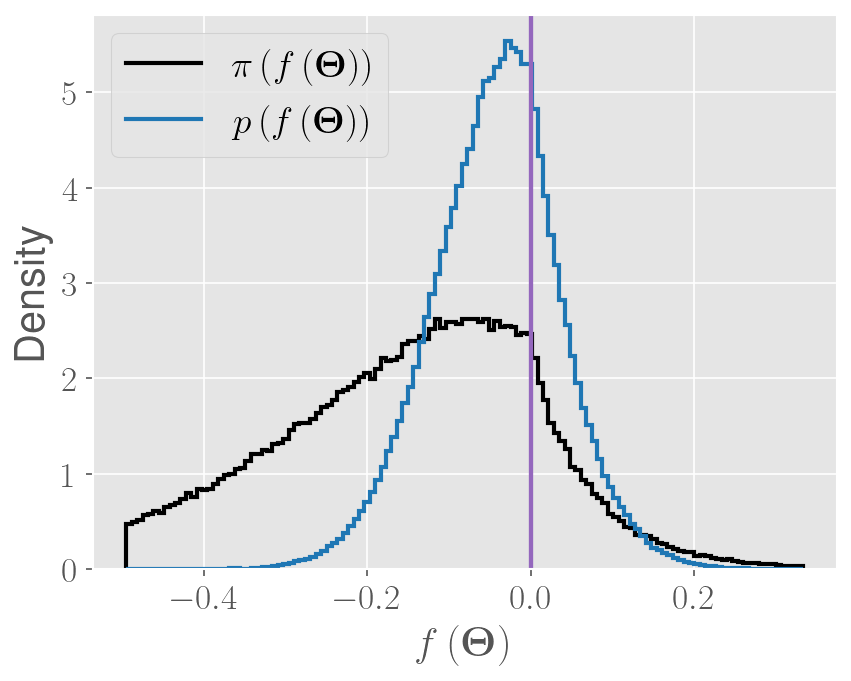}
  \caption{}
  \label{fig:example_1_case_1_full_output_posterior}
\end{subfigure}%
\begin{subfigure}{.48\textwidth}
  \centering
  \includegraphics[width=\textwidth]{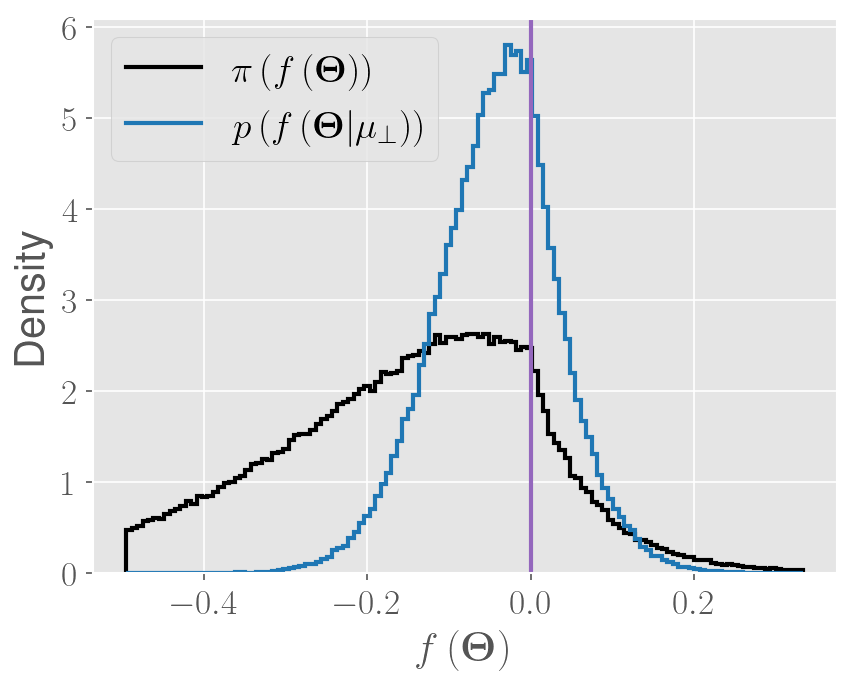}
  \caption{}
  \label{fig:example_1_case_1_conditional_output_posterior}
\end{subfigure}
\caption{Calibration results for Case 1 of example 1. Figures (a) and (b) are heatmaps comparing the full marginal posteriors and the conditional active marginal posteriors, respectively, of each $ \theta_i $ with the prior. The first column in each heatmap is the prior, which is identical for all parameter components, and the subsequent columns are the calibrated posteriors of the parameters in order. Figures (c) and (d) plot the probability density of output for samples from the full posterior and the conditional active posterior, respectively, in blue. Also in (c) and (d), the probability density of the output corresponding to samples from $ \pi \left( \boldsymbol{\Theta} \right) $ is drawn in black, and the observed $ f_{\boldsymbol{\Theta}}^* $ is in purple.}
\label{fig:example_1_case_1_calibration_results}
\end{figure}

\begin{figure}[t!bhp]
\centering
\begin{subfigure}{.48\textwidth}
  \centering
  \includegraphics[width=\textwidth]{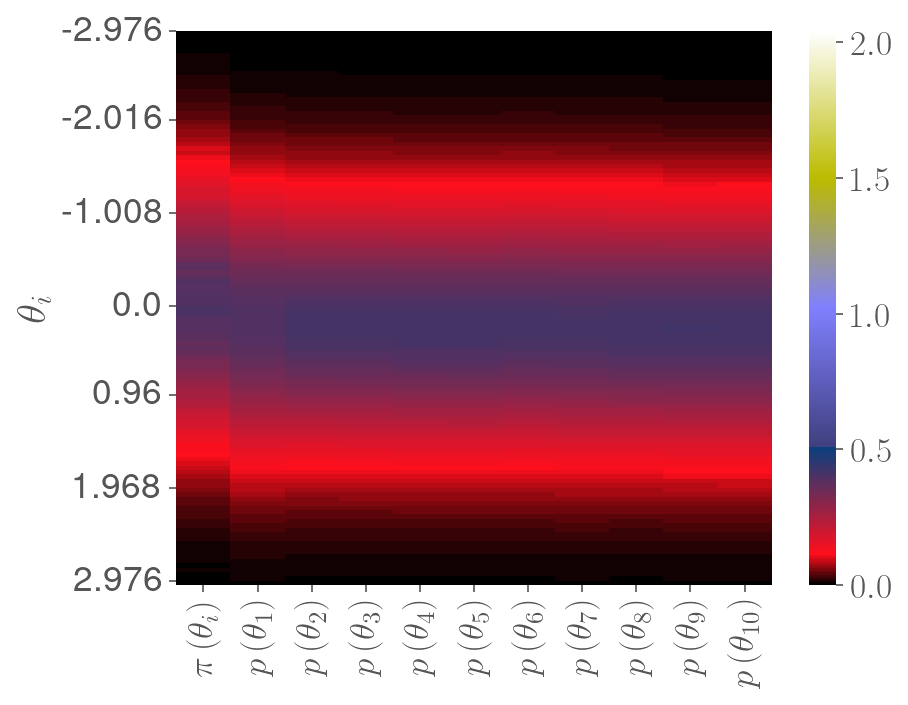}
  \caption{}
  \label{fig:example_1_case_2_full_posterior_heatmap}
\end{subfigure}%
\begin{subfigure}{.48\textwidth}
  \centering
  \includegraphics[width=\textwidth]{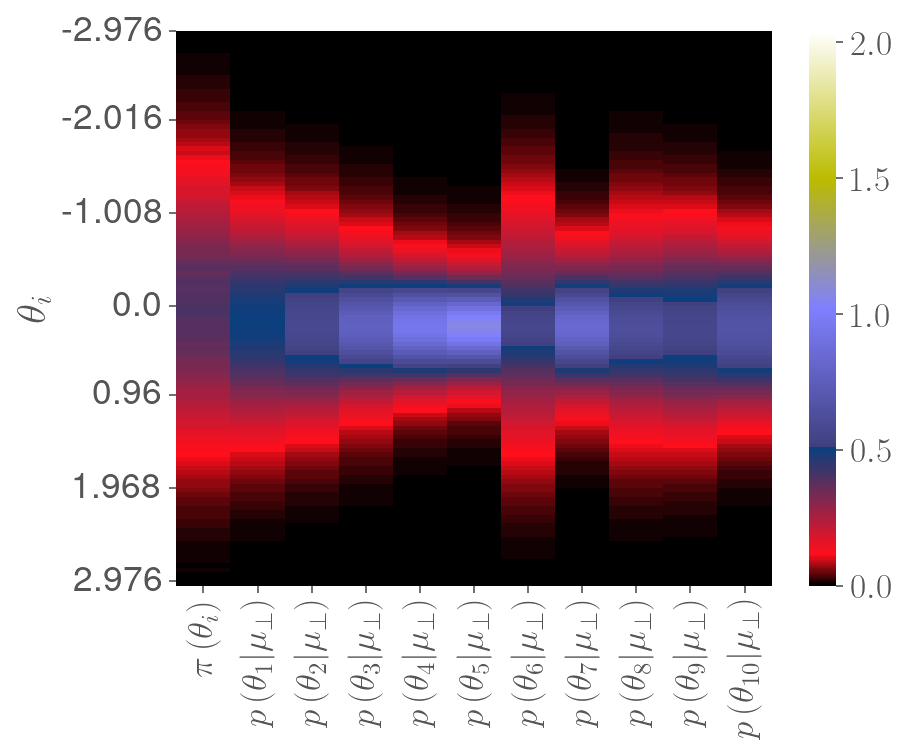}
  \caption{}
  \label{fig:example_1_case_2_conditional_posterior_heatmap}
\end{subfigure}
\begin{subfigure}{.48\textwidth}
  \centering
  \includegraphics[width=\textwidth]{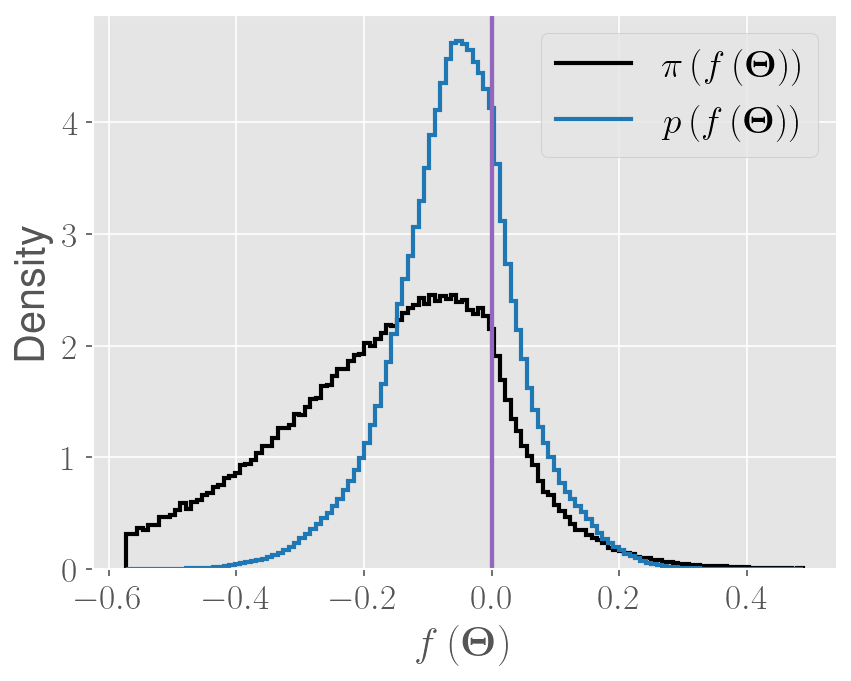}
  \caption{}
  \label{fig:example_1_case_2_full_output_posterior}
\end{subfigure}%
\begin{subfigure}{.48\textwidth}
  \centering
  \includegraphics[width=\textwidth]{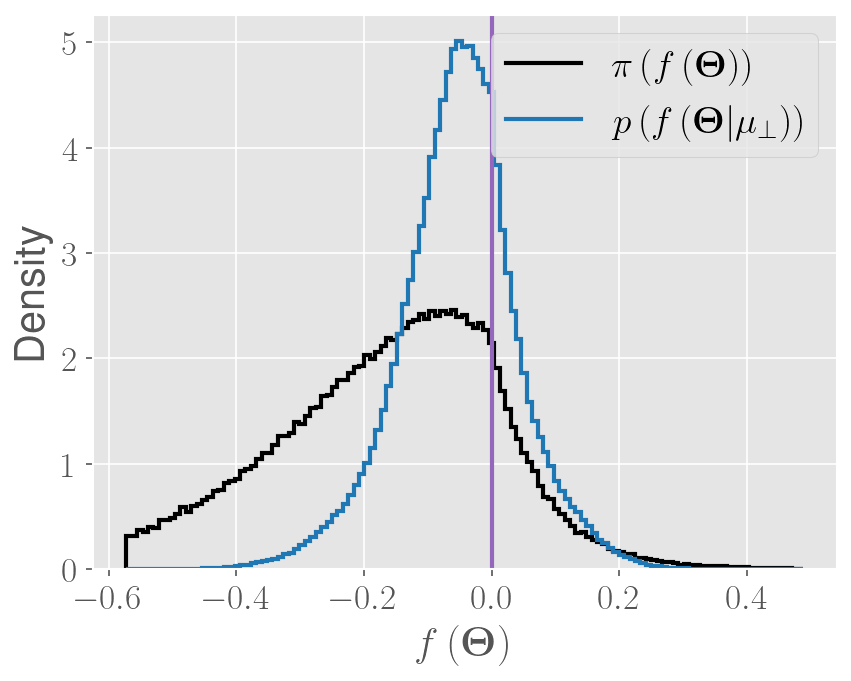}
  \caption{}
  \label{fig:example_1_case_2_conditional_output_posterior}
\end{subfigure}
\caption{Calibration results for Case 2 of example 1. Figures (a) and (b) are heatmaps comparing the full marginal posteriors and the conditional active marginal posteriors, respectively, of each $ \theta_i $ with the prior. The first column in each heatmap is the prior, which is identical for all parameter components, and the subsequent columns are the calibrated posteriors of the parameters in order. Figures (c) and (d) plot the probability density of output for samples from the full posterior and the conditional active posterior, respectively, in blue. Also in (c) and (d), the probability density of the output corresponding to samples from $ \pi \left( \boldsymbol{\Theta} \right) $ is drawn in black, and the observed $ f_{\boldsymbol{\Theta}}^* $ is in purple.}
\label{fig:example_1_case_2_calibration_results}
\end{figure}

\begin{figure}[t!bhp]
\centering
\begin{subfigure}{0.8\textwidth}
  \centering
  \includegraphics[width=\textwidth]{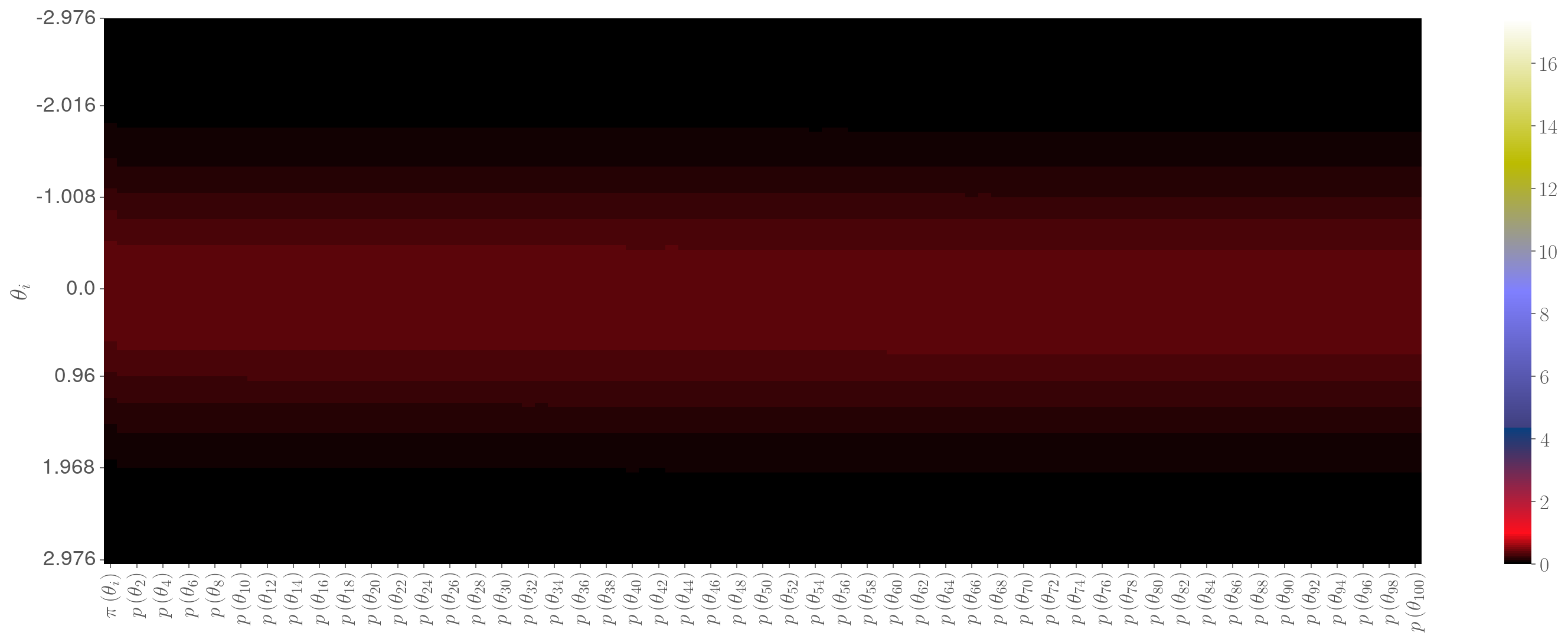}
  \caption{}
  \label{fig:example_1_case_3_full_posterior_heatmap}
\end{subfigure}
\begin{subfigure}{0.8\textwidth}
  \centering
  \includegraphics[width=\textwidth]{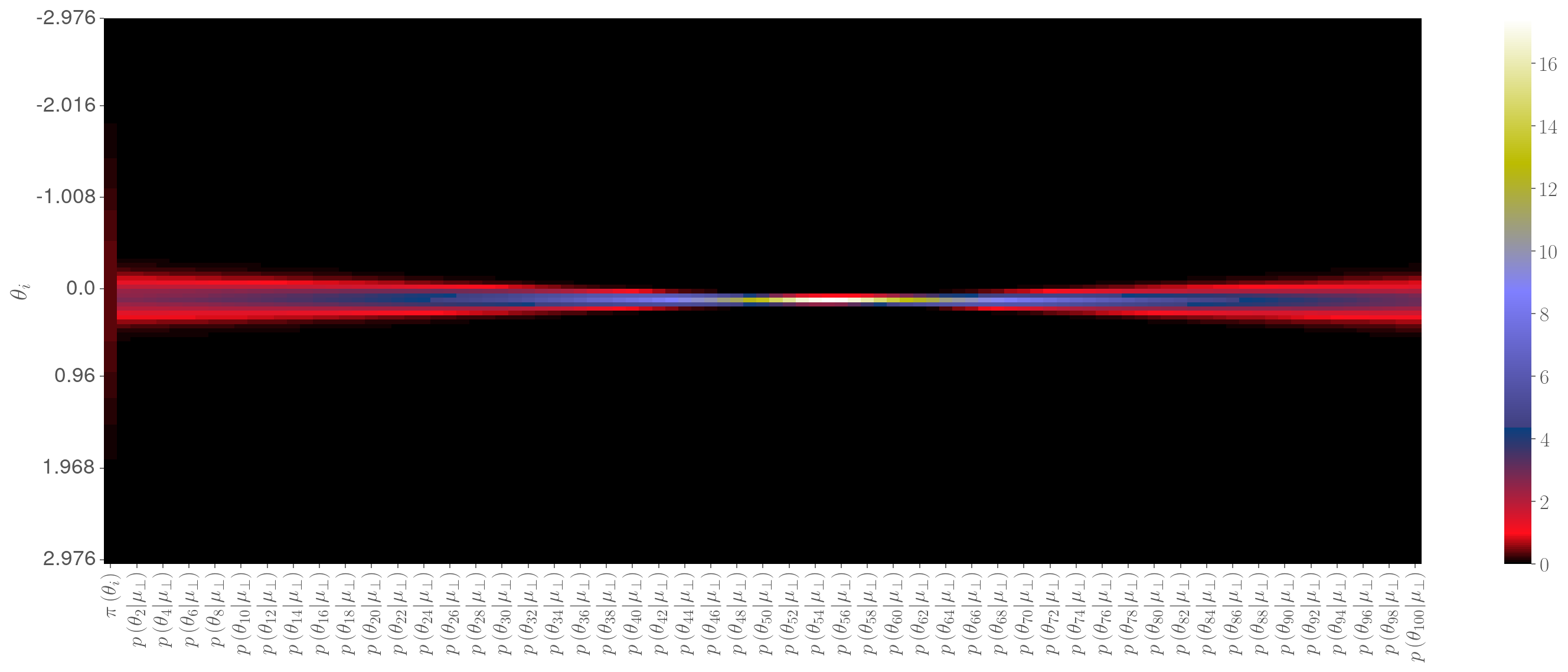}
  \caption{}
  \label{fig:example_1_case_3_conditional_posterior_heatmap}
\end{subfigure}
\begin{subfigure}{.48\textwidth}
  \centering
  \includegraphics[width=\textwidth]{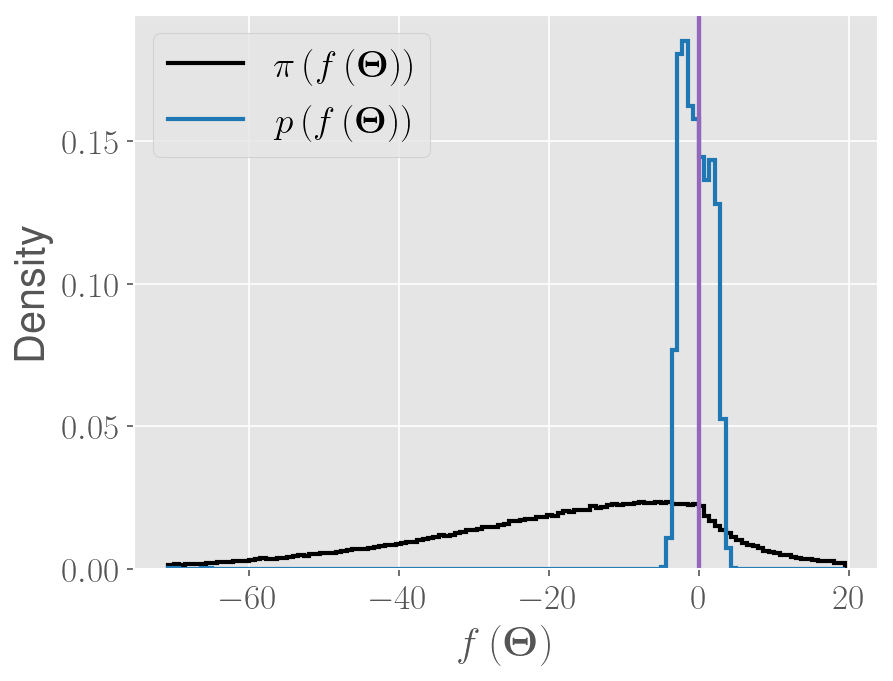}
  \caption{}
  \label{fig:example_1_case_3_full_output_posterior}
\end{subfigure}%
\begin{subfigure}{.48\textwidth}
  \centering
  \includegraphics[width=\textwidth]{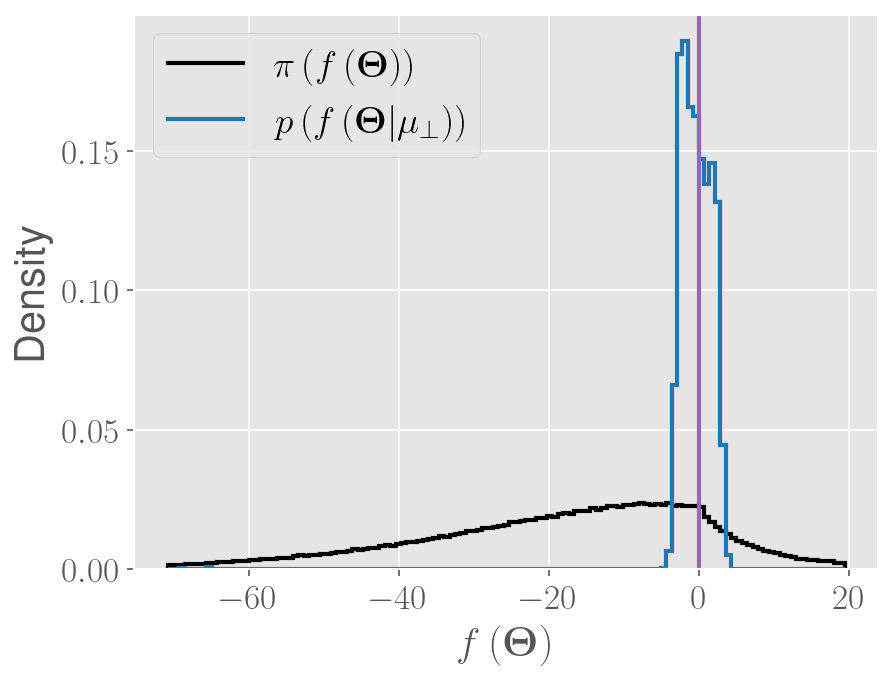}
  \caption{}
  \label{fig:example_1_case_3_conditional_output_posterior}
\end{subfigure}
\caption{Calibration results for Case 3 of example 1. Figures (a) and (b) are heatmaps comparing the full marginal posteriors and the conditional active marginal posteriors, respectively, of each $ \theta_i $ with the prior. The first column in each heatmap is the prior, which is identical for all parameter components, and the subsequent columns are the calibrated posteriors of the parameters in order. Figures (c) and (d) plot the probability density of output for samples from the full posterior and the conditional active posterior, respectively, in blue. Also in (c) and (d), the probability density of the output corresponding to samples from $ \pi \left( \boldsymbol{\Theta} \right) $ is drawn in black, and the observed $ f_{\boldsymbol{\Theta}}^* $ is in purple.}
\label{fig:example_1_case_3_calibration_results}
\end{figure}

Following the steps of our framework, first the LIS is constructed as in Section~\ref{section:likelihood_informed_subspace} for the misfit using $ 69 $ gradient evaluations for Cases 1 and 2, and $ 1381 $ for Case 3, at parameter values sampled from the standard Normal prior $ \pi \left( \boldsymbol{\Theta} \right) $; the number of gradient evaluations were chosen based on Eq.~\eqref{eqn:gradient_sample_size_heuristic_AS} with $ \alpha = 3 $ and $ m = d_p $. (In practical cases, where the analyst may have a guess for the latent-space dimension $ d_l $ based on expert judgment, or some external upper bound on $ d_l $, $ m $ should be chosen as this guess or constraint, as suggested by Constantine~\cite{ConstantineActiveSubspacesBook2015}. Here, we opt to implement the most conservative case, where we want all eigenvectors to be as accurately estimated as possible.) In each case, $ d_l $ is selected as the smallest integer such that the cumulative sum of the leading eigenvalues exceeds $ 90 \% $, which ensures that $ 90 \% $ of the variance of the misfit gradient is captured by the active subspace. Figures~\ref{fig:example_1_case_1_cumulative_variance_gradient} - ~\ref{fig:example_1_case_3_cumulative_variance_gradient} plot these variance fractions for Cases 1, 2, and 3, respectively. As expected, we get $ d_l = 2 $ for Cases 1 and 3, but a larger $ d_l $ for Case 2 where the non-linear third term in Eq.~\eqref{eqn:example_1_function} has a larger relative contribution. Figure~\ref{fig:active_subspace_results_for_example_1} also visualizes the first two columns $ \hat{\mathbf{w}}_1 $ and $ \hat{\mathbf{w}}_2 $ of the rotation matrix $ \hat{\mathbf{W}} $ (the directions in the original parameter space that capture the largest contribution of the misfit gradient variance) in terms of the direction cosines associated with each parameter $ \theta_i $ (which quantifies the contribution of $ \theta_i $ to the vector $ \hat{\mathbf{w}}_1 $ or $ \hat{\mathbf{w}}_2 $). The orange line in each subfigure corresponds to the nominal vector $ \hat{\mathbf{w}}_1 $ or $ \hat{\mathbf{w}}_2 $, while $ N = 1000 $ bootstrap replicates of these vectors (i.e., $ \hat{\mathbf{w}}_1^{(i)} $ and $ \hat{\mathbf{w}}_2^{(i)} $, $ i = 1, \dots, 1000 $) are drawn in blue, showcasing a band of uncertainty. 
The rest of the columns $ \hat{\mathbf{w}}_3 $ - $ \hat{\mathbf{w}}_{10} $ for Case 1 are provided in Appendix~\ref{appendix:additional_figures_example_1} for reference.

It is clear to see from Figures~\ref{fig:example_1_case_1_projection_vector_1} to~\ref{fig:example_1_case_3_projection_vector_2} that the uncertainty in the estimated columns of the rotation vector increases as their corresponding eigenvalue (i.e., gradient variance fraction) decreases. First, in Cases 1 and 3, the initial eigenvectors have low uncertainty while the later ones have high uncertainty. Secondly, comparing Case 2 against 1 and 3, where the initial eigenvectors themselves correspond to lower eigenvalues due to a comparatively weaker active subspace, the uncertainty in even the first two columns is high compared to those of Cases 1 and 3. This increasing estimation uncertainty with decreasing variance contribution of the directions serves as another incentive to use the conditional active posterior (Eq.~\eqref{eqn:conditional_active_posterior_general_definition}), which depends only on $ \hat{\mathbf{W}}_{d_l} $, i.e., the first $ d_l $ columns of the rotation vector. The full posterior as defined in Eq.~\eqref{eqn:system_parameter_posterior_marginalized_over_rotation_matrix}, meanwhile, requires all the columns of $ \hat{\mathbf{W}} $, including the later ones with high uncertainty. 


Next, a surrogate for the misfit is constructed in the LIS following the procedure in Section~\ref{section:surrogate_and_generalized_likelihood}. As noted in Section~\ref{section:projection_surrogate}, the model and gradient evaluations used for building the LIS were insufficient to train this surrogate, since they are too concentrated around the origin. Instead, $ 500 $ samples are selected uniformly at random from the set $ \left[ -3, 3 \right]^{d_p} $, which sufficiently covers the parameter domain. Again, a more efficient implementation may use active learning to minimize the number of additional model evaluations. However, as this manuscript focuses on exploring the behavior of the framework and not on an optimal implementation, the above naive sample selection is sufficient for our purposes. We see from Figure~\ref{fig:example_1_surrogate_quality} that the misfit is well predicted in each of the three cases considered, with nearly unit $ R^2 $. (Note that the concentration of the samples around the mode is not an issue for the LIS construction, since the theory of active subspaces requires samples to be drawn according to a specified distribution, which is the prior distribution $ \pi \left( \boldsymbol{\Theta} \right) $ in our case.)

Finally, Bayesian inference is carried out using the proposed framework. $ N = 1000 $ bootstrap replicates of the projection matrix $ \hat{\mathbf{W}} $ are used for Cases 1 and 2, while $ 250 $ are used for Case 3. The results of the calibration are collected in Figures~\ref{fig:example_1_case_1_calibration_results} -~\ref{fig:example_1_case_3_calibration_results} for Cases 1 - 3, respectively. Consistently, we see that while the full posteriors shift perceptibly towards the true value of $ \theta_j = \nicefrac{1}{\sqrt{d_p}} $, the uncertainty in the posteriors remains comparable to that of the prior (Figures~\ref{fig:example_1_case_1_full_posterior_heatmap} -~\ref{fig:example_1_case_3_full_posterior_heatmap}). This is because the QoI function, and hence the misfit, is only sensitive to $ 2 - 5 $ directions in the domain. Consequently, the priors along the inactive directions are retained completely, which contributes a large amount of the high uncertainty seen in the full posteriors. Despite this, when comparing the prior probability density of $ f \left( \mathbf{X}, \boldsymbol{\Theta} \right) $ with the density corresponding to the full posterior of $ f \left( \mathbf{X}, \boldsymbol{\Theta} \right) $ (Figures~\ref{fig:example_1_case_1_full_output_posterior} -~\ref{fig:example_1_case_3_full_output_posterior}), we see a clear reduction in the uncertainty, signaling effective calibration. On the other hand, the conditional active posteriors (Figures~\ref{fig:example_1_case_1_conditional_posterior_heatmap} -~\ref{fig:example_1_case_3_conditional_posterior_heatmap}) have a comparably much lower spread than the full posteriors, since the parameters are not varied along the inactive directions, thereby removing their contribution to the calibrated posteriors. Simultaneously, the probability density of $ f \left( \mathbf{X}, \boldsymbol{\Theta} \right) $ corresponding to the conditional active posterior (Figures~\ref{fig:example_1_case_1_conditional_output_posterior} -~\ref{fig:example_1_case_3_conditional_output_posterior}) is indistinguishable from that corresponding to the full posterior mentioned above (Figures~\ref{fig:example_1_case_1_full_output_posterior} -~\ref{fig:example_1_case_3_full_output_posterior}). Clearly, both the full posterior and the conditional active posterior result in equivalent uncertainty in the QoIs, despite the conditional active posterior being significantly more confident about the inferred parameters for the same computational cost as the full posterior. It is also noteworthy that the conditional active posteriors for Case 2 are wider than those for Case 1, which is a direct consequence of the weaker active subspace and thus fewer discarded inactive variables in Case 2.

Estimating high-dimensional kernel densities from samples is usually intractable, making it challenging to compute the joint posteriors defined in Eqs.~\eqref{eqn:system_parameter_posterior_marginalized_over_rotation_matrix} or~\eqref{eqn:conditional_active_posterior_general_definition} from $ \mathfrak{C}_{\boldsymbol{\Theta}}^{(i)} $, $ i = 1, \dots, N $. However, low-dimensional marginals can be obtained by first computing the relevant marginal for each individual sample set replicate $ \mathfrak{C}_{\boldsymbol{\Theta}}^{(i)} $, and then averaging over the $ N $ replicates. This is the procedure used to generate the marginals in this and the subsequent sections. Appendix~\ref{appendix:additional_figures_example_1} collects figures that show all one-dimensional marginals for both the full posterior and the conditional active posterior for Cases 1 and 2, plotted in terms of the nominal marginal constructed using $ \mathfrak{C}_{\boldsymbol{\Theta}} $ along with uncertainty bands corresponding to the $ 90 \% $ confidence interval evaluated using $ \mathfrak{C}_{\boldsymbol{\Theta}}^{(i)} $, $ i = 1, \dots, N $.

\subsection{Example 2: Vehicle Side Impact}
\label{section:example_2_vehicle_side_impact}

For the second numerical illustration, we adapt a case study on vehicle crashworthiness from Gu et al.~\cite{gu2001optimisation}. The vehicle's safety is assessed using a side-impact test that mimics the behavior of a passenger vehicle struck on its side by a light-duty truck (e.g., an SUV) at $ \SI{30}{mph} $. At the moment of impact, the response of the crash test dummy within the test vehicle, as well as the vehicle itself, is measured and characterized using ten quantities of interest (whose descriptions and symbols are listed in Table~\ref{tab:example_2_qois}). A finite element model developed at Ford Motor Company was used to simulate this test and construct response surfaces for these ten output quantities using eleven design variables, corresponding to the sizes and material properties of several key components (as described in Table~\ref{tab:example_2_parameters}).

The case study has been used numerous times in the literature, although primarily in the context of reliability-based design optimization~\cite{youn2004reliability, du2004sequential, zou2006direct, liang2017pareto}, robust design optimization~\cite{gu2001optimisation, sinha2007multi}, and reliability prediction~\cite{bichon2011efficient}. In this work, we perform Bayesian inference with it, adopting the specific polynomial response surfaces used by Bichon et al.~\cite{bichon2011efficient} and expressed as
\begin{equation}
    \label{eqn:example_2_output_quantities}
    \begin{aligned}
        L \left( \boldsymbol{\Theta} \right) = &1.16 - 0.3717 v_2 v_4 - 0.00931 v_2 v_{10} - 0.484 v_3 v_9 + 0.01343 v_6 v_{10} \\
        F \left( \boldsymbol{\Theta} \right) = &4.72 - 0.5 v_4 - 0.19 v_2 v_3 - 0.0122 v_4 v_{10} + 0.009325 v_6 v_{10} + 0.000191 v_{11}^2 \\
        D_u \left( \boldsymbol{\Theta} \right) = &28.98 + 3.818 v_3 - 4.2 v_1 v_2 + 0.0207 v_5 v_{10} + 6.63 v_6 v_9 - 7.7 v_7 v_8 + 0.32 v_9 v_{10} \\
        D_m \left( \boldsymbol{\Theta} \right) = &33.86 + 2.95 v_3 + 0.1792 v_{10} - 5.057 v_1 v_2 - 11.0 v_2 v_8 - 0.0215 v_5 v_{10} - 9.98 v_7 v_8 + 22.0 v_8 v_9 \\
        D_l \left( \boldsymbol{\Theta} \right) = &46.36 - 9.9 v_2 - 12.9 v_1 v_8 + 0.1107 v_3 v_{10} \\
        VC_u \left( \boldsymbol{\Theta} \right) = &0.261 - 0.0159 v_1 v_2 - 0.188 v_1 v_8 - 0.019 v_2 v_7 + 0.0144 v_3 v_5 + 0.0008757 v_5 v_{10} + 0.08045 v_6 v_9 \\
         &+ 0.00139 v_8 v_{11} + 0.00001575 v_{10} v_{11} \\
        VC_m \left( \boldsymbol{\Theta} \right) = &0.214 + 0.00817 v_5 - 0.131 v_1 v_8 - 0.0704 v_1 v_9 + 0.03099 v_2 v_6 - 0.018 v_2 v_7 + 0.0208 v_3 v_8 \\ 
         &+ 0.121 v_3 v_9 - 0.00364 v_5 v_6 + 0.0007715 v_5 v_{10} - 0.0005354 v_6 v_{10} + 0.00121 v_8 v_{11} \\
        VC_l \left( \boldsymbol{\Theta} \right) = &0.74 - 0.61 v_2 -0.163 v_3 v_8 + 0.001232 v_3 v_{10} - 0.166 v_7 v_9 + 0.227 v_2^2 \\
        V_B \left( \boldsymbol{\Theta} \right) = &10.58 - 0.674 v_1 v_2 - 1.95 v_2 v_8 + 0.02054 v_3 v_{10} - 0.0198 v_4 v_{10} + 0.028 v_6 v_{10} \\
        V_D \left( \boldsymbol{\Theta} \right) = &16.45 - 0.489 v_3 v_7 - 0.843 v_5 v_6 + 0.0432 v_9 v_{10} - 0.0556 v_9 v_{11} - 0.000786 v_{11}^2
    \end{aligned}
\end{equation}
where $ v_j $, $ j = 1, \dots, d_p = 11 $ denote the design variables. These are then normalized to produce the parameter vector $ \boldsymbol{\Theta} = \begin{bmatrix} \theta_1 & \dots & \theta_{11} \end{bmatrix} $ using Eq.~\eqref{eqn:example_2_random_variable_transformations}, such that $ \theta_j $, $ \forall j = 1, \dots, 11 $ have the same scale and are all centered at $ 0 $. (Consequently, we notate the ten QoIs as functions of $ \boldsymbol{\Theta} $.)
\begin{equation}
    \label{eqn:example_2_random_variable_transformations}
    \begin{aligned}
        v_1 &= 0.03 \theta_1 + 0.5 \qquad \qquad &
        v_2 &= 0.03 \theta_2 + 1.310 \qquad \qquad &
        v_3 &= 0.03 \theta_3 + 0.5 \\
        v_4 &= 0.03 \theta_4 + 1.395 \qquad \qquad &
        v_5 &= 0.03 \theta_5 + 0.875 \qquad \qquad &
        v_6 &= 0.03 \theta_6 + 1.2 \\
        v_7 &= 0.03 \theta_7 + 0.4 \qquad \qquad &
        v_8 &= 0.006 \theta_8 + 0.345 \qquad \qquad &
        v_9 &= 0.006 \theta_9 + 0.192 \\
        v_{10} &= 10.0 \theta_{10} \qquad \qquad &
        v_{11} &= 10.0 \theta_{11} 
    \end{aligned}
\end{equation}

The calibration problem for this section is constructed by conducting a ``hypothetical experiment'' with $ \theta_j = -0.6745 $ ($ j = 1, \dots 11 $); when the QoIs are evaluated at this $ \boldsymbol{\Theta} $ using the response surfaces from Eq.~\eqref{eqn:example_2_output_quantities}, the ``observed'' values listed in the third column of Table~\ref{tab:example_2_qois} are obtained. We then infer the values of the design variables that produce the listed QoI outputs using the proposed framework. For ease of study, we focus on the calibration results for the parameters $ \theta_j $ instead of the original design variables $ v_j $, since the $ \theta_j $'s have the same prior and can be meaningfully visualized together. The prior distribution of $ \boldsymbol{\Theta} $ is assumed to be the $ 11 $-dimensional standard normal distribution, which results in the design variables $ v_j $ having priors that match their specified distributions from~\cite{bichon2011efficient} (reproduced for reference in Table~\ref{tab:example_2_parameters}). (As an aside, we remark that $ \theta_j = -0.6745 $ was chosen as it is the $ 25$-th percentile of a standard normal distribution.) Two cases are explored here: in Case 1, only the viscous criterion at the middle location $ VC_m $ is considered in the QoI function $ f \left( \boldsymbol{\Theta} \right) $, while Case 2 involves all ten observation quantities (suitably transformed to ensure that the QoI function has uncorrelated components). In both cases, the observation noise is assumed to have a variance equal to $ 0.001 \operatorname{\mathbb{V}ar} \left[ f \left( \boldsymbol{\Theta} \right) \right] $.

\begin{table}[t!bhp]
    \centering
    \resizebox{\columnwidth}{!}{%
    \begin{tabular}{c|c|c|c|c}
        Symbol & Name & Mean & Standard Deviation & Distribution \\
        \hline \hline
        $ v_1 $ & B-pillar inner thickness (in $ \SI{}{mm} $) & $ 0.5 $ & $ 0.03 $ & Gaussian \\
        $ v_2 $ & B-pillar reinforcement thickness (in $ \SI{}{mm} $) & $ 1.31 $ & $ 0.03 $ & Gaussian \\
        $ v_3 $ & Floor side inner thickness (in $ \SI{}{mm} $) & $ 0.5 $ & $ 0.03 $ & Gaussian \\
        $ v_4 $ & Cross members thickness (in $ \SI{}{mm} $) & $ 1.395 $ & $ 0.03 $ & Gaussian \\
        $ v_5 $ & Door beam thickness (in $ \SI{}{mm} $) & $ 0.875 $ & $ 0.03 $ & Gaussian \\
        $ v_6 $ & Door belt line reinforcement thickness (in $ \SI{}{mm} $) & $ 1.2 $ & $ 0.03 $ & Gaussian \\
        $ v_7 $ & Roof rail thickness (in $ \SI{}{mm} $) & $ 0.4 $ & $ 0.03 $ & Gaussian \\
        $ v_8 $ & B-pillar inner material property (in $ \SI{}{GPa} $) & $ 0.345 $ & $ 0.006 $ & Gaussian \\
        $ v_9 $ & Floor side inner material property (in $ \SI{}{GPa} $) & $ 0.192 $ & $ 0.006 $ & Gaussian \\
        $ v_{10} $ & Deviation of impact location from barrier height (in $ \SI{}{mm} $) & $ 0.0 $ & $ 10.0 $ & Gaussian \\
        $ v_{10} $ & Deviation of impact location from barrier hitting position (in $ \SI{}{mm} $) & $ 0.0 $ & $ 10.0 $ & Gaussian \\
        \hline
    \end{tabular}
    }
    \caption{List of design variables for the vehicle side-impact test problem, and their assumed prior distributions.}
    \label{tab:example_2_parameters}
\end{table}

\begin{table}[t!bhp]
    \centering
    \begin{tabular}{c|c|c}
        Symbol & Name & Calibration values \\
        \hline \hline
        $ L $ & Abdomen load & $ \SI{0.431}{kN} $ \\
        $ F $ & Pubic symphysis force & $ \SI{3.963}{kN} $\\
        $ D_u $ & Rib deflection at upper location & $ \SI{28.161}{mm} $ \\
        $ D_m $ & Rib deflection at middle location & $ \SI{26.342}{mm} $ \\
        $ D_l $ & Rib deflection at lower location & $ \SI{31.123}{mm} $ \\
        $ VC_u $ & Viscous criteria at upper location & $ \SI{0.227}{ms^{-1}} $ \\
        $ VC_m $ & Viscous criteria at middle location & $ \SI{0.239}{ms^{-1}} $ \\
        $ VC_l $ & Viscous criteria at lower location & $ \SI{0.288}{ms^{-1}} $ \\
        $ V_B $ & Velocity at B-pillar & $ \SI{9.200}{ms^{-1}} $ \\
        $ V_D $ & Velocity at door & $ \SI{5.491}{ms^{-1}} $ \\
        \hline
    \end{tabular}
    \caption{List of measured quantities of interest for the vehicle side-impact test problem. The symbolic notation for each quantity is itemized in the first column, and the corresponding physical quantities being measured are described in the second column. The third column lists the observed values for the hypothetical test experiment, as described in Section~\ref{section:example_2_vehicle_side_impact}.}
    \label{tab:example_2_qois}
\end{table}

\begin{figure}[t!bhp]
\centering
\begin{subfigure}{.30\textwidth}
  \centering
  \includegraphics[width=\textwidth]{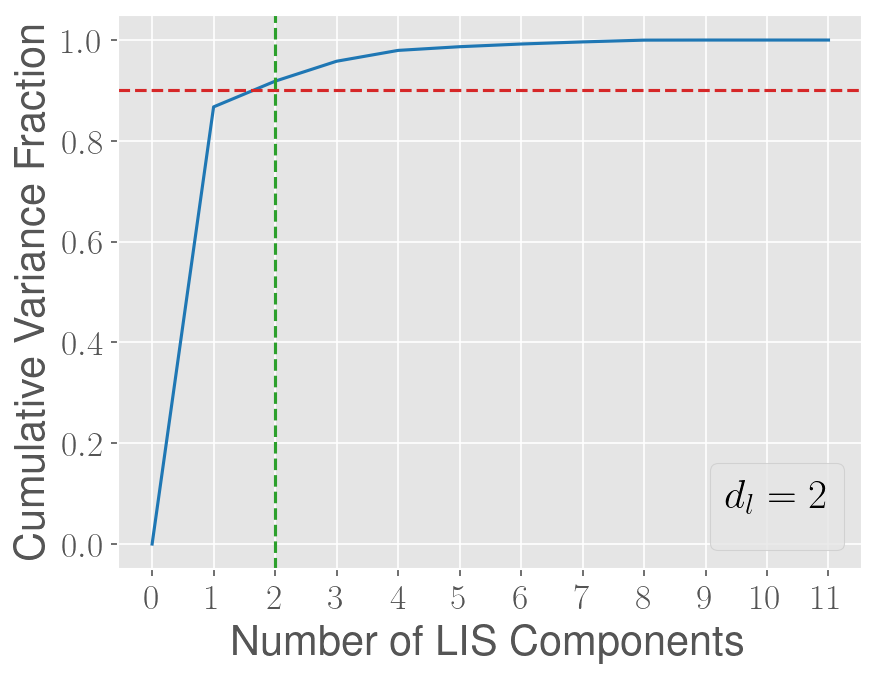}
  \caption{}
  \label{fig:example_2_case_1_cumulative_variance_gradient}
\end{subfigure}%
\begin{subfigure}{.30\textwidth}
  \centering
  \includegraphics[width=\textwidth]{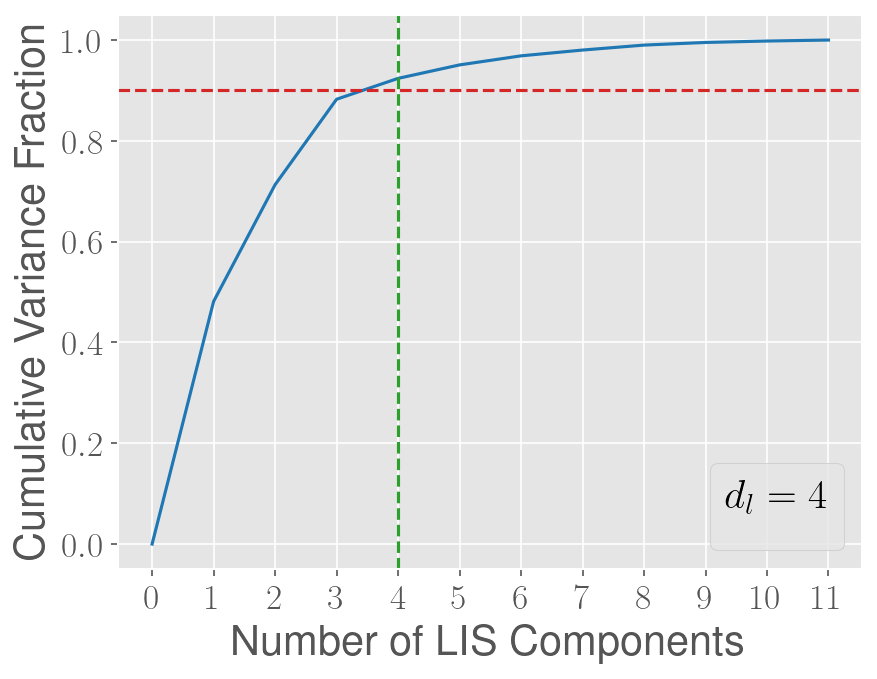}
  \caption{}
  \label{fig:example_2_case_2_cumulative_variance_gradient}
\end{subfigure}
\begin{subfigure}{.30\textwidth}
  \centering
  \includegraphics[width=\textwidth]{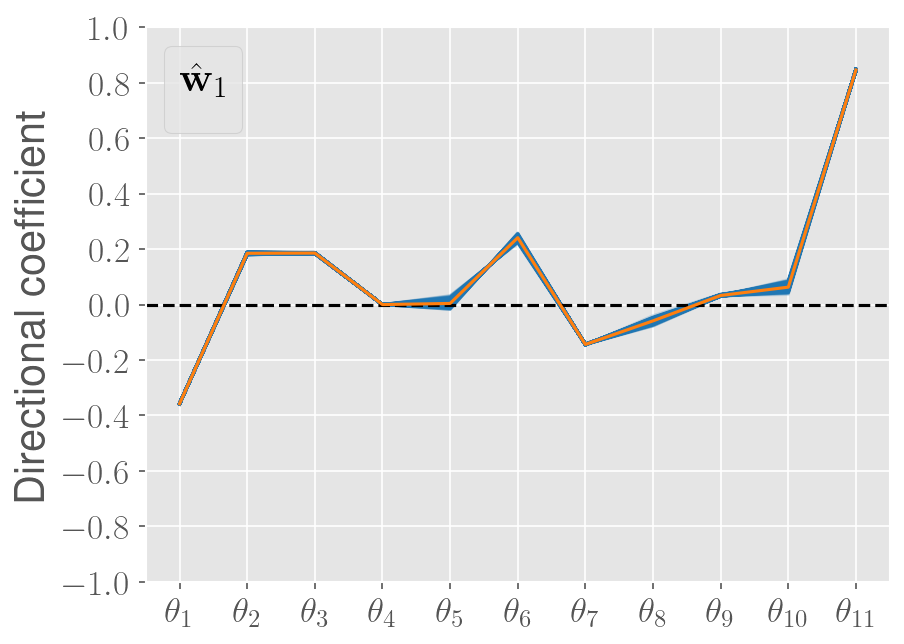}
  \caption{}
  \label{fig:example_2_case_1_projection_vector_1}
\end{subfigure}%
\begin{subfigure}{.30\textwidth}
  \centering
  \includegraphics[width=\linewidth]{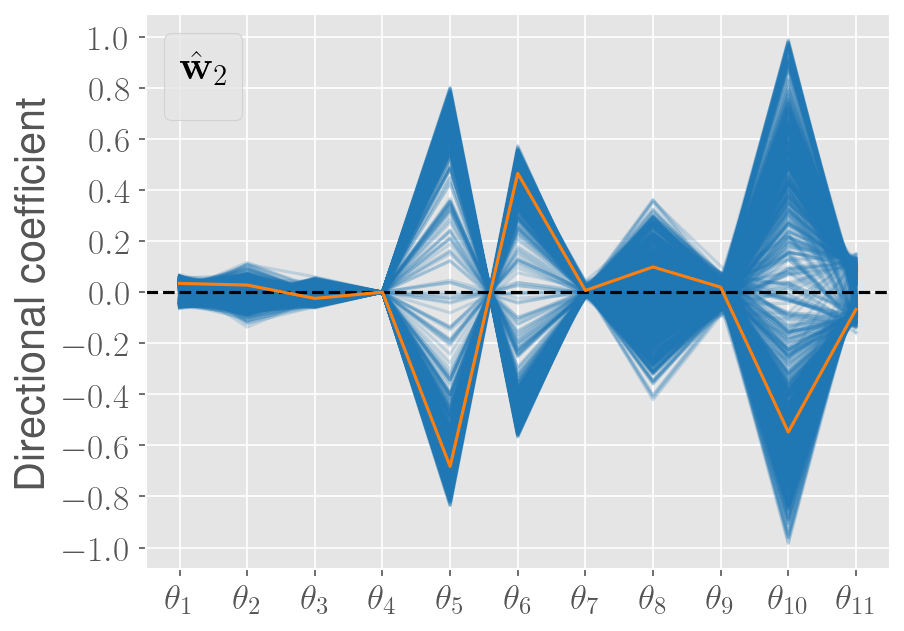}
  \caption{}
  \label{fig:example_2_case_1_projection_vector_2}
\end{subfigure}%
\begin{subfigure}{.30\textwidth}
  \centering
  \includegraphics[width=\linewidth]{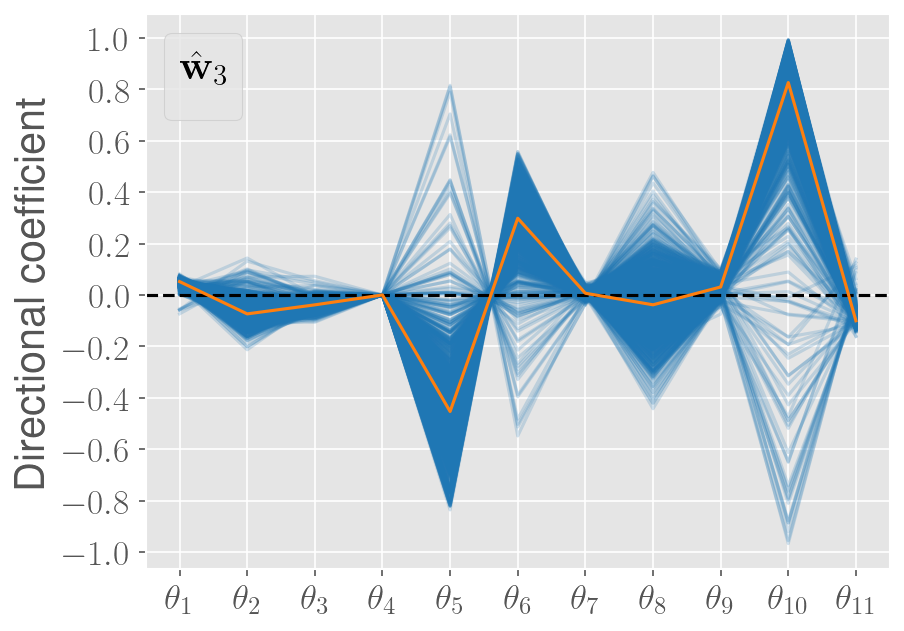}
  \caption{}
  \label{fig:example_2_case_1_projection_vector_3}
\end{subfigure}
\begin{subfigure}{.30\textwidth}
  \centering
  \includegraphics[width=\textwidth]{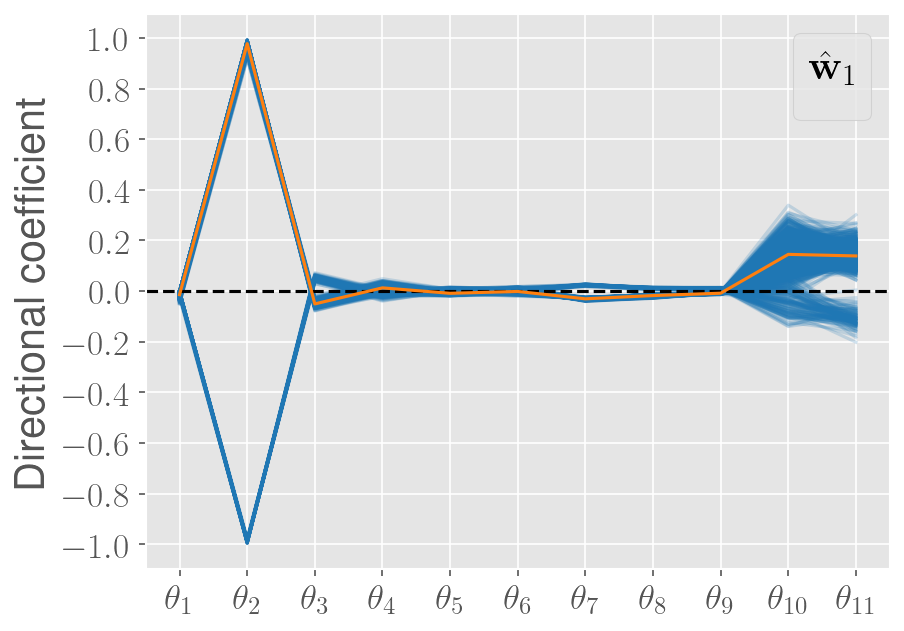}
  \caption{}
  \label{fig:example_2_case_2_projection_vector_1}
\end{subfigure}%
\begin{subfigure}{.30\textwidth}
  \centering
  \includegraphics[width=\linewidth]{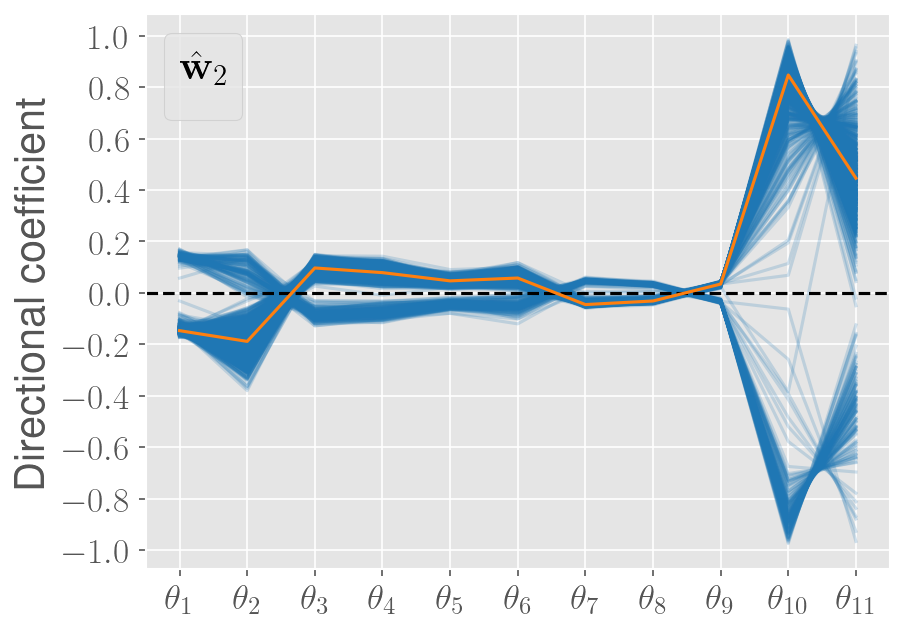}
  \caption{}
  \label{fig:example_2_case_2_projection_vector_2}
\end{subfigure}%
\begin{subfigure}{.30\textwidth}
  \centering
  \includegraphics[width=\linewidth]{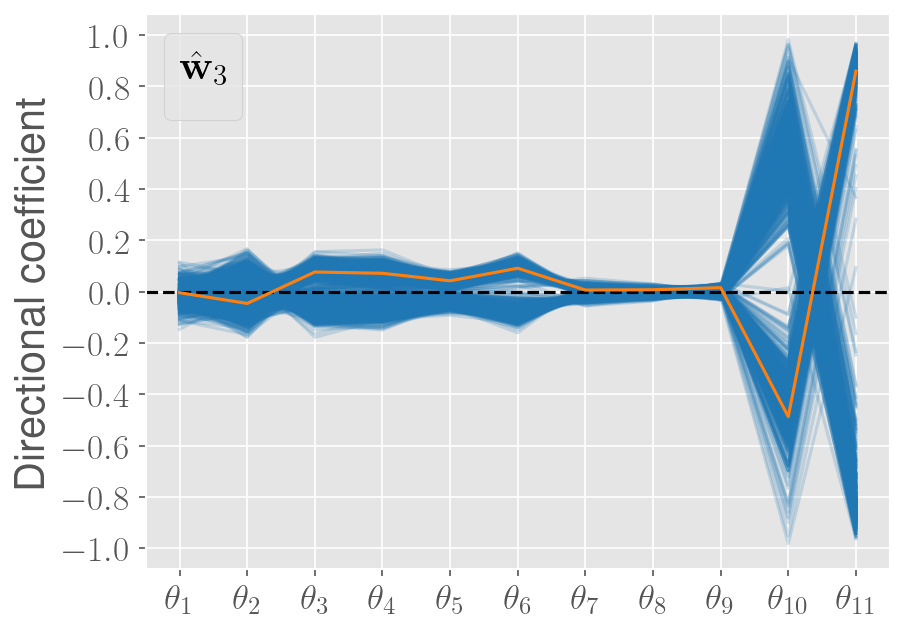}
  \caption{}
  \label{fig:example_2_case_2_projection_vector_3}
\end{subfigure}
\caption{Active subspace results for the vehicle side-impact test problem in example 2. Figures (a), and (b) plot the cumulative gradient variance explained as a function of the number of eigenvectors included for Cases 1 and 2, respectively. In these figures, the red dashed line represents the $ 90 \% $ variance fraction threshold, which is used to select $ d_l $, while the green dashed line marks the chosen value of $ d_l $. Figures (c), (d), and (e) visualize $ \hat{\mathbf{w}}_1 $, $ \hat{\mathbf{w}}_2 $, and $ \hat{\mathbf{w}}_3 $, respectively, for Case 1, with the nominal vector in orange and $ 1000 $ bootstrapped replicates in blue. Figures (f), (g), and (h) visualize $ \hat{\mathbf{w}}_1 $, $ \hat{\mathbf{w}}_2 $, and $ \hat{\mathbf{w}}_3 $ for Case 2, following the same scheme.}
\label{fig:active_subspace_results_for_example_2}
\end{figure}

\begin{figure}[t!bhp]
\centering
\begin{subfigure}{.30\textwidth}
  \centering
  \includegraphics[width=\textwidth]{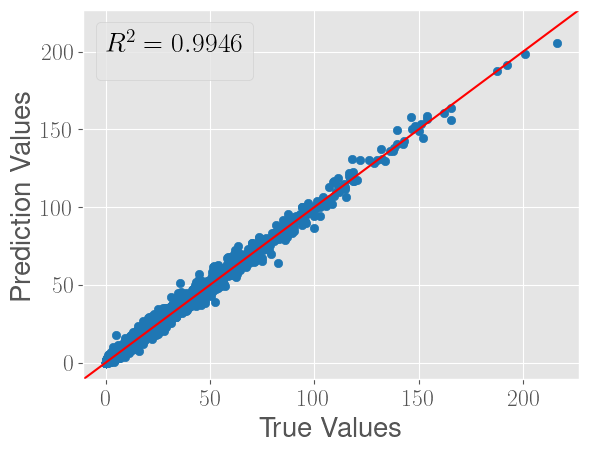}
  \caption{}
  \label{fig:example_2_case_1_surrogate_quality}
\end{subfigure}%
\begin{subfigure}{.30\textwidth}
  \centering
  \includegraphics[width=\linewidth]{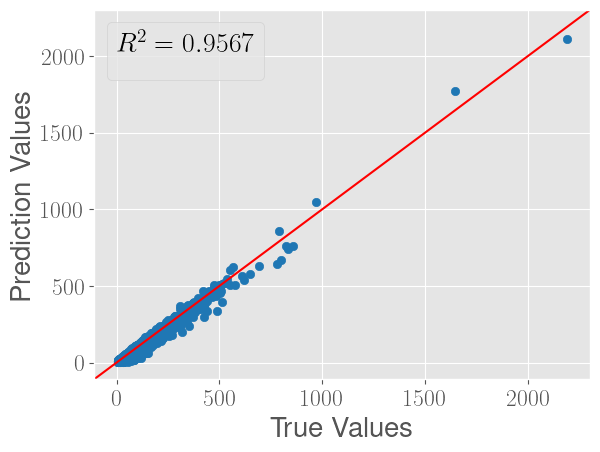}
  \caption{}
  \label{fig:example_2_case_2_surrogate_quality}
\end{subfigure}
\caption{Misfit surrogate quality results for example 2. Figure (a) plots the predicted misfit values vs. the true misfit values for Case 1 and (b) for Case 2. Each figure includes the $ R^2 $ value, and the $ x=y $ line (i.e., $ \SI{45}{\degree} $ line) highlighted in red.}
\label{fig:example_2_surrogate_quality}
\end{figure}

We once again begin by constructing the likelihood-informed subspace (LIS) for both cases, as summarized in Figure~\ref{fig:active_subspace_results_for_example_2}, obtaining a two-dimensional active subspace for Case 1 and a four-dimensional active subspace for Case 2. The first three columns of the projection vector, i.e., $ \hat{\mathbf{w}}_1 $ - $ \hat{\mathbf{w}}_3 $ are included in Figure~\ref{fig:active_subspace_results_for_example_2}, with the rest provided in Appendix~\ref{appendix:additional_figures_example_2}. Trends similar to those described in Section~\ref{section:example_1_polynomial} are observed here regarding the estimation uncertainty of these projection vectors. The misfit surrogates constructed in the latent space once again require an additional set of $ 500 $ training samples, but are highly accurate for both cases, as evidenced by Figure~\ref{fig:example_2_surrogate_quality}.

\begin{figure}[t!bhp]
\centering
\begin{subfigure}{.48\textwidth}
  \centering
  \includegraphics[width=\textwidth]{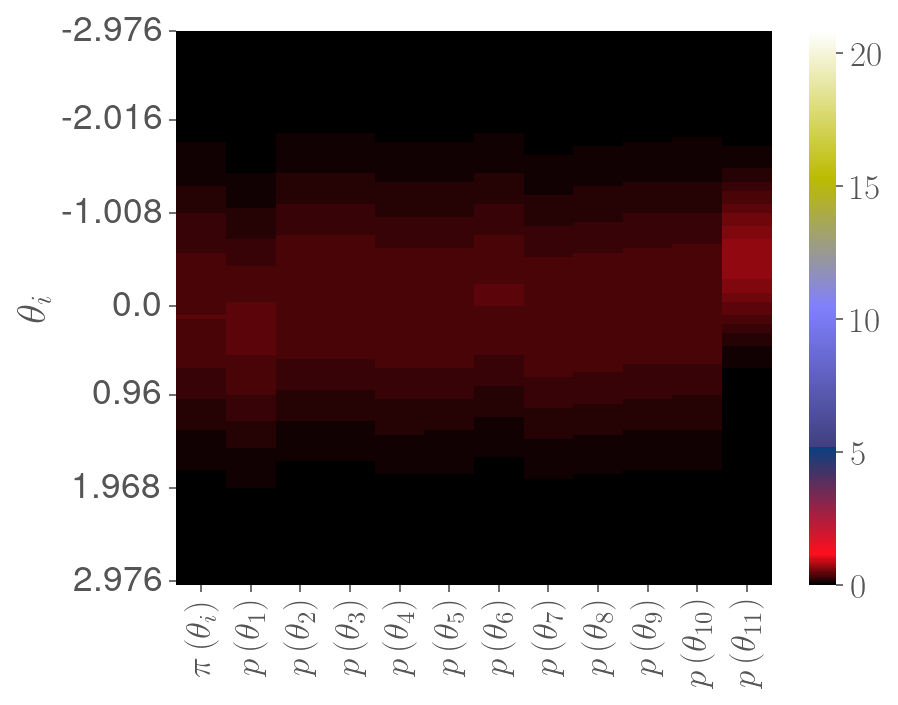}
  \caption{}
  \label{fig:example_2_case_1_full_posterior_heatmap}
\end{subfigure}%
\begin{subfigure}{.48\textwidth}
  \centering
  \includegraphics[width=\textwidth]{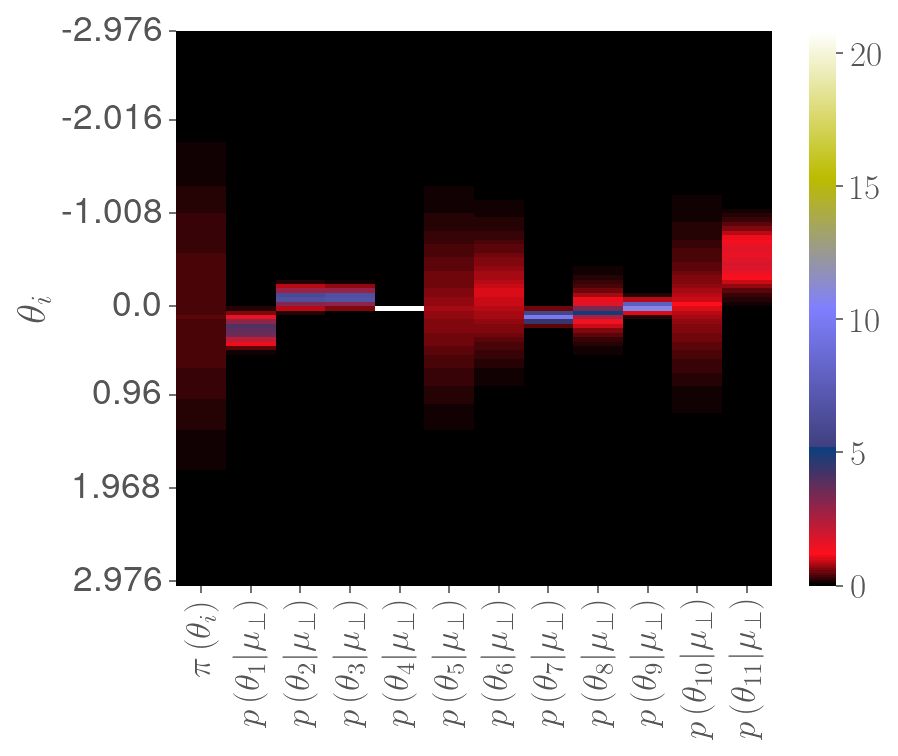}
  \caption{}
  \label{fig:example_2_case_1_conditional_posterior_heatmap}
\end{subfigure}
\begin{subfigure}{.48\textwidth}
  \centering
  \includegraphics[width=\textwidth]{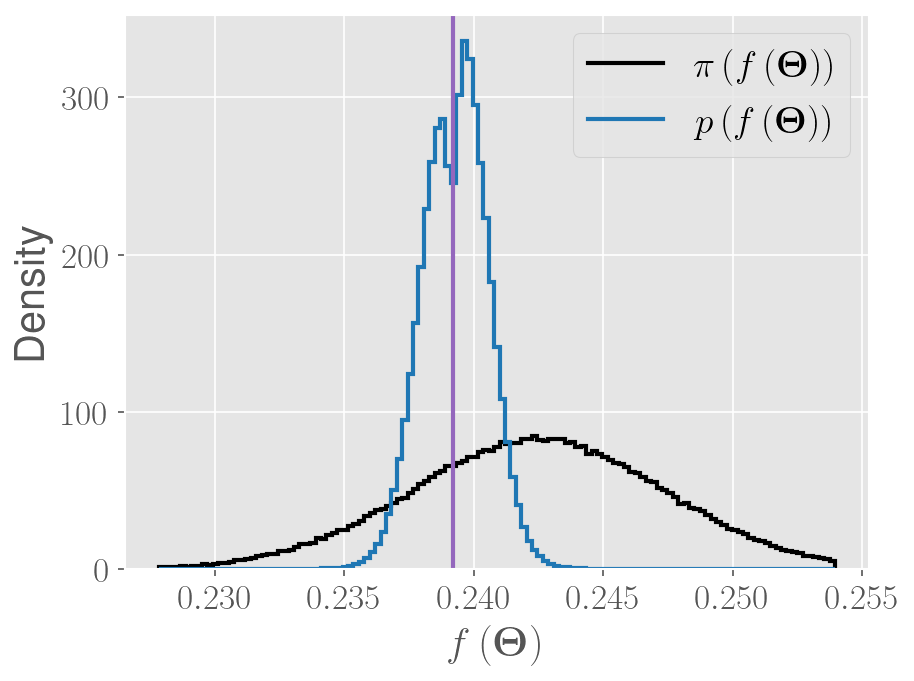}
  \caption{}
  \label{fig:example_2_case_1_full_output_posterior}
\end{subfigure}%
\begin{subfigure}{.48\textwidth}
  \centering
  \includegraphics[width=\textwidth]{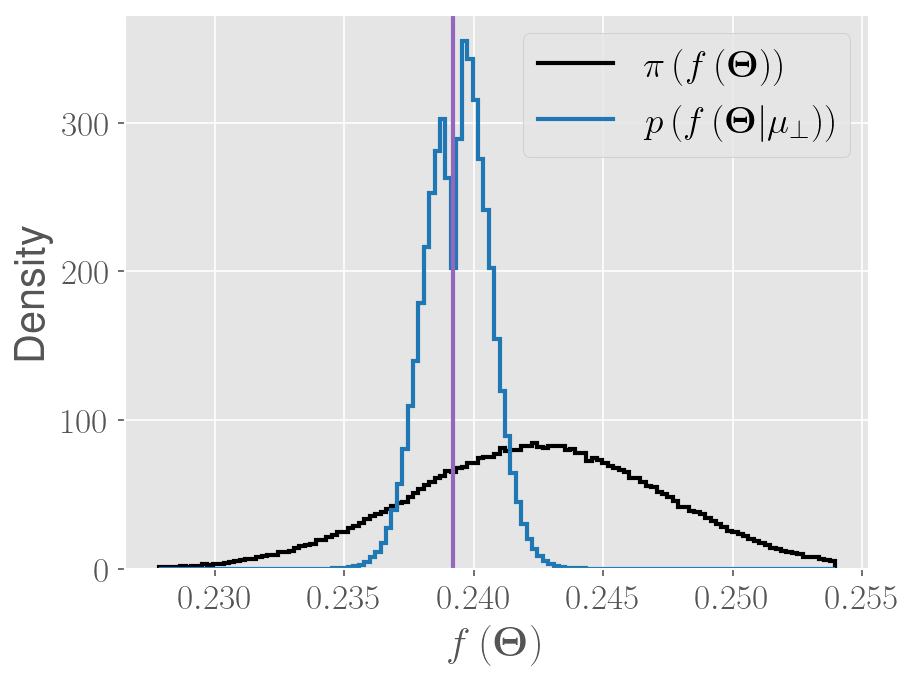}
  \caption{}
  \label{fig:example_2_case_1_conditional_output_posterior}
\end{subfigure}
\caption{Calibration results for Case 1 of example 2. Figures (a) and (b) are heatmaps comparing the full marginal posteriors and the conditional active marginal posteriors, respectively, of each $ \theta_i $ with the prior. The first column in each heatmap is the prior, which is identical for all parameter components, and the subsequent columns are the calibrated posteriors of the parameters in order. Figures (c) and (d) plot the probability density of output for samples from the full posterior and the conditional active posterior, respectively, in blue. Also in (c) and (d), the probability density of the output corresponding to samples from $ \pi \left( \boldsymbol{\Theta} \right) $ is drawn in black, and the observed $ f_{\boldsymbol{\Theta}}^* $ is in purple.}
\label{fig:example_2_case_1_calibration_results}
\end{figure}

Inspecting the calibration results for Case 1 illustrated in Figure~\ref{fig:example_2_case_1_calibration_results}, we notice that only the posterior of $ \theta_{11} $ is centered around the chosen ``true'' value of $ -0.6745 $. However, this is to be expected, as the QoI function value at $ \theta_j = -0.6745 $, $ j = 1, \dots, 11 $ is not unique. There are a large number of points in the parameter space that produce the same value of $ f \left( \boldsymbol{\Theta} \right) = VC_m \left( \boldsymbol{\Theta} \right) $. The likelihood has peaks of similar height at all such coordinates, so the point with the highest prior probability density is chosen as the mode of the calibrated posterior. And indeed, we see in Figures~\ref{fig:example_2_case_1_full_output_posterior} and~\ref{fig:example_2_case_1_conditional_output_posterior} that the probability density of the QoI function under the calibrated posteriors is concentrated around the observed data, implying that samples from the calibrated posterior result in QoI values in the vicinity of the observation. Further, the dimension reduction increases this smearing effect, since all points in the original parameter space that share the same low-dimensional representation will have the same predicted likelihood under our framework, even if they have different QoI values. $ \theta_{11} $, which dominates the behavior of the likelihood by virtue of having by far the highest contribution in the most important (by a large margin) projection vector, is thus the only parameter that shows a strong preference for the chosen ``true'' value. In terms of posterior spread, we once again see that the conditional active posteriors infer parameter values with much higher confidence without a meaningful difference in the posterior probability density of the QoI function.

\begin{figure}[t!bhp]
\centering
\begin{subfigure}{.48\textwidth}
  \centering
  \includegraphics[width=\textwidth]{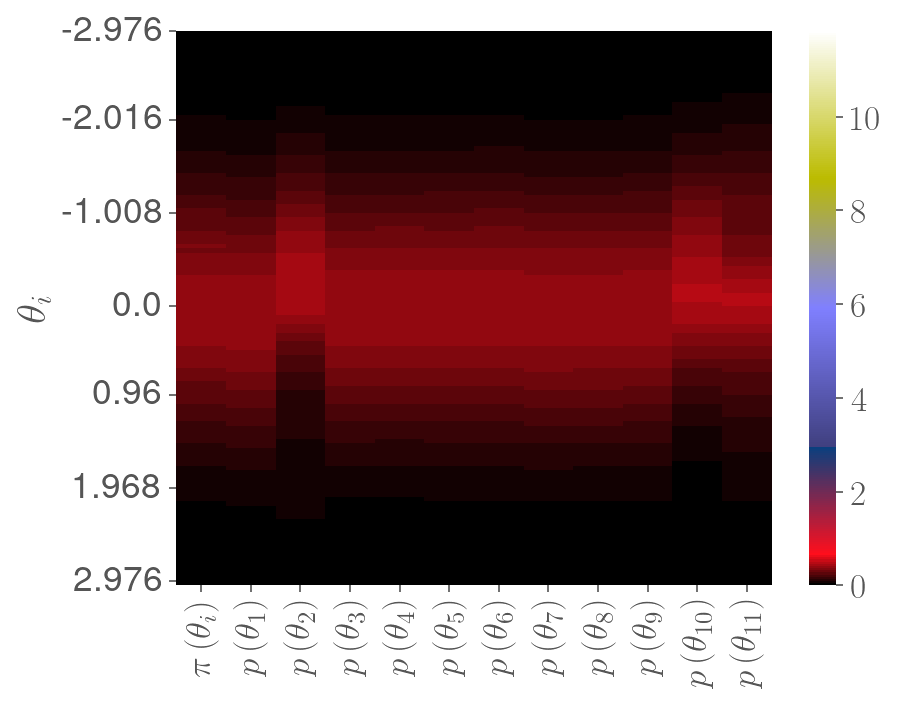}
  \caption{}
  \label{fig:example_2_case_2_full_posterior_heatmap}
\end{subfigure}%
\begin{subfigure}{.48\textwidth}
  \centering
  \includegraphics[width=\textwidth]{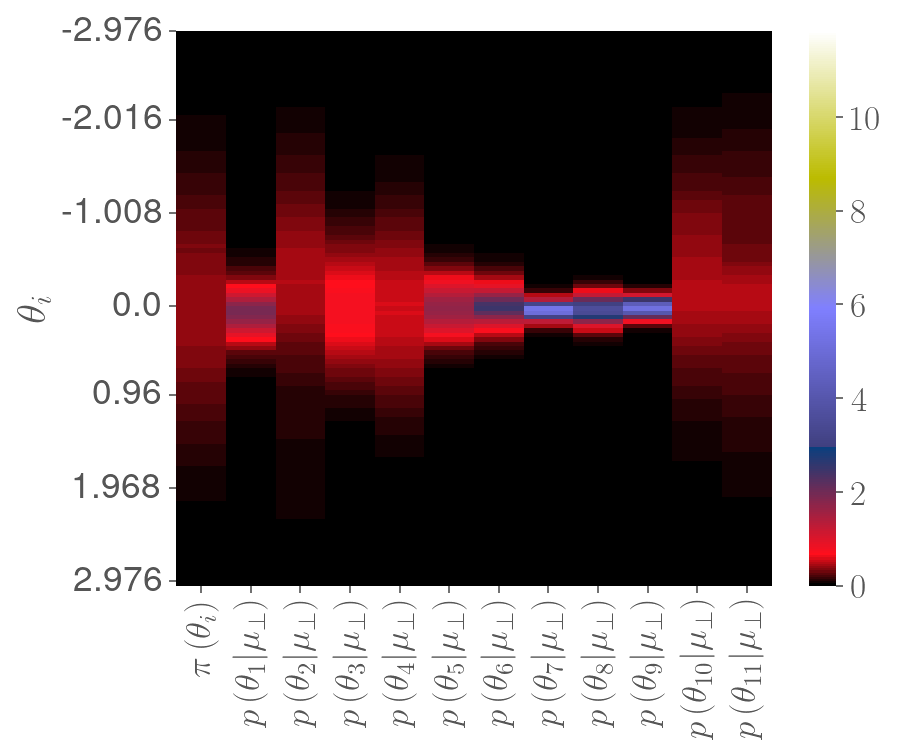}
  \caption{}
  \label{fig:example_2_case_2_conditional_posterior_heatmap}
\end{subfigure}
\begin{subfigure}{.48\textwidth}
  \centering
  \includegraphics[width=\textwidth]{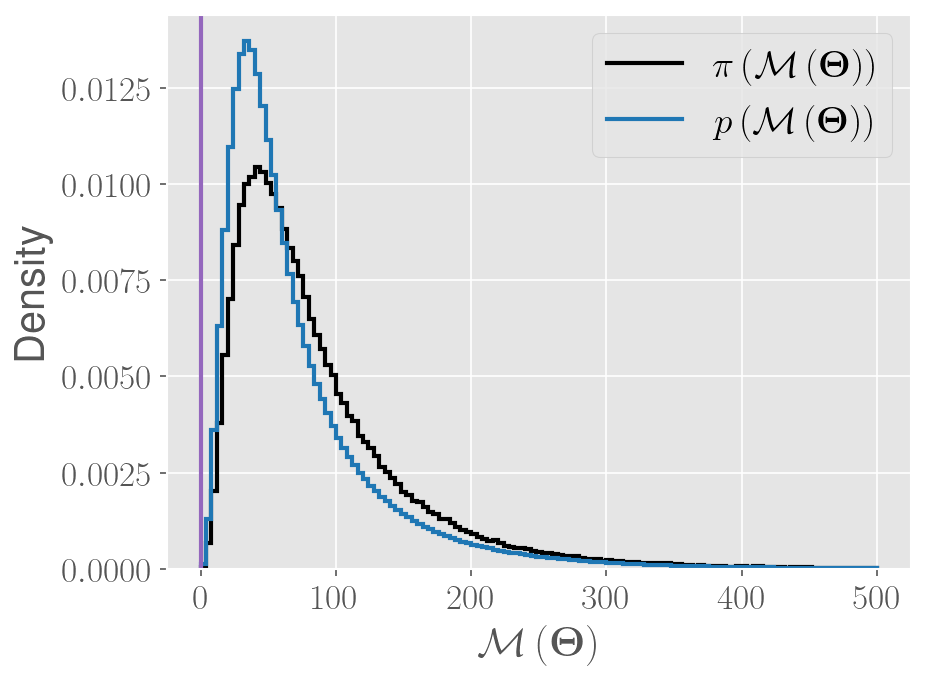}
  \caption{}
  \label{fig:example_2_case_2_full_misfit_posterior}
\end{subfigure}%
\begin{subfigure}{.48\textwidth}
  \centering
  \includegraphics[width=\textwidth]{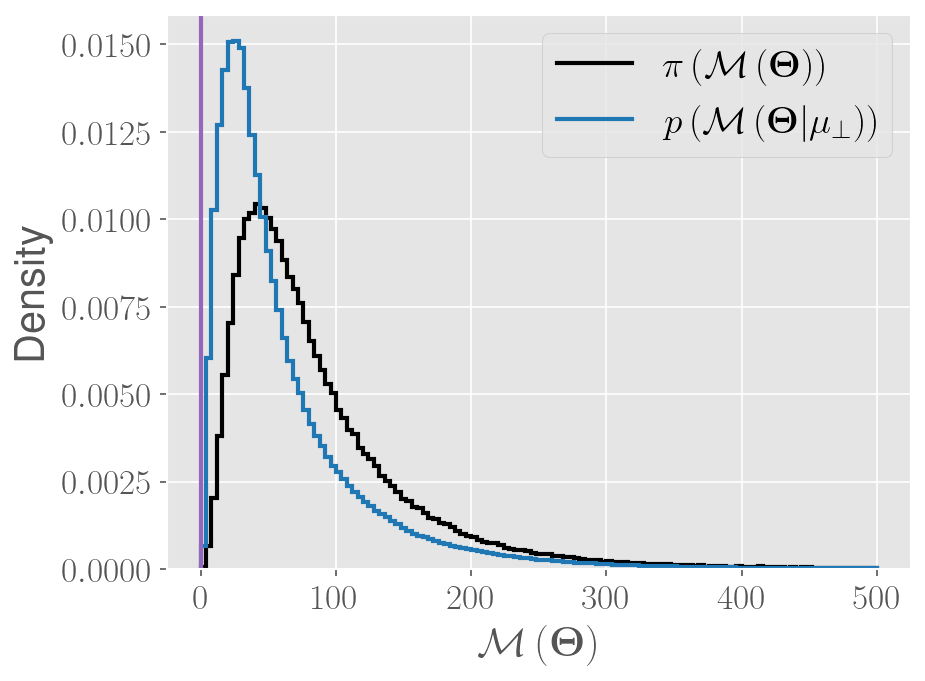}
  \caption{}
  \label{fig:example_2_case_2_conditional_misfit_posterior}
\end{subfigure}
\caption{Calibration results for Case 2 of example 2. Figures (a) and (b) are heatmaps comparing the full marginal posteriors and the conditional active marginal posteriors, respectively, of each $ \theta_i $ with the prior. The first column in each heatmap is the prior, which is identical for all parameter components, and the subsequent columns are the calibrated posteriors of the parameters in order. Figures (c) and (d) plot the probability density of the misfit function for samples from the full posterior and the conditional active posterior, respectively, in blue. Also in (c) and (d), the probability density of the output corresponding to samples from $ \pi \left( \boldsymbol{\Theta} \right) $ is drawn in black, and $ \mathcal{M} \left( \boldsymbol{\Theta} \right) = 0 $ (when the QoIs match the observed outputs) is in purple.}
\label{fig:example_2_case_2_calibration_results}
\end{figure}

For Case 2, the calibration results are more nuanced (Figure~\ref{fig:example_2_case_2_calibration_results}). Here, the QoI function is ten-dimensional, with the individual quantities of interest pulling in different directions. Parameter values that make one QoI move close to the observation may cause others to deviate markedly from the observation. The misfit then averages out these discrepancies by considering the Euclidean norm of the ten-dimensional QoI vector. Furthermore, due to the information loss caused by the low-dimensional projection, it is possible for two points in the parameter space to have different QoI function - hence, misfit function - values while having the same low-dimensional representation. 
Consequently, only a tiny region of the parameter space corresponds to all ten response quantities having minimal deviations from the observation, which is necessary to have a near-zero misfit value, and this region is further reduced due to the averaging effect of the LIS-plus-surrogate approximation. The distribution of the misfit also necessarily narrows towards a non-zero mean value with increasing dimensionality of the QoI vector as a consequence of the Law of Large Numbers. In combination, these two effects make it exceedingly difficult for the calibration procedure to reduce the spread of the misfit's posterior probability density relative to the prior as $ d_{\text{out}} $ increases. Figures~\ref{fig:example_2_case_2_full_misfit_posterior} and~\ref{fig:example_2_case_2_conditional_misfit_posterior} highlight this issue clearly; there is only a small difference between the prior and posterior misfit distributions. (However, an inspection of the marginal posteriors of the QoI components provided in Appendix~\ref{appendix:additional_figures_example_2} reveals a clear change in the QoI function posterior as compared to the prior, verifying that some calibration has occurred.)

The above discussion underscores the importance of optimizing the construction of the QoI function based on a careful selection of measurable system outputs that are the most informative for the problem of interest. Although the proposed framework can handle vector-valued QoI functions, the construction of the misfit function itself (which is not unique to our framework) imposes practical limitations on the quality of calibration that can be achieved when the QoI function is multi-dimensional. The more quantities of interest that need to be balanced, the more difficult it becomes for the likelihood function to appropriately incorporate all the often conflicting information. In practice, it is best to select a small number of important observable quantities, or implement some other dimension reduction method on the QoI function itself, to improve the quality of Bayesian inference that can be achieved with the proposed (or any) calibration algorithm. Works by Guo et al.~\cite{guo2024active} and others provide a roadmap for how high-dimensional output functions can be reduced to a small number of highly informative features. Aside from the complexities introduced by the vector-valued nature of the QoI function, the rest of the observations regarding the calibrated parameter posterior distributions remain as before. 

\section{Conclusion}
\label{section:conclusion}

In this manuscript, we proposed a robust and efficient framework for the Bayesian calibration of many-parameter system models that incorporates dimension reduction through active subspaces and couples it with surrogate modeling. Following prior literature on \textit{Likelihood-informed Subspaces}~\cite{cui2014likelihood, constantine2014active}, we construct a low-dimensional representation of the parameter space focused on the misfit function and build our surrogate to predict the misfit. Special emphasis is placed on ensuring compatibility between the mathematical assumptions governing the active subspace dimension reduction and those underlying the surrogate model construction. In this pursuit, an appropriate form for the misfit surrogate is derived (a Gamma random field when the noise in the observed data is assumed Gaussian as per Kennedy \& O'Hagan \cite{kennedy2001bayesian}), and a generalized likelihood is formulated which can account for the surrogate prediction error when quantities other than QoIs are approximated by the surrogate (e.g., the misfit value). The algorithm also accounts for the uncertainty introduced by the active subspace dimension reduction procedure and includes an explicit method for quantifying this uncertainty and incorporating it into the calibrated posteriors. Finally, we discuss the practical merits of the \textit{conditional active posterior density}, which ignores the prior uncertainty of the inactive parts of the parameter vector, resulting in a significantly more confident inference of the parameters compared to the full posterior for a negligible change in the posterior uncertainty of the QoI and misfit functions. The framework is shown to perform well for a closed-form polynomial function in $ 10 $ and $ 100 $ dimensions, and in a $ 11 $-dimensional vehicle side-impact case study.

A key assumption in our work is that the likelihood-informed subspace is sufficiently low-dimensional for the misfit surrogate to be successfully trained and MCMC to be effective. Provided an efficient low-dimensional representation exists, the proposed method can handle a large number of parameters simultaneously. The primary limitations of the framework, therefore, stem from the construction of the likelihood-informed subspace, many of which are discussed in~\cite{constantine2016accelerating}; for example, functions with high localized variability or large gradients with small oscillatory behavior can lead to inappropriate directions being selected as the active subspace. Another significant limitation involves the dimensionality of the QoI vector, as highlighted in Section~\ref{section:example_2_vehicle_side_impact}. As the number of output quantities increases, the misfit function becomes less able to home in on the true parameter values by balancing deviations from the observed data across all QoI components. Hence, for some practical problems (e.g., engineering systems where the output is a time series or an image matrix, etc.), a dimension reduction step may be needed for the QoI function as well, to select a small number of features that capture the relevant details of the output.

Although this manuscript details a specific algorithm for Bayesian calibration, the discussions throughout can also be used as guidance for constructing alternate frameworks. Other dimension reduction methods for function inputs/parameters exist that operate analogously to active subspaces, such as inverse regression-based methods, which can be employed to construct the LIS instead of active subspaces. A variety of surrogate modeling tools can be used as well, as long as prediction variance is provided along with the mean prediction. An analyst can construct the misfit function based on alternative formulations of the likelihood, rather than the Gaussian likelihood present in this manuscript. Or functions other than the misfit can be approximated using the surrogate model. Nearly every piece of the proposed algorithm can be exchanged for other preferred tools in a plug-and-play manner, as long as the skeleton of the framework is respected, as follows: (1) The dimension reduction and surrogate modeling are applied on the same function (misfit/QoI/etc.); (2) the form of the surrogate is derived from the assumed form of the observation noise and QoI uncertainty, if any; (3) the original (i.e., model-vs.-observation) likelihood (e.g., Eq.~\eqref{eqn:misfit_to_likelihood}) can be determined uniquely and deterministically from the quantity predicted by the surrogate; (4) the generalized likelihood (Eq.~\eqref{eqn:generalized_likelihood}) is used to incorporate the surrogate prediction uncertainty; and (5) the projection matrix is either obtained without uncertainty, or its uncertainty can be averaged over when constructing the posterior distributions for the parameters. As long as the above steps are followed, the mathematical derivation of the proposed framework will be applicable even if alternate tools are applied. Future work should focus on incorporating more robust dimension reduction for both the output vector and the parameter space to alleviate the limitations listed above, while following the aforementioned skeleton. This will allow for the creation of even more robust and comprehensive algorithms to calibrate many parameter models for complex engineering systems that cannot be satisfactorily handled by the current iteration of the framework as presented in this manuscript.

\section{Acknowledgement}

The authors acknowledge financial support from the National Aeronautics and Space Administration-Space Technology Research Institute (NASA STRI project 35, Award No. 80NSSC23K1342. Technical monitor: Timothy Poe).

\appendix

\section{Deriving the Kennedy-O'Hagan framework from the generalized likelihood formulation}
\label{appendix:deriving_KOH_from_generalized_likelihood}

In Section~\ref{section:generalized_likelihood}, we derived the generalized likelihood (Eq.~\eqref{eqn:generalized_likelihood}) that allows nearly any function of the system parameters to be approximated by a surrogate and subsequently used within Bayes' theorem. Here, we show that the form of the likelihood in the original Kennedy-O'Hagan calibration framework - where a Gaussian process is used to capture the behavior of the QoI function - can be seen as a special case of the generalized likelihood formulation.

The specific assumptions of the Kennedy-O'Hagan framework are listed in Section~\ref{section:notation}. We additionally assume $ \kappa = 1 $, which is a common addition adopted in the literature. We also note that there is no provision for a separate surrogate modeling error term. Then, replacing the computer model with a Gaussian process surrogate as per~\cite{kennedy2001bayesian}, we can write
\begin{gather}
    f_{\boldsymbol{\Theta}}^* = \mathcal{G}_{\text{KOH}} \left( \boldsymbol{\Theta} \right) \label{eqn:gaussian_QoI_model_surrogate_KOH} \\
    \text{where } \mathcal{G}_{\text{KOH}} \left( \boldsymbol{\Theta} \right) \sim \mathcal{N} \left( \boldsymbol{\mu}_{\text{KOH}} \left( \mathbf{X}^*, \boldsymbol{\Theta} \right), \boldsymbol{\Sigma}_{\text{KOH}} \left( \mathbf{X}^*, \boldsymbol{\Theta} \right) \right) \\
    \Rightarrow \rho \left( f_{\boldsymbol{\Theta}}^* | \boldsymbol{\Theta} \right) = \phi_{\text{KOH}} \left( f_{\boldsymbol{\Theta}}^* - \boldsymbol{\mu}_{\text{KOH}} \left( \mathbf{X}^*, \boldsymbol{\Theta} \right) \right) \label{eqn:KOH_density_f_given_surrogate}
\end{gather}
where $ \phi_{\text{KOH}} $ is the density function of the Gaussian distribution of $ \mathcal{G}_{\text{KOH}} $. From Eqs.~\eqref{eqn:observation_to_model} and~\eqref{eqn:general_model_distribution_assumptions_KOH_framework}, we get
\begin{gather}
    \mathbf{Y}^* = f_{\boldsymbol{\Theta}}^* + \eta \left( \mathbf{X}^* \right) + \varepsilon_{\mathbf{Y}} \\
    \text{where } \eta \left( \mathbf{X}^* \right) \sim \mathcal{N} \left( \boldsymbol{\mu}_\eta \left( \mathbf{X}^*, \boldsymbol{\Theta} \right), \boldsymbol{\Sigma}_\eta \left( \mathbf{X}^*, \boldsymbol{\Theta} \right) \right) \; \text{, and } \; \varepsilon_{\mathbf{Y}} \sim \mathcal{N} \left( \boldsymbol{0}, \boldsymbol{\Sigma}_\varepsilon \left( \boldsymbol{\Theta} \right) \right) \\
    \Rightarrow \rho \left( \mathbf{Y}^* | f_{\boldsymbol{\Theta}}^* \right) = \phi_\mathbf{Y} \left( \mathbf{Y}^* - \boldsymbol{\mu}_\eta \left( \mathbf{X}^*, \boldsymbol{\Theta} \right) - f_{\boldsymbol{\Theta}}^* \right) \label{eqn:KOH_density_Y_given_f}
\end{gather}
where $ \phi_{\mathbf{Y}} \left( \mathbf{Y} - \eta \left( \mathbf{X}^* \right) - f_{\boldsymbol{\Theta}}^* \right) $ is the density function of the distribution $ \mathcal{N} \left( \boldsymbol{\mu}_\eta \left( \mathbf{X}^*, \boldsymbol{\Theta} \right) + f_{\boldsymbol{\Theta}}^* , \boldsymbol{\Sigma}_\mathbf{Y} \left( \mathbf{X}^*, \boldsymbol{\Theta} \right) \right) $, with $ \boldsymbol{\Sigma}_\mathbf{Y} \left( \mathbf{X}^*, \boldsymbol{\Theta} \right) = \boldsymbol{\Sigma}_\eta \left( \mathbf{X}^*, \boldsymbol{\Theta} \right) + \boldsymbol{\Sigma}_\varepsilon \left( \boldsymbol{\Theta} \right) $.

Substituting Eqs.~\eqref{eqn:KOH_density_Y_given_f} and~\eqref{eqn:KOH_density_f_given_surrogate} into Eq.~\eqref{eqn:generalized_likelihood}, we get
\begin{align}
    \mathcal{L} \left( \boldsymbol{\Theta} \right) &= \int_{f_{\boldsymbol{\Theta}}^*} \rho \left( \mathbf{Y}^* | f_{\boldsymbol{\Theta}}^* \right) \cdot \rho \left( f_{\boldsymbol{\Theta}}^* | \boldsymbol{\Theta} \right) d f_{\boldsymbol{\Theta}}^* \\
    &= \int_{f_{\boldsymbol{\Theta}}^*} \phi_\mathbf{Y} \left( \mathbf{Y}^* - \boldsymbol{\mu}_\eta \left( \mathbf{X}^*, \boldsymbol{\Theta} \right) - f_{\boldsymbol{\Theta}}^* \right) \cdot \phi_{\text{KOH}} \left( f_{\boldsymbol{\Theta}}^* - \boldsymbol{\mu}_{\text{KOH}} \left( \mathbf{X}^*, \boldsymbol{\Theta} \right) \right) d f_{\boldsymbol{\Theta}}^* \label{eqn:KOH_likelihood_convolution_form}
\end{align}
The above equation is clearly a convolution of Gaussian density functions, which is well known to be equivalent to the density of the sum of the Gaussian distributions involved in the convolution. Thus,
\begin{align}
    \mathcal{L} \left( \boldsymbol{\Theta} \right) &= \phi_{\left( \mathbf{Y} + \text{KOH} \right)} \left( \mathbf{Y}^* - \boldsymbol{\mu}_\eta \left( \mathbf{X}^*, \boldsymbol{\Theta} \right) - \boldsymbol{\mu}_{\text{KOH}} \left( \mathbf{X}^*, \boldsymbol{\Theta} \right) \right) \label{eqn:KOH_likelihood_gaussian_density_form}
\end{align}

Usually, the likelihood for the Kennedy-O'Hagan framework with the Gaussian assumptions is reached by using the additive property of Gaussians after substituting Eq.~\eqref{eqn:gaussian_QoI_model_surrogate_KOH} into Eq.~\eqref{eqn:observation_to_model}, i.e.,
\begin{gather}
    \mathbf{Y}^* = \mathcal{G}_{\text{KOH}} + \eta \left( \mathbf{X}^* \right) + \varepsilon_{\mathbf{Y}} \\
    \Rightarrow \mathcal{L} \left( \boldsymbol{\Theta} \right) = \phi_{\left( \mathbf{Y} + \text{KOH} \right)} \left( \mathbf{Y}^* - \boldsymbol{\mu}_{\text{KOH}} \left( \mathbf{X}^*, \boldsymbol{\Theta} \right) - \boldsymbol{\mu}_\eta \left( \mathbf{X}^*, \boldsymbol{\Theta} \right) \right) 
\end{gather}
which is the same expression as the one derived using the generalized likelihood.

\section{Additional Figures}
\label{appendix:additional_figures}

In this section, we include additional visualizations of the active subspace eigenvectors and calibrated posteriors for the test problems discussed in Section~\ref{section:numerical_examples}.

\subsection{Supplementary Figures for the Polynomial Test Function}
\label{appendix:additional_figures_example_1}

\begin{figure}[t!bhp]
\centering
\begin{subfigure}{.24\textwidth}
  \centering
  \includegraphics[width=\textwidth]{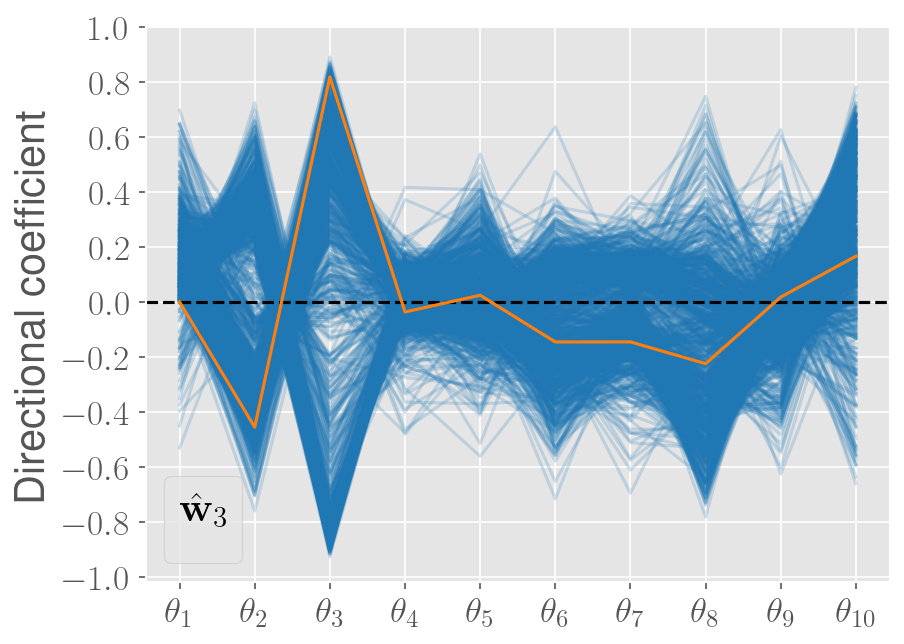}
  \caption{}
  \label{fig:example_1_case_1_projection_vector_dim_3}
\end{subfigure}%
\begin{subfigure}{.24\textwidth}
  \centering
  \includegraphics[width=\textwidth]{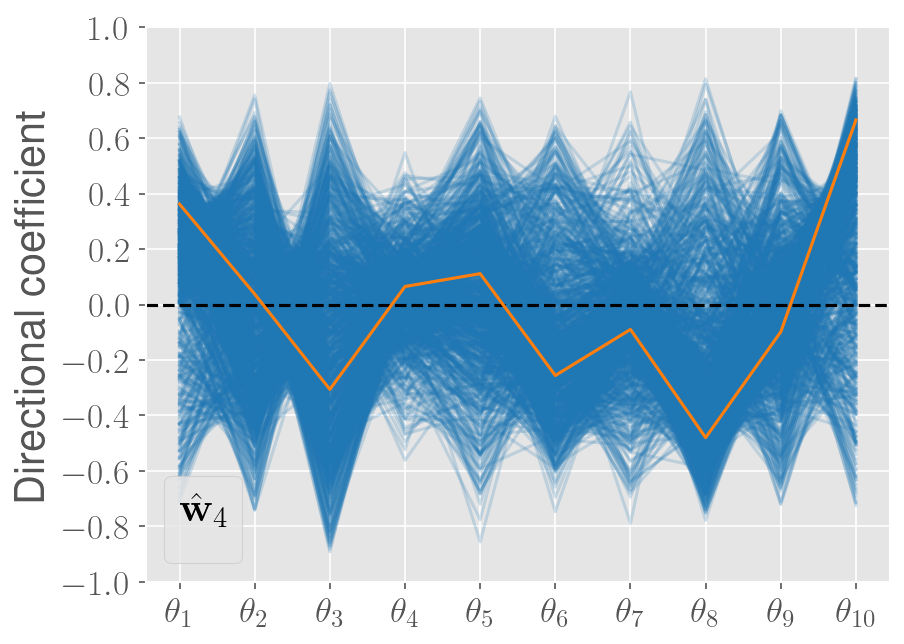}
  \caption{}
  \label{fig:example_1_case_1_projection_vector_dim_4}
\end{subfigure}%
\begin{subfigure}{.24\textwidth}
  \centering
  \includegraphics[width=\textwidth]{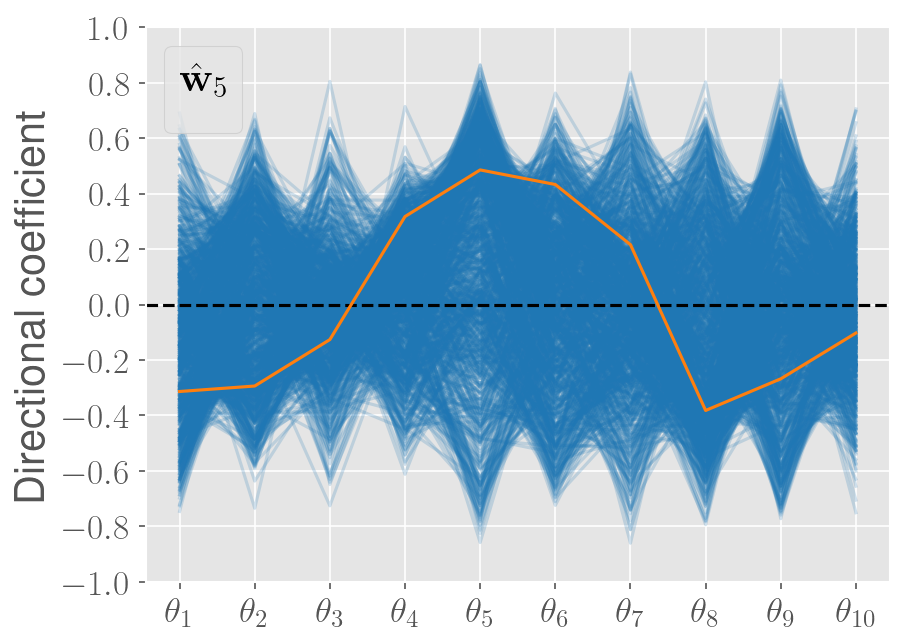}
  \caption{}
  \label{fig:example_1_case_1_projection_vector_dim_5}
\end{subfigure}%
\begin{subfigure}{.24\textwidth}
  \centering
  \includegraphics[width=\textwidth]{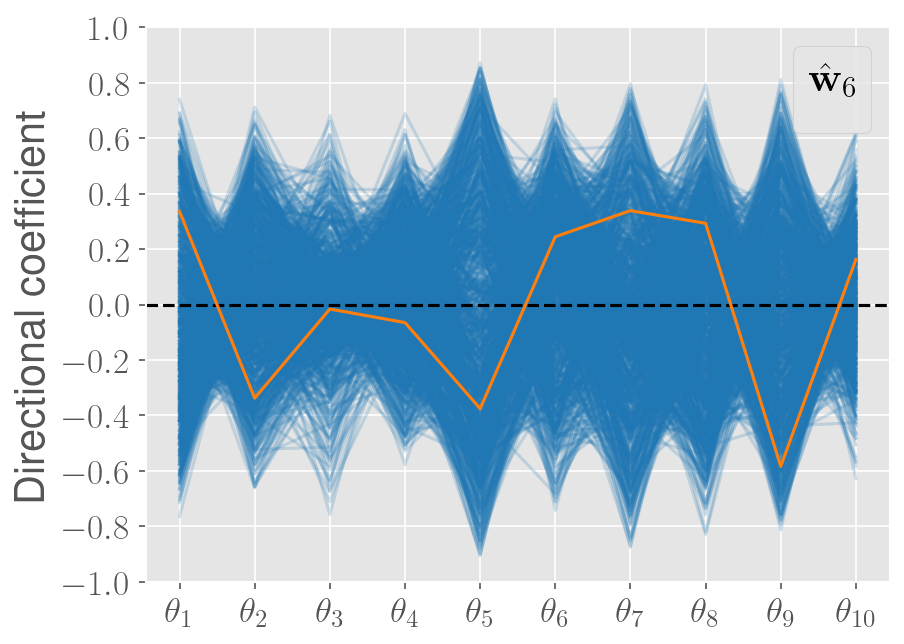}
  \caption{}
  \label{fig:example_1_case_1_projection_vector_dim_6}
\end{subfigure}
\begin{subfigure}{.24\textwidth}
  \centering
  \includegraphics[width=\textwidth]{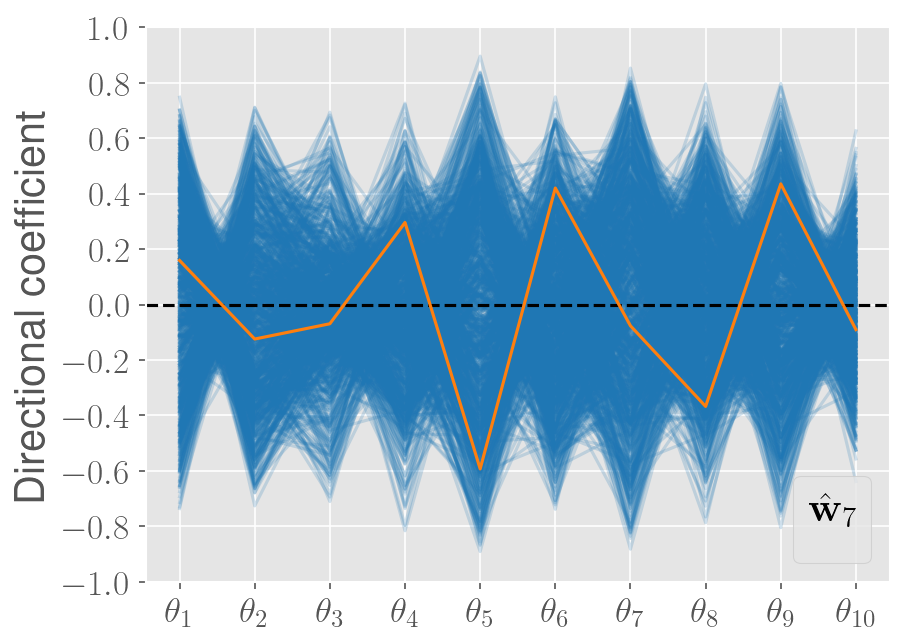}
  \caption{}
  \label{fig:example_1_case_1_projection_vector_dim_7}
\end{subfigure}%
\begin{subfigure}{.24\textwidth}
  \centering
  \includegraphics[width=\textwidth]{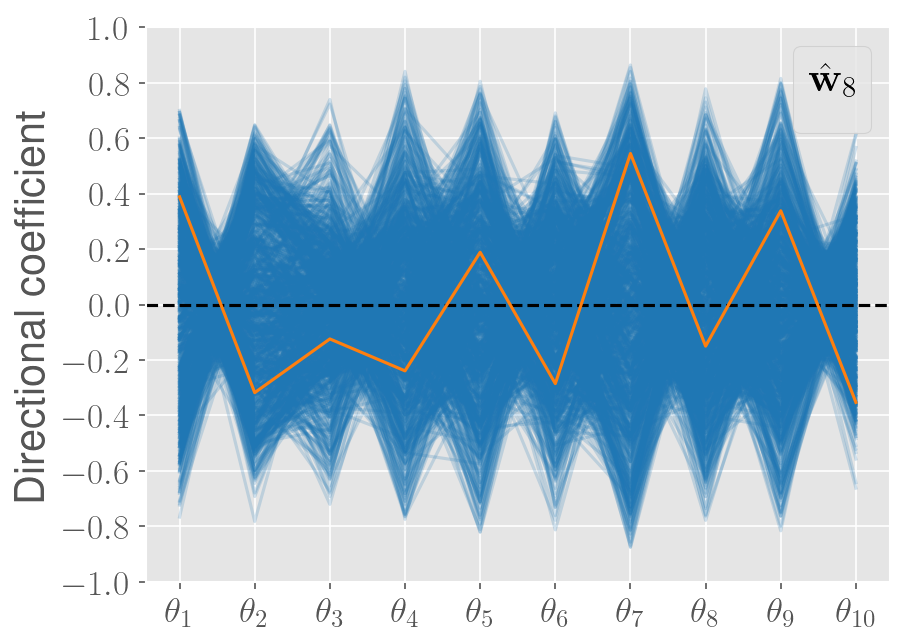}
  \caption{}
  \label{fig:example_1_case_1_projection_vector_dim_8}
\end{subfigure}%
\begin{subfigure}{.24\textwidth}
  \centering
  \includegraphics[width=\textwidth]{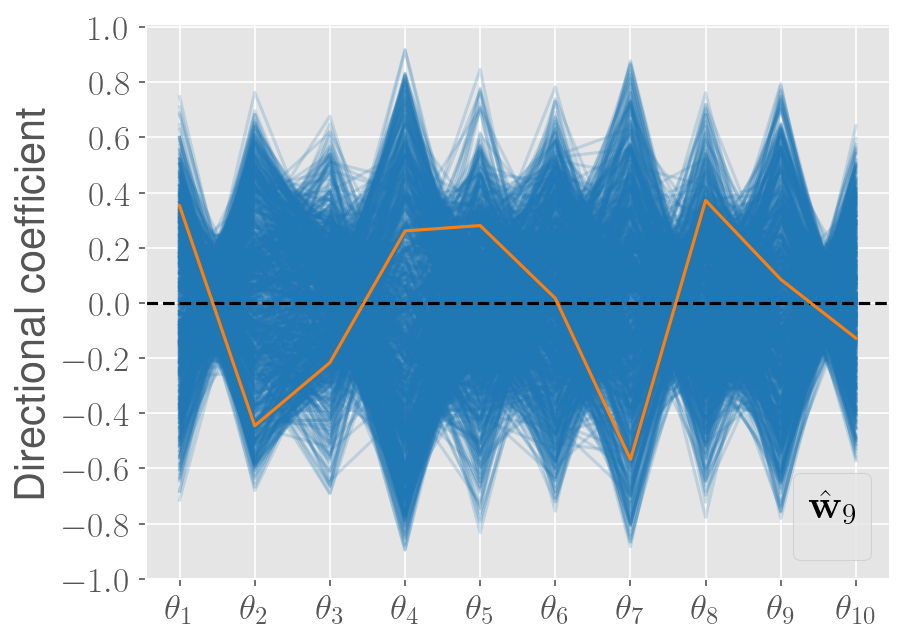}
  \caption{}
  \label{fig:example_1_case_1_projection_vector_dim_9}
\end{subfigure}%
\begin{subfigure}{.24\textwidth}
  \centering
  \includegraphics[width=\textwidth]{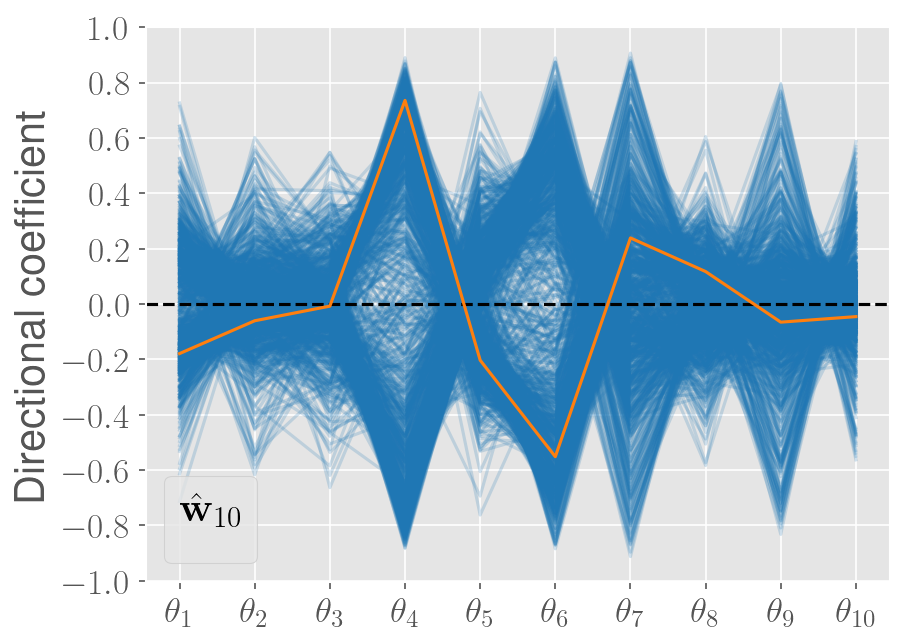}
  \caption{}
  \label{fig:example_1_case_1_projection_vector_dim_10}
\end{subfigure}
\caption{Plots of $ \hat{\mathbf{w}}_i $, $ i = 3, \dots, 10 $ for Case 1 of the polynomial test function (see Section~\ref{section:example_1_polynomial}). The nominal vector is in orange, and 1000 bootstrapped replicates are in blue. $ \hat{\mathbf{w}}_1 $ and $ \hat{\mathbf{w}}_2 $ are in Figure~\ref{fig:active_subspace_results_for_example_1}.}
\label{fig:example_1_case_1_all_projection_vectors}
\end{figure}

\begin{figure}[t!bhp]
\centering
\begin{subfigure}{.24\textwidth}
  \centering
  \includegraphics[width=\textwidth]{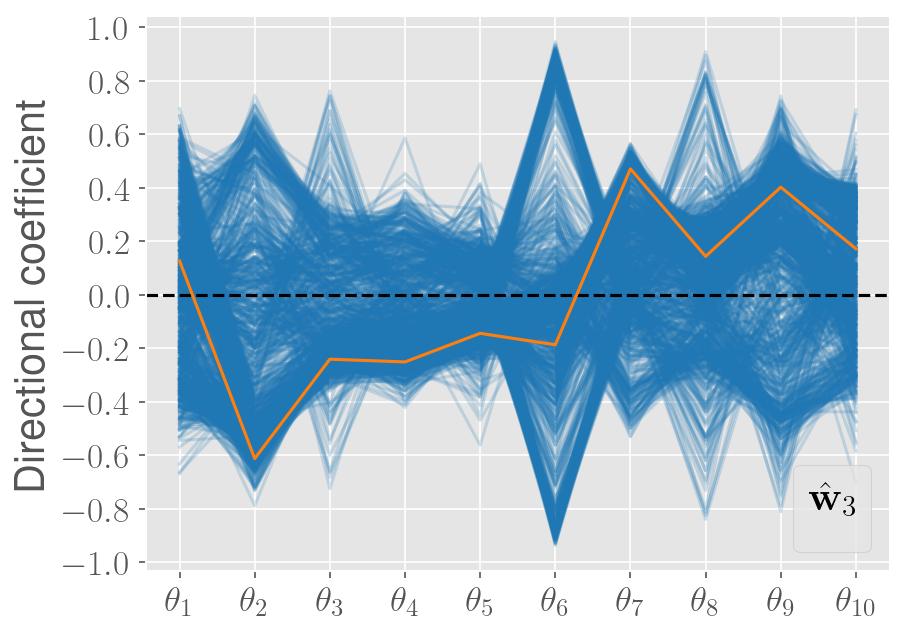}
  \caption{}
  \label{fig:example_1_case_2_projection_vector_dim_3}
\end{subfigure}%
\begin{subfigure}{.24\textwidth}
  \centering
  \includegraphics[width=\textwidth]{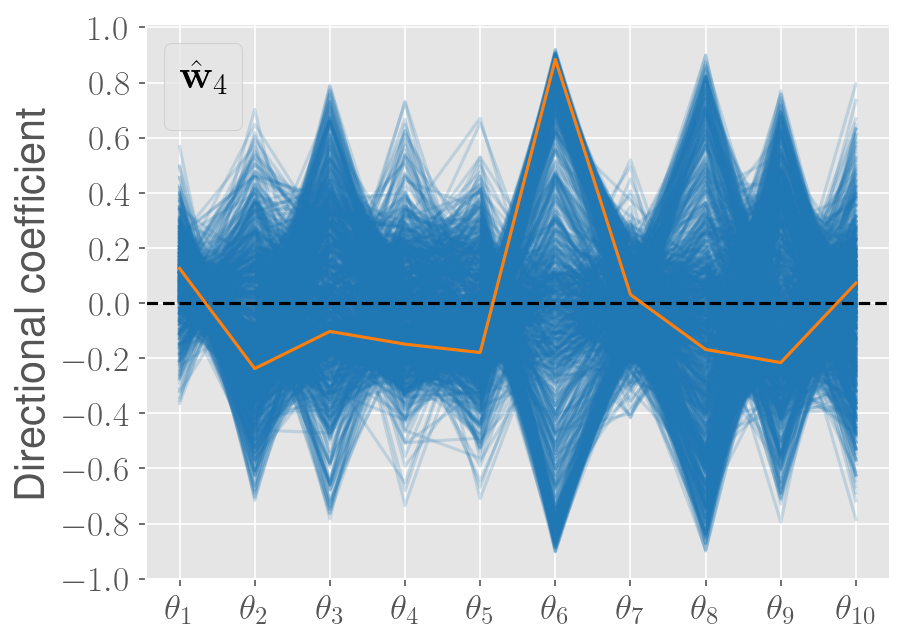}
  \caption{}
  \label{fig:example_1_case_2_projection_vector_dim_4}
\end{subfigure}%
\begin{subfigure}{.24\textwidth}
  \centering
  \includegraphics[width=\textwidth]{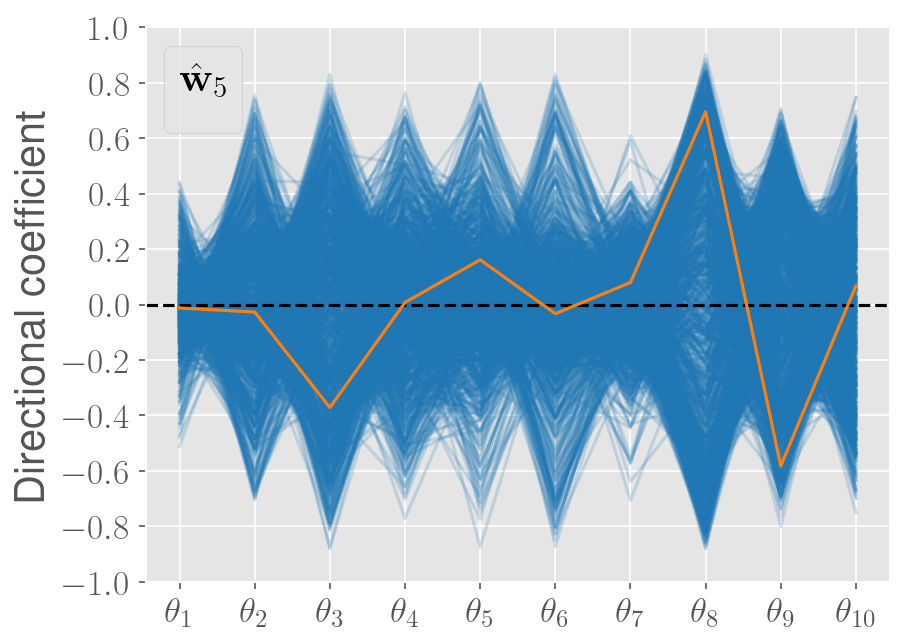}
  \caption{}
  \label{fig:example_1_case_2_projection_vector_dim_5}
\end{subfigure}%
\begin{subfigure}{.24\textwidth}
  \centering
  \includegraphics[width=\textwidth]{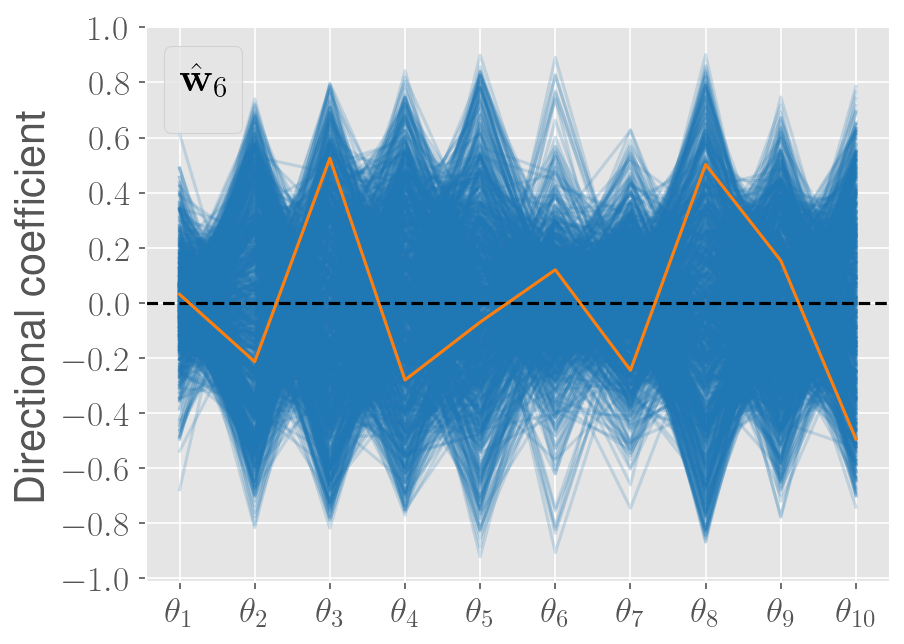}
  \caption{}
  \label{fig:example_1_case_2_projection_vector_dim_6}
\end{subfigure}
\begin{subfigure}{.24\textwidth}
  \centering
  \includegraphics[width=\textwidth]{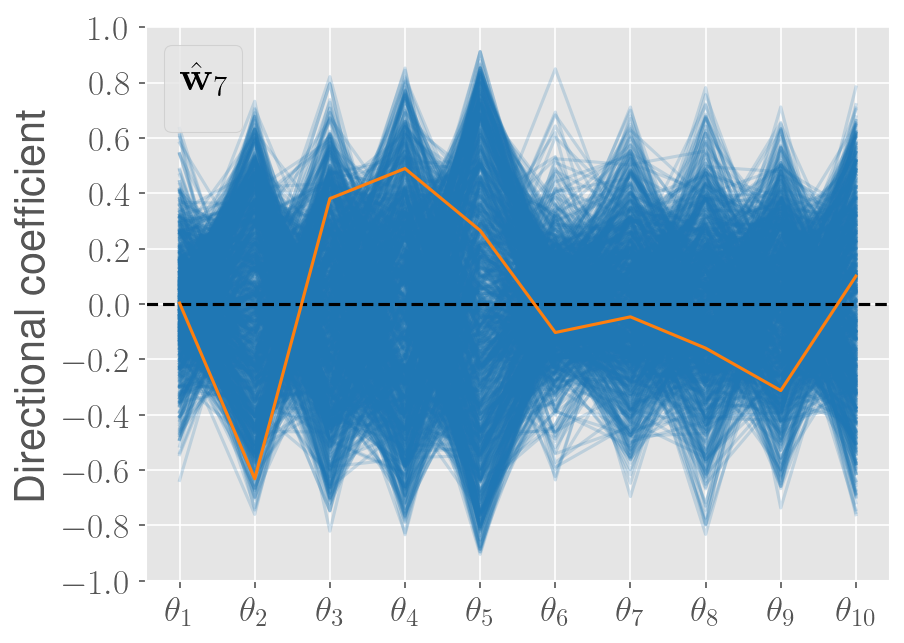}
  \caption{}
  \label{fig:example_1_case_2_projection_vector_dim_7}
\end{subfigure}%
\begin{subfigure}{.24\textwidth}
  \centering
  \includegraphics[width=\textwidth]{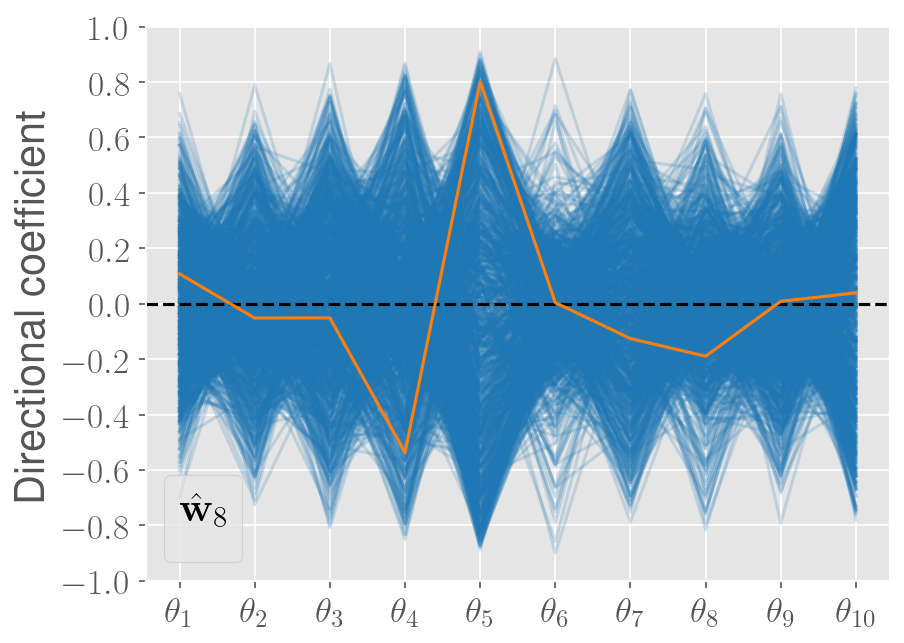}
  \caption{}
  \label{fig:example_1_case_2_projection_vector_dim_8}
\end{subfigure}%
\begin{subfigure}{.24\textwidth}
  \centering
  \includegraphics[width=\textwidth]{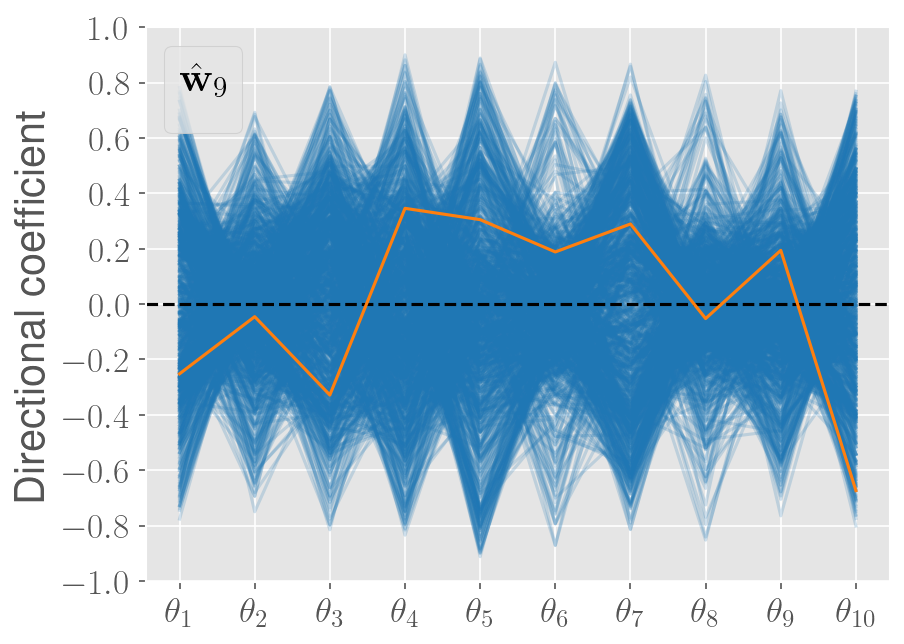}
  \caption{}
  \label{fig:example_1_case_2_projection_vector_dim_9}
\end{subfigure}%
\begin{subfigure}{.24\textwidth}
  \centering
  \includegraphics[width=\textwidth]{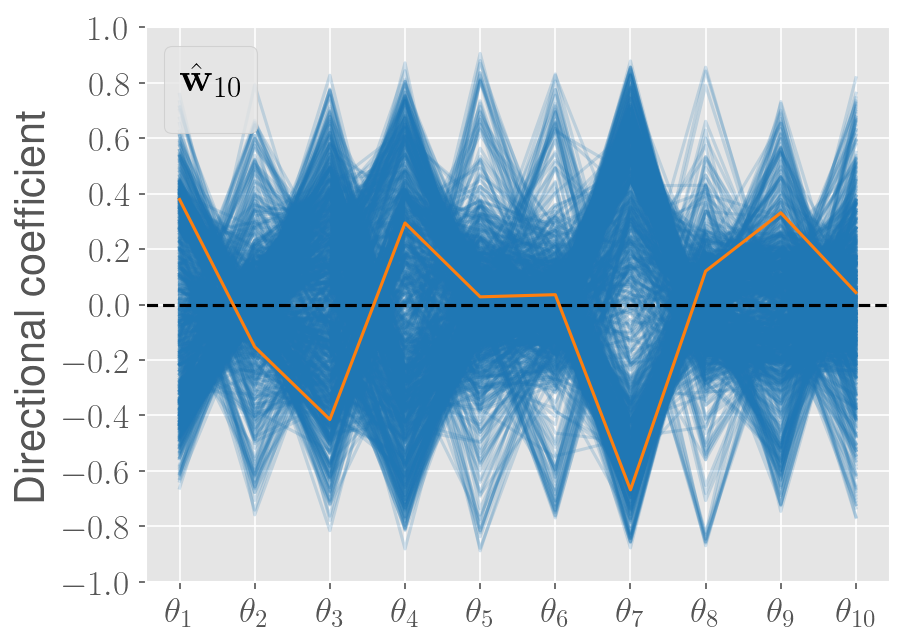}
  \caption{}
  \label{fig:example_1_case_2_projection_vector_dim_10}
\end{subfigure}
\caption{Plots of $ \hat{\mathbf{w}}_i $, $ i = 3, \dots, 10 $ for Case 2 of the polynomial test function (see Section~\ref{section:example_1_polynomial}). The nominal vector is in orange, and 1000 bootstrapped replicates are in blue. $ \hat{\mathbf{w}}_1 $ and $ \hat{\mathbf{w}}_2 $ are in Figure~\ref{fig:active_subspace_results_for_example_1}.}
\label{fig:example_1_case_2_all_projection_vectors}
\end{figure}

\begin{figure}[t!bhp]
\centering
\begin{subfigure}{.24\textwidth}
  \centering
  \includegraphics[width=\textwidth]{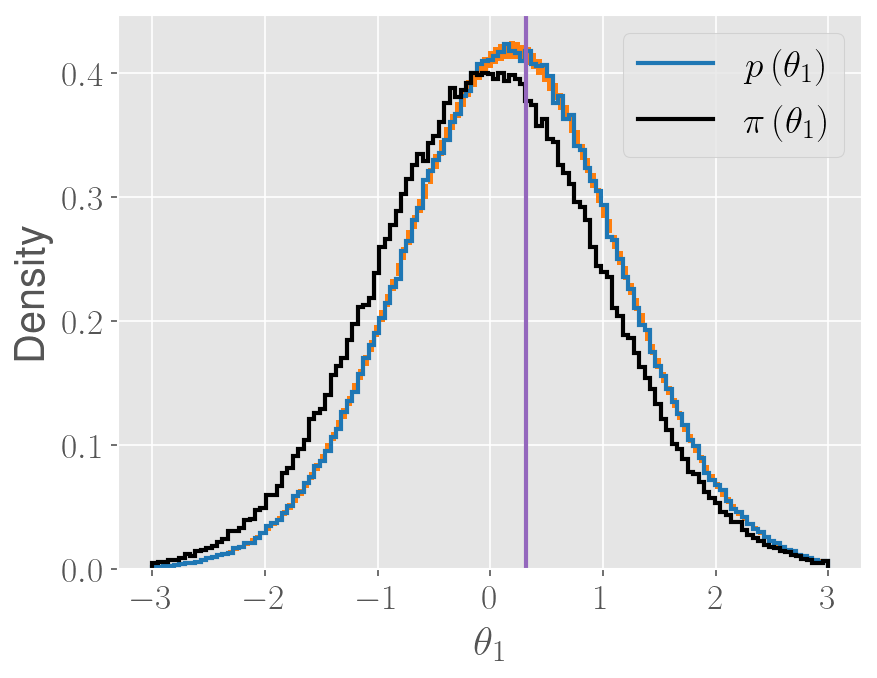}
  \caption{}
  \label{fig:example_1_case_1_full_posterior_nominal_and_errorbars_dim_1}
\end{subfigure}%
\begin{subfigure}{.24\textwidth}
  \centering
  \includegraphics[width=\textwidth]{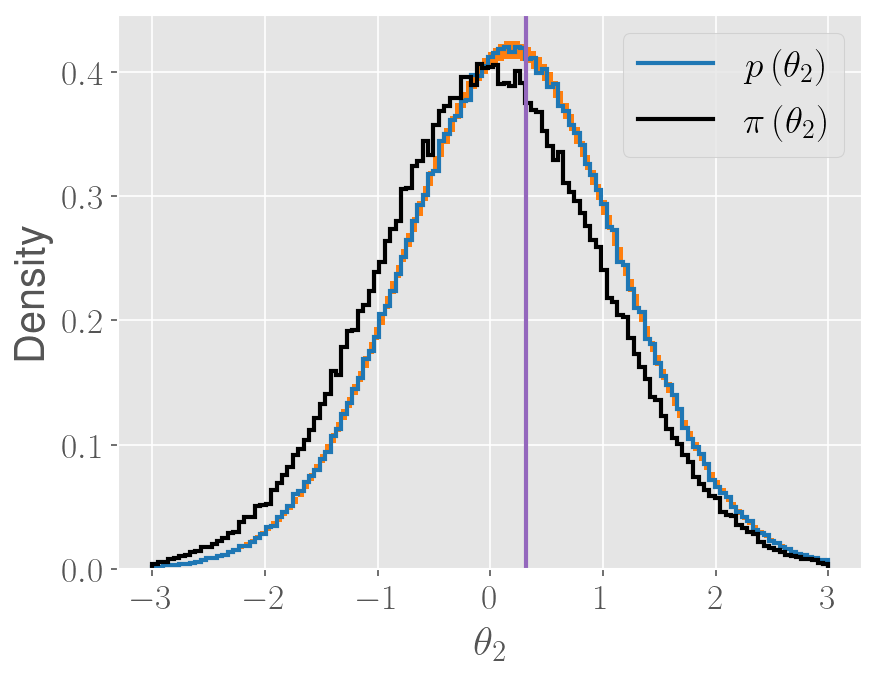}
  \caption{}
  \label{fig:example_1_case_1_full_posterior_nominal_and_errorbars_dim_2}
\end{subfigure}%
\begin{subfigure}{.24\textwidth}
  \centering
  \includegraphics[width=\textwidth]{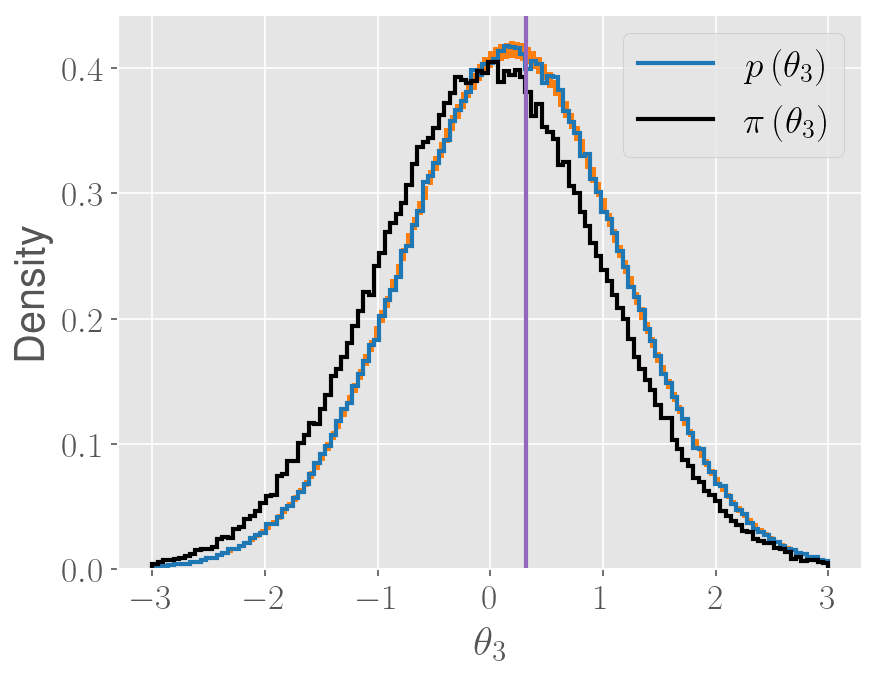}
  \caption{}
  \label{fig:example_1_case_1_full_posterior_nominal_and_errorbars_dim_3}
\end{subfigure}%
\begin{subfigure}{.24\textwidth}
  \centering
  \includegraphics[width=\textwidth]{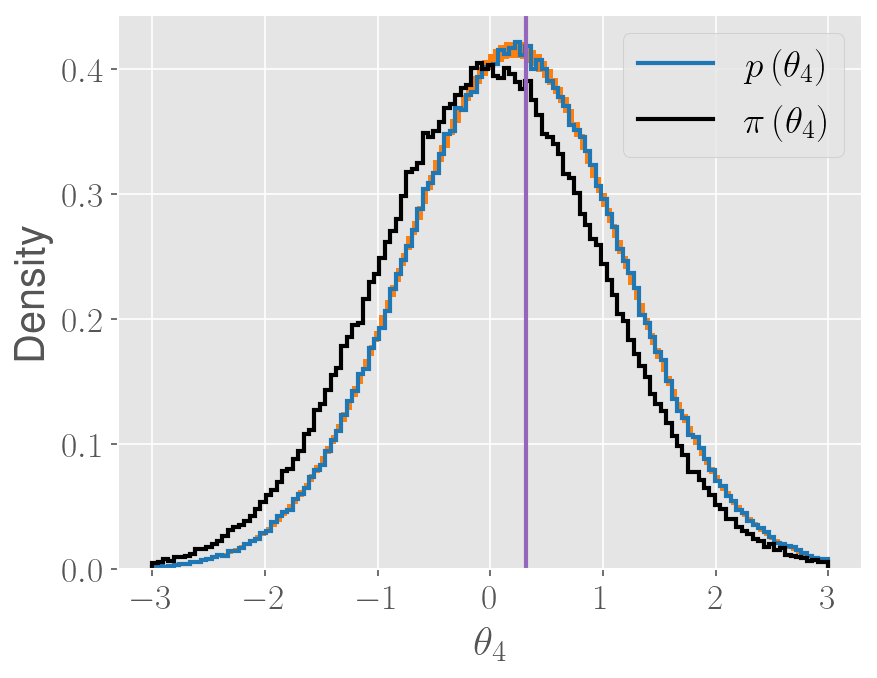}
  \caption{}
  \label{fig:example_1_case_1_full_posterior_nominal_and_errorbars_dim_4}
\end{subfigure}
\begin{subfigure}{.24\textwidth}
  \centering
  \includegraphics[width=\textwidth]{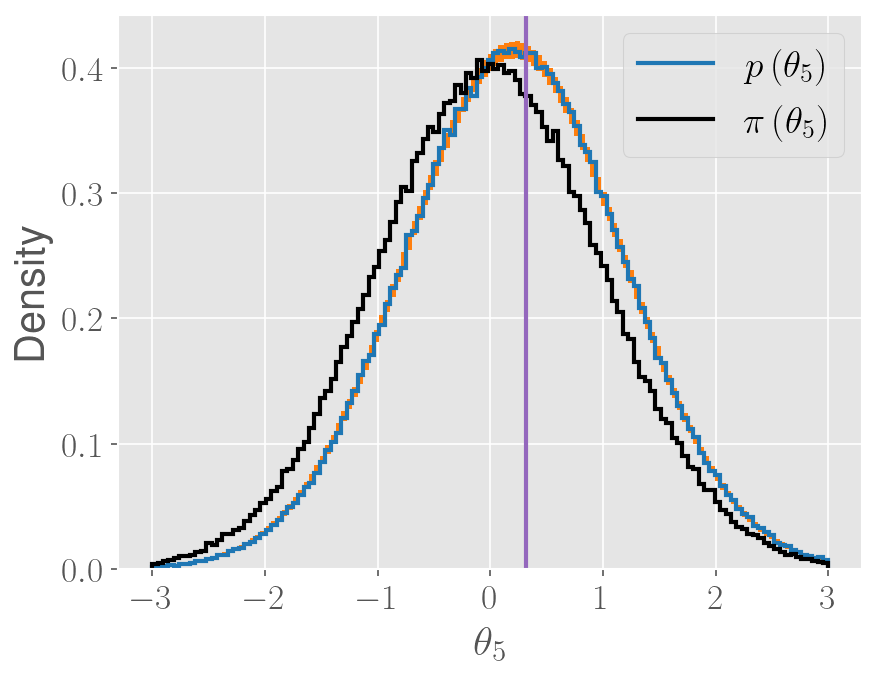}
  \caption{}
  \label{fig:example_1_case_1_full_posterior_nominal_and_errorbars_dim_5}
\end{subfigure}%
\begin{subfigure}{.24\textwidth}
  \centering
  \includegraphics[width=\textwidth]{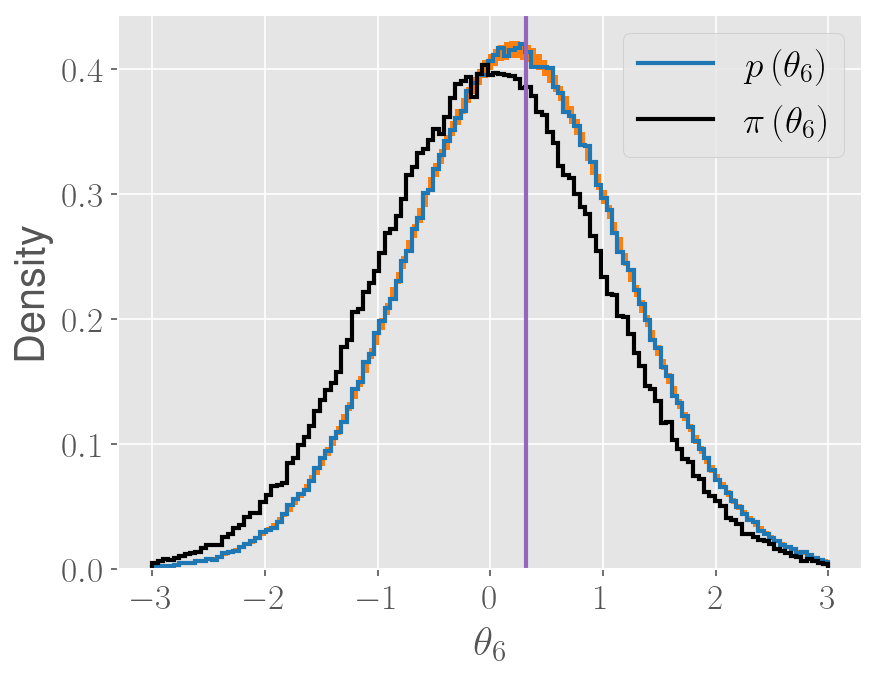}
  \caption{}
  \label{fig:example_1_case_1_full_posterior_nominal_and_errorbars_dim_6}
\end{subfigure}%
\begin{subfigure}{.24\textwidth}
  \centering
  \includegraphics[width=\textwidth]{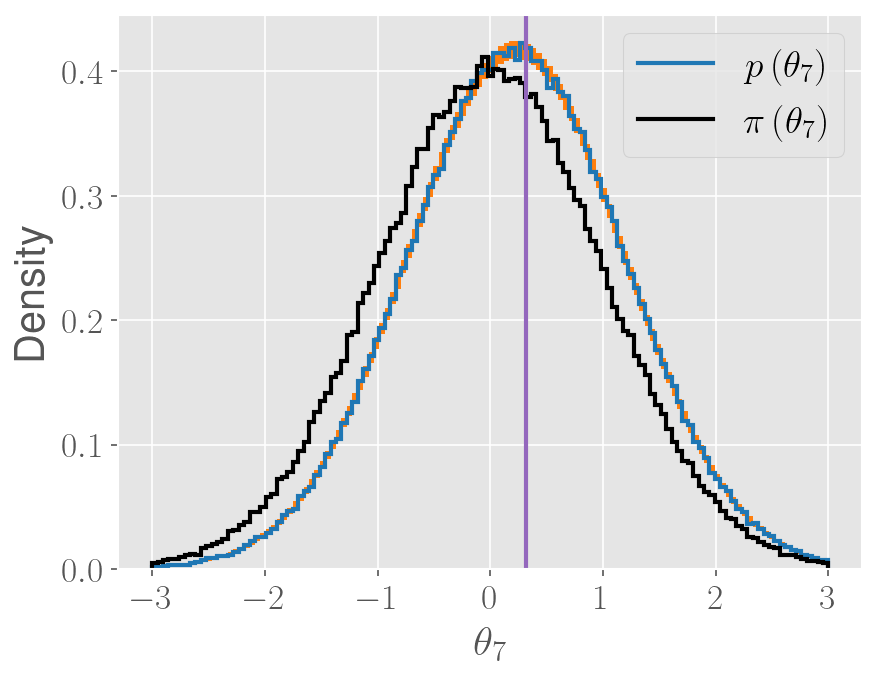}
  \caption{}
  \label{fig:example_1_case_1_full_posterior_nominal_and_errorbars_dim_7}
\end{subfigure}%
\begin{subfigure}{.24\textwidth}
  \centering
  \includegraphics[width=\textwidth]{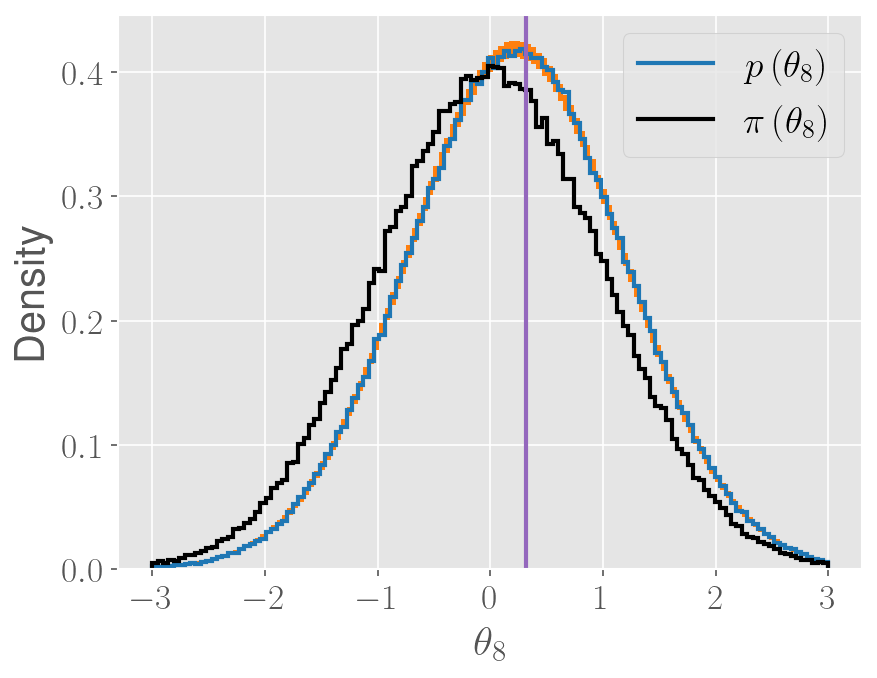}
  \caption{}
  \label{fig:example_1_case_1_full_posterior_nominal_and_errorbars_dim_8}
\end{subfigure}
\begin{subfigure}{.24\textwidth}
  \centering
  \includegraphics[width=\textwidth]{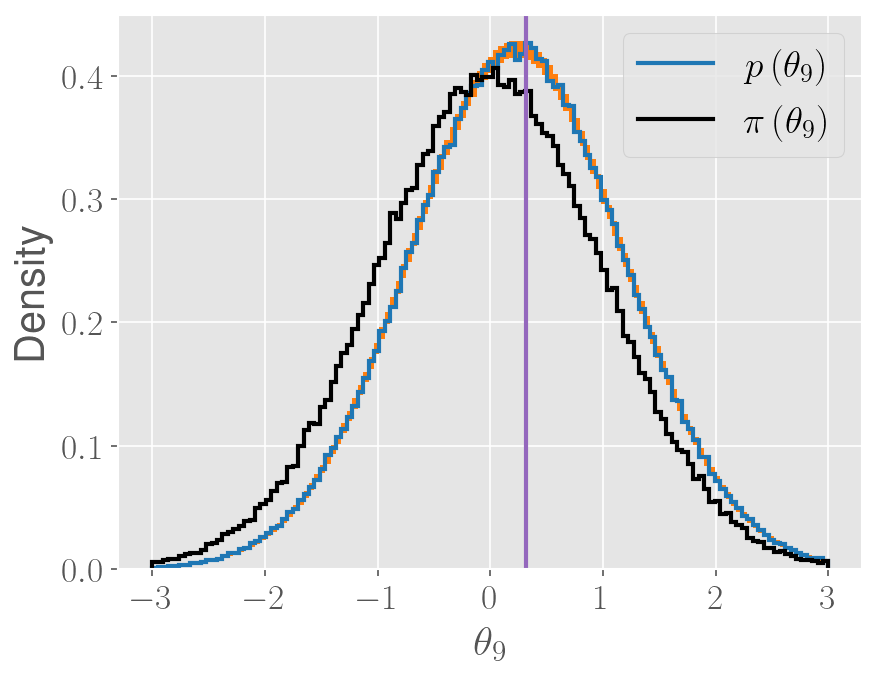}
  \caption{}
  \label{fig:example_1_case_1_full_posterior_nominal_and_errorbars_dim_9}
\end{subfigure}%
\begin{subfigure}{.24\textwidth}
  \centering
  \includegraphics[width=\textwidth]{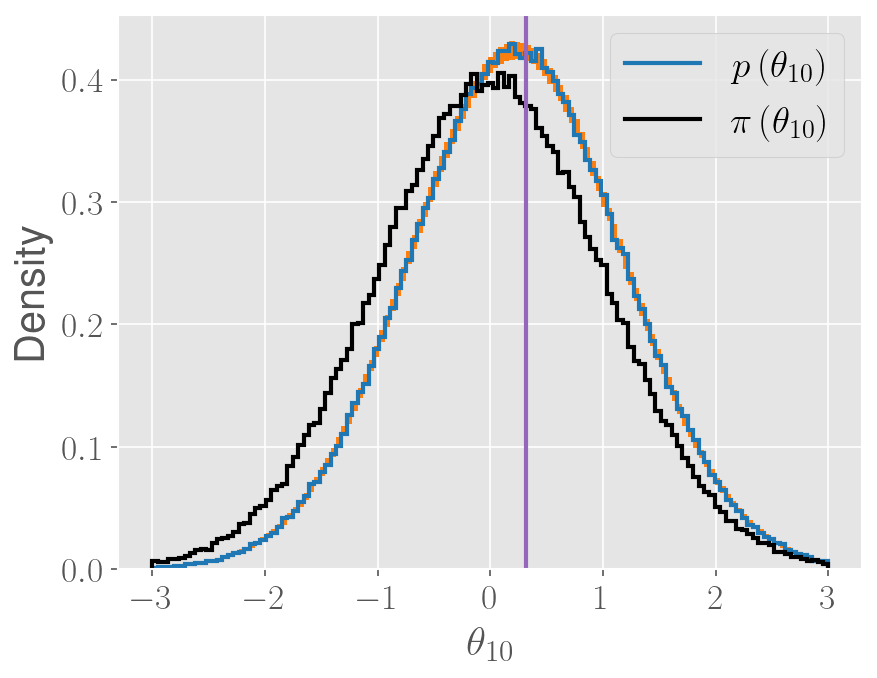}
  \caption{}
  \label{fig:example_1_case_1_full_posterior_nominal_and_errorbars_dim_10}
\end{subfigure}%
\begin{subfigure}{.24\textwidth}
  \centering
  \includegraphics[width=\textwidth]{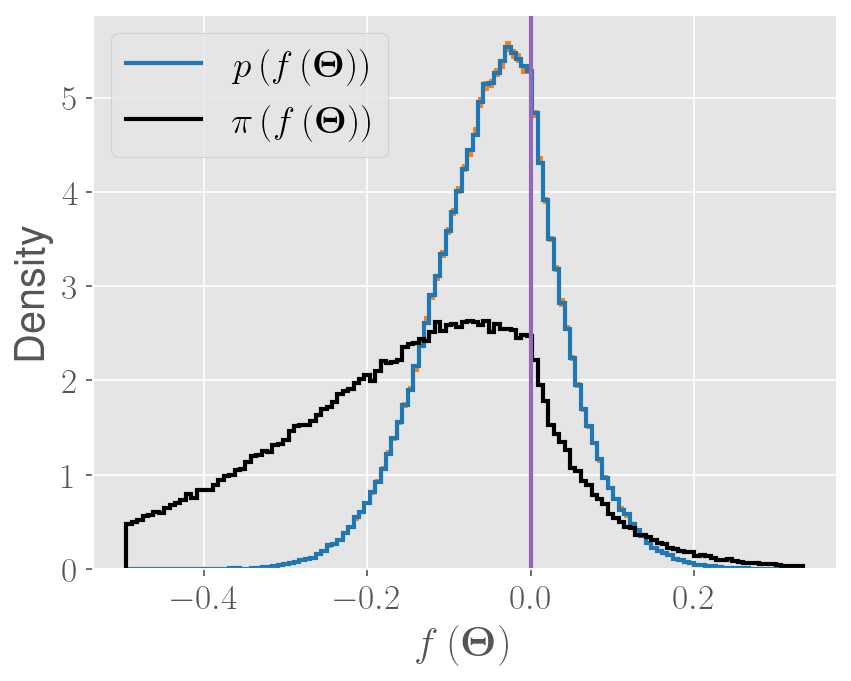}
  \caption{}
  \label{fig:example_1_case_1_full_output_nominal_and_errorbars}
\end{subfigure}%
\begin{subfigure}{.24\textwidth}
  \centering
  \includegraphics[width=\textwidth]{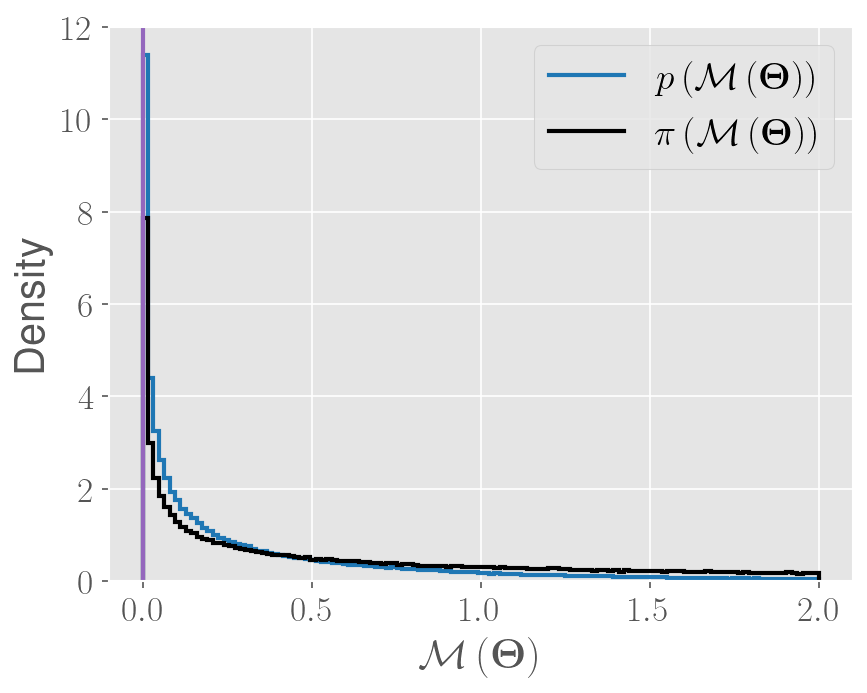}
  \caption{}
  \label{fig:example_1_case_1_full_misfit_nominal_and_errorbars}
\end{subfigure}
\caption{Plots of full posteriors for Case 1 of the polynomial test function (see Section~\ref{section:example_1_polynomial}). Figures (a) - (j) plot the posteriors for $ \theta_1 $ - $ \theta_{10} $, figure (k) plots the density of $ f \left( \boldsymbol{\Theta} \right) $ corresponding to the full posterior $ p \left( \boldsymbol{\Theta} \right) $, and figure (l) plots the density of $ \mathcal{M} \left( \boldsymbol{\Theta} \right) $ corresponding to the full posterior $ p \left( \boldsymbol{\Theta} \right) $. In each figure, the nominal calibrated posterior constructed using $ \mathfrak{C}_{\boldsymbol{\Theta}} $ is in blue, while the orange bands depict the $ 90 \% $ confidence interval of the posteriors constructed using the sample set replicates $ \mathfrak{C}_{\boldsymbol{\Theta}}^{(i)} $ is in orange. Additionally, the prior density is drawn in black, with the purple line indicating the observed value used for calibration.}
\label{fig:calibrated_full_posteriors_results_for_example_1_case_1}
\end{figure}

\begin{figure}[t!bhp]
\centering
\begin{subfigure}{.24\textwidth}
  \centering
  \includegraphics[width=\textwidth]{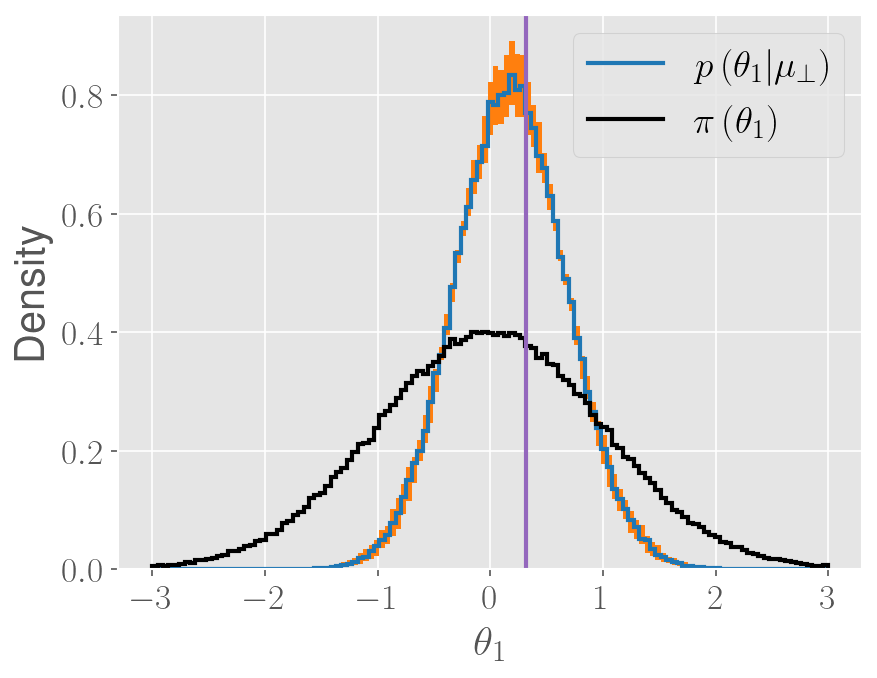}
  \caption{}
  \label{fig:example_1_case_1_conditional_posterior_nominal_and_errorbars_dim_1}
\end{subfigure}%
\begin{subfigure}{.24\textwidth}
  \centering
  \includegraphics[width=\textwidth]{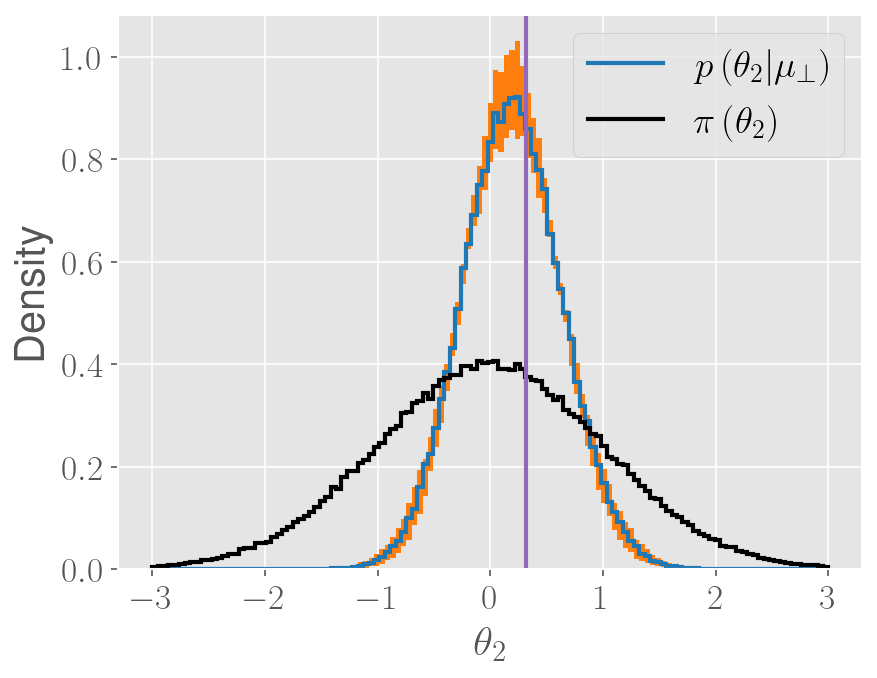}
  \caption{}
  \label{fig:example_1_case_1_conditional_posterior_nominal_and_errorbars_dim_2}
\end{subfigure}%
\begin{subfigure}{.24\textwidth}
  \centering
  \includegraphics[width=\textwidth]{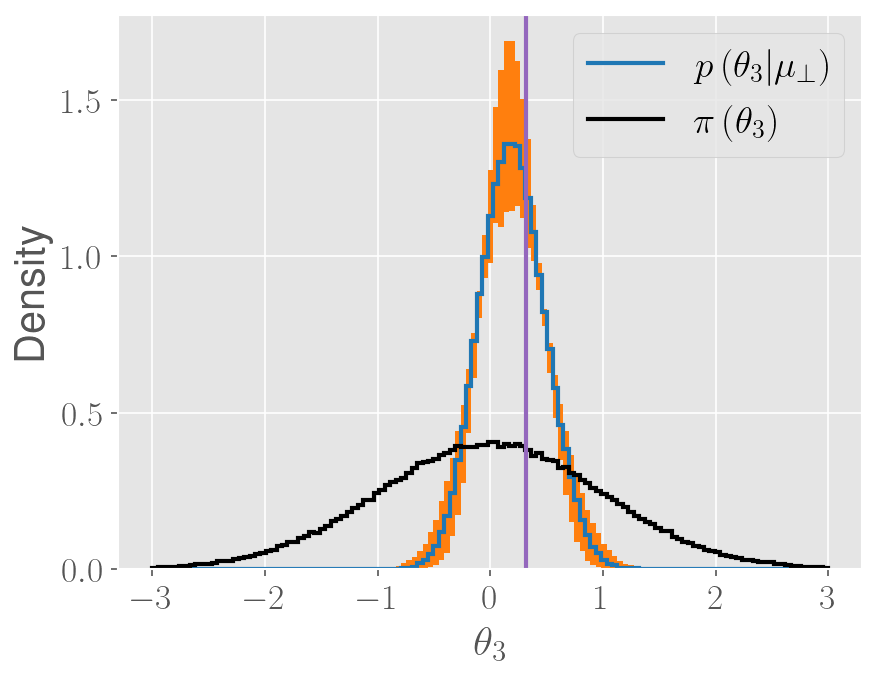}
  \caption{}
  \label{fig:example_1_case_1_conditional_posterior_nominal_and_errorbars_dim_3}
\end{subfigure}%
\begin{subfigure}{.24\textwidth}
  \centering
  \includegraphics[width=\textwidth]{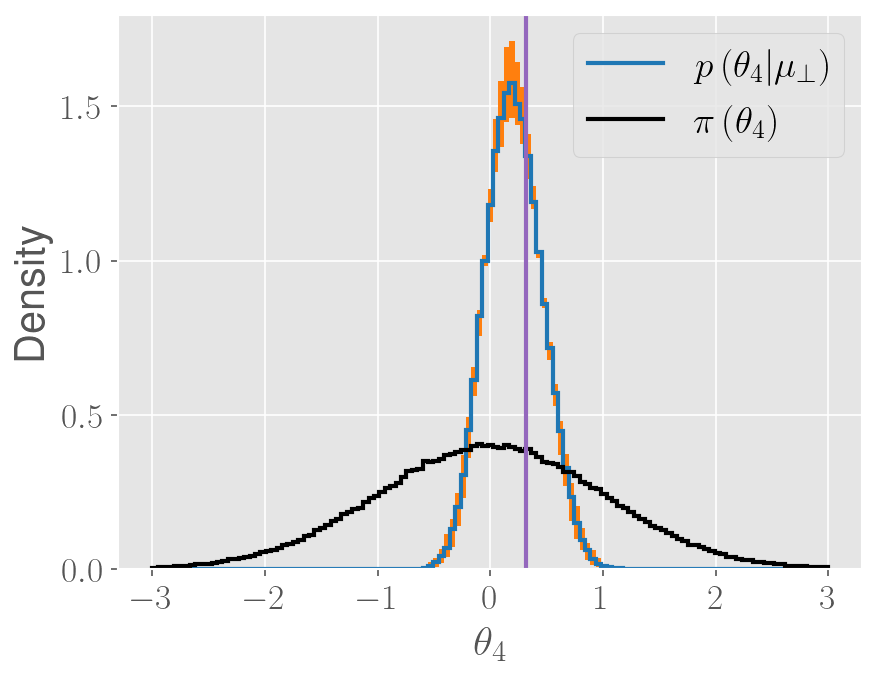}
  \caption{}
  \label{fig:example_1_case_1_conditional_posterior_nominal_and_errorbars_dim_4}
\end{subfigure}
\begin{subfigure}{.24\textwidth}
  \centering
  \includegraphics[width=\textwidth]{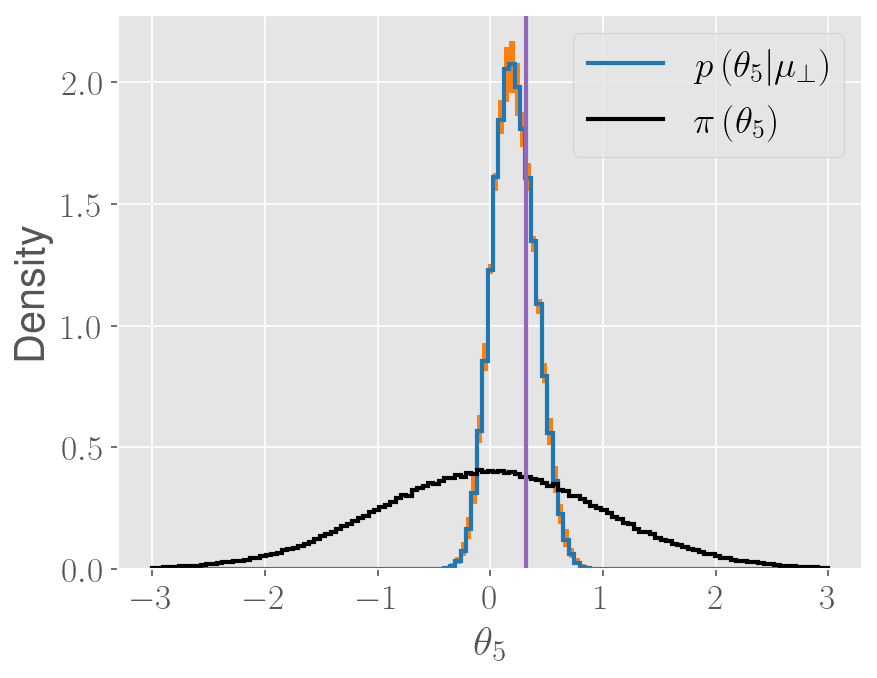}
  \caption{}
  \label{fig:example_1_case_1_conditional_posterior_nominal_and_errorbars_dim_5}
\end{subfigure}%
\begin{subfigure}{.24\textwidth}
  \centering
  \includegraphics[width=\textwidth]{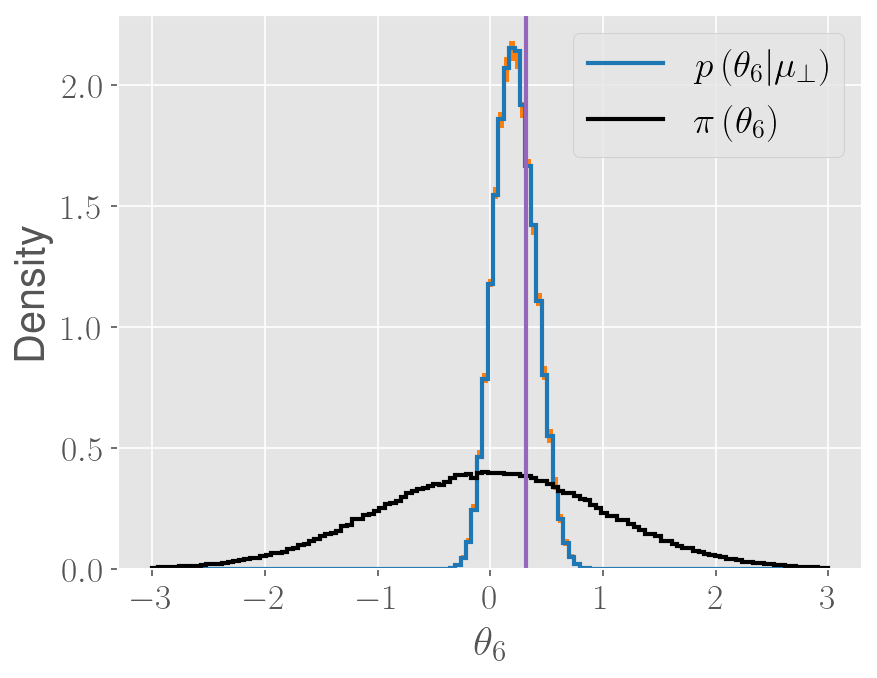}
  \caption{}
  \label{fig:example_1_case_1_conditional_posterior_nominal_and_errorbars_dim_6}
\end{subfigure}%
\begin{subfigure}{.24\textwidth}
  \centering
  \includegraphics[width=\textwidth]{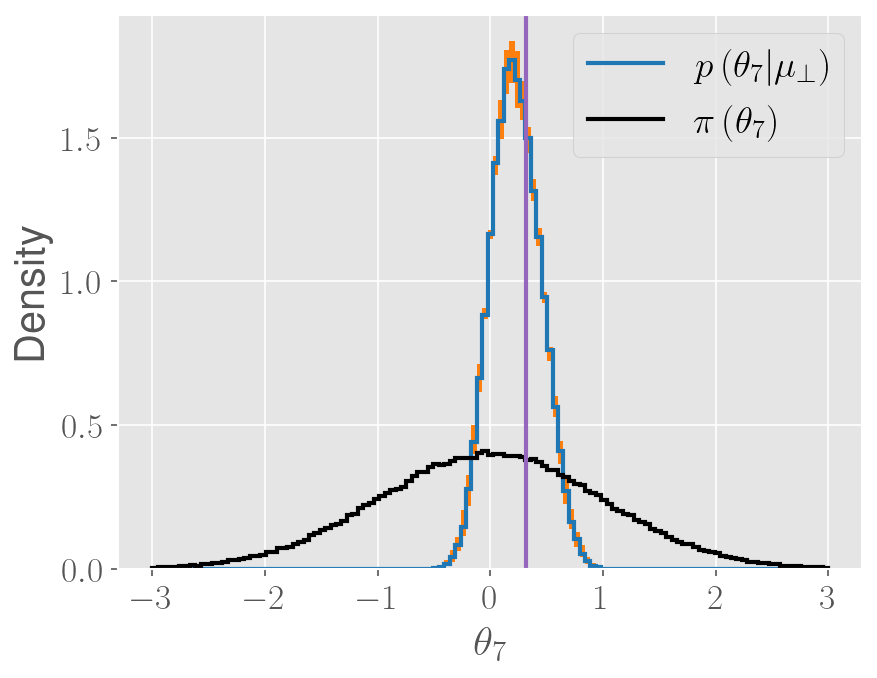}
  \caption{}
  \label{fig:example_1_case_1_conditional_posterior_nominal_and_errorbars_dim_7}
\end{subfigure}%
\begin{subfigure}{.24\textwidth}
  \centering
  \includegraphics[width=\textwidth]{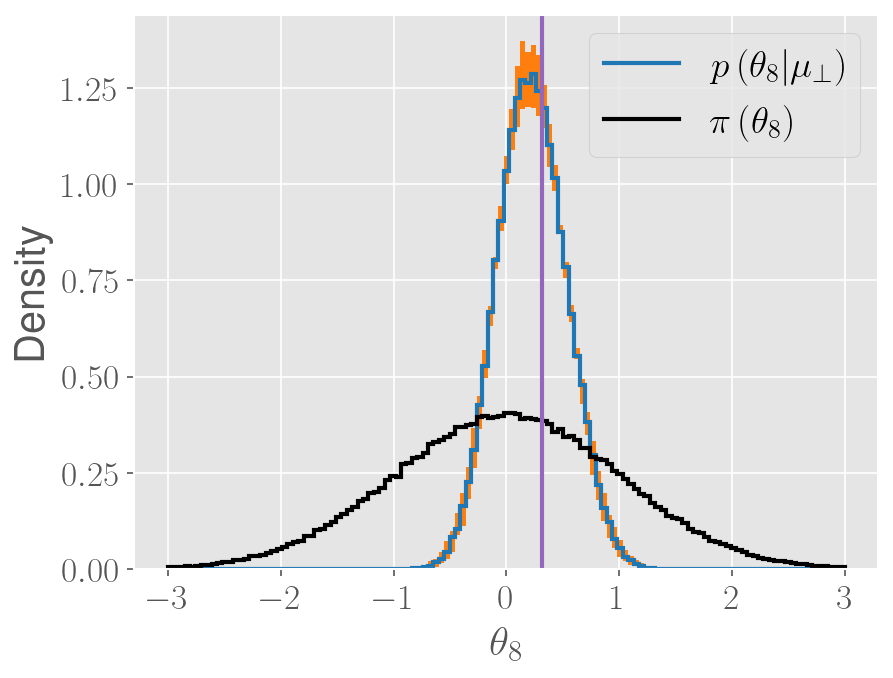}
  \caption{}
  \label{fig:example_1_case_1_conditional_posterior_nominal_and_errorbars_dim_8}
\end{subfigure}
\begin{subfigure}{.24\textwidth}
  \centering
  \includegraphics[width=\textwidth]{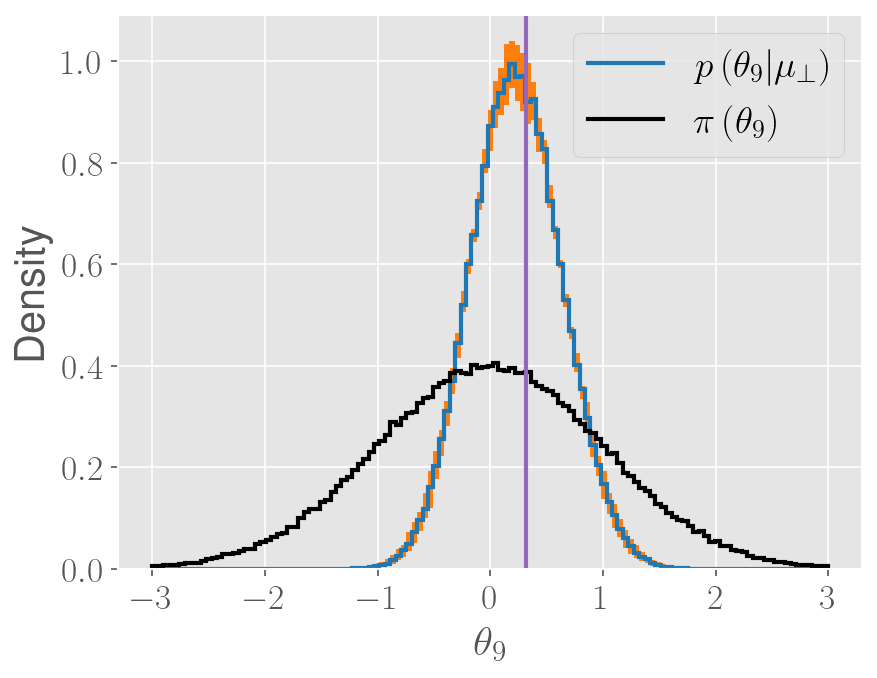}
  \caption{}
  \label{fig:example_1_case_1_conditional_posterior_nominal_and_errorbars_dim_9}
\end{subfigure}%
\begin{subfigure}{.24\textwidth}
  \centering
  \includegraphics[width=\textwidth]{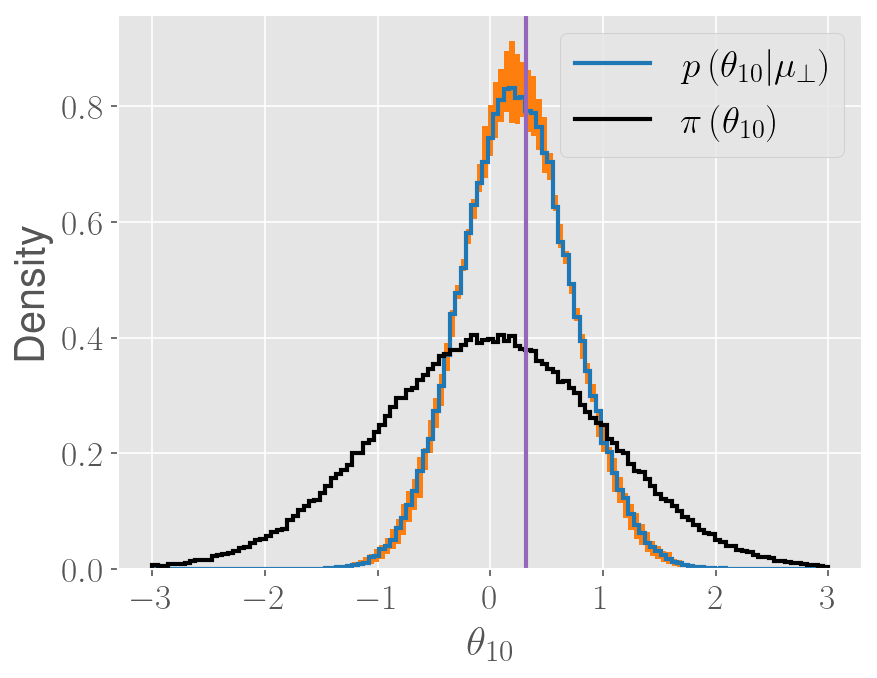}
  \caption{}
  \label{fig:example_1_case_1_conditional_posterior_nominal_and_errorbars_dim_10}
\end{subfigure}%
\begin{subfigure}{.24\textwidth}
  \centering
  \includegraphics[width=\textwidth]{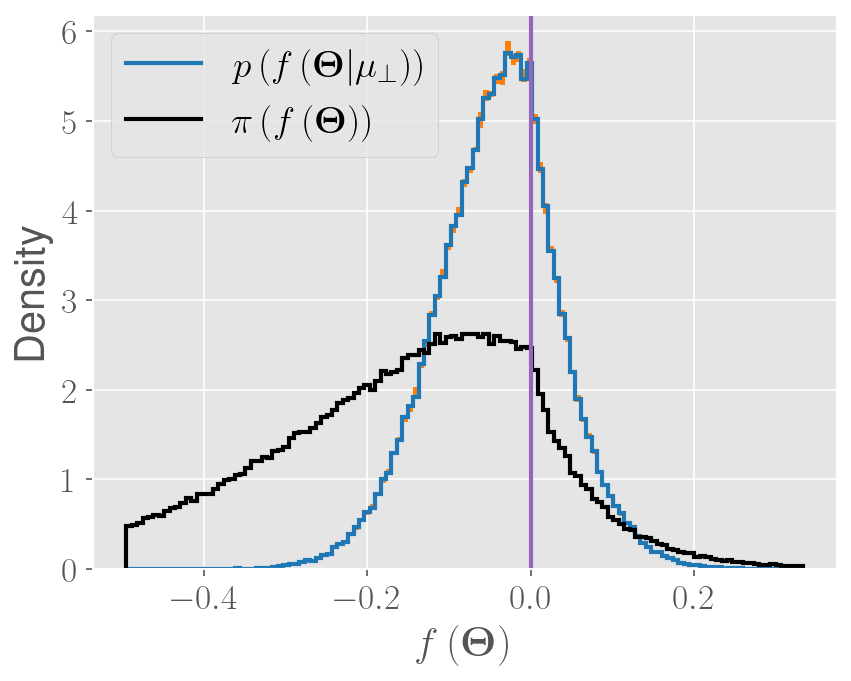}
  \caption{}
  \label{fig:example_1_case_1_conditional_output_nominal_and_errorbars}
\end{subfigure}%
\begin{subfigure}{.24\textwidth}
  \centering
  \includegraphics[width=\textwidth]{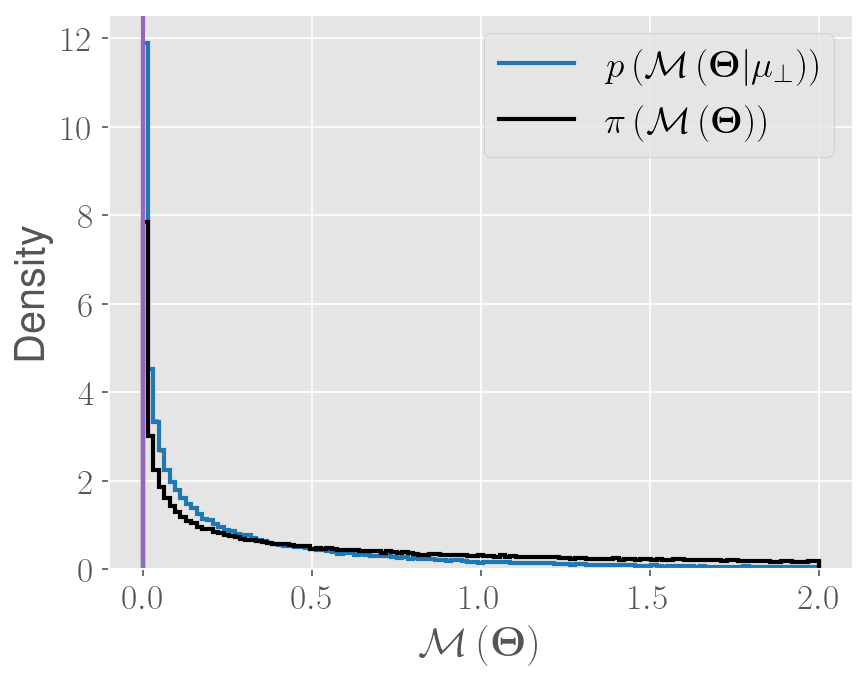}
  \caption{}
  \label{fig:example_1_case_1_conditional_misfit_nominal_and_errorbars}
\end{subfigure}
\caption{Plots of conditional active calibrated posteriors for Case 1 of the polynomial test function (see Section~\ref{section:example_1_polynomial}). Figures (a) - (j) plot the posteriors for $ \theta_1 $ - $ \theta_{10} $, figure (k) plots the density of $ f \left( \boldsymbol{\Theta} \right) $ corresponding to the conditional active calibrated posterior $ p \left( \boldsymbol{\Theta}_{\mu_{\perp}} \right) $, and figure (l) plots the density of $ \mathcal{M} \left( \boldsymbol{\Theta} \right) $ corresponding to the conditional active calibrated posterior $ p \left( \boldsymbol{\Theta}_{\mu_{\perp}} \right) $. In each figure, the nominal calibrated posterior constructed using $ \mathfrak{C}_{\boldsymbol{\Theta}} $ is in blue, while the orange bands depict the $ 90 \% $ confidence interval of the posteriors constructed using the sample set replicates $ \mathfrak{C}_{\boldsymbol{\Theta}}^{(i)} $ is in orange. Additionally, the prior density is drawn in black, with the purple line indicating the observed value used for calibration.}
\label{fig:calibrated_conditional_posteriors_results_for_example_1_case_1}
\end{figure}

\begin{figure}[t!bhp]
\centering
\begin{subfigure}{.24\textwidth}
  \centering
  \includegraphics[width=\textwidth]{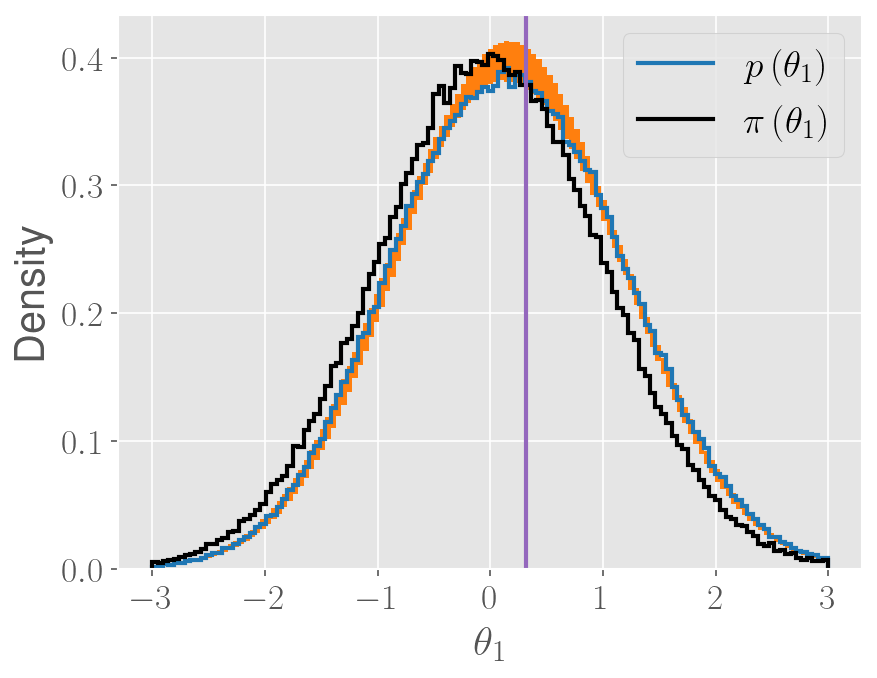}
  \caption{}
  \label{fig:example_1_case_2_full_posterior_nominal_and_errorbars_dim_1}
\end{subfigure}%
\begin{subfigure}{.24\textwidth}
  \centering
  \includegraphics[width=\textwidth]{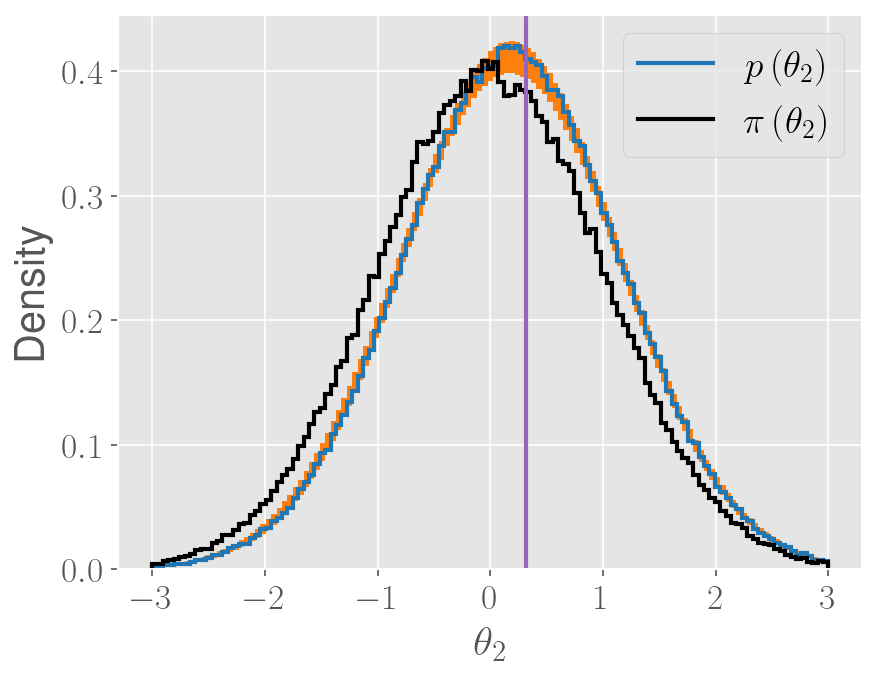}
  \caption{}
  \label{fig:example_1_case_2_full_posterior_nominal_and_errorbars_dim_2}
\end{subfigure}%
\begin{subfigure}{.24\textwidth}
  \centering
  \includegraphics[width=\textwidth]{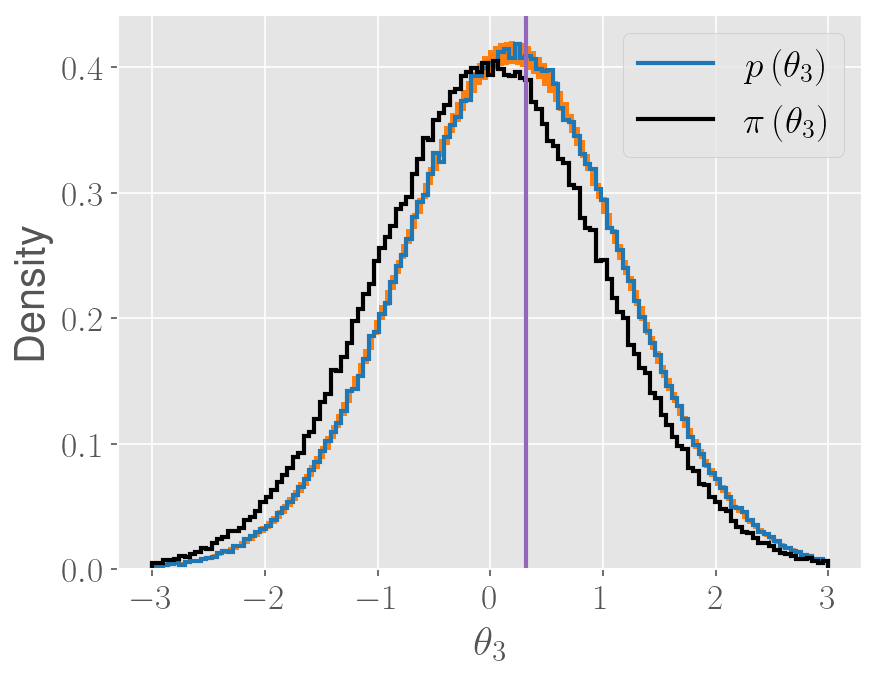}
  \caption{}
  \label{fig:example_1_case_2_full_posterior_nominal_and_errorbars_dim_3}
\end{subfigure}%
\begin{subfigure}{.24\textwidth}
  \centering
  \includegraphics[width=\textwidth]{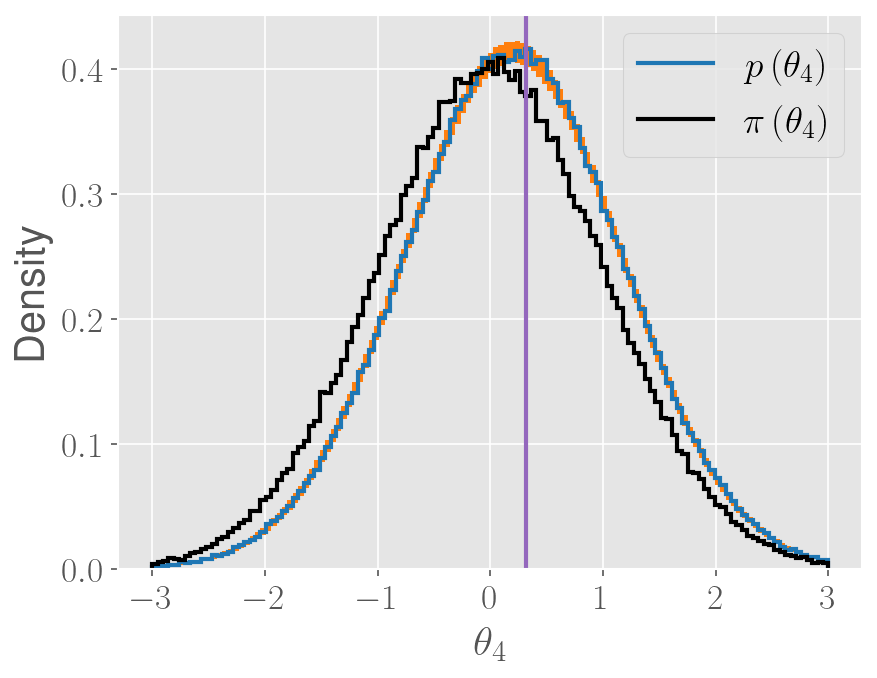}
  \caption{}
  \label{fig:example_1_case_2_full_posterior_nominal_and_errorbars_dim_4}
\end{subfigure}
\begin{subfigure}{.24\textwidth}
  \centering
  \includegraphics[width=\textwidth]{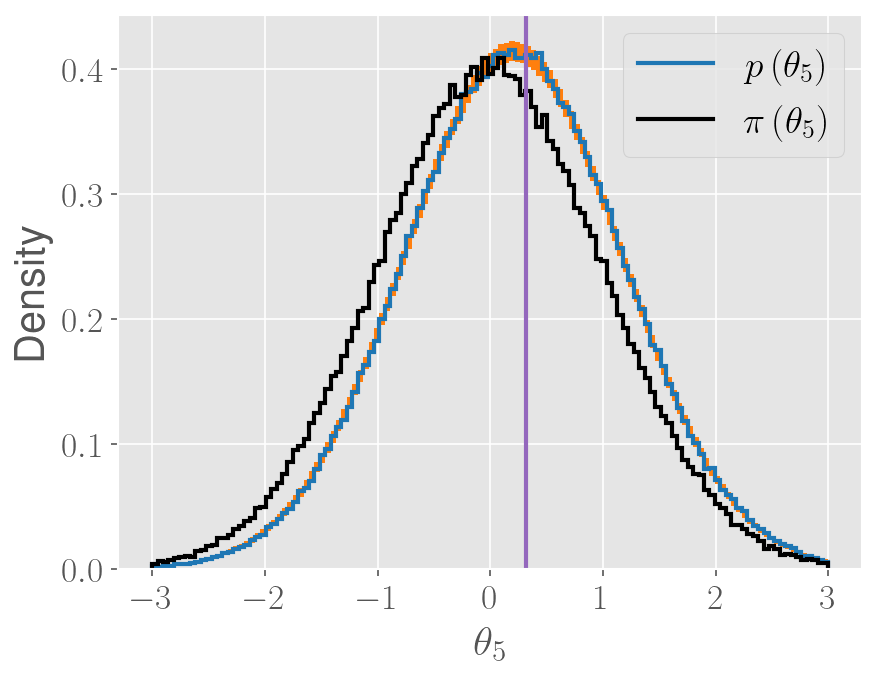}
  \caption{}
  \label{fig:example_1_case_2_full_posterior_nominal_and_errorbars_dim_5}
\end{subfigure}%
\begin{subfigure}{.24\textwidth}
  \centering
  \includegraphics[width=\textwidth]{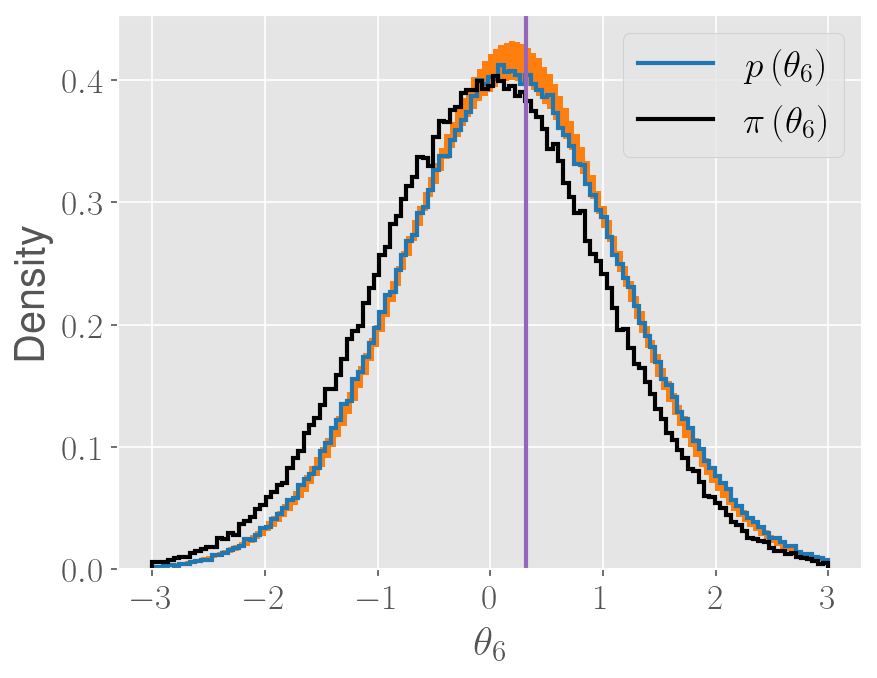}
  \caption{}
  \label{fig:example_1_case_2_full_posterior_nominal_and_errorbars_dim_6}
\end{subfigure}%
\begin{subfigure}{.24\textwidth}
  \centering
  \includegraphics[width=\textwidth]{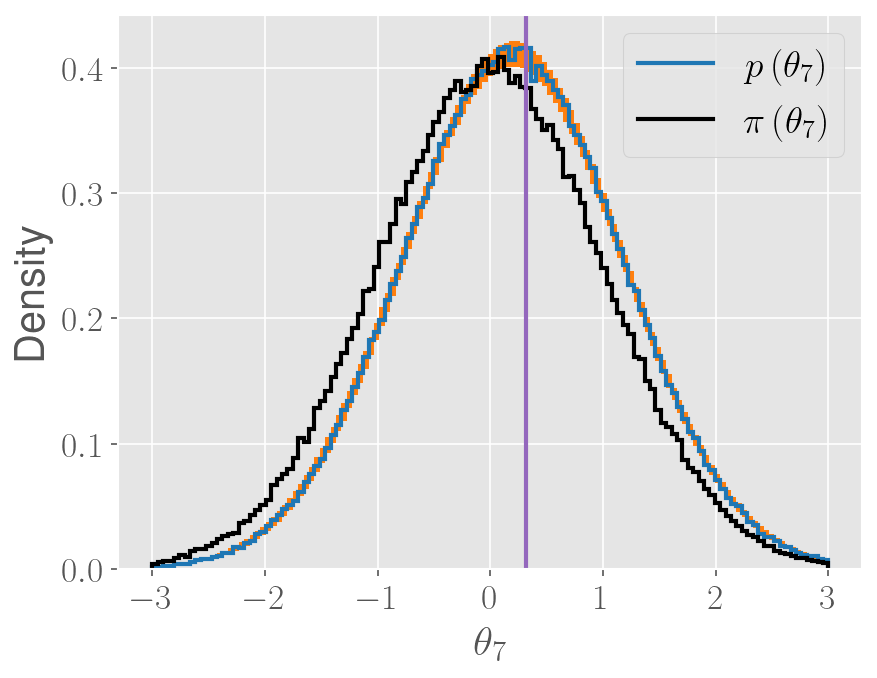}
  \caption{}
  \label{fig:example_1_case_2_full_posterior_nominal_and_errorbars_dim_7}
\end{subfigure}%
\begin{subfigure}{.24\textwidth}
  \centering
  \includegraphics[width=\textwidth]{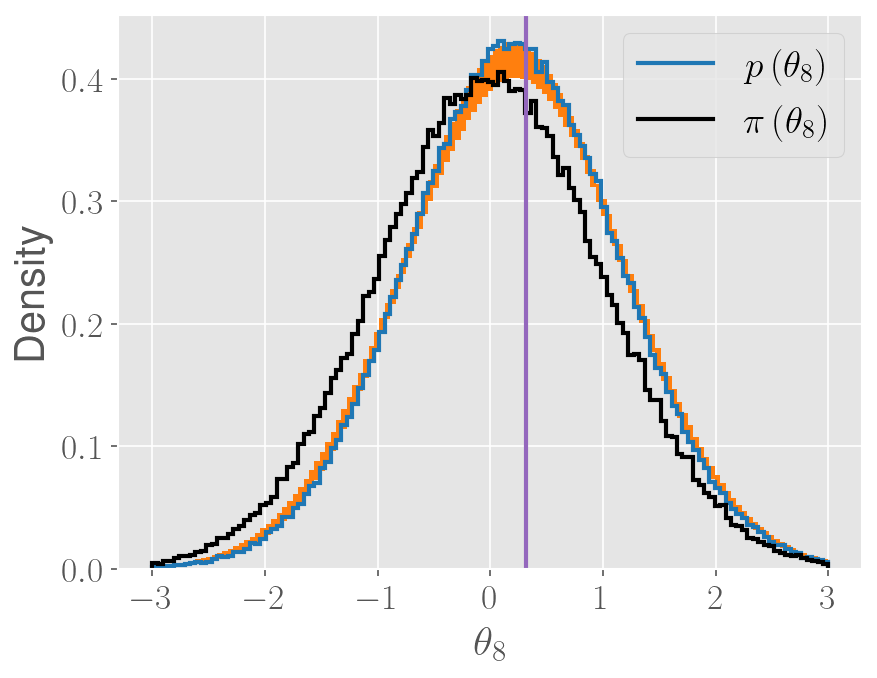}
  \caption{}
  \label{fig:example_1_case_2_full_posterior_nominal_and_errorbars_dim_8}
\end{subfigure}
\begin{subfigure}{.24\textwidth}
  \centering
  \includegraphics[width=\textwidth]{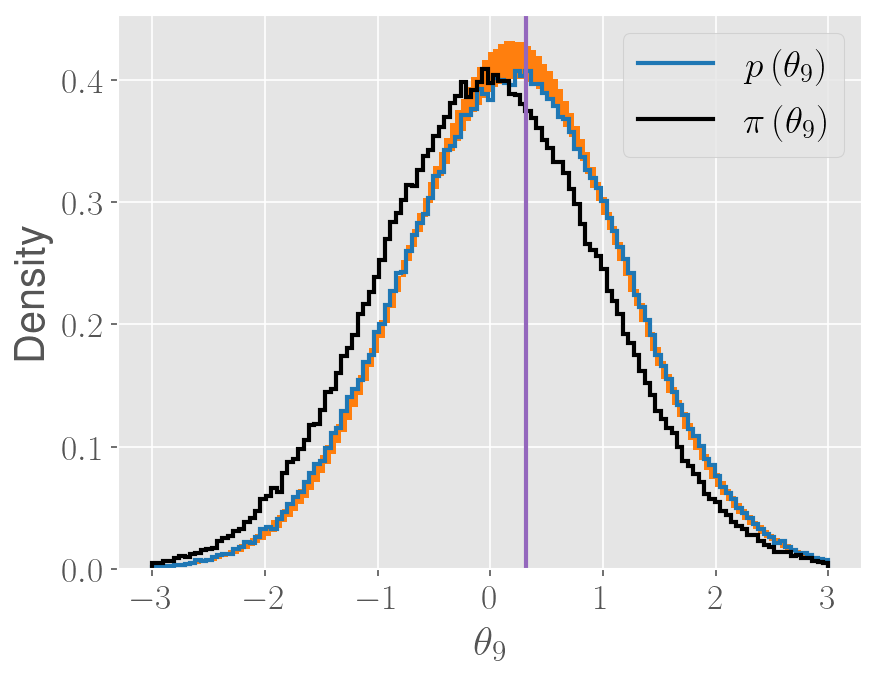}
  \caption{}
  \label{fig:example_1_case_2_full_posterior_nominal_and_errorbars_dim_9}
\end{subfigure}%
\begin{subfigure}{.24\textwidth}
  \centering
  \includegraphics[width=\textwidth]{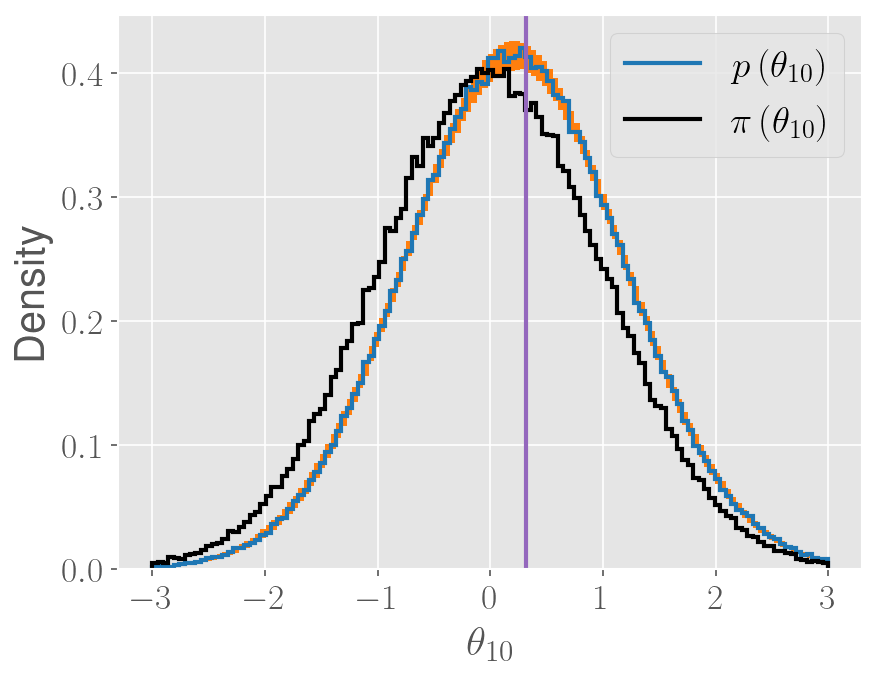}
  \caption{}
  \label{fig:example_1_case_2_full_posterior_nominal_and_errorbars_dim_10}
\end{subfigure}%
\begin{subfigure}{.24\textwidth}
  \centering
  \includegraphics[width=\textwidth]{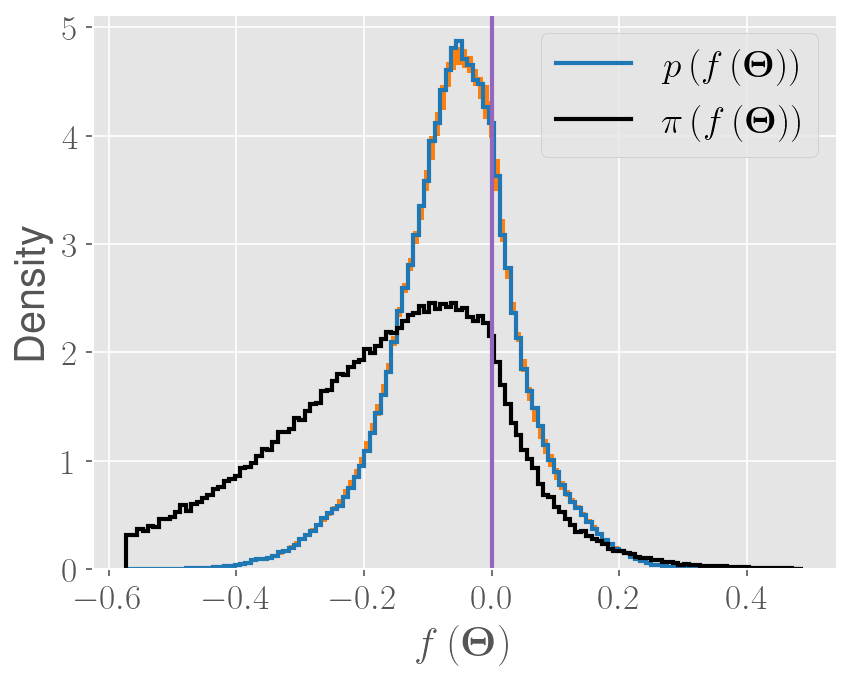}
  \caption{}
  \label{fig:example_1_case_2_full_output_nominal_and_errorbars}
\end{subfigure}%
\begin{subfigure}{.24\textwidth}
  \centering
  \includegraphics[width=\textwidth]{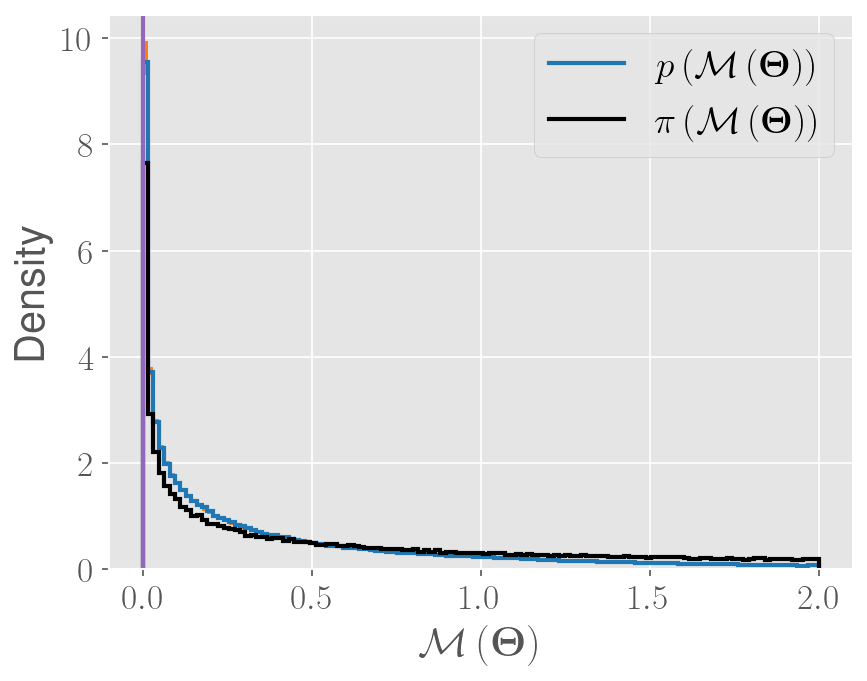}
  \caption{}
  \label{fig:example_1_case_2_full_misfit_nominal_and_errorbars}
\end{subfigure}
\caption{Plots of full posteriors for Case 2 of the polynomial test function (see Section~\ref{section:example_1_polynomial}). Figures (a) - (j) plot the posteriors for $ \theta_1 $ - $ \theta_{10} $, figure (k) plots the density of $ f \left( \boldsymbol{\Theta} \right) $ corresponding to the full posterior $ p \left( \boldsymbol{\Theta} \right) $, and figure (l) plots the density of $ \mathcal{M} \left( \boldsymbol{\Theta} \right) $ corresponding to the full posterior $ p \left( \boldsymbol{\Theta} \right) $. In each figure, the nominal calibrated posterior constructed using $ \mathfrak{C}_{\boldsymbol{\Theta}} $ is in blue, while the orange bands depict the $ 90 \% $ confidence interval of the posteriors constructed using the sample set replicates $ \mathfrak{C}_{\boldsymbol{\Theta}}^{(i)} $ is in orange. Additionally, the prior density is drawn in black, with the purple line indicating the observed value used for calibration.}
\label{fig:calibrated_full_posteriors_results_for_example_1_case_2}
\end{figure}

\begin{figure}[t!bhp]
\centering
\begin{subfigure}{.24\textwidth}
  \centering
  \includegraphics[width=\textwidth]{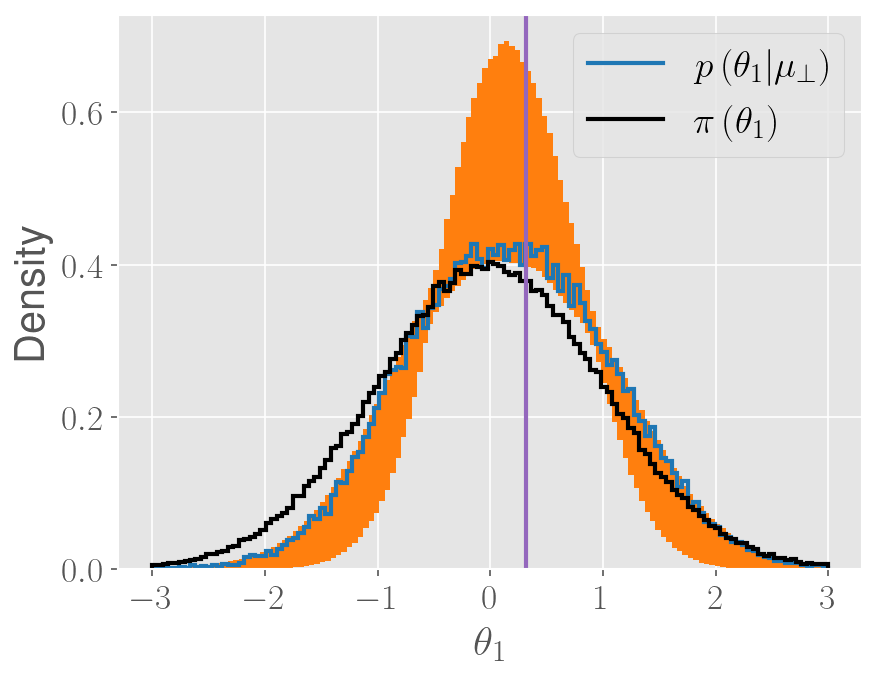}
  \caption{}
  \label{fig:example_1_case_2_conditional_posterior_nominal_and_errorbars_dim_1}
\end{subfigure}%
\begin{subfigure}{.24\textwidth}
  \centering
  \includegraphics[width=\textwidth]{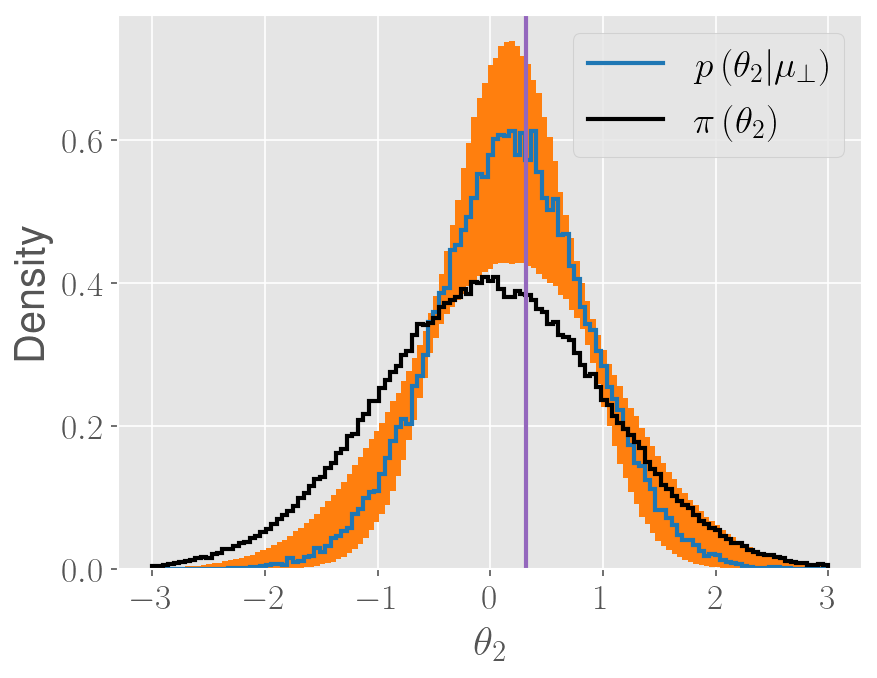}
  \caption{}
  \label{fig:example_1_case_2_conditional_posterior_nominal_and_errorbars_dim_2}
\end{subfigure}%
\begin{subfigure}{.24\textwidth}
  \centering
  \includegraphics[width=\textwidth]{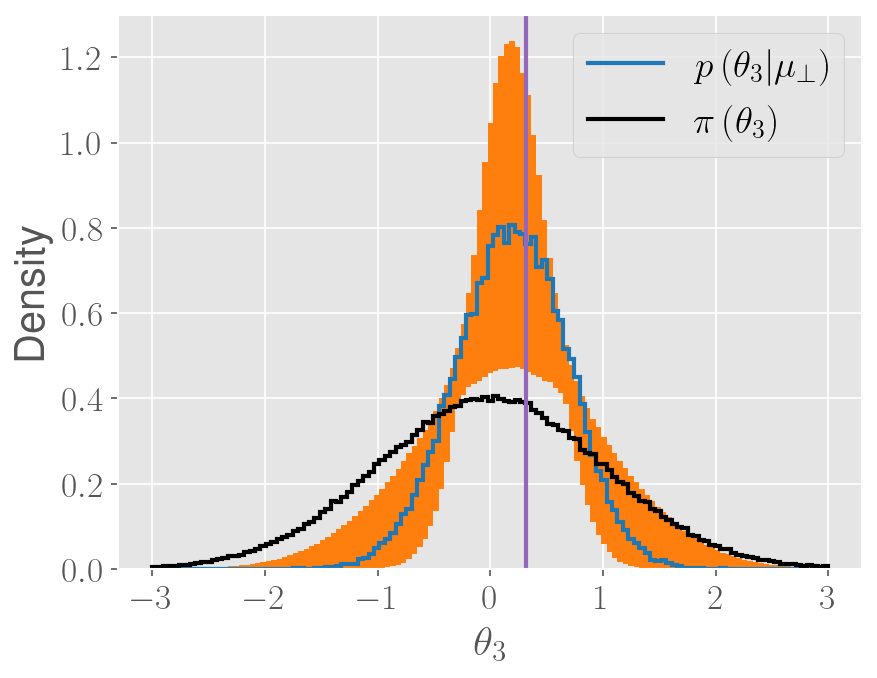}
  \caption{}
  \label{fig:example_1_case_2_conditional_posterior_nominal_and_errorbars_dim_3}
\end{subfigure}%
\begin{subfigure}{.24\textwidth}
  \centering
  \includegraphics[width=\textwidth]{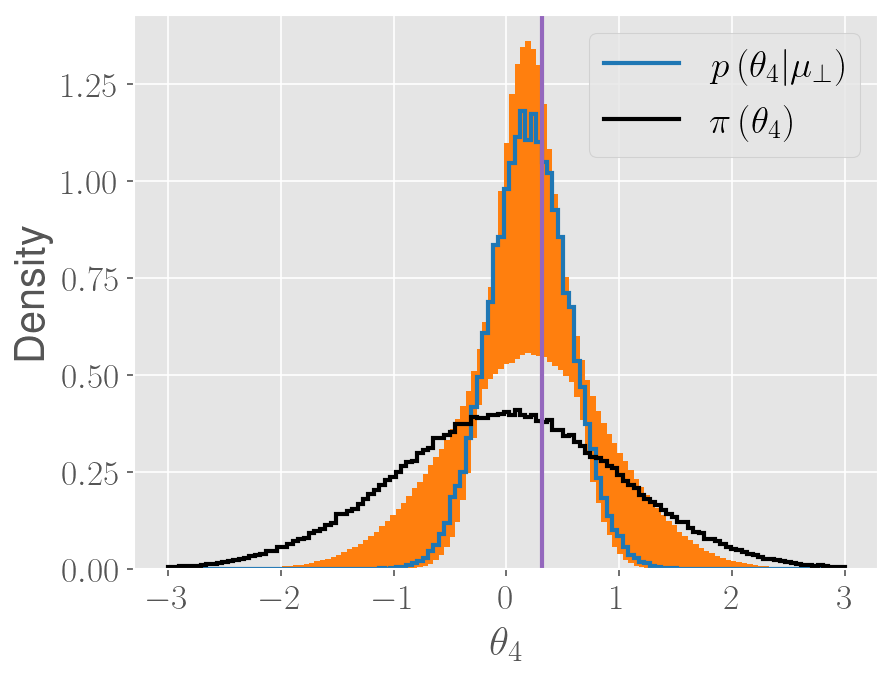}
  \caption{}
  \label{fig:example_1_case_2_conditional_posterior_nominal_and_errorbars_dim_4}
\end{subfigure}
\begin{subfigure}{.24\textwidth}
  \centering
  \includegraphics[width=\textwidth]{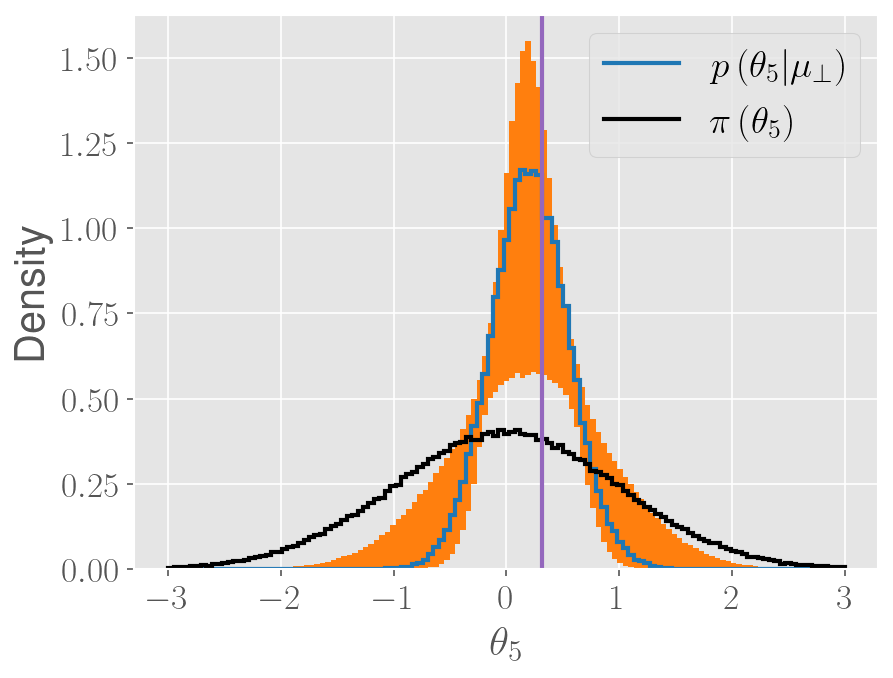}
  \caption{}
  \label{fig:example_1_case_2_conditional_posterior_nominal_and_errorbars_dim_5}
\end{subfigure}%
\begin{subfigure}{.24\textwidth}
  \centering
  \includegraphics[width=\textwidth]{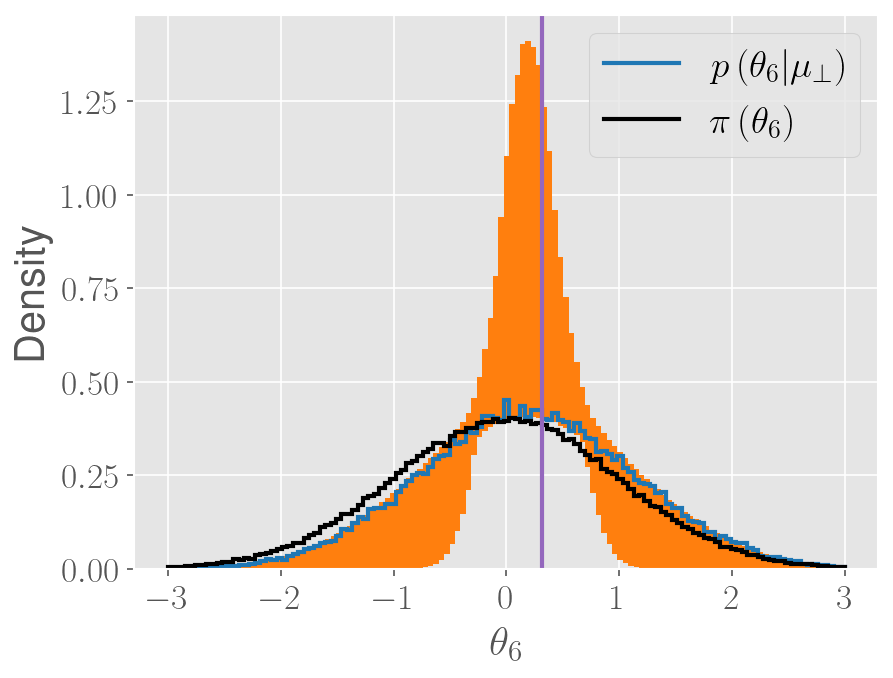}
  \caption{}
  \label{fig:example_1_case_2_conditional_posterior_nominal_and_errorbars_dim_6}
\end{subfigure}%
\begin{subfigure}{.24\textwidth}
  \centering
  \includegraphics[width=\textwidth]{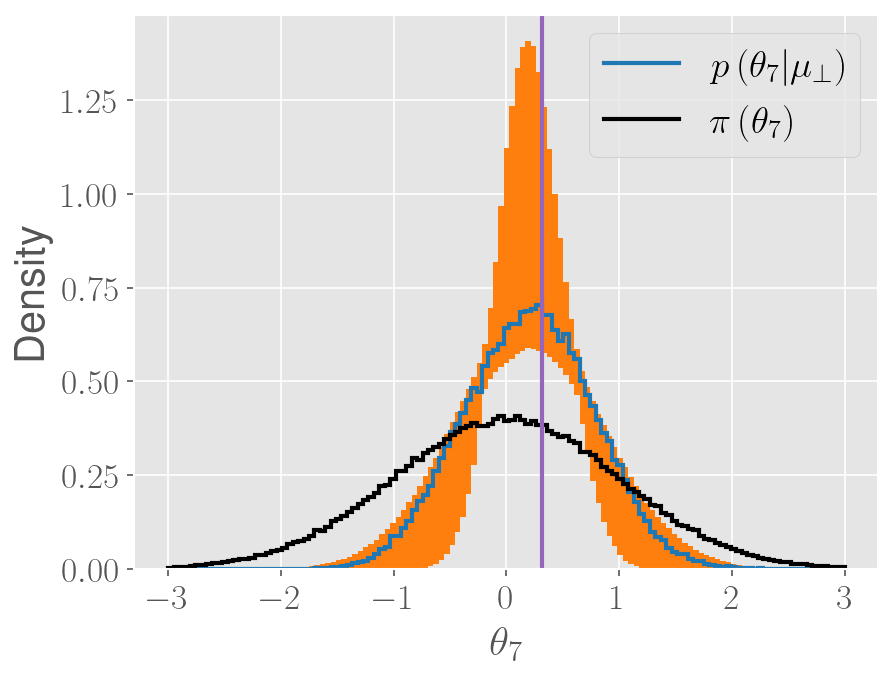}
  \caption{}
  \label{fig:example_1_case_2_conditional_posterior_nominal_and_errorbars_dim_7}
\end{subfigure}%
\begin{subfigure}{.24\textwidth}
  \centering
  \includegraphics[width=\textwidth]{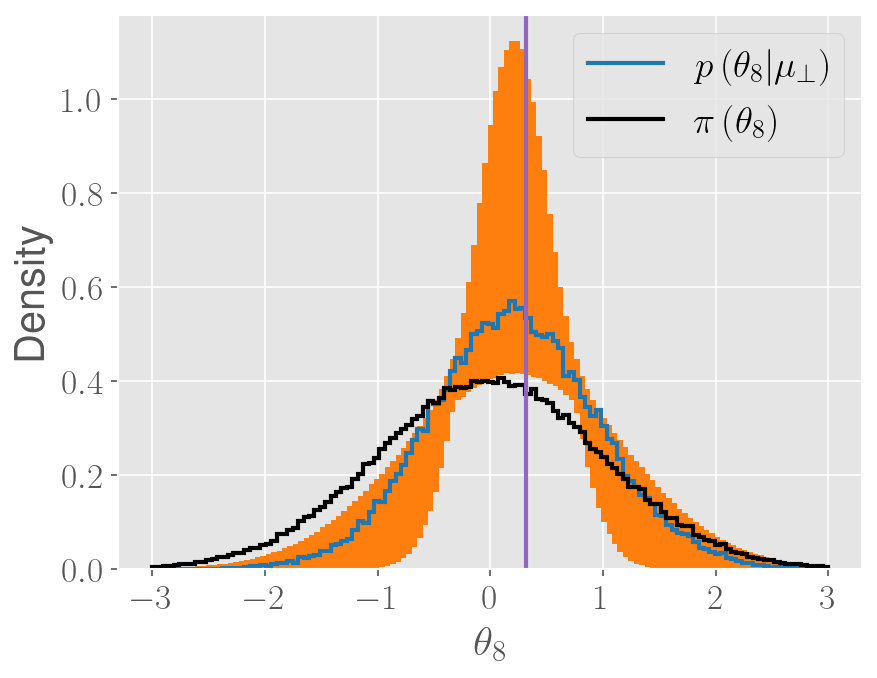}
  \caption{}
  \label{fig:example_1_case_2_conditional_posterior_nominal_and_errorbars_dim_8}
\end{subfigure}
\begin{subfigure}{.24\textwidth}
  \centering
  \includegraphics[width=\textwidth]{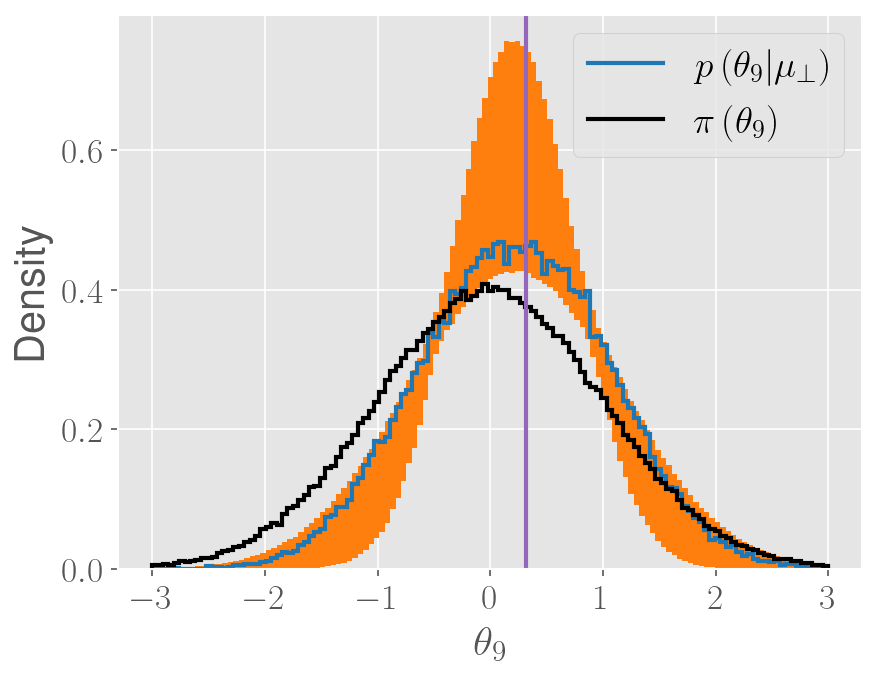}
  \caption{}
  \label{fig:example_1_case_2_conditional_posterior_nominal_and_errorbars_dim_9}
\end{subfigure}%
\begin{subfigure}{.24\textwidth}
  \centering
  \includegraphics[width=\textwidth]{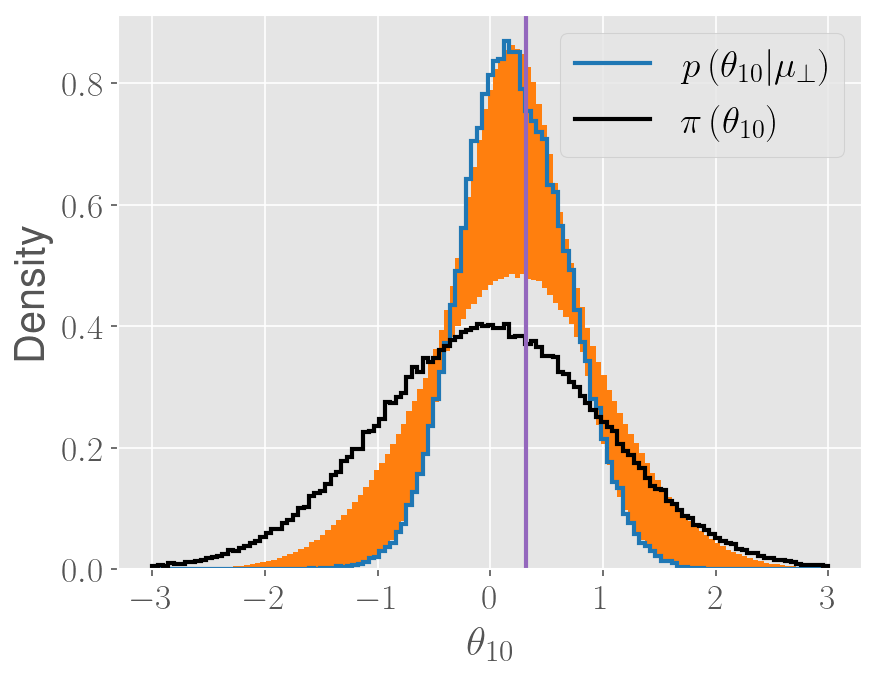}
  \caption{}
  \label{fig:example_1_case_2_conditional_posterior_nominal_and_errorbars_dim_10}
\end{subfigure}%
\begin{subfigure}{.24\textwidth}
  \centering
  \includegraphics[width=\textwidth]{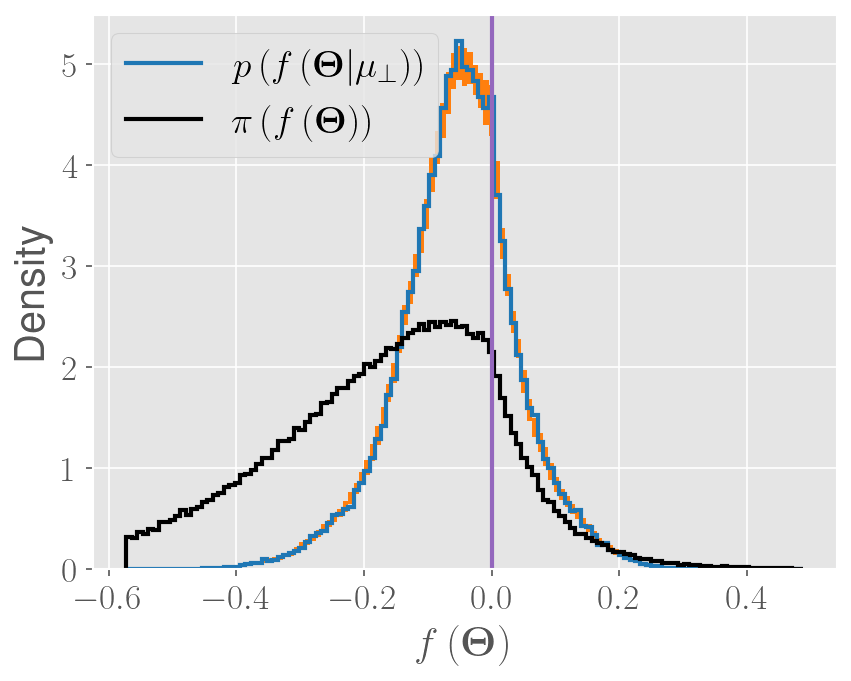}
  \caption{}
  \label{fig:example_1_case_2_conditional_output_nominal_and_errorbars}
\end{subfigure}%
\begin{subfigure}{.24\textwidth}
  \centering
  \includegraphics[width=\textwidth]{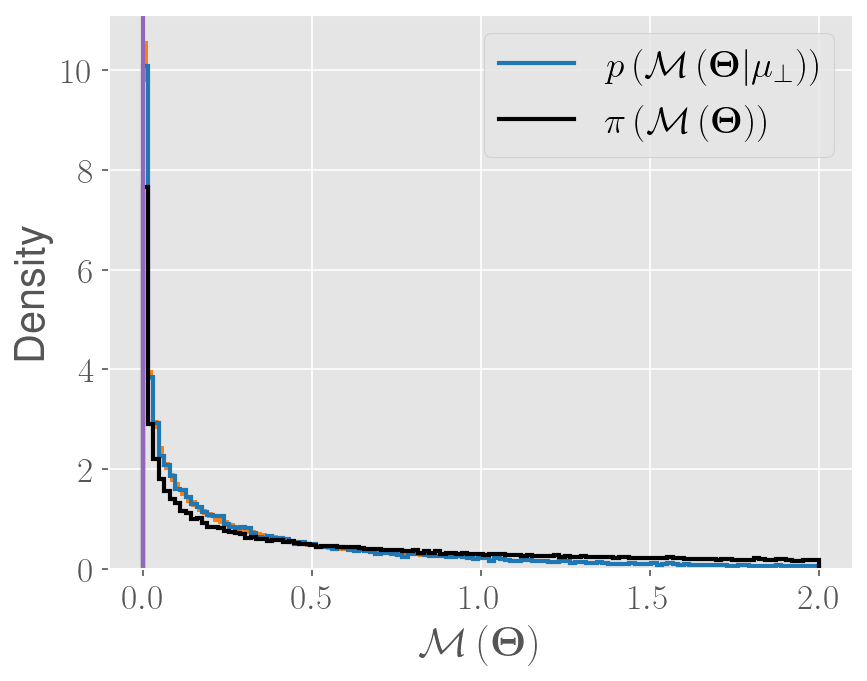}
  \caption{}
  \label{fig:example_1_case_2_conditional_misfit_nominal_and_errorbars}
\end{subfigure}
\caption{Plots of conditional active calibrated posteriors for Case 2 of the polynomial test function (see Section~\ref{section:example_1_polynomial}). Figures (a) - (j) plot the posteriors for $ \theta_1 $ - $ \theta_{10} $, figure (k) plots the density of $ f \left( \boldsymbol{\Theta} \right) $ corresponding to the conditional active calibrated posterior $ p \left( \boldsymbol{\Theta}_{\mu_{\perp}} \right) $, and figure (l) plots the density of $ \mathcal{M} \left( \boldsymbol{\Theta} \right) $ corresponding to the conditional active calibrated posterior $ p \left( \boldsymbol{\Theta}_{\mu_{\perp}} \right) $. In each figure, the nominal calibrated posterior constructed using $ \mathfrak{C}_{\boldsymbol{\Theta}} $ is in blue, while the orange bands depict the $ 90 \% $ confidence interval of the posteriors constructed using the sample set replicates $ \mathfrak{C}_{\boldsymbol{\Theta}}^{(i)} $ is in orange. Additionally, the prior density is drawn in black, with the purple line indicating the observed value used for calibration.}
\label{fig:calibrated_conditional_posteriors_results_for_example_1_case_2}
\end{figure}

Supplementary figures for the polynomial test function from Section~\ref{section:example_1_polynomial} are included here. The LIS projection vectors not included in Figure~\ref{fig:active_subspace_results_for_example_1} (i.e., eigenvectors 3 - 10) are shown in Figure~\ref{fig:example_1_case_1_all_projection_vectors} for Case 1 and Figure~\ref{fig:example_1_case_2_all_projection_vectors} for Case 2. Figures~\ref{fig:calibrated_full_posteriors_results_for_example_1_case_1} and~\ref{fig:calibrated_conditional_posteriors_results_for_example_1_case_1} collect all one-dimensional marginals as well as the output and misfit probability density functions corresponding to the full posterior and the conditional active posterior for Case 1, respectively. Figures~\ref{fig:calibrated_full_posteriors_results_for_example_1_case_2} and~\ref{fig:calibrated_conditional_posteriors_results_for_example_1_case_2} do the same for Case 2. Similar plots for Case 3 are omitted, since the problem is too high dimensional.

\subsection{Supplementary Figures for the Vehicle Side Impact Problem}
\label{appendix:additional_figures_example_2}

\begin{figure}[t!bhp]
\centering
\begin{subfigure}{.24\textwidth}
  \centering
  \includegraphics[width=\textwidth]{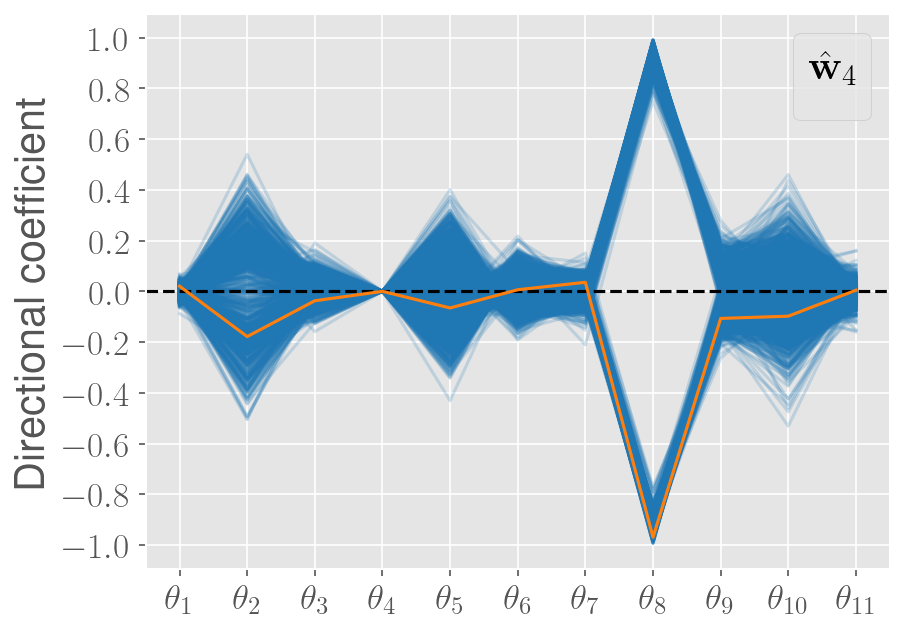}
  \caption{}
  \label{fig:example_2_case_1_projection_vector_dim_4}
\end{subfigure}%
\begin{subfigure}{.24\textwidth}
  \centering
  \includegraphics[width=\textwidth]{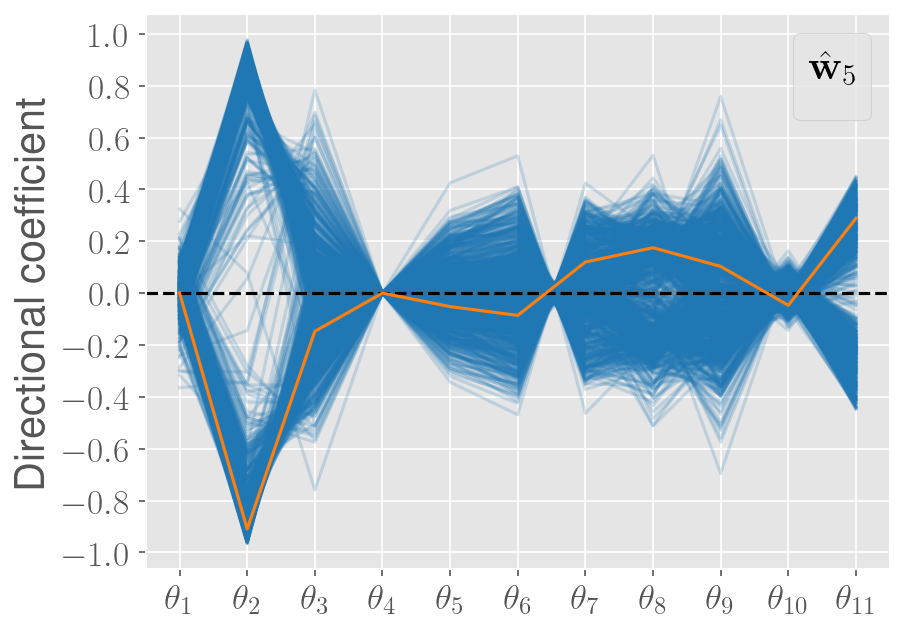}
  \caption{}
  \label{fig:example_2_case_1_projection_vector_dim_5}
\end{subfigure}%
\begin{subfigure}{.24\textwidth}
  \centering
  \includegraphics[width=\textwidth]{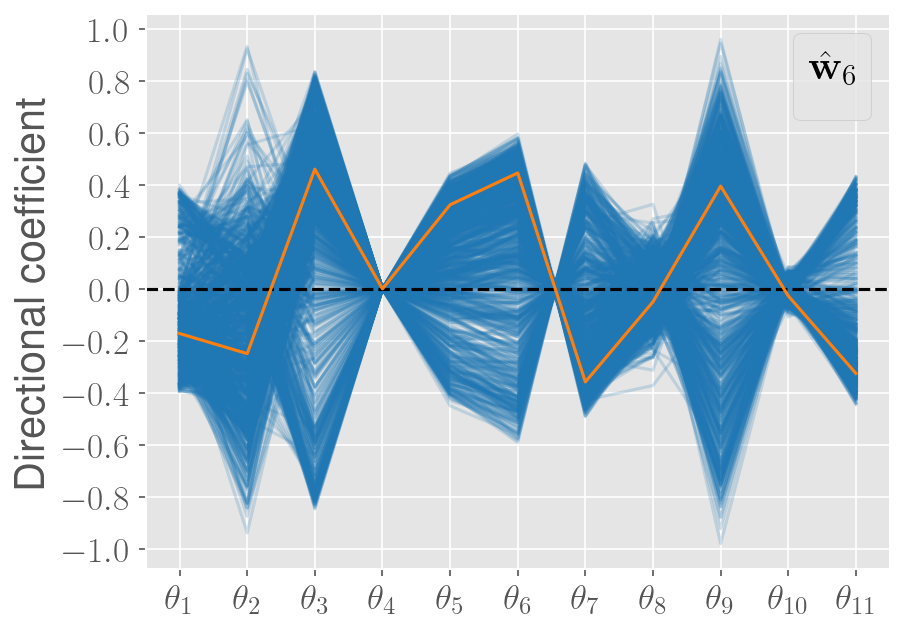}
  \caption{}
  \label{fig:example_2_case_1_projection_vector_dim_6}
\end{subfigure}%
\begin{subfigure}{.24\textwidth}
  \centering
  \includegraphics[width=\textwidth]{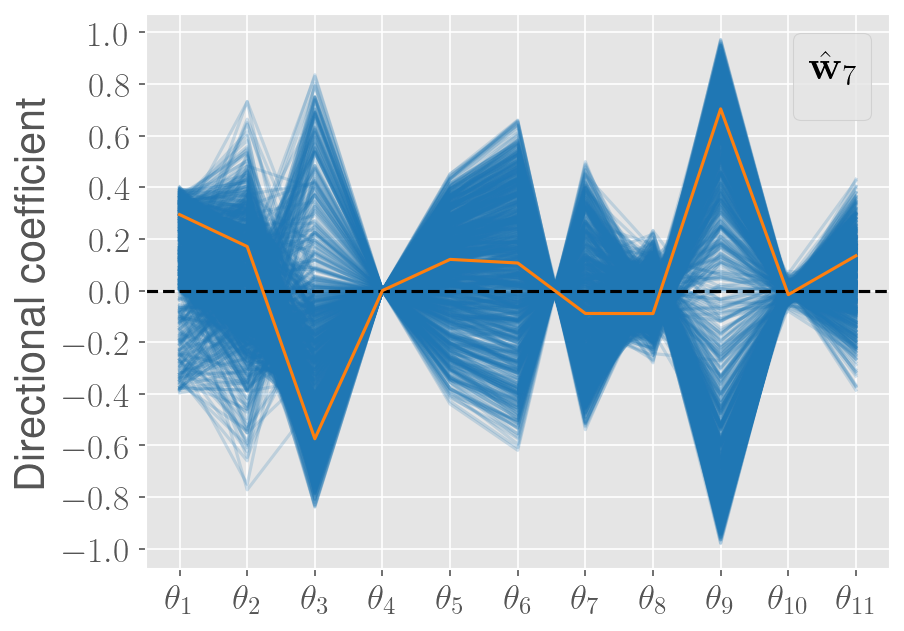}
  \caption{}
  \label{fig:example_2_case_1_projection_vector_dim_7}
\end{subfigure}
\begin{subfigure}{.24\textwidth}
  \centering
  \includegraphics[width=\textwidth]{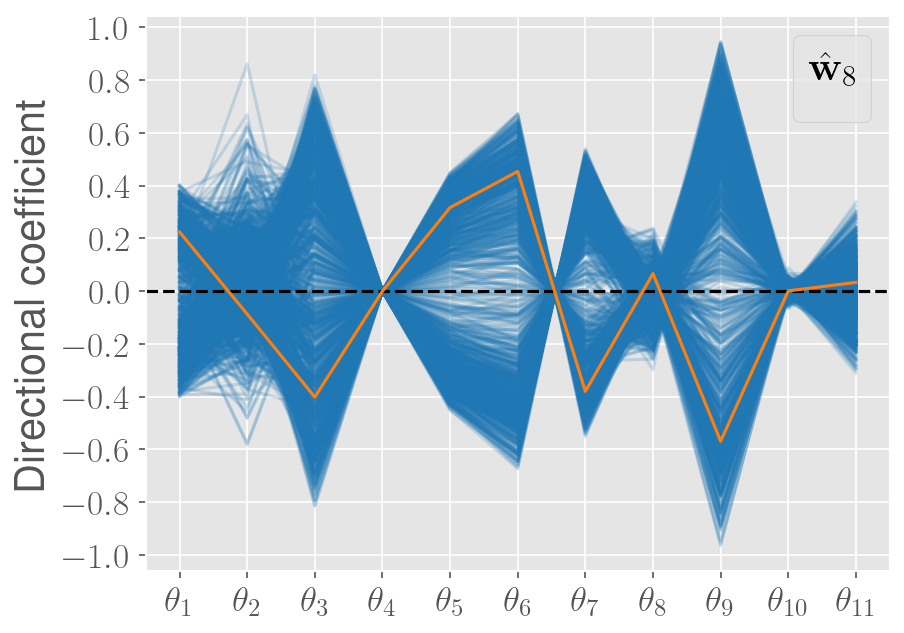}
  \caption{}
  \label{fig:example_2_case_1_projection_vector_dim_8}
\end{subfigure}%
\begin{subfigure}{.24\textwidth}
  \centering
  \includegraphics[width=\textwidth]{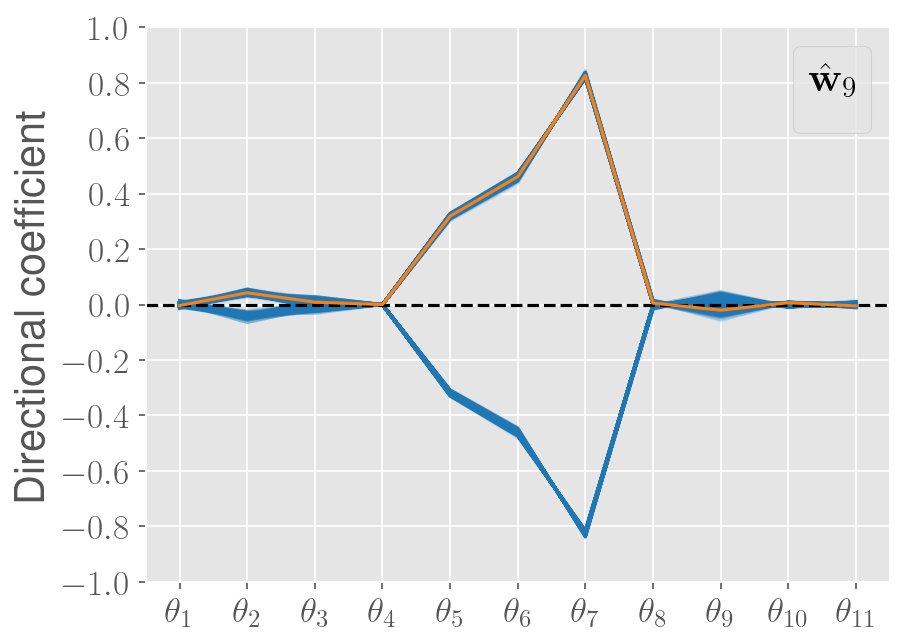}
  \caption{}
  \label{fig:example_2_case_1_projection_vector_dim_9}
\end{subfigure}%
\begin{subfigure}{.24\textwidth}
  \centering
  \includegraphics[width=\textwidth]{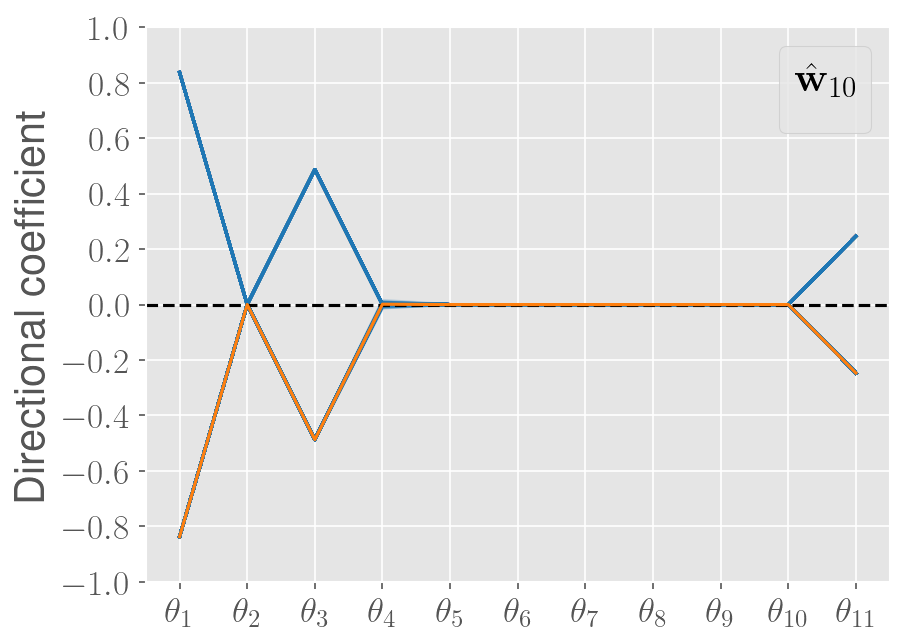}
  \caption{}
  \label{fig:example_2_case_1_projection_vector_dim_10}
\end{subfigure}%
\begin{subfigure}{.24\textwidth}
  \centering
  \includegraphics[width=\textwidth]{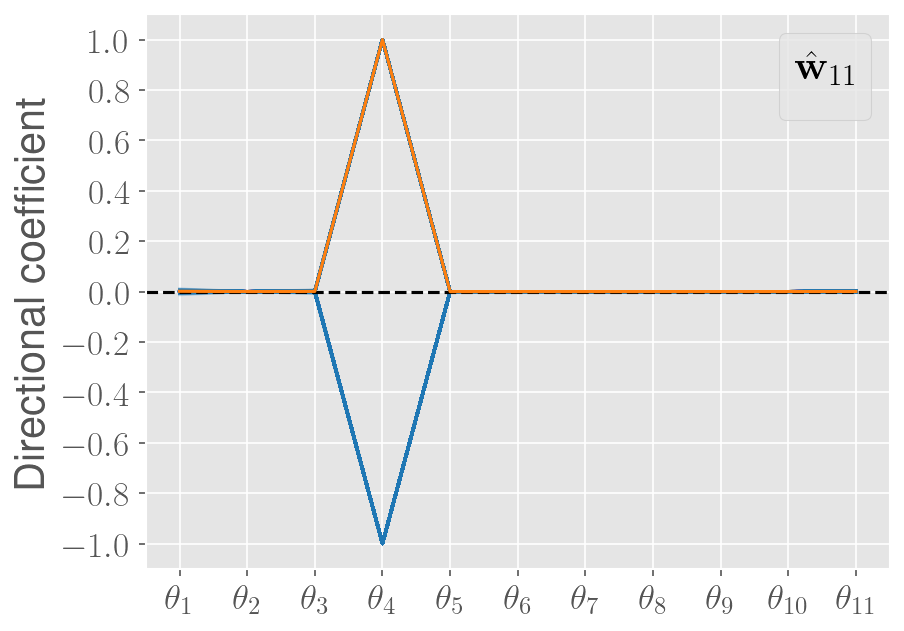}
  \caption{}
  \label{fig:example_2_case_1_projection_vector_dim_11}
\end{subfigure}
\caption{Plots of $ \hat{\mathbf{w}}_i $, $ i = 4, \dots, 11 $ for Case 1 of the vehicle side impact test problem (see Section~\ref{section:example_2_vehicle_side_impact}). The nominal vector is in orange, and 1000 bootstrapped replicates are in blue. $ \hat{\mathbf{w}}_1 $, $ \hat{\mathbf{w}}_2 $, and $ \hat{\mathbf{w}}_3 $ are in Figure~\ref{fig:active_subspace_results_for_example_2}.}
\label{fig:example_2_case_1_all_projection_vectors}
\end{figure}

\begin{figure}[t!bhp]
\centering
\begin{subfigure}{.24\textwidth}
  \centering
  \includegraphics[width=\textwidth]{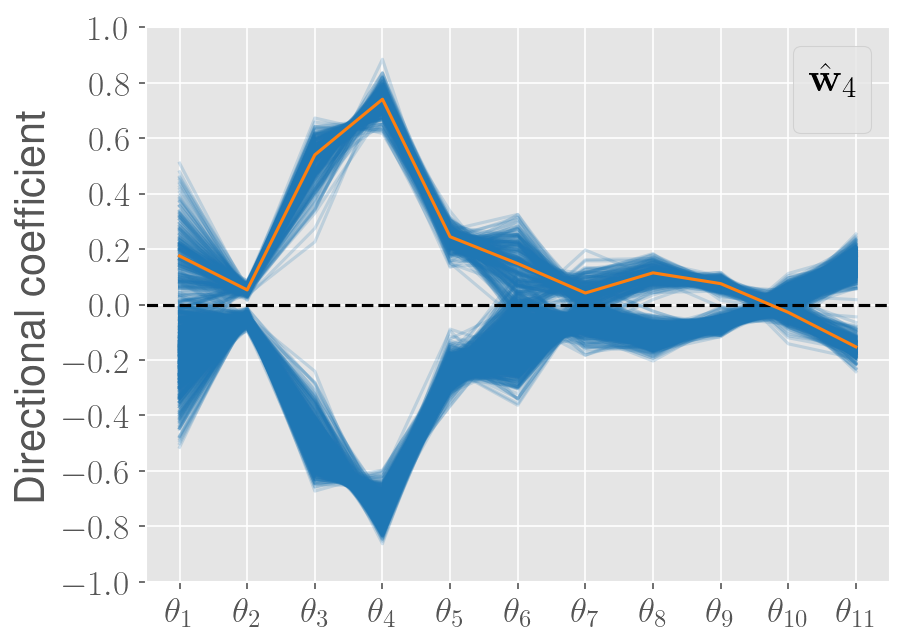}
  \caption{}
  \label{fig:example_2_case_2_projection_vector_dim_4}
\end{subfigure}%
\begin{subfigure}{.24\textwidth}
  \centering
  \includegraphics[width=\textwidth]{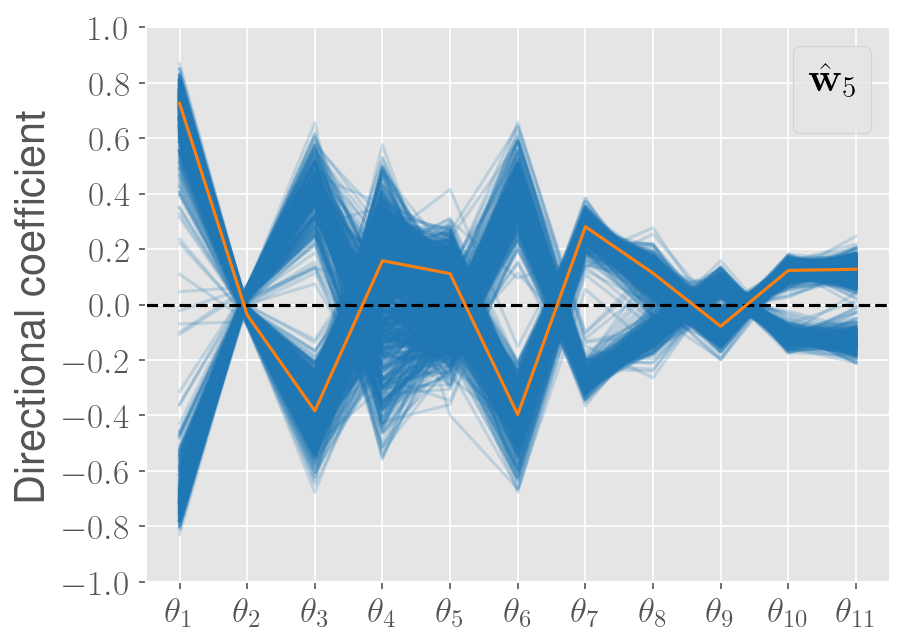}
  \caption{}
  \label{fig:example_2_case_2_projection_vector_dim_5}
\end{subfigure}%
\begin{subfigure}{.24\textwidth}
  \centering
  \includegraphics[width=\textwidth]{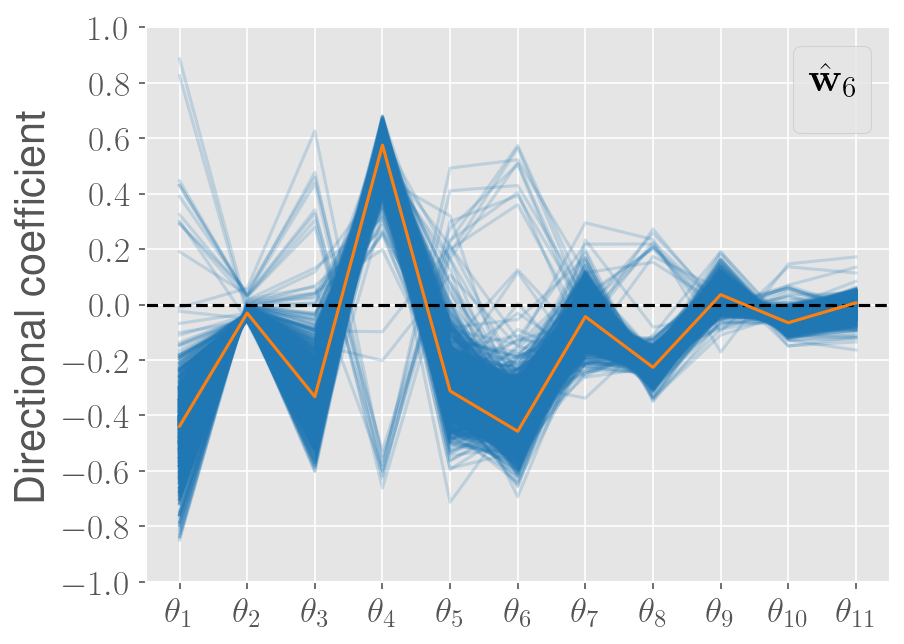}
  \caption{}
  \label{fig:example_2_case_2_projection_vector_dim_6}
\end{subfigure}%
\begin{subfigure}{.24\textwidth}
  \centering
  \includegraphics[width=\textwidth]{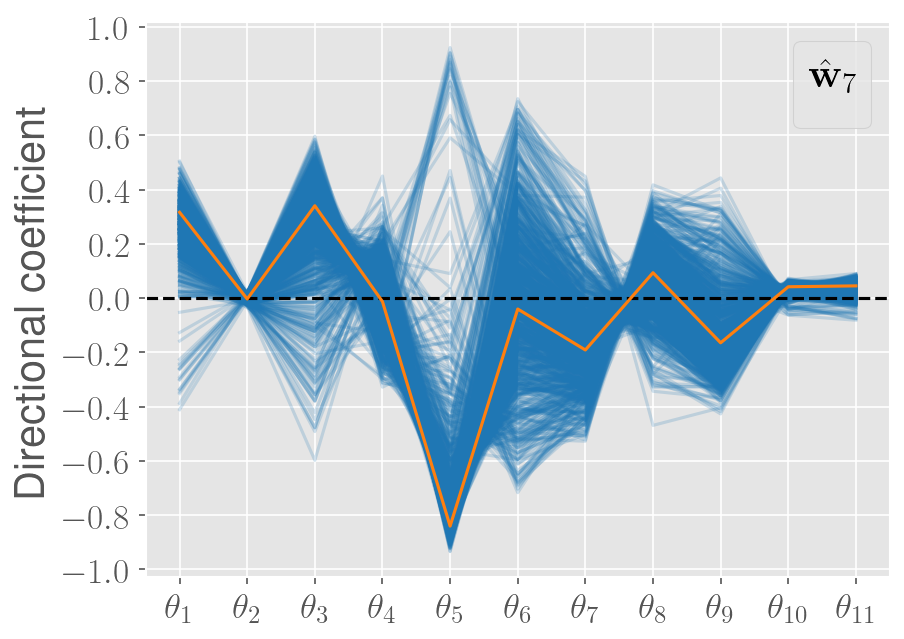}
  \caption{}
  \label{fig:example_2_case_2_projection_vector_dim_7}
\end{subfigure}
\begin{subfigure}{.24\textwidth}
  \centering
  \includegraphics[width=\textwidth]{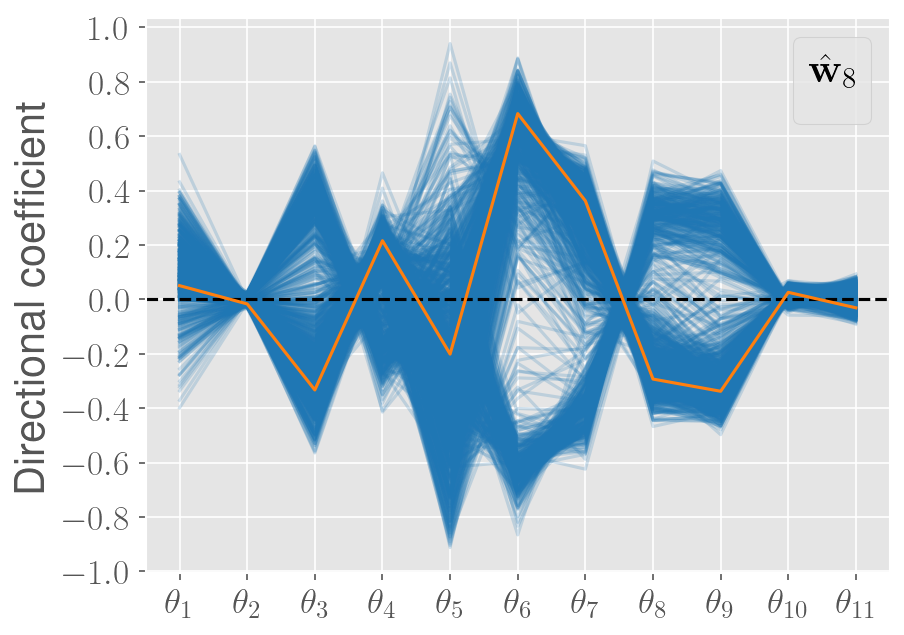}
  \caption{}
  \label{fig:example_2_case_2_projection_vector_dim_8}
\end{subfigure}%
\begin{subfigure}{.24\textwidth}
  \centering
  \includegraphics[width=\textwidth]{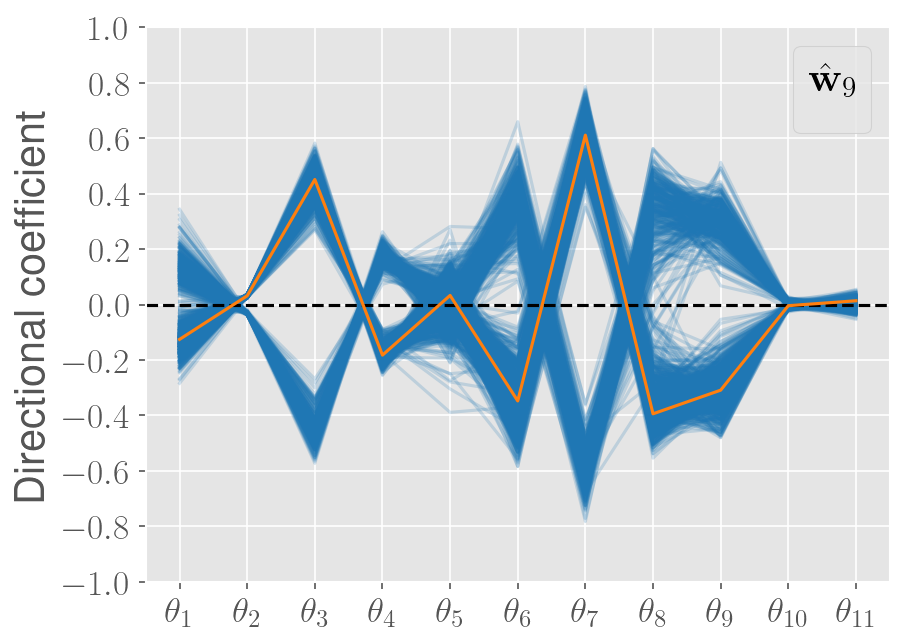}
  \caption{}
  \label{fig:example_2_case_2_projection_vector_dim_9}
\end{subfigure}%
\begin{subfigure}{.24\textwidth}
  \centering
  \includegraphics[width=\textwidth]{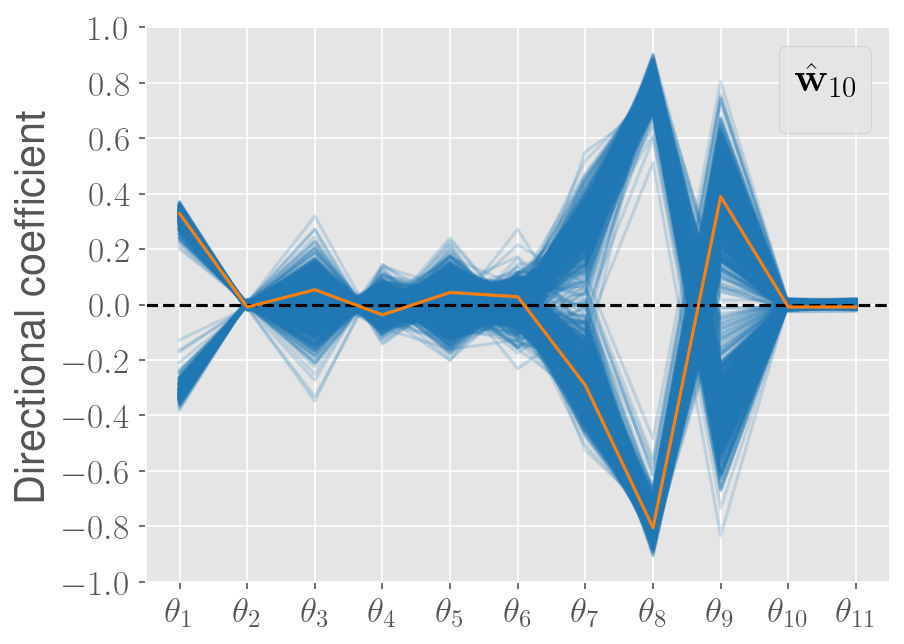}
  \caption{}
  \label{fig:example_2_case_2_projection_vector_dim_10}
\end{subfigure}%
\begin{subfigure}{.24\textwidth}
  \centering
  \includegraphics[width=\textwidth]{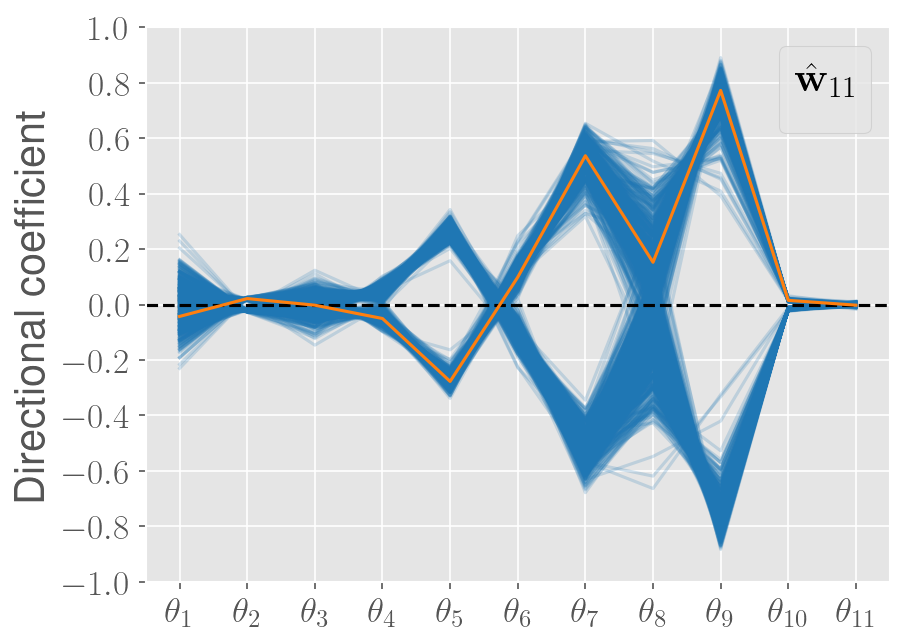}
  \caption{}
  \label{fig:example_2_case_2_projection_vector_dim_11}
\end{subfigure}
\caption{Plots of $ \hat{\mathbf{w}}_i $, $ i = 4, \dots, 11 $ for Case 2 of the vehicle side impact test problem (see Section~\ref{section:example_2_vehicle_side_impact}). The nominal vector is in orange, and 1000 bootstrapped replicates are in blue. $ \hat{\mathbf{w}}_1 $ and $ \hat{\mathbf{w}}_2 $ are in Figure~\ref{fig:active_subspace_results_for_example_2}.}
\label{fig:example_2_case_2_all_projection_vectors}
\end{figure}

\begin{figure}[t!bhp]
\centering
\begin{subfigure}{.24\textwidth}
  \centering
  \includegraphics[width=\textwidth]{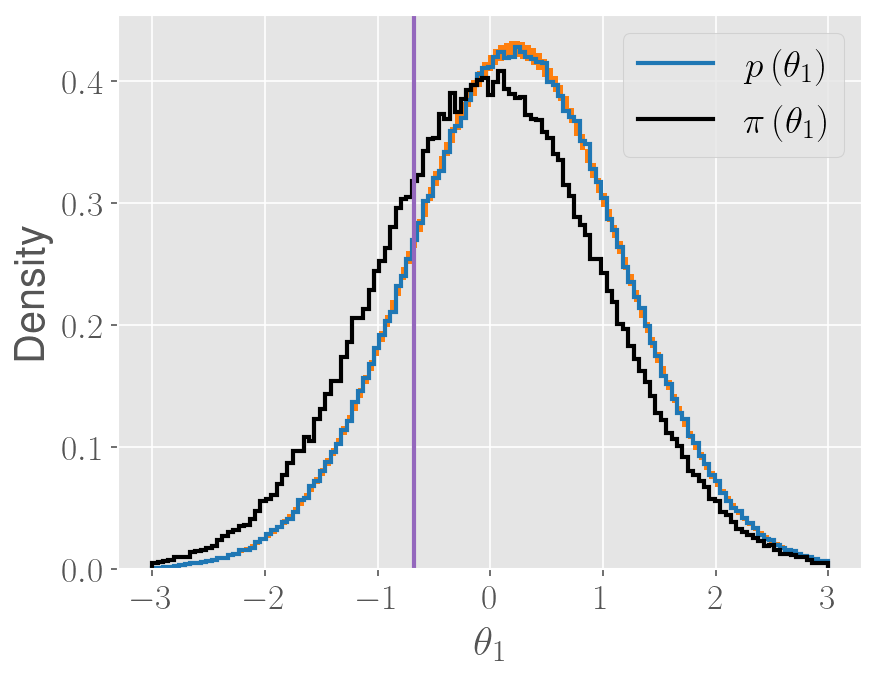}
  \caption{}
  \label{fig:example_2_case_1_full_posterior_nominal_and_errorbars_dim_1}
\end{subfigure}%
\begin{subfigure}{.24\textwidth}
  \centering
  \includegraphics[width=\textwidth]{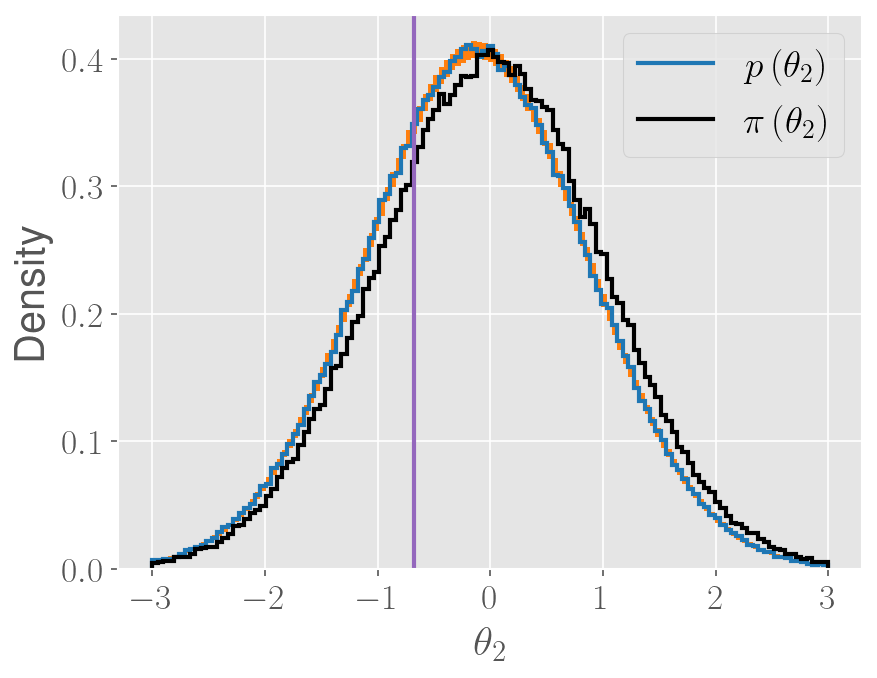}
  \caption{}
  \label{fig:example_2_case_1_full_posterior_nominal_and_errorbars_dim_2}
\end{subfigure}%
\begin{subfigure}{.24\textwidth}
  \centering
  \includegraphics[width=\textwidth]{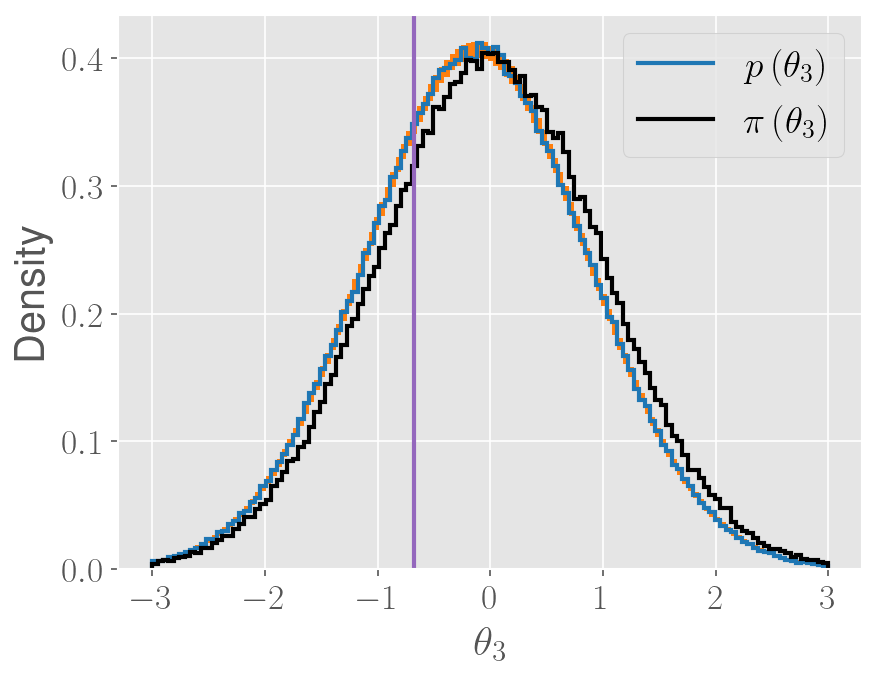}
  \caption{}
  \label{fig:example_2_case_1_full_posterior_nominal_and_errorbars_dim_3}
\end{subfigure}%
\begin{subfigure}{.24\textwidth}
  \centering
  \includegraphics[width=\textwidth]{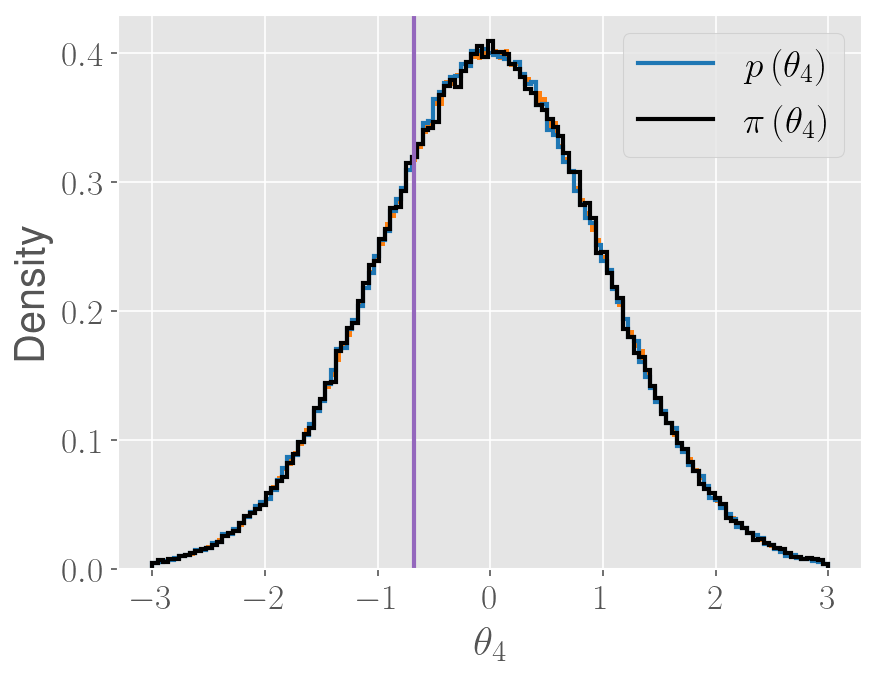}
  \caption{}
  \label{fig:example_2_case_1_full_posterior_nominal_and_errorbars_dim_4}
\end{subfigure}
\begin{subfigure}{.24\textwidth}
  \centering
  \includegraphics[width=\textwidth]{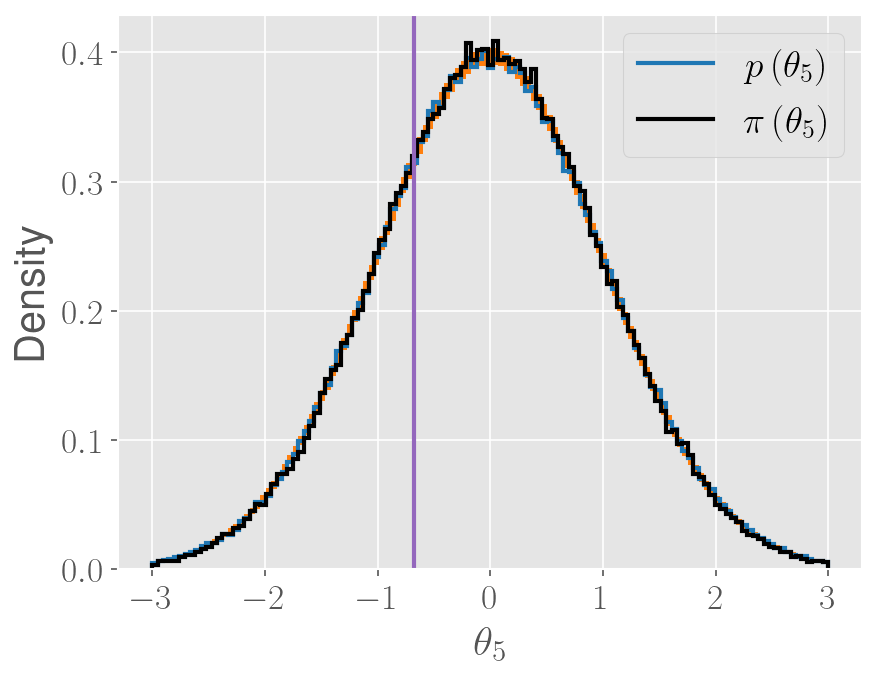}
  \caption{}
  \label{fig:example_2_case_1_full_posterior_nominal_and_errorbars_dim_5}
\end{subfigure}%
\begin{subfigure}{.24\textwidth}
  \centering
  \includegraphics[width=\textwidth]{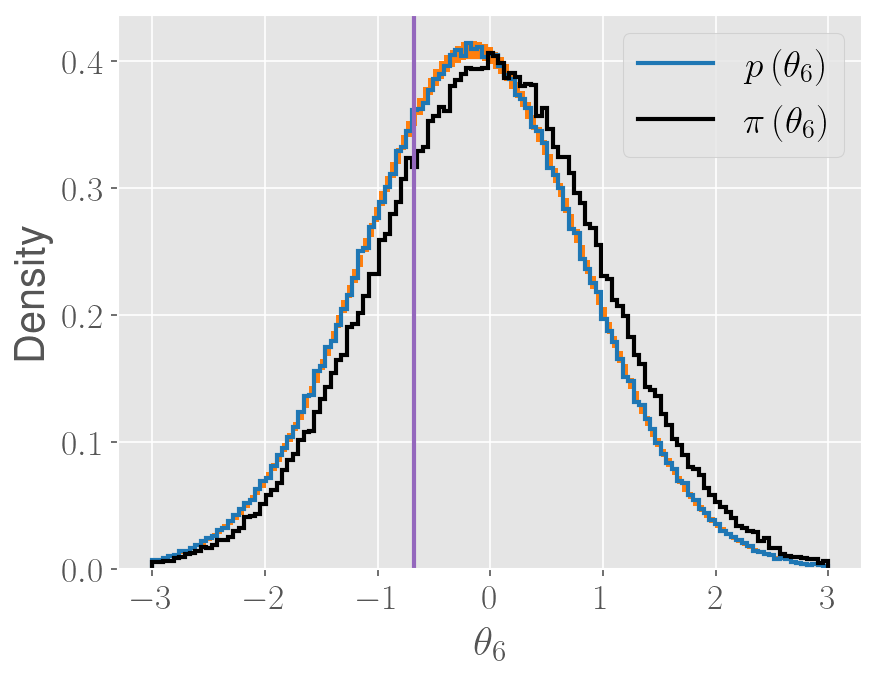}
  \caption{}
  \label{fig:example_2_case_1_full_posterior_nominal_and_errorbars_dim_6}
\end{subfigure}%
\begin{subfigure}{.24\textwidth}
  \centering
  \includegraphics[width=\textwidth]{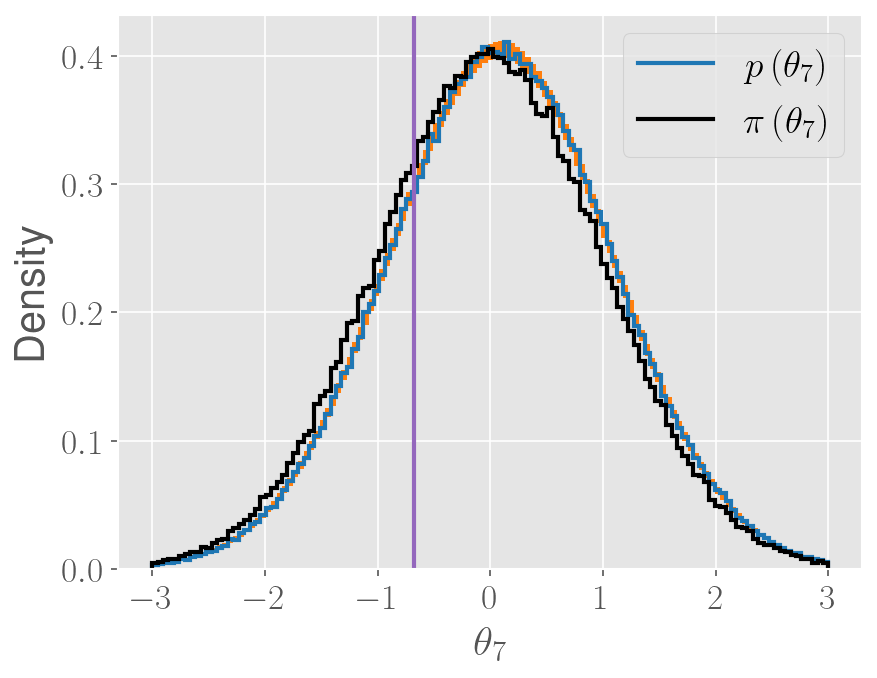}
  \caption{}
  \label{fig:example_2_case_1_full_posterior_nominal_and_errorbars_dim_7}
\end{subfigure}%
\begin{subfigure}{.24\textwidth}
  \centering
  \includegraphics[width=\textwidth]{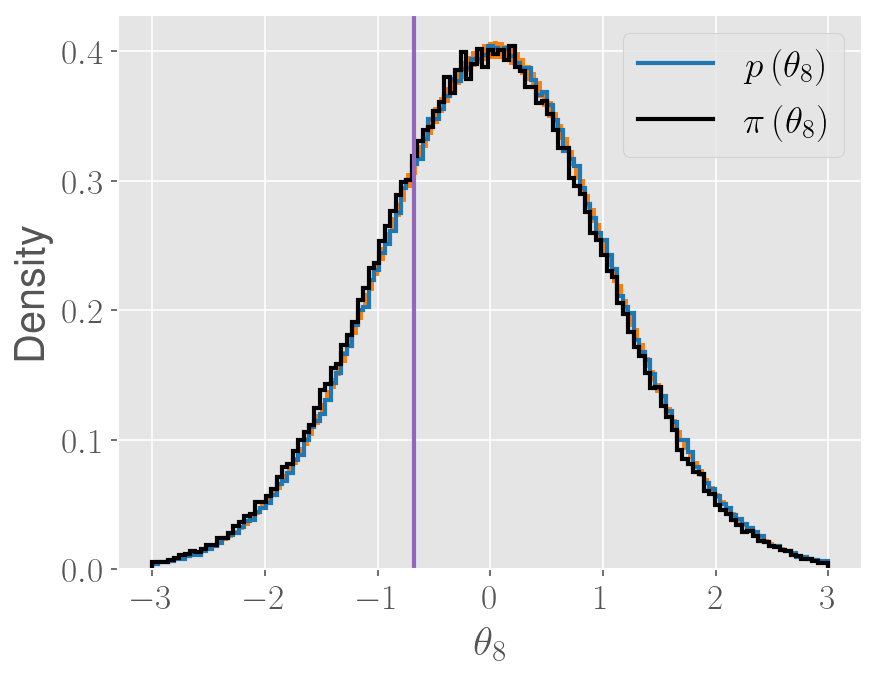}
  \caption{}
  \label{fig:example_2_case_1_full_posterior_nominal_and_errorbars_dim_8}
\end{subfigure}
\begin{subfigure}{.24\textwidth}
  \centering
  \includegraphics[width=\textwidth]{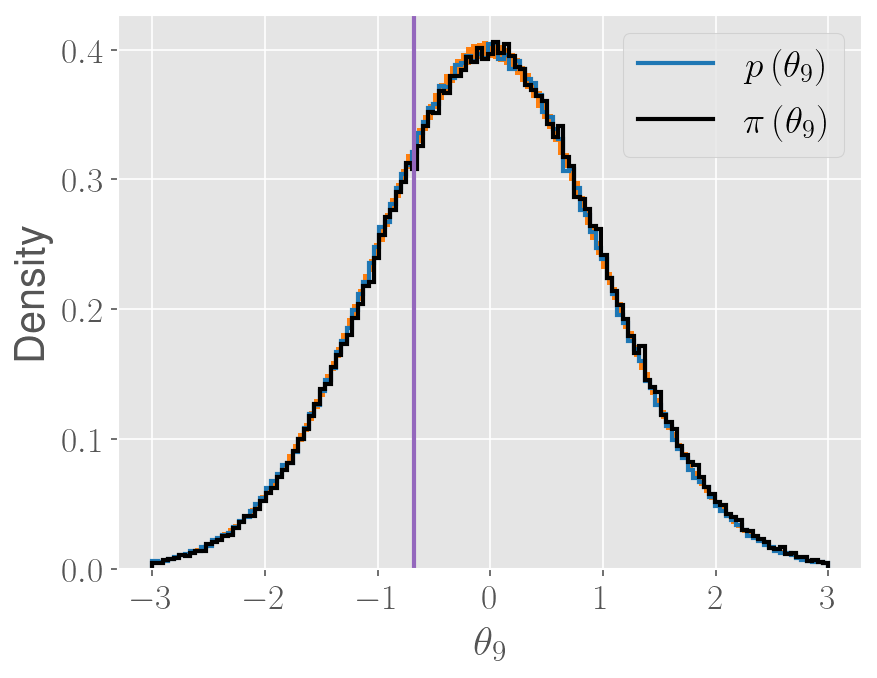}
  \caption{}
  \label{fig:example_2_case_1_full_posterior_nominal_and_errorbars_dim_9}
\end{subfigure}%
\begin{subfigure}{.24\textwidth}
  \centering
  \includegraphics[width=\textwidth]{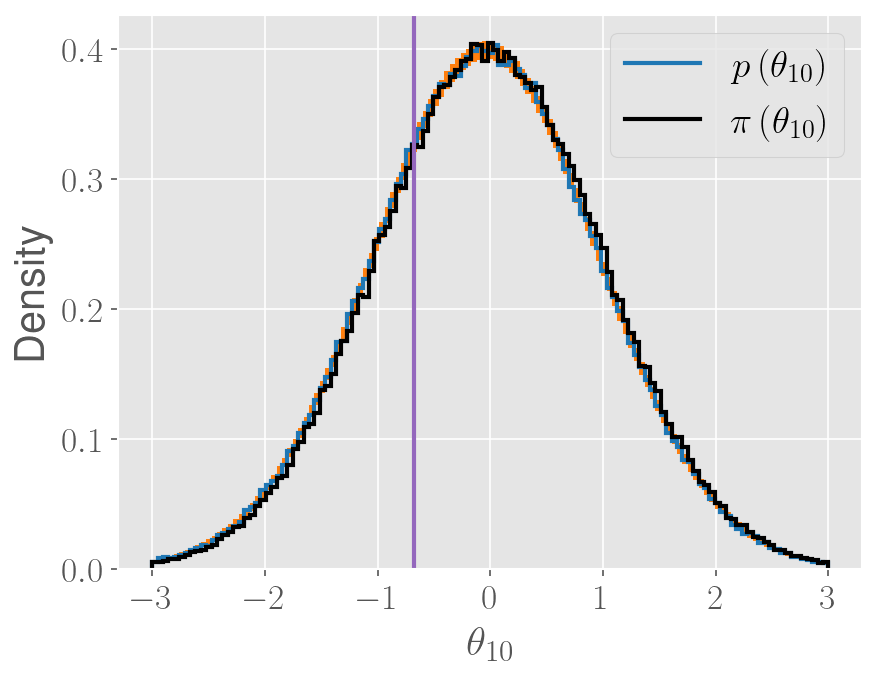}
  \caption{}
  \label{fig:example_2_case_1_full_posterior_nominal_and_errorbars_dim_10}
\end{subfigure}%
\begin{subfigure}{.24\textwidth}
  \centering
  \includegraphics[width=\textwidth]{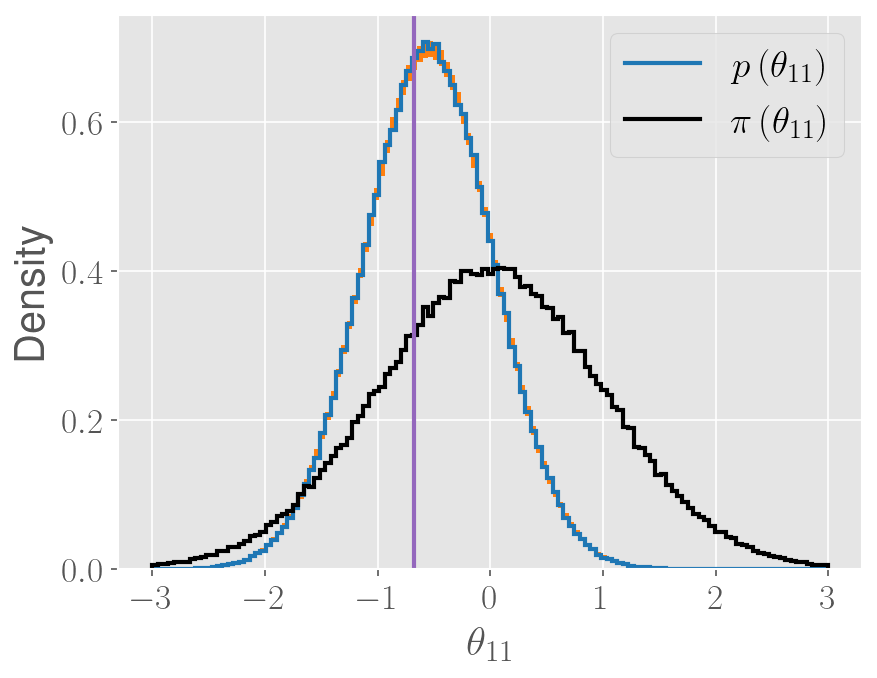}
  \caption{}
  \label{fig:example_2_case_1_full_posterior_nominal_and_errorbars_dim_11}
\end{subfigure}%
\begin{subfigure}{.24\textwidth}
  \centering
  \includegraphics[width=\textwidth]{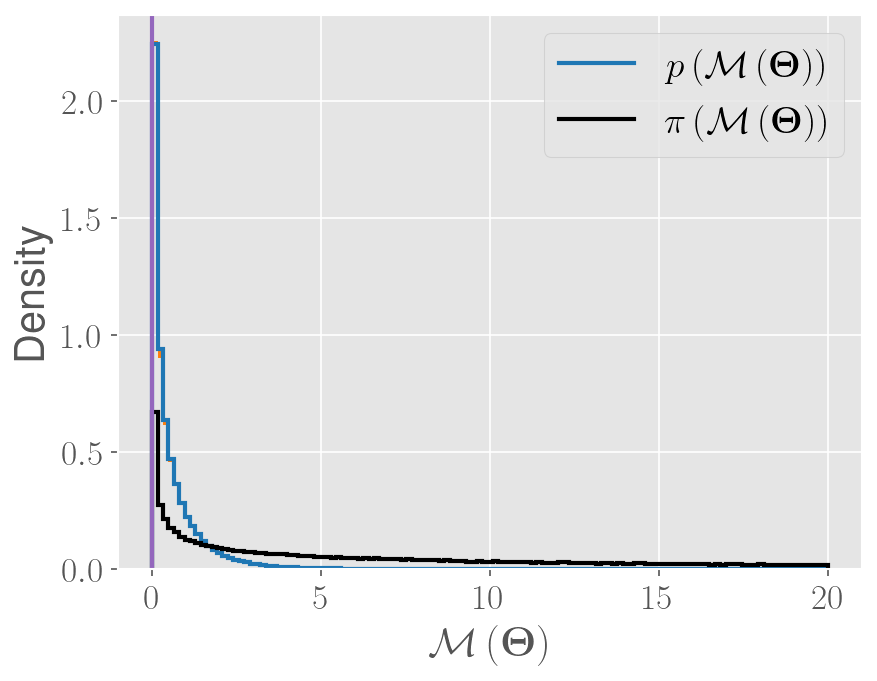}
  \caption{}
  \label{fig:example_2_case_1_full_misfit_nominal_and_errorbars}
\end{subfigure}
\caption{Plots of full posteriors for Case 1 of the vehicle side impact test problem (see Section~\ref{section:example_2_vehicle_side_impact}). Figures (a) - (k) plot the posteriors for $ \theta_1 $ - $ \theta_{11} $, and figure (l) plots the density of $ \mathcal{M} \left( \boldsymbol{\Theta} \right) $ corresponding to the full posterior $ p \left( \boldsymbol{\Theta} \right) $. In each figure, the nominal calibrated posterior constructed using $ \mathfrak{C}_{\boldsymbol{\Theta}} $ is in blue, while the orange bands depict the $ 90 \% $ confidence interval of the posteriors constructed using the sample set replicates $ \mathfrak{C}_{\boldsymbol{\Theta}}^{(i)} $ is in orange. Additionally, the prior density is drawn in black, with the purple line indicating the observed value used for calibration.}
\label{fig:calibrated_full_posteriors_results_for_example_2_case_1}
\end{figure}

\begin{figure}[t!bhp]
\centering
\begin{subfigure}{.24\textwidth}
  \centering
  \includegraphics[width=\textwidth]{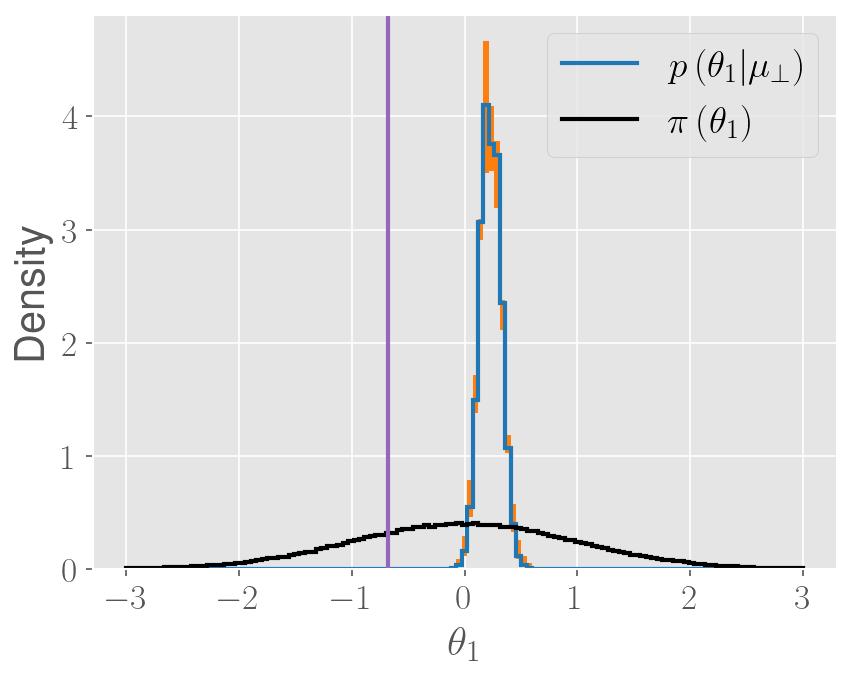}
  \caption{}
  \label{fig:example_2_case_1_conditional_posterior_nominal_and_errorbars_dim_1}
\end{subfigure}%
\begin{subfigure}{.24\textwidth}
  \centering
  \includegraphics[width=\textwidth]{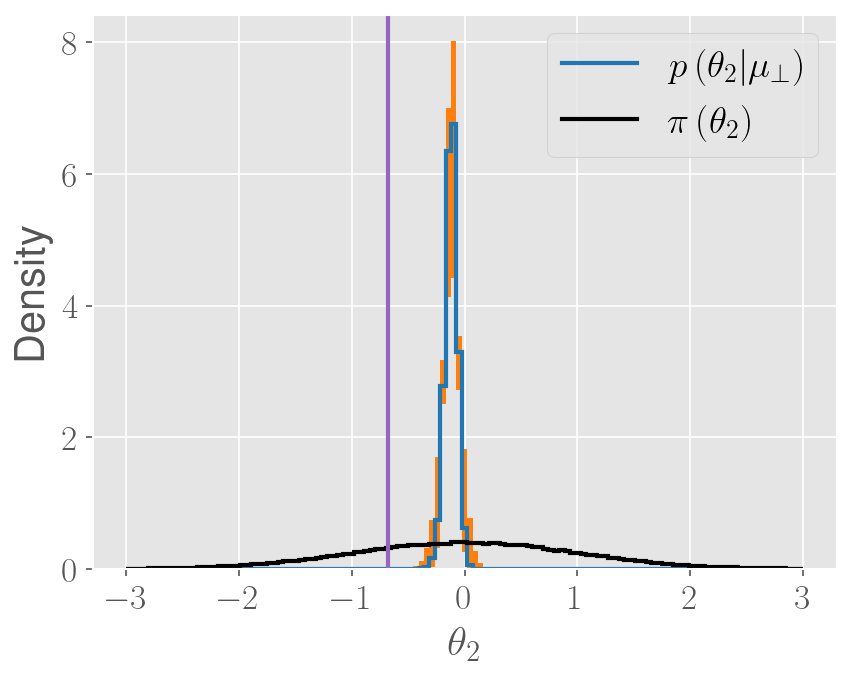}
  \caption{}
  \label{fig:example_2_case_1_conditional_posterior_nominal_and_errorbars_dim_2}
\end{subfigure}%
\begin{subfigure}{.24\textwidth}
  \centering
  \includegraphics[width=\textwidth]{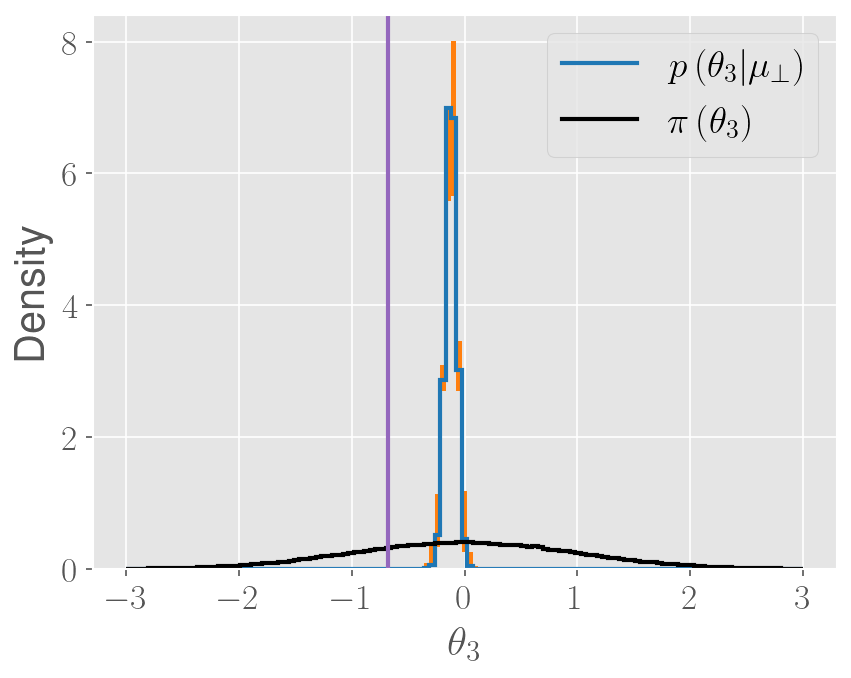}
  \caption{}
  \label{fig:example_2_case_1_conditional_posterior_nominal_and_errorbars_dim_3}
\end{subfigure}%
\begin{subfigure}{.24\textwidth}
  \centering
  \includegraphics[width=\textwidth]{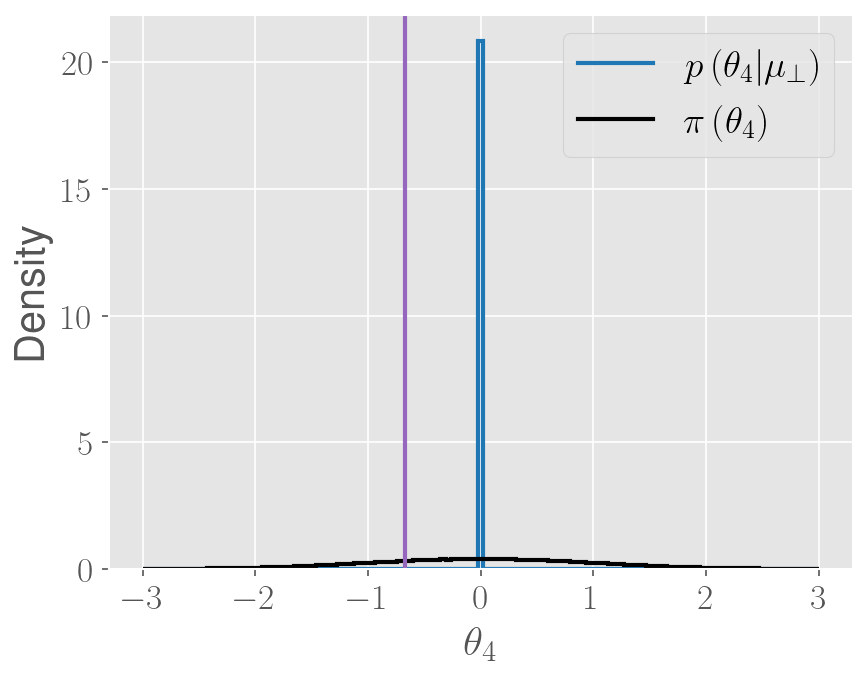}
  \caption{}
  \label{fig:example_2_case_1_conditional_posterior_nominal_and_errorbars_dim_4}
\end{subfigure}
\begin{subfigure}{.24\textwidth}
  \centering
  \includegraphics[width=\textwidth]{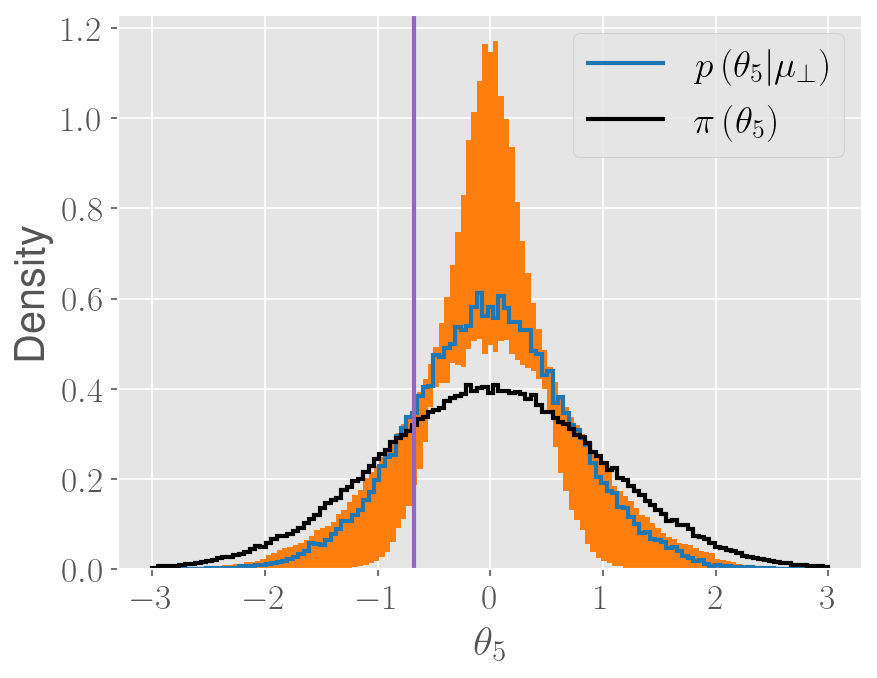}
  \caption{}
  \label{fig:example_2_case_1_conditional_posterior_nominal_and_errorbars_dim_5}
\end{subfigure}%
\begin{subfigure}{.24\textwidth}
  \centering
  \includegraphics[width=\textwidth]{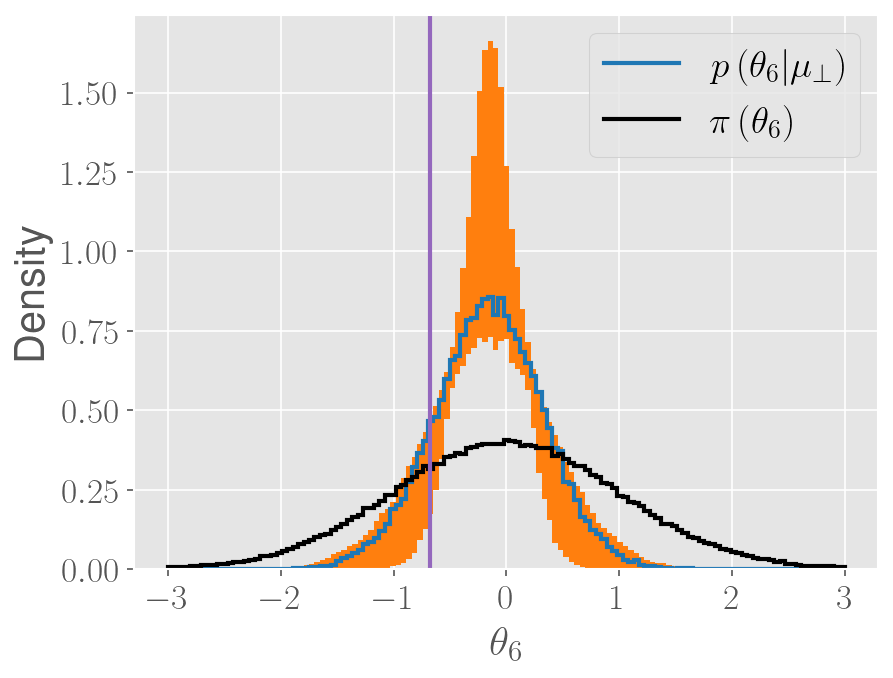}
  \caption{}
  \label{fig:example_2_case_1_conditional_posterior_nominal_and_errorbars_dim_6}
\end{subfigure}%
\begin{subfigure}{.24\textwidth}
  \centering
  \includegraphics[width=\textwidth]{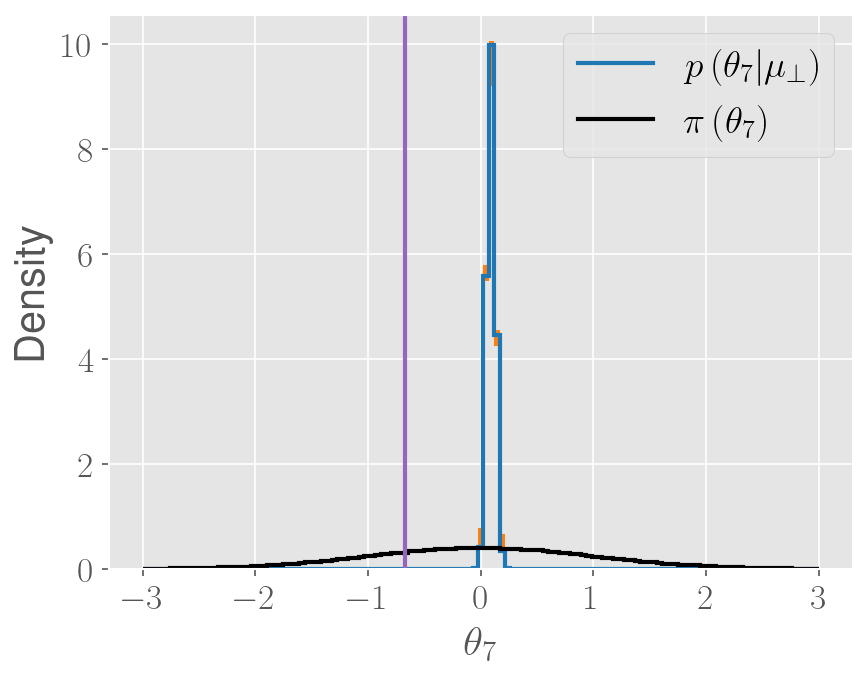}
  \caption{}
  \label{fig:example_2_case_1_conditional_posterior_nominal_and_errorbars_dim_7}
\end{subfigure}%
\begin{subfigure}{.24\textwidth}
  \centering
  \includegraphics[width=\textwidth]{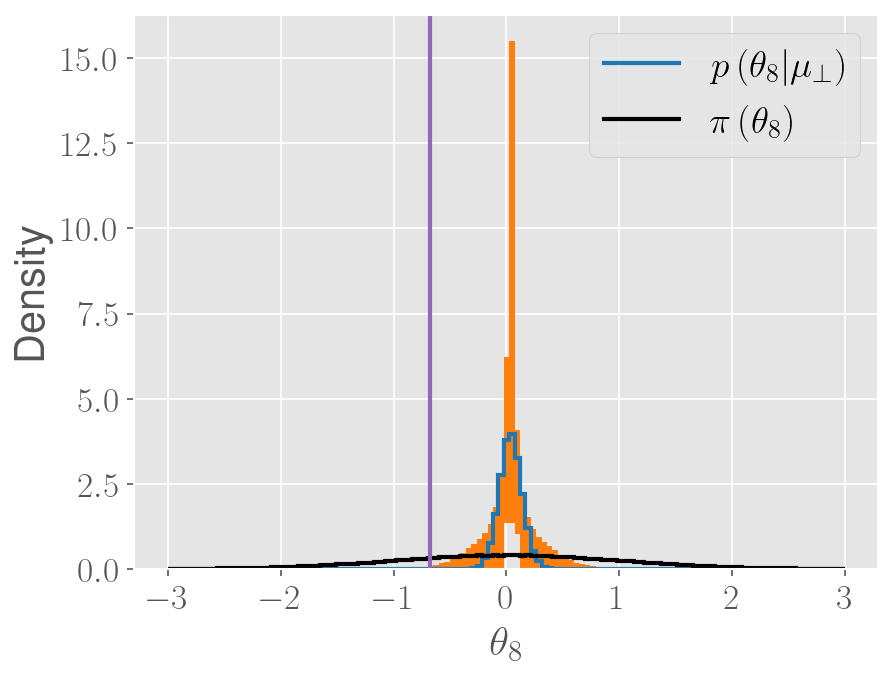}
  \caption{}
  \label{fig:example_2_case_1_conditional_posterior_nominal_and_errorbars_dim_8}
\end{subfigure}
\begin{subfigure}{.24\textwidth}
  \centering
  \includegraphics[width=\textwidth]{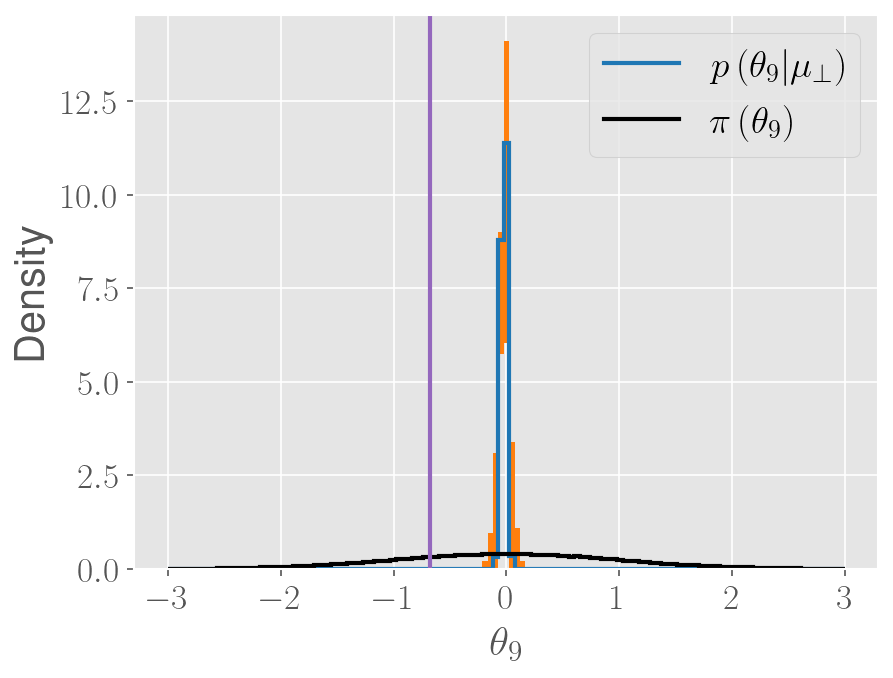}
  \caption{}
  \label{fig:example_2_case_1_conditional_posterior_nominal_and_errorbars_dim_9}
\end{subfigure}%
\begin{subfigure}{.24\textwidth}
  \centering
  \includegraphics[width=\textwidth]{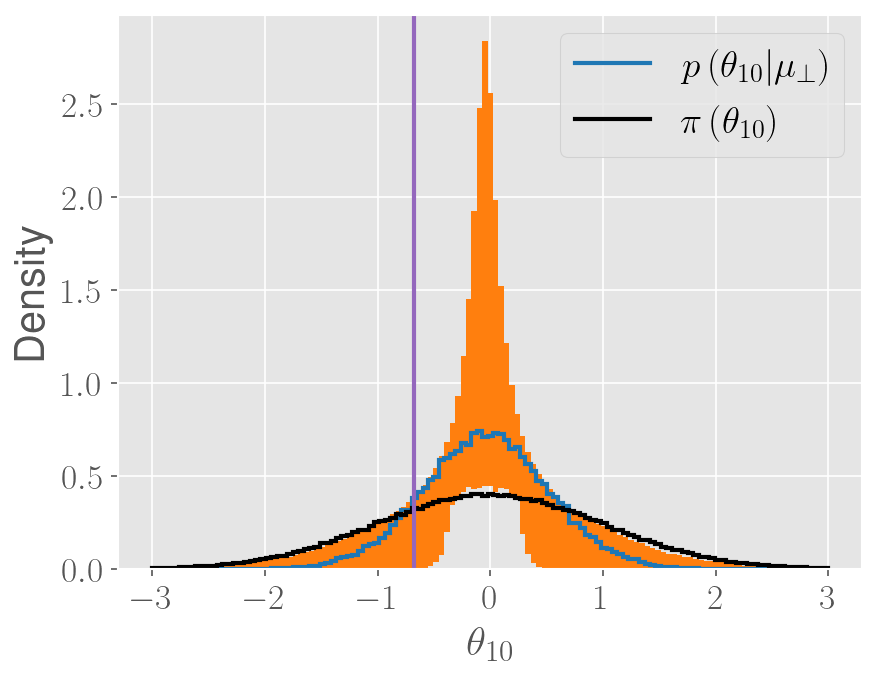}
  \caption{}
  \label{fig:example_2_case_1_conditional_posterior_nominal_and_errorbars_dim_10}
\end{subfigure}%
\begin{subfigure}{.24\textwidth}
  \centering
  \includegraphics[width=\textwidth]{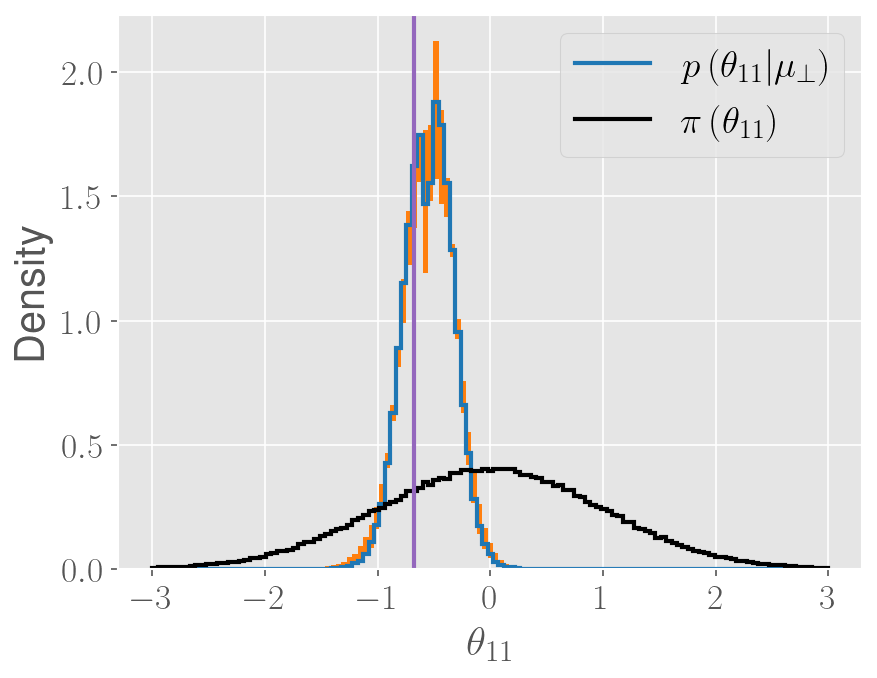}
  \caption{}
  \label{fig:example_2_case_1_conditional_posterior_nominal_and_errorbars_dim_11}
\end{subfigure}%
\begin{subfigure}{.24\textwidth}
  \centering
  \includegraphics[width=\textwidth]{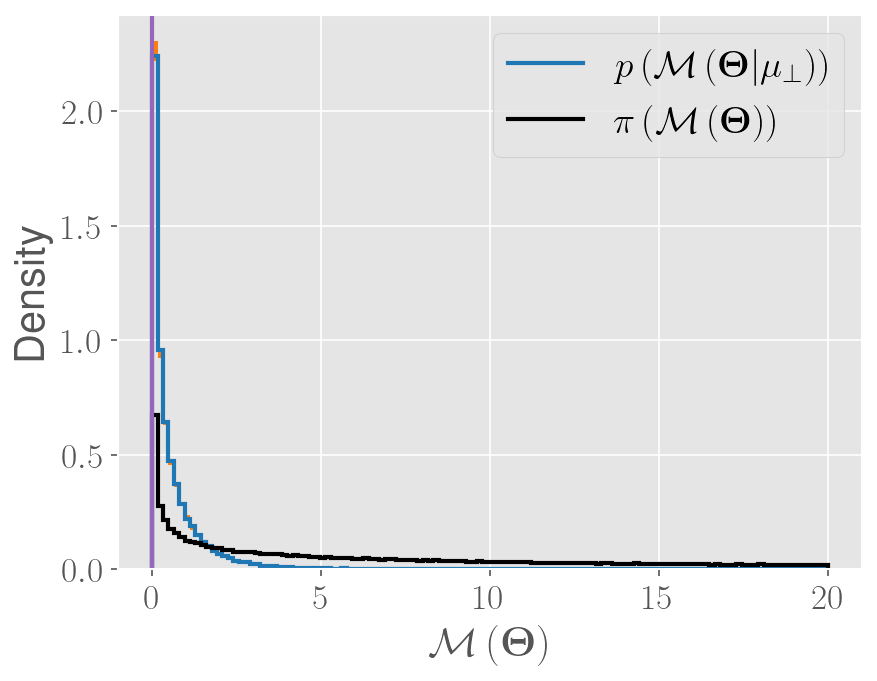}
  \caption{}
  \label{fig:example_2_case_1_conditional_misfit_nominal_and_errorbars}
\end{subfigure}
\caption{Plots of conditional active calibrated posteriors for Case 1 of the vehicle side impact test problem (see Section~\ref{section:example_2_vehicle_side_impact}). Figures (a) - (k) plot the posteriors for $ \theta_1 $ - $ \theta_{11} $, and figure (l) plots the density of $ \mathcal{M} \left( \boldsymbol{\Theta} \right) $ corresponding to the full posterior $ p \left( \boldsymbol{\Theta}_{\mu_{\perp}} \right) $. In each figure, the nominal calibrated posterior constructed using $ \mathfrak{C}_{\boldsymbol{\Theta}} $ is in blue, while the orange bands depict the $ 90 \% $ confidence interval of the posteriors constructed using the sample set replicates $ \mathfrak{C}_{\boldsymbol{\Theta}}^{(i)} $ is in orange. Additionally, the prior density is drawn in black, with the purple line indicating the observed value used for calibration.}
\label{fig:calibrated_conditional_posteriors_results_for_example_2_case_1}
\end{figure}

\begin{figure}[t!bhp]
\centering
\begin{subfigure}{.24\textwidth}
  \centering
  \includegraphics[width=\textwidth]{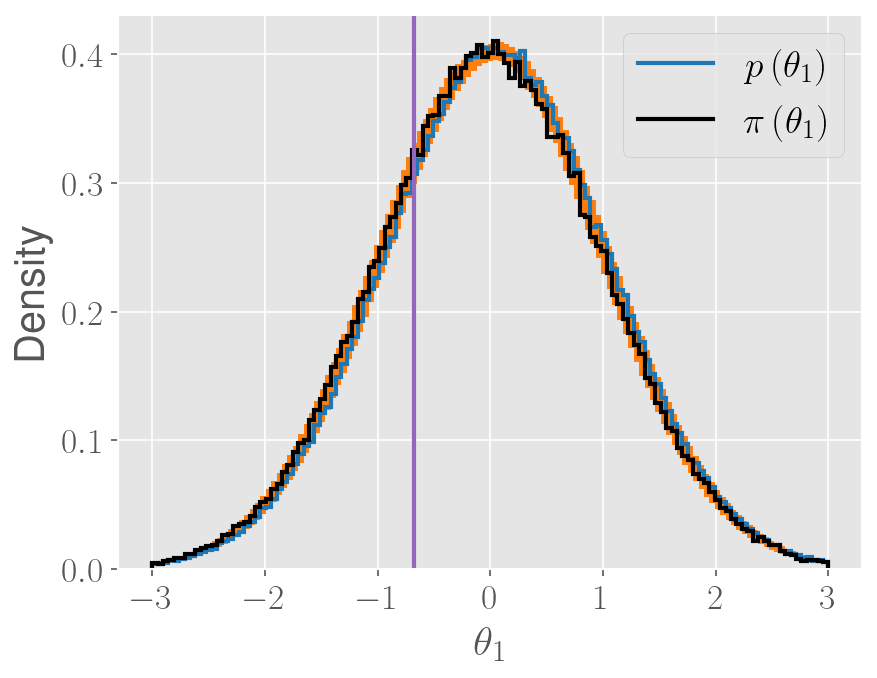}
  \caption{}
  \label{fig:example_2_case_2_full_posterior_nominal_and_errorbars_dim_1}
\end{subfigure}%
\begin{subfigure}{.24\textwidth}
  \centering
  \includegraphics[width=\textwidth]{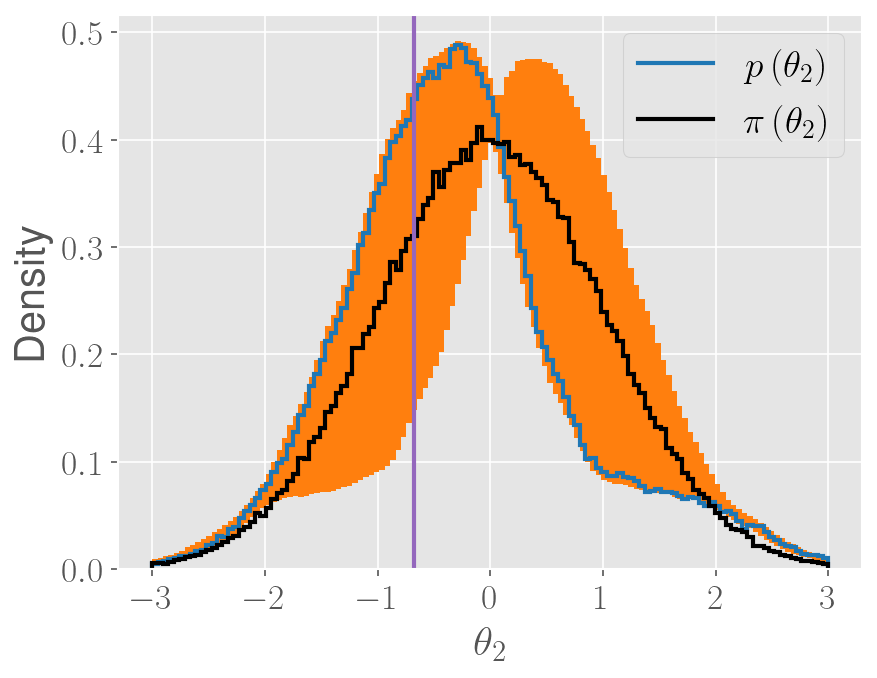}
  \caption{}
  \label{fig:example_2_case_2_full_posterior_nominal_and_errorbars_dim_2}
\end{subfigure}%
\begin{subfigure}{.24\textwidth}
  \centering
  \includegraphics[width=\textwidth]{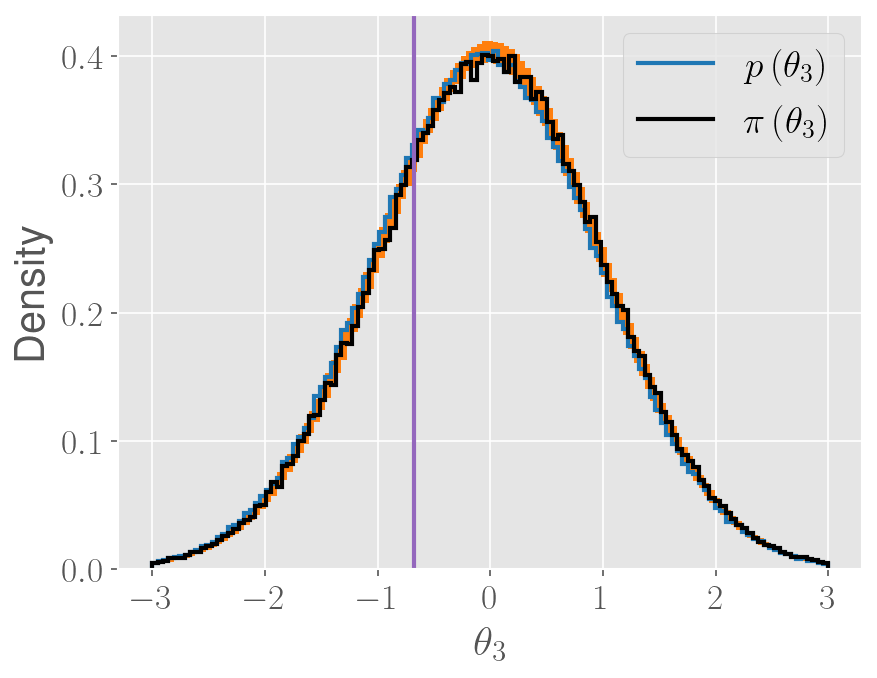}
  \caption{}
  \label{fig:example_2_case_2_full_posterior_nominal_and_errorbars_dim_3}
\end{subfigure}%
\begin{subfigure}{.24\textwidth}
  \centering
  \includegraphics[width=\textwidth]{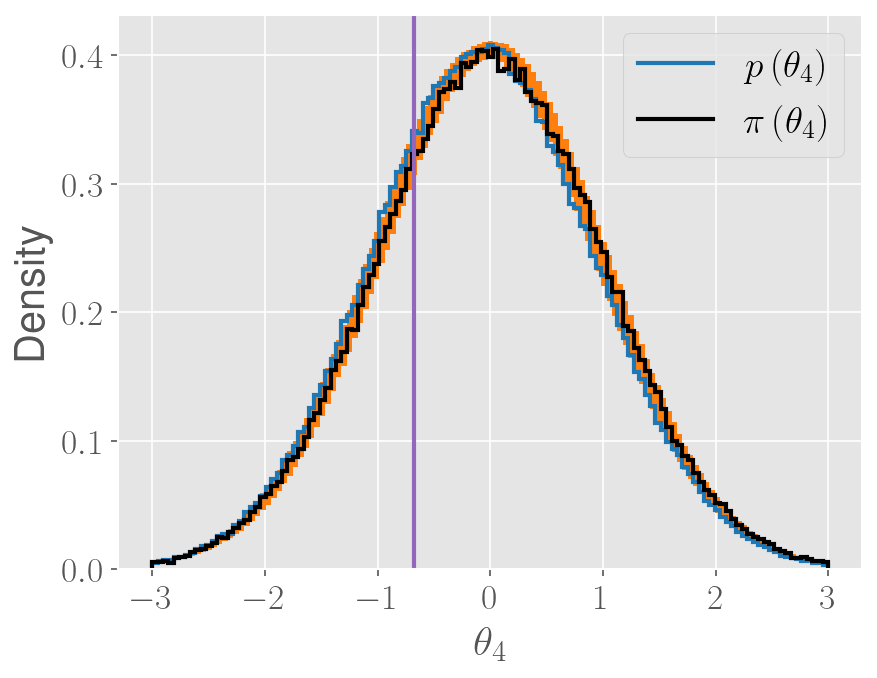}
  \caption{}
  \label{fig:example_2_case_2_full_posterior_nominal_and_errorbars_dim_4}
\end{subfigure}
\begin{subfigure}{.24\textwidth}
  \centering
  \includegraphics[width=\textwidth]{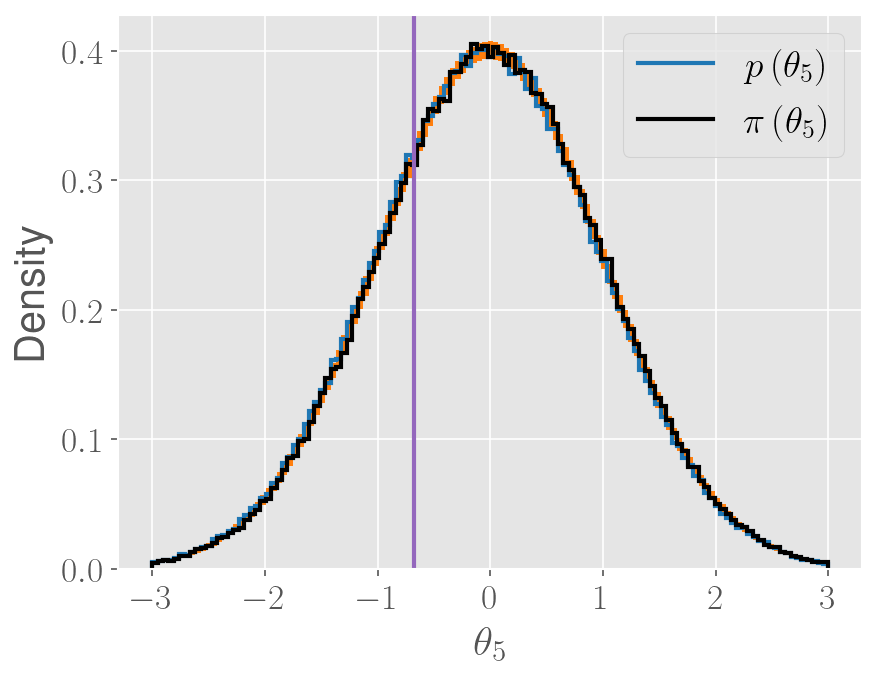}
  \caption{}
  \label{fig:example_2_case_2_full_posterior_nominal_and_errorbars_dim_5}
\end{subfigure}%
\begin{subfigure}{.24\textwidth}
  \centering
  \includegraphics[width=\textwidth]{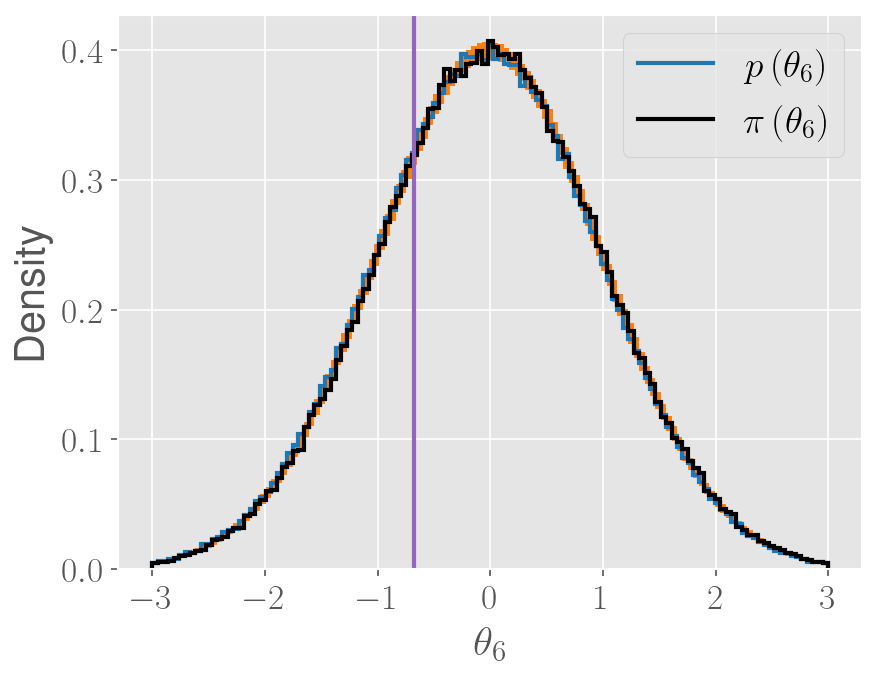}
  \caption{}
  \label{fig:example_2_case_2_full_posterior_nominal_and_errorbars_dim_6}
\end{subfigure}%
\begin{subfigure}{.24\textwidth}
  \centering
  \includegraphics[width=\textwidth]{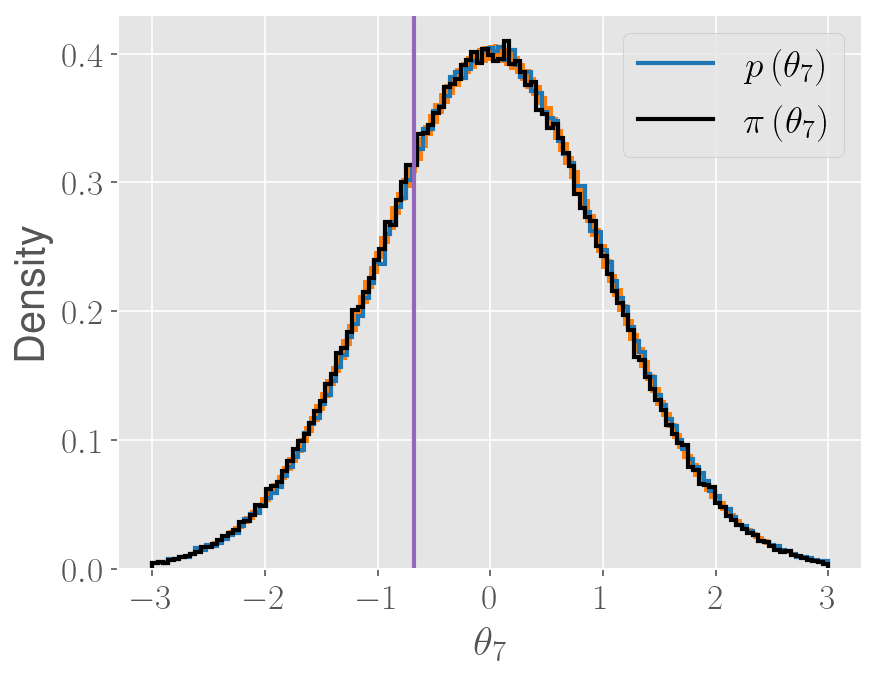}
  \caption{}
  \label{fig:example_2_case_2_full_posterior_nominal_and_errorbars_dim_7}
\end{subfigure}%
\begin{subfigure}{.24\textwidth}
  \centering
  \includegraphics[width=\textwidth]{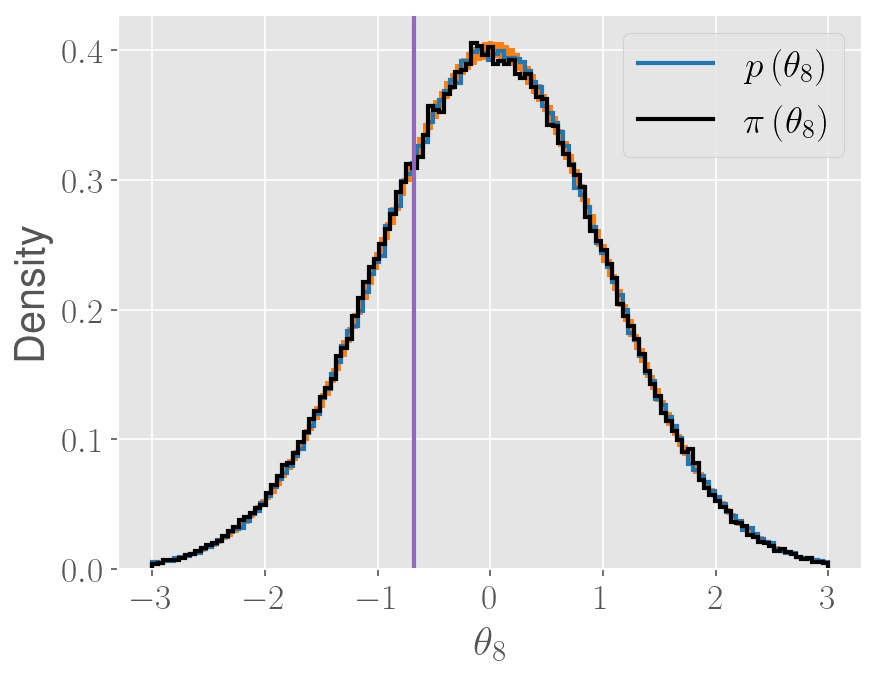}
  \caption{}
  \label{fig:example_2_case_2_full_posterior_nominal_and_errorbars_dim_8}
\end{subfigure}
\begin{subfigure}{.24\textwidth}
  \centering
  \includegraphics[width=\textwidth]{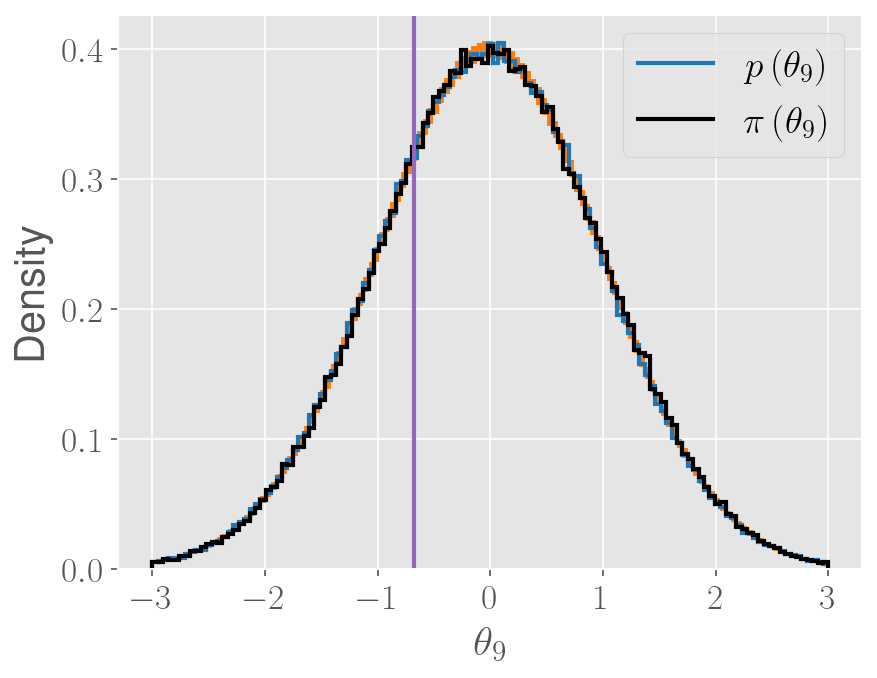}
  \caption{}
  \label{fig:example_2_case_2_full_posterior_nominal_and_errorbars_dim_9}
\end{subfigure}%
\begin{subfigure}{.24\textwidth}
  \centering
  \includegraphics[width=\textwidth]{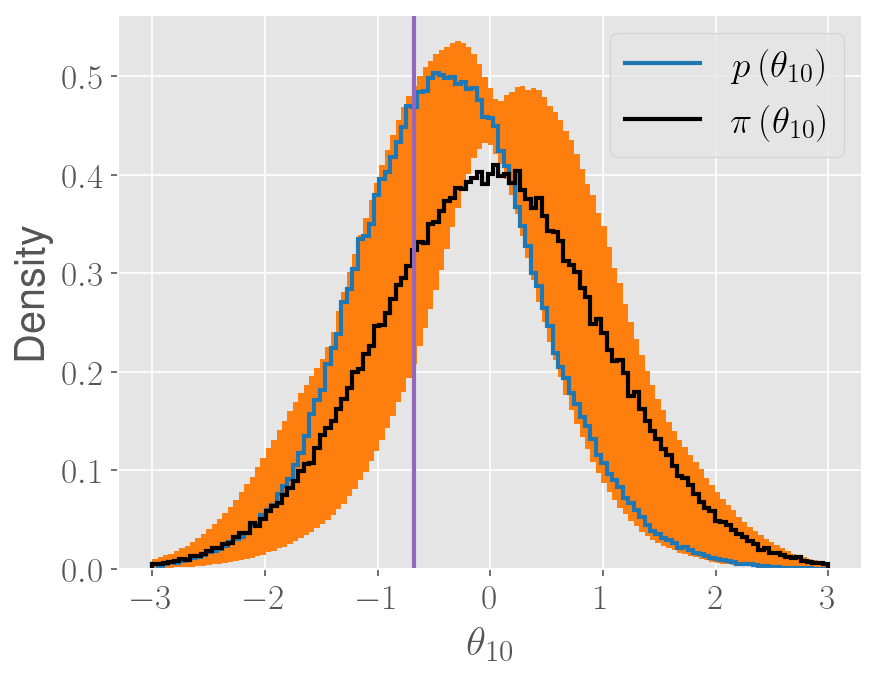}
  \caption{}
  \label{fig:example_2_case_2_full_posterior_nominal_and_errorbars_dim_10}
\end{subfigure}%
\begin{subfigure}{.24\textwidth}
  \centering
  \includegraphics[width=\textwidth]{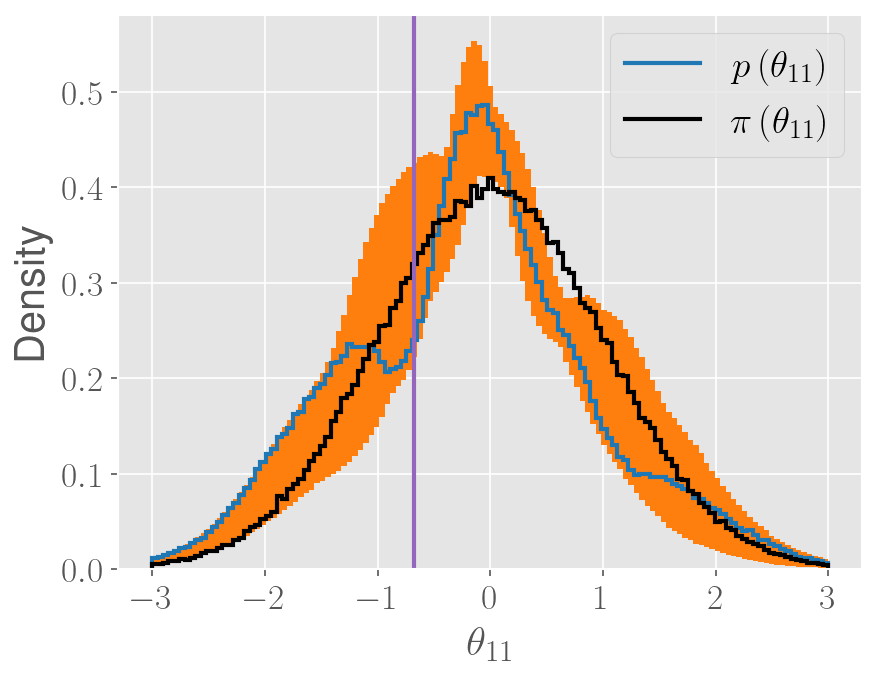}
  \caption{}
  \label{fig:example_2_case_2_full_posterior_nominal_and_errorbars_dim_11}
\end{subfigure}%
\begin{subfigure}{.24\textwidth}
  \centering
  \includegraphics[width=\textwidth]{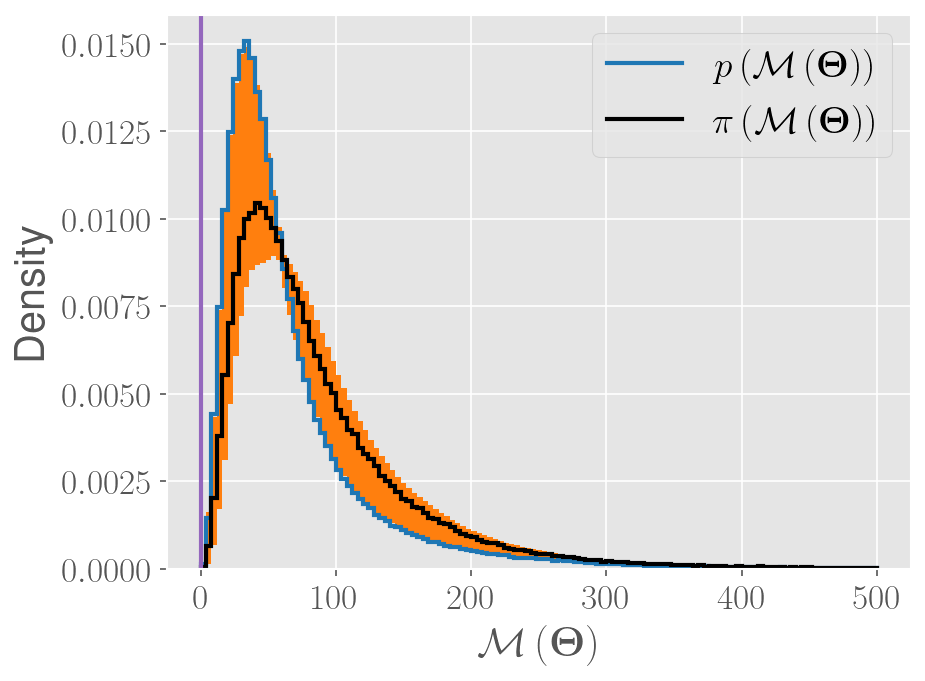}
  \caption{}
  \label{fig:example_2_case_2_full_misfit_nominal_and_errorbars}
\end{subfigure}
\caption{Plots of full posteriors for Case 2 of the vehicle side impact test problem (see Section~\ref{section:example_2_vehicle_side_impact}). Figures (a) - (k) plot the posteriors for $ \theta_1 $ - $ \theta_{11} $, and figure (l) plots the density of $ \mathcal{M} \left( \boldsymbol{\Theta} \right) $ corresponding to the full posterior $ p \left( \boldsymbol{\Theta} \right) $. In each figure, the nominal calibrated posterior constructed using $ \mathfrak{C}_{\boldsymbol{\Theta}} $ is in blue, while the orange bands depict the $ 90 \% $ confidence interval of the posteriors constructed using the sample set replicates $ \mathfrak{C}_{\boldsymbol{\Theta}}^{(i)} $ is in orange. Additionally, the prior density is drawn in black, with the purple line indicating the observed value used for calibration.}
\label{fig:calibrated_full_posteriors_results_for_example_2_case_2}
\end{figure}

\begin{figure}[t!bhp]
\centering
\begin{subfigure}{.24\textwidth}
  \centering
  \includegraphics[width=\textwidth]{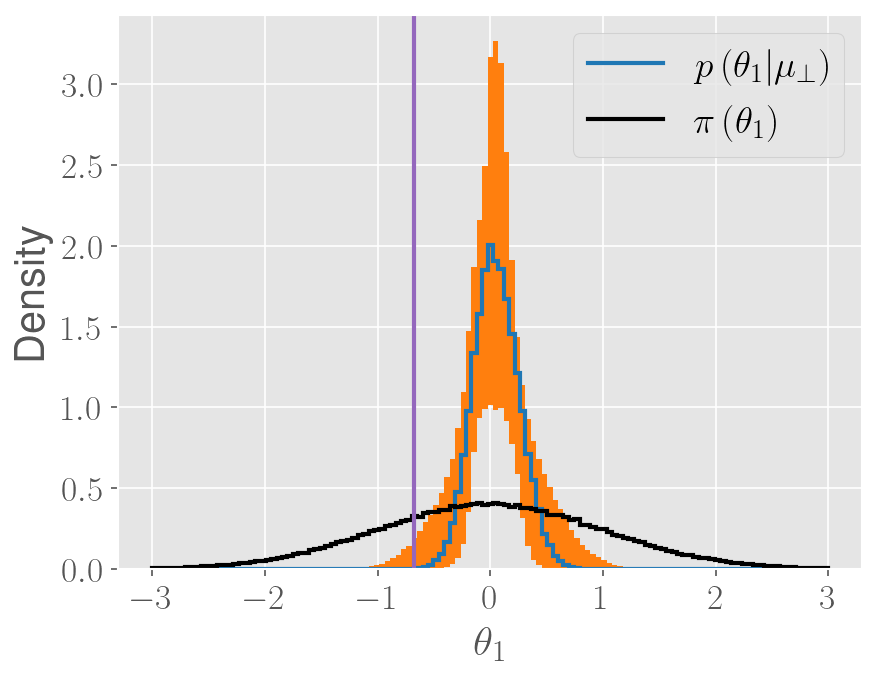}
  \caption{}
  \label{fig:example_2_case_2_conditional_posterior_nominal_and_errorbars_dim_1}
\end{subfigure}%
\begin{subfigure}{.24\textwidth}
  \centering
  \includegraphics[width=\textwidth]{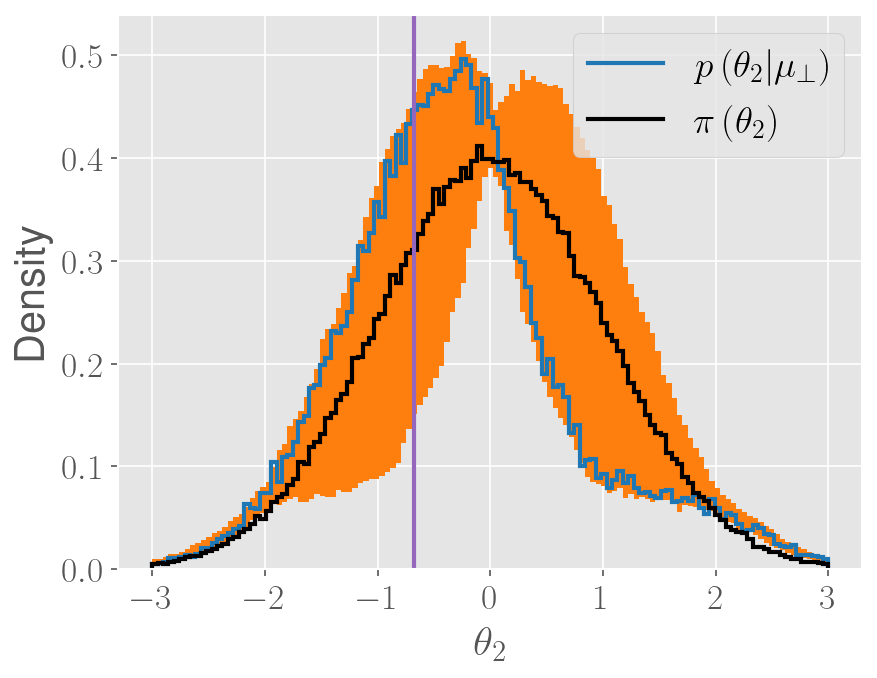}
  \caption{}
  \label{fig:example_2_case_2_conditional_posterior_nominal_and_errorbars_dim_2}
\end{subfigure}%
\begin{subfigure}{.24\textwidth}
  \centering
  \includegraphics[width=\textwidth]{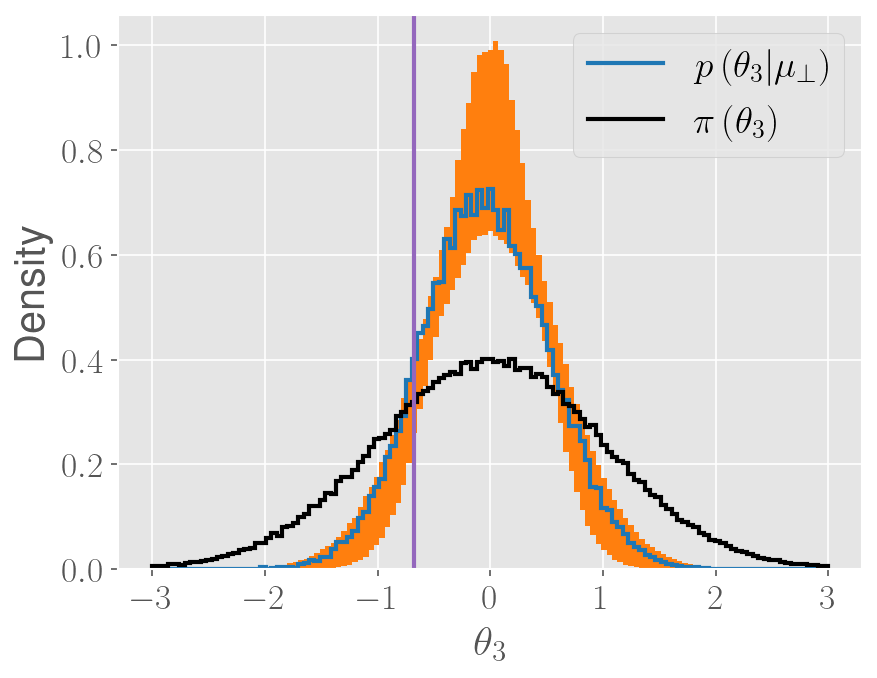}
  \caption{}
  \label{fig:example_2_case_2_conditional_posterior_nominal_and_errorbars_dim_3}
\end{subfigure}%
\begin{subfigure}{.24\textwidth}
  \centering
  \includegraphics[width=\textwidth]{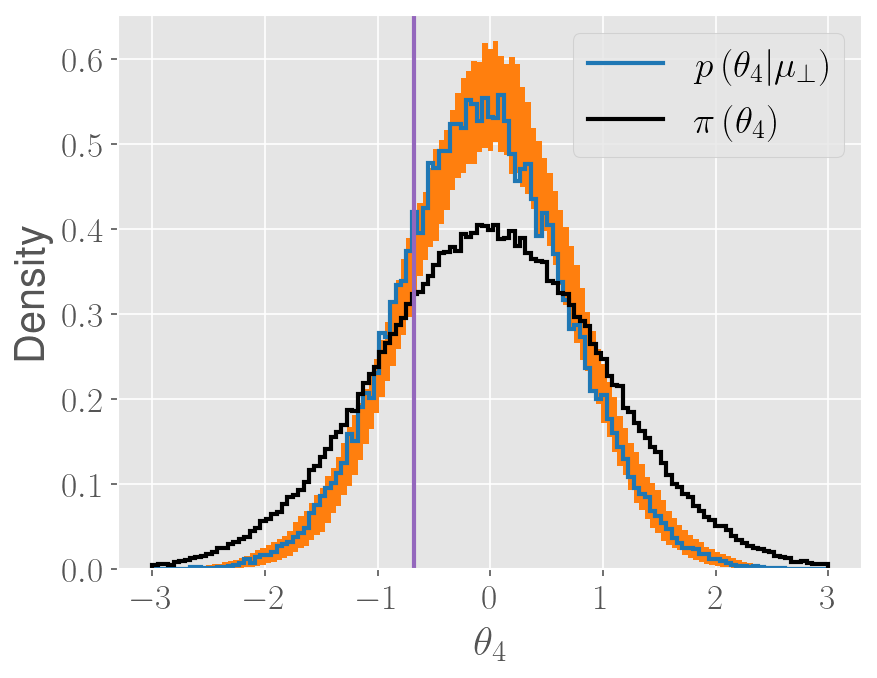}
  \caption{}
  \label{fig:example_2_case_2_conditional_posterior_nominal_and_errorbars_dim_4}
\end{subfigure}
\begin{subfigure}{.24\textwidth}
  \centering
  \includegraphics[width=\textwidth]{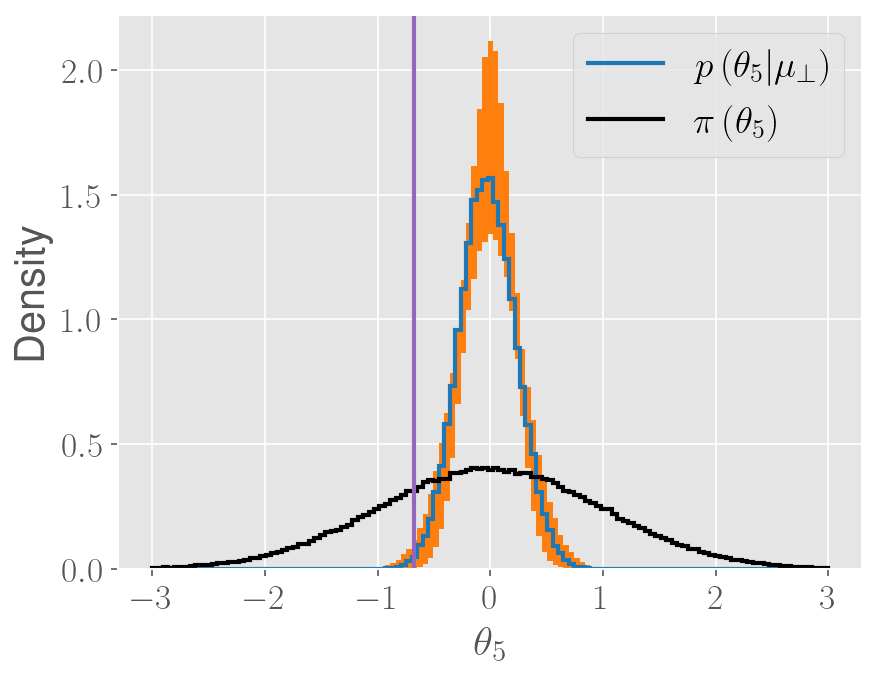}
  \caption{}
  \label{fig:example_2_case_2_conditional_posterior_nominal_and_errorbars_dim_5}
\end{subfigure}%
\begin{subfigure}{.24\textwidth}
  \centering
  \includegraphics[width=\textwidth]{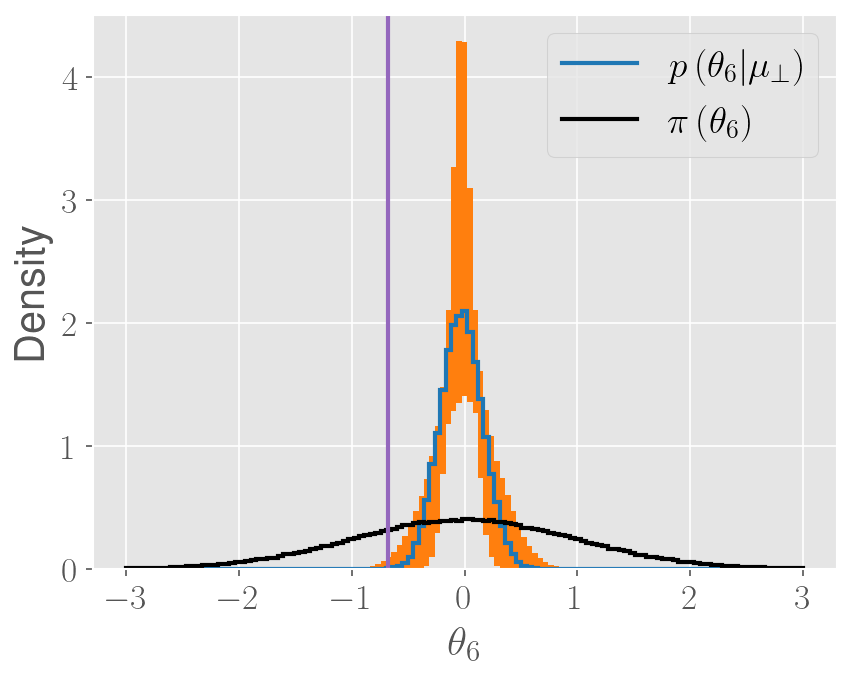}
  \caption{}
  \label{fig:example_2_case_2_conditional_posterior_nominal_and_errorbars_dim_6}
\end{subfigure}%
\begin{subfigure}{.24\textwidth}
  \centering
  \includegraphics[width=\textwidth]{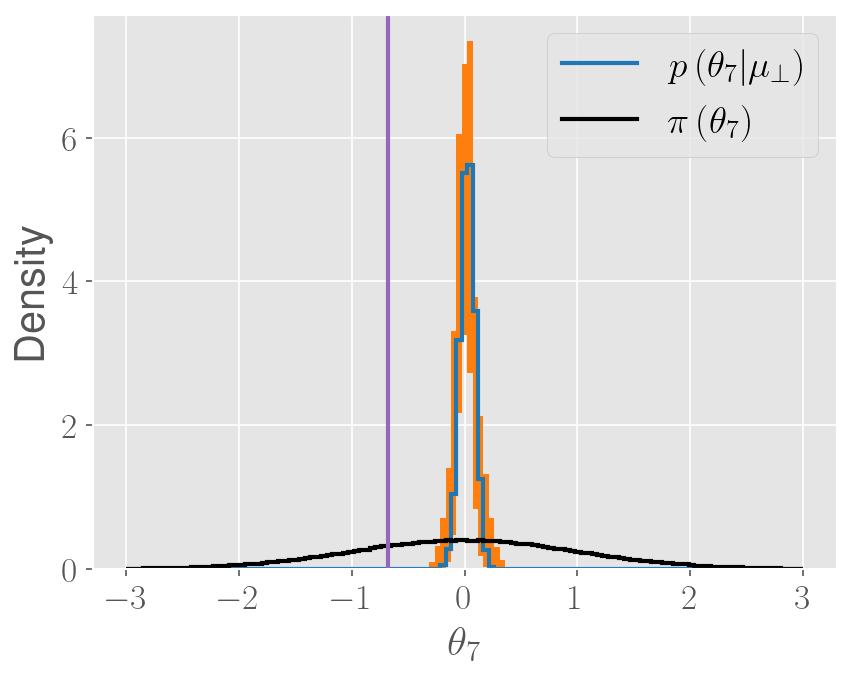}
  \caption{}
  \label{fig:example_2_case_2_conditional_posterior_nominal_and_errorbars_dim_7}
\end{subfigure}%
\begin{subfigure}{.24\textwidth}
  \centering
  \includegraphics[width=\textwidth]{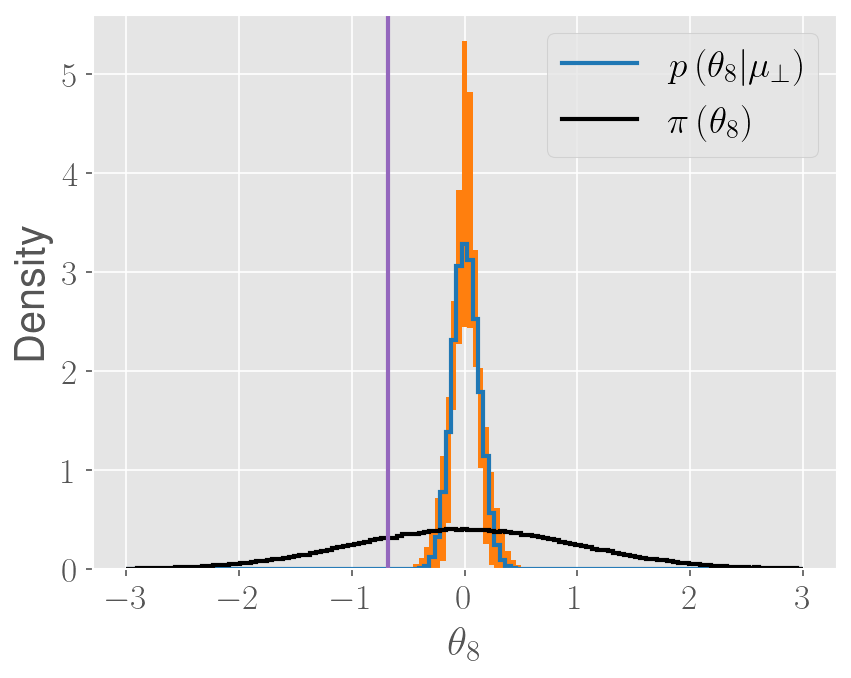}
  \caption{}
  \label{fig:example_2_case_2_conditional_posterior_nominal_and_errorbars_dim_8}
\end{subfigure}
\begin{subfigure}{.24\textwidth}
  \centering
  \includegraphics[width=\textwidth]{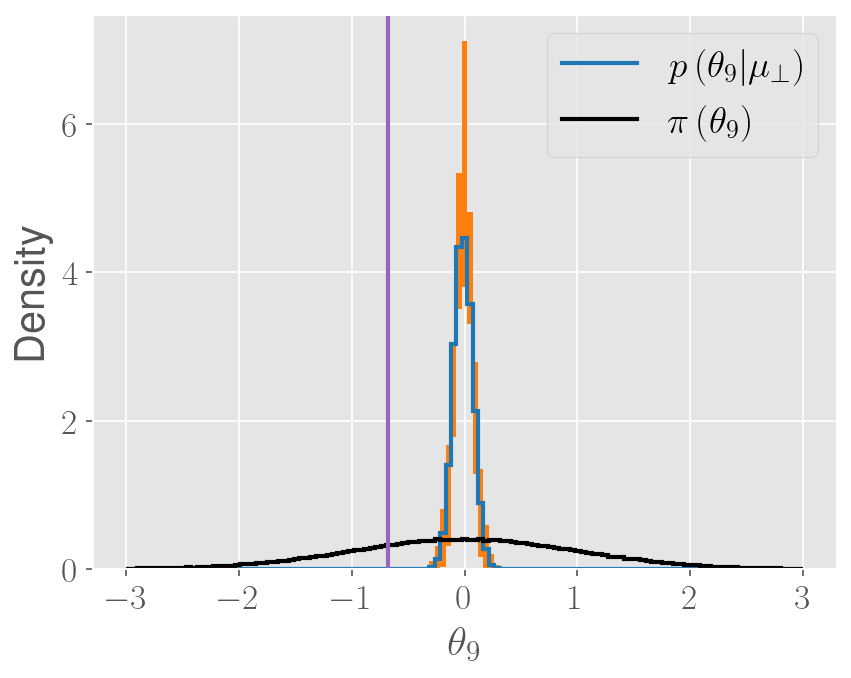}
  \caption{}
  \label{fig:example_2_case_2_conditional_posterior_nominal_and_errorbars_dim_9}
\end{subfigure}%
\begin{subfigure}{.24\textwidth}
  \centering
  \includegraphics[width=\textwidth]{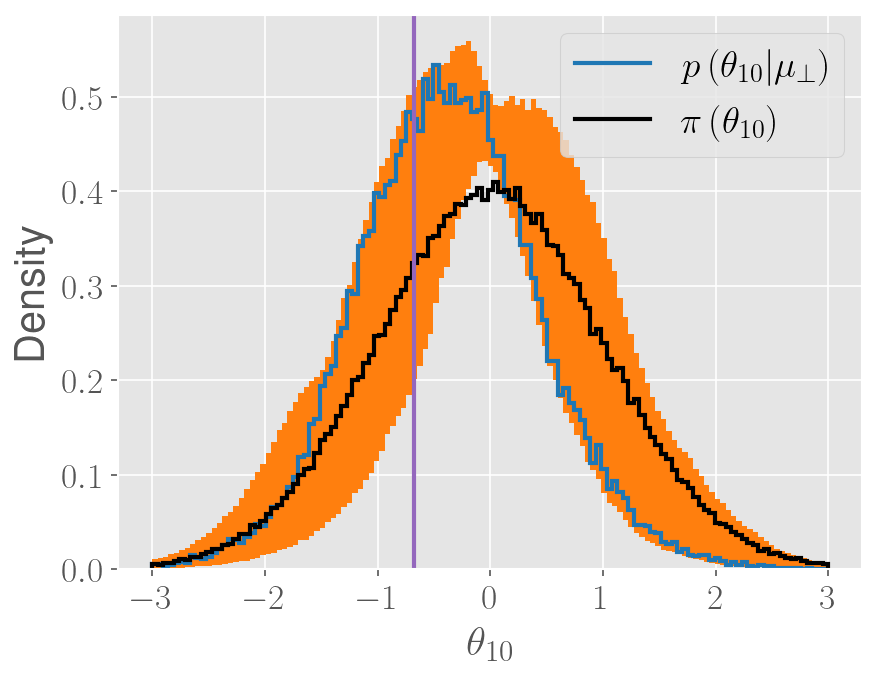}
  \caption{}
  \label{fig:example_2_case_2_conditional_posterior_nominal_and_errorbars_dim_10}
\end{subfigure}%
\begin{subfigure}{.24\textwidth}
  \centering
  \includegraphics[width=\textwidth]{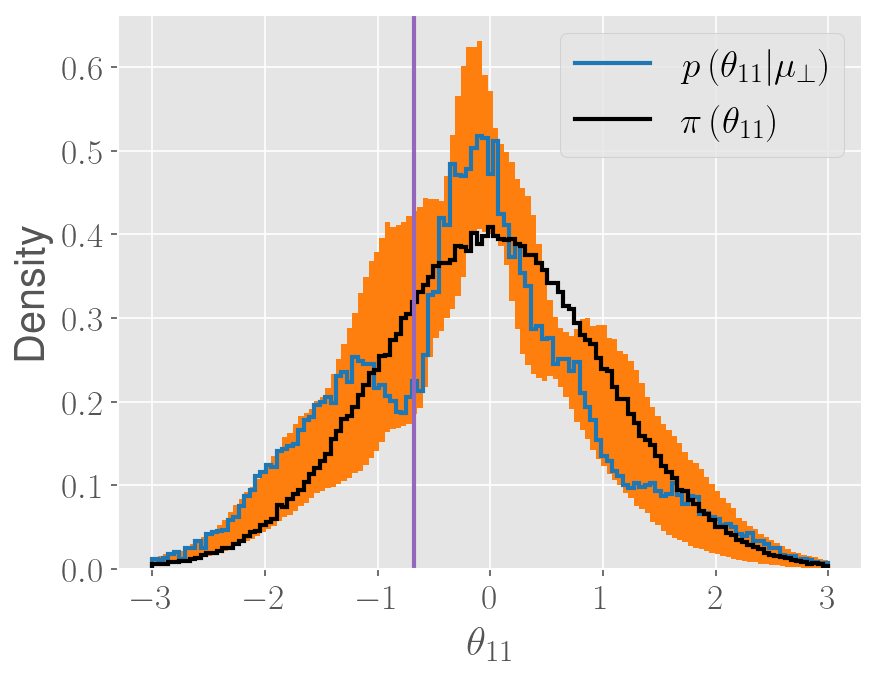}
  \caption{}
  \label{fig:example_2_case_2_conditional_posterior_nominal_and_errorbars_dim_11}
\end{subfigure}%
\begin{subfigure}{.24\textwidth}
  \centering
  \includegraphics[width=\textwidth]{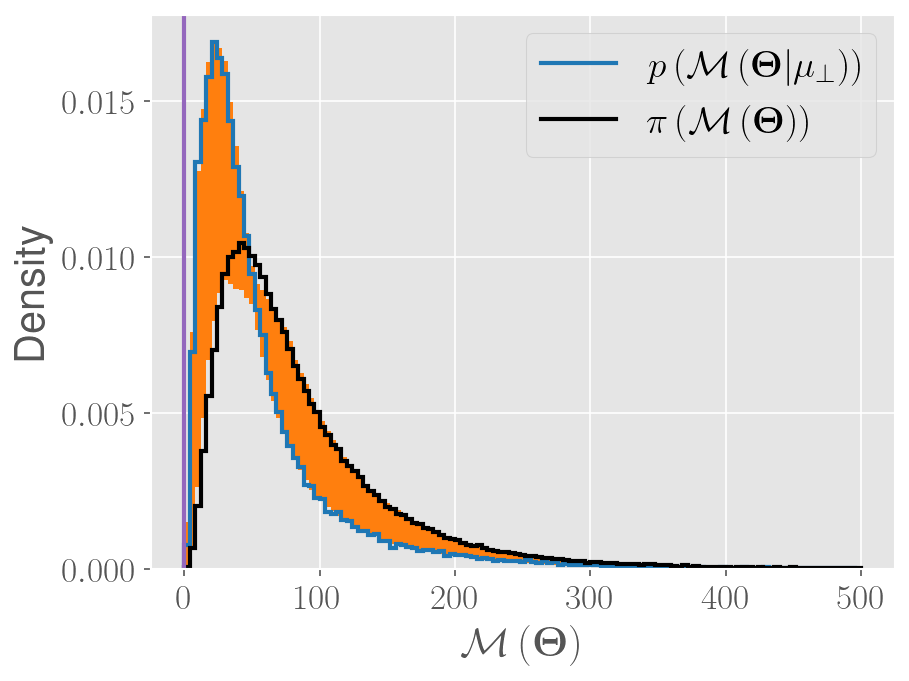}
  \caption{}
  \label{fig:example_2_case_2_conditional_misfit_nominal_and_errorbars}
\end{subfigure}
\caption{Plots of conditional active calibrated posteriors for Case 2 of the vehicle side impact test problem (see Section~\ref{section:example_2_vehicle_side_impact}). Figures (a) - (k) plot the posteriors for $ \theta_1 $ - $ \theta_{11} $, and figure (l) plots the density of $ \mathcal{M} \left( \boldsymbol{\Theta} \right) $ corresponding to the full posterior $ p \left( \boldsymbol{\Theta}_{\mu_{\perp}} \right) $. In each figure, the nominal calibrated posterior constructed using $ \mathfrak{C}_{\boldsymbol{\Theta}} $ is in blue, while the orange bands depict the $ 90 \% $ confidence interval of the posteriors constructed using the sample set replicates $ \mathfrak{C}_{\boldsymbol{\Theta}}^{(i)} $ is in orange. Additionally, the prior density is drawn in black, with the purple line indicating the observed value used for calibration.}
\label{fig:calibrated_conditional_posteriors_results_for_example_2_case_2}
\end{figure}

\begin{figure}[t!bhp]
\centering
\begin{subfigure}{.24\textwidth}
  \centering
  \includegraphics[width=\textwidth]{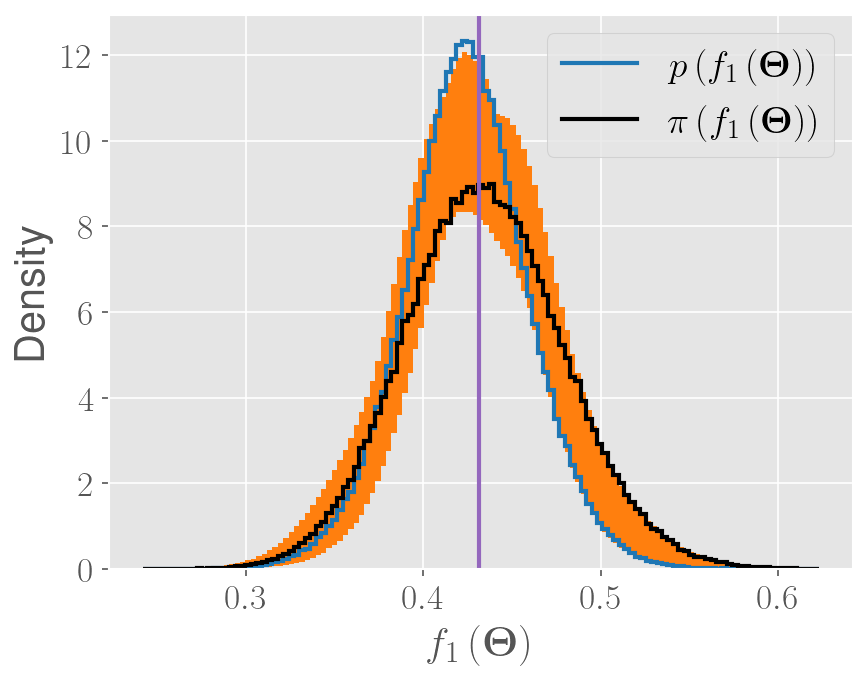}
  \caption{}
  \label{fig:example_2_case_2_marginal_posterior_output_nominal_and_errorbars_dim_1}
\end{subfigure}%
\begin{subfigure}{.24\textwidth}
  \centering
  \includegraphics[width=\textwidth]{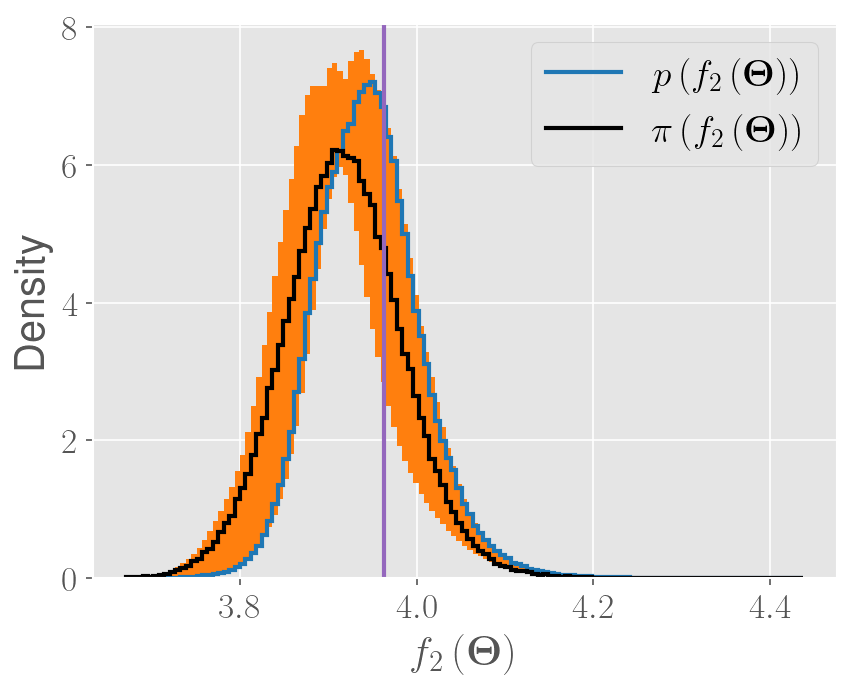}
  \caption{}
  \label{fig:example_2_case_2_marginal_posterior_output_nominal_and_errorbars_dim_2}
\end{subfigure}%
\begin{subfigure}{.24\textwidth}
  \centering
  \includegraphics[width=\textwidth]{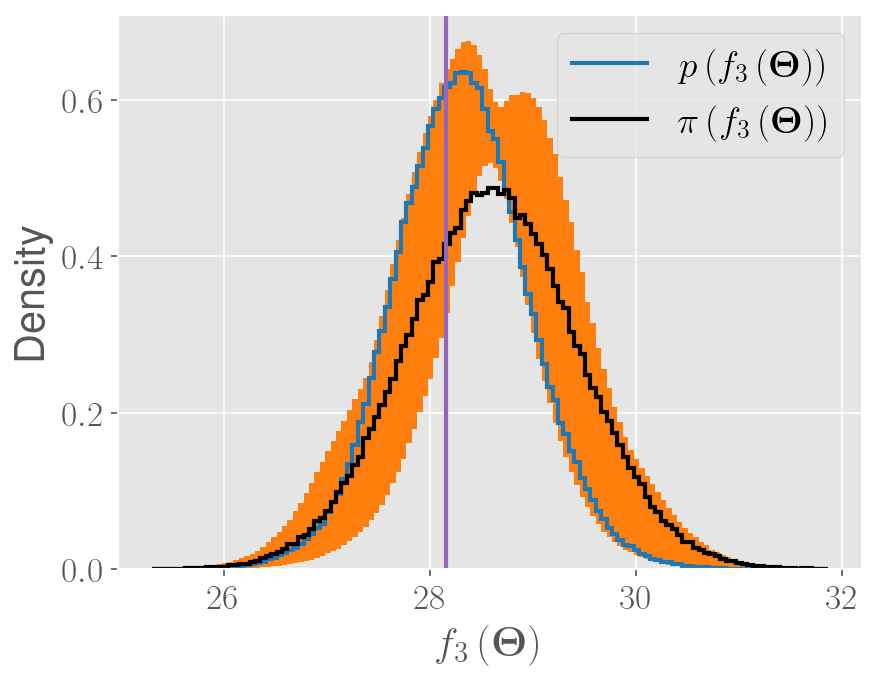}
  \caption{}
  \label{fig:example_2_case_2_marginal_posterior_output_nominal_and_errorbars_dim_3}
\end{subfigure}%
\begin{subfigure}{.24\textwidth}
  \centering
  \includegraphics[width=\textwidth]{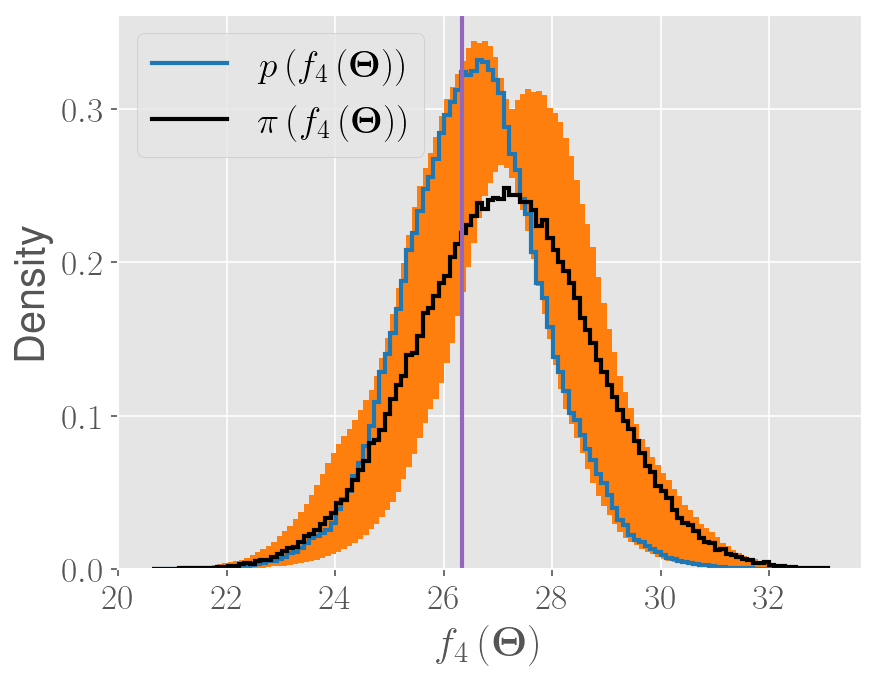}
  \caption{}
  \label{fig:example_2_case_2_marginal_posterior_output_nominal_and_errorbars_dim_4}
\end{subfigure}
\begin{subfigure}{.24\textwidth}
  \centering
  \includegraphics[width=\textwidth]{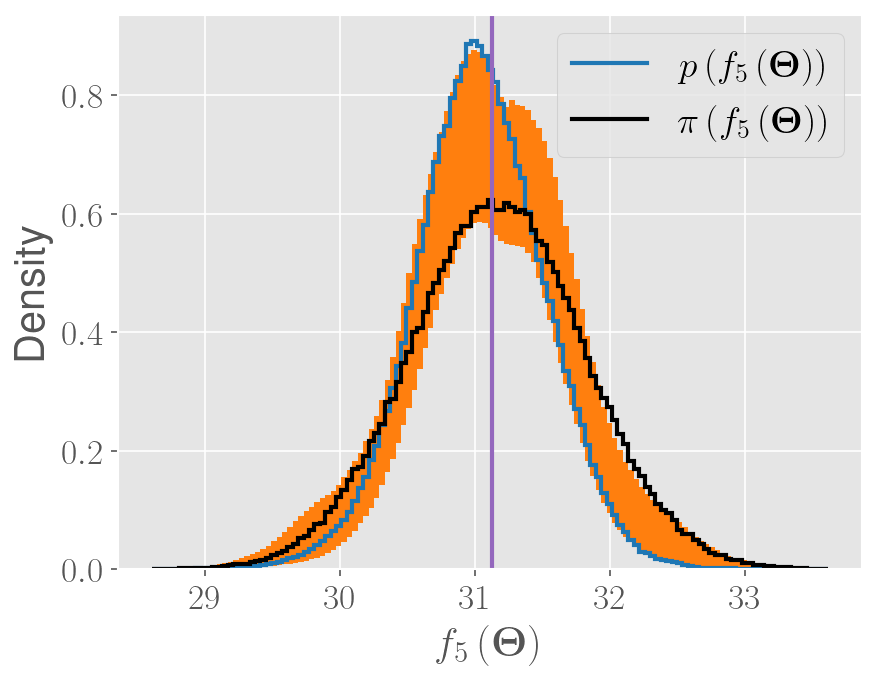}
  \caption{}
  \label{fig:example_2_case_2_marginal_posterior_output_nominal_and_errorbars_dim_5}
\end{subfigure}%
\begin{subfigure}{.24\textwidth}
  \centering
  \includegraphics[width=\textwidth]{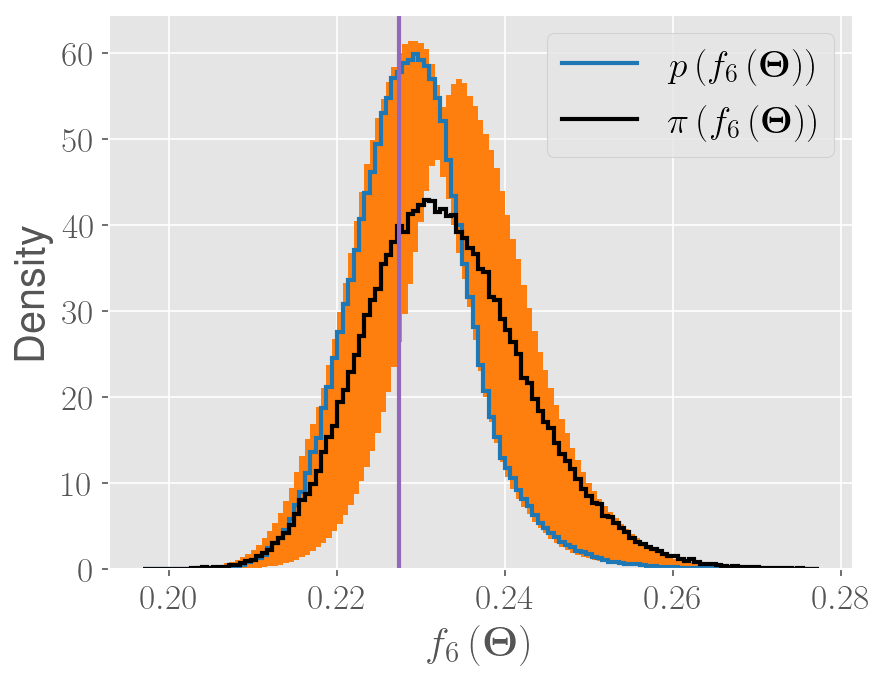}
  \caption{}
  \label{fig:example_2_case_2_marginal_posterior_output_nominal_and_errorbars_dim_6}
\end{subfigure}%
\begin{subfigure}{.24\textwidth}
  \centering
  \includegraphics[width=\textwidth]{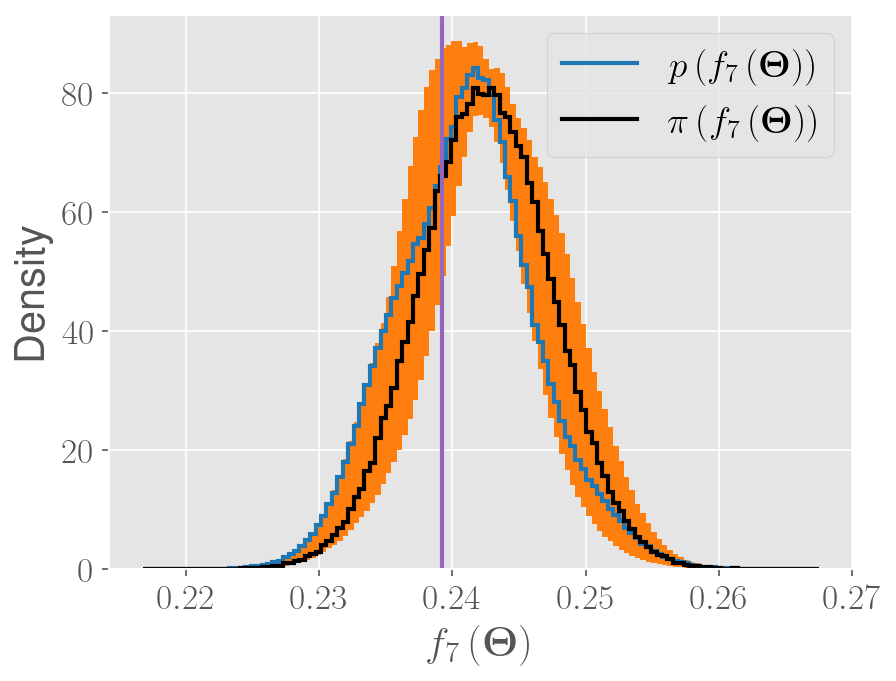}
  \caption{}
  \label{fig:example_2_case_2_marginal_posterior_output_nominal_and_errorbars_dim_7}
\end{subfigure}%
\begin{subfigure}{.24\textwidth}
  \centering
  \includegraphics[width=\textwidth]{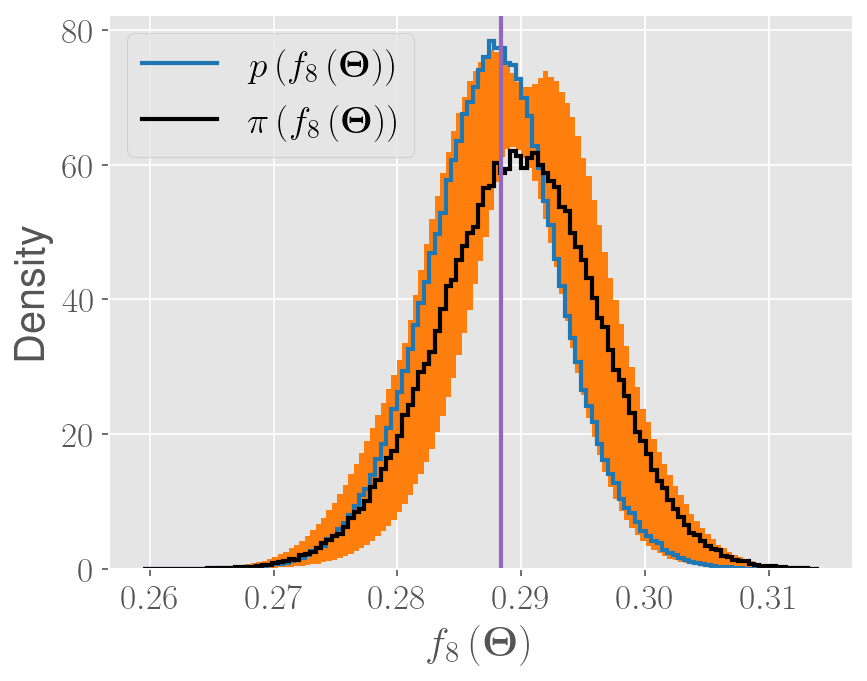}
  \caption{}
  \label{fig:example_2_case_2_marginal_posterior_output_nominal_and_errorbars_dim_8}
\end{subfigure}
\begin{subfigure}{.24\textwidth}
  \centering
  \includegraphics[width=\textwidth]{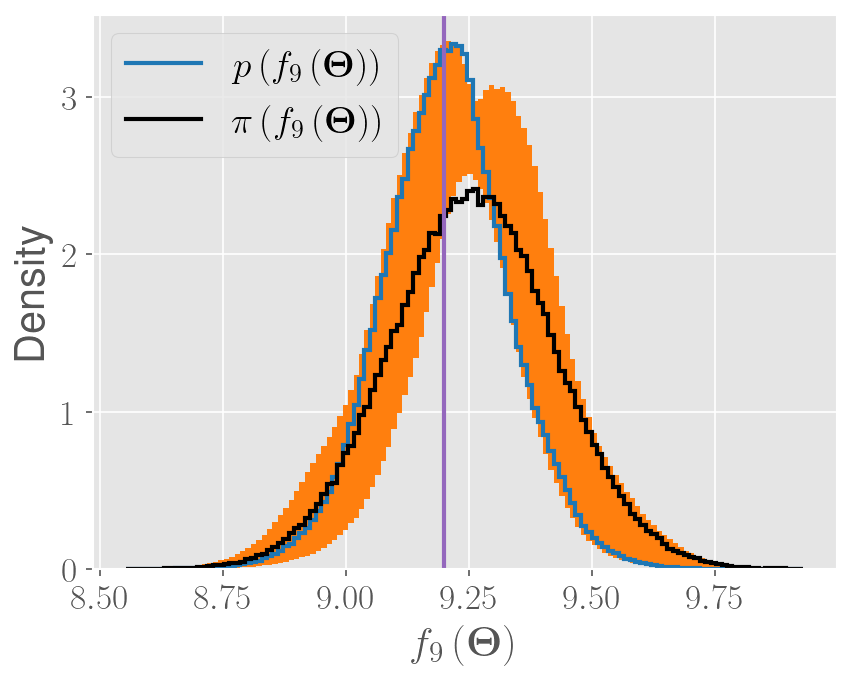}
  \caption{}
  \label{fig:example_2_case_2_marginal_posterior_output_nominal_and_errorbars_dim_9}
\end{subfigure}%
\begin{subfigure}{.24\textwidth}
  \centering
  \includegraphics[width=\textwidth]{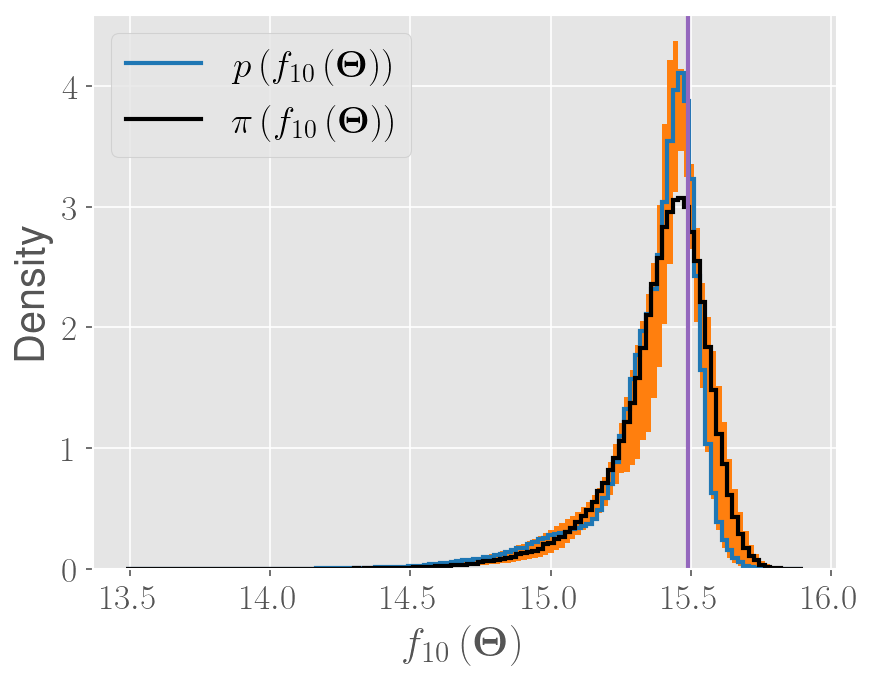}
  \caption{}
  \label{fig:example_2_case_2_marginal_posterior_output_nominal_and_errorbars_dim_10}
\end{subfigure}%
\begin{subfigure}{.24\textwidth}
  \centering
  \includegraphics[width=\textwidth]{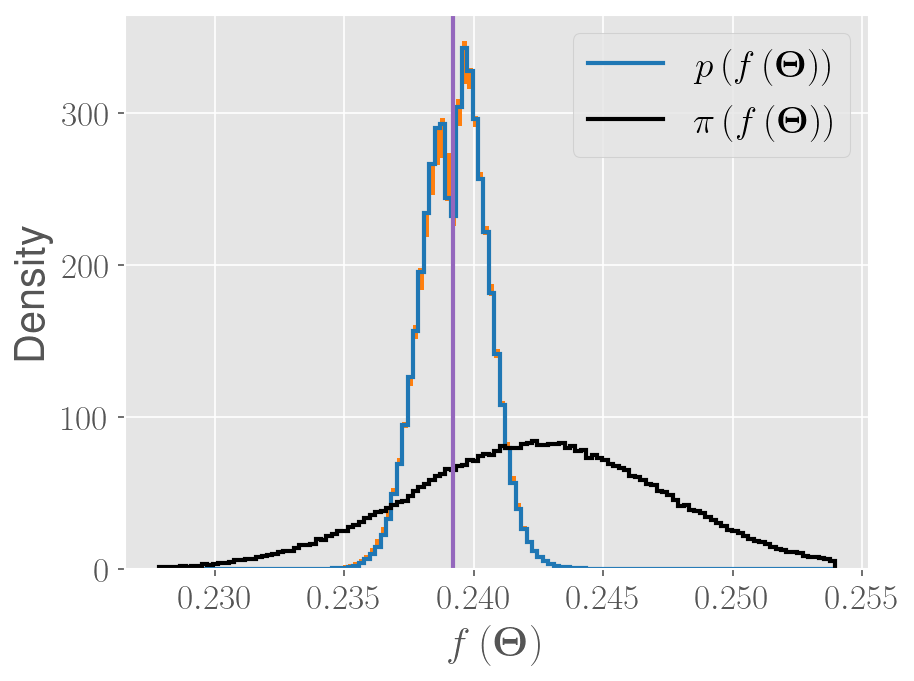}
  \caption{}
  \label{fig:example_2_case_1_marginal_posterior_output_nominal_and_errorbars}
\end{subfigure}
\caption{Plots of the density of $ f \left( \boldsymbol{\Theta} \right) $ corresponding to the full posterior $ p \left( \boldsymbol{\Theta} \right) $ for the vehicle side impact test problem (see Section~\ref{section:example_2_vehicle_side_impact}). Figures (a) - (j) depict the marginal output densities of the 10 output components of $ f \left( \boldsymbol{\Theta} \right) $ (i.e., $ f_i \left( \boldsymbol{\Theta} \right) $, $ i = 1, \dots, 10 $) for Case 2, while figure (k) depicts the density of $ f \left( \boldsymbol{\Theta} \right) $ for Case 1. In each figure, the nominal calibrated posterior constructed using $ \mathfrak{C}_{\boldsymbol{\Theta}} $ is in blue, while the orange bands depict the $ 90 \% $ confidence interval of the posteriors constructed using the sample set replicates $ \mathfrak{C}_{\boldsymbol{\Theta}}^{(i)} $ is in orange. Additionally, the prior density is drawn in black, with the purple line indicating the observed value used for calibration.}
\label{fig:calibrated_marginal_output_posteriors_results_for_example_2}
\end{figure}

\begin{figure}[t!bhp]
\centering
\begin{subfigure}{.24\textwidth}
  \centering
  \includegraphics[width=\textwidth]{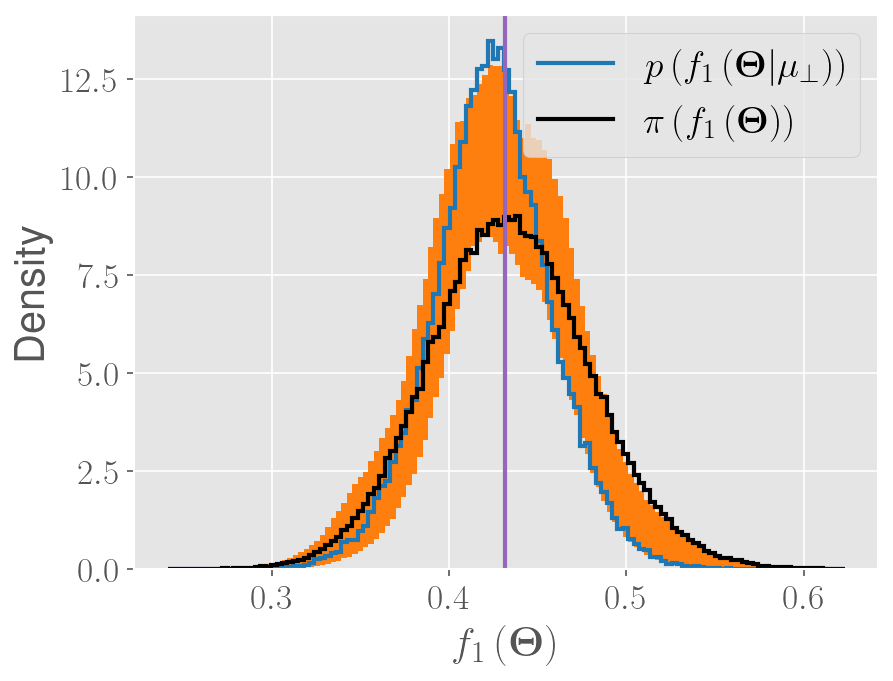}
  \caption{}
  \label{fig:example_2_case_2_conditional_posterior_output_nominal_and_errorbars_dim_1}
\end{subfigure}%
\begin{subfigure}{.24\textwidth}
  \centering
  \includegraphics[width=\textwidth]{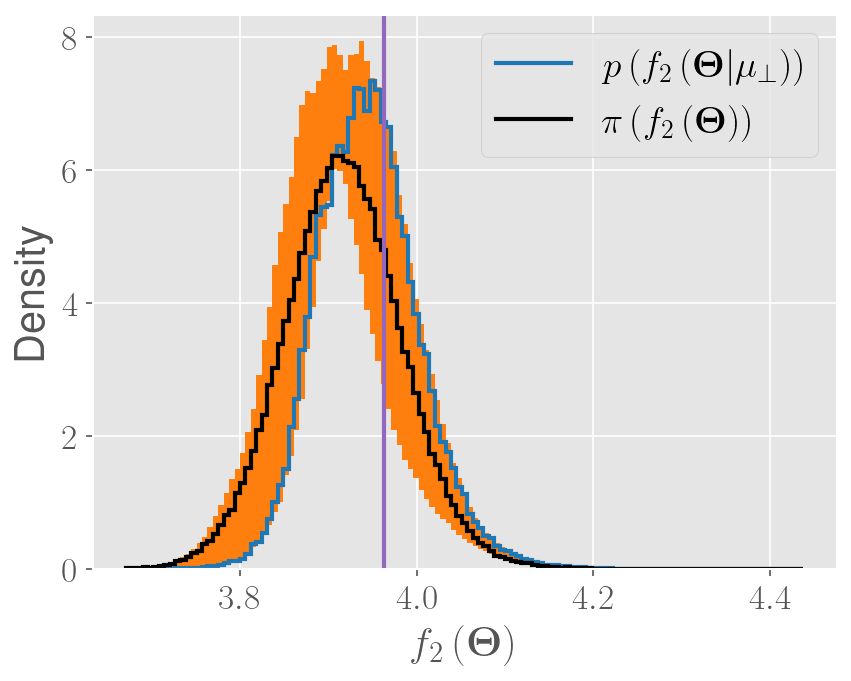}
  \caption{}
  \label{fig:example_2_case_2_conditional_posterior_output_nominal_and_errorbars_dim_2}
\end{subfigure}%
\begin{subfigure}{.24\textwidth}
  \centering
  \includegraphics[width=\textwidth]{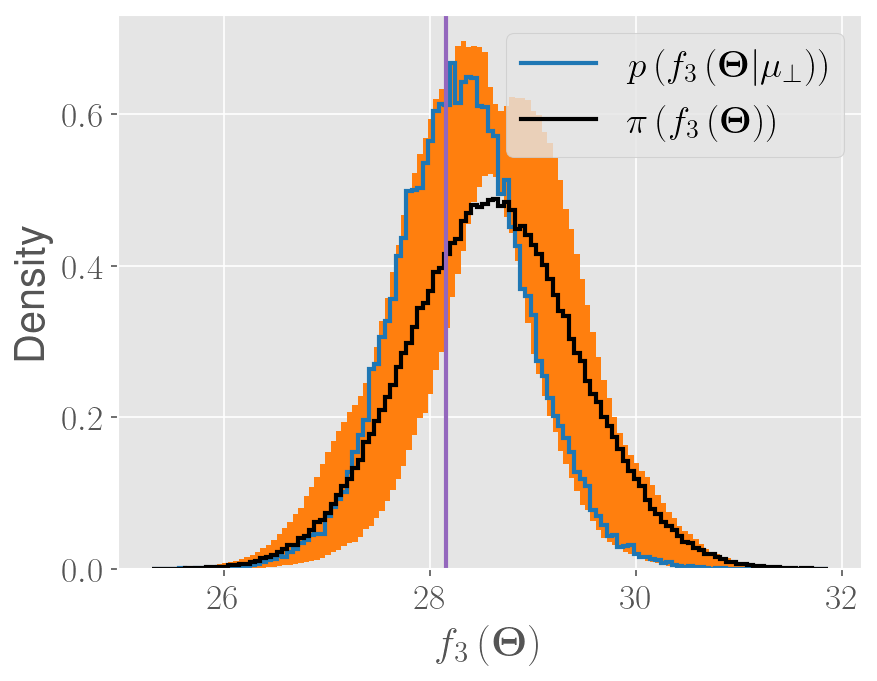}
  \caption{}
  \label{fig:example_2_case_2_conditional_posterior_output_nominal_and_errorbars_dim_3}
\end{subfigure}%
\begin{subfigure}{.24\textwidth}
  \centering
  \includegraphics[width=\textwidth]{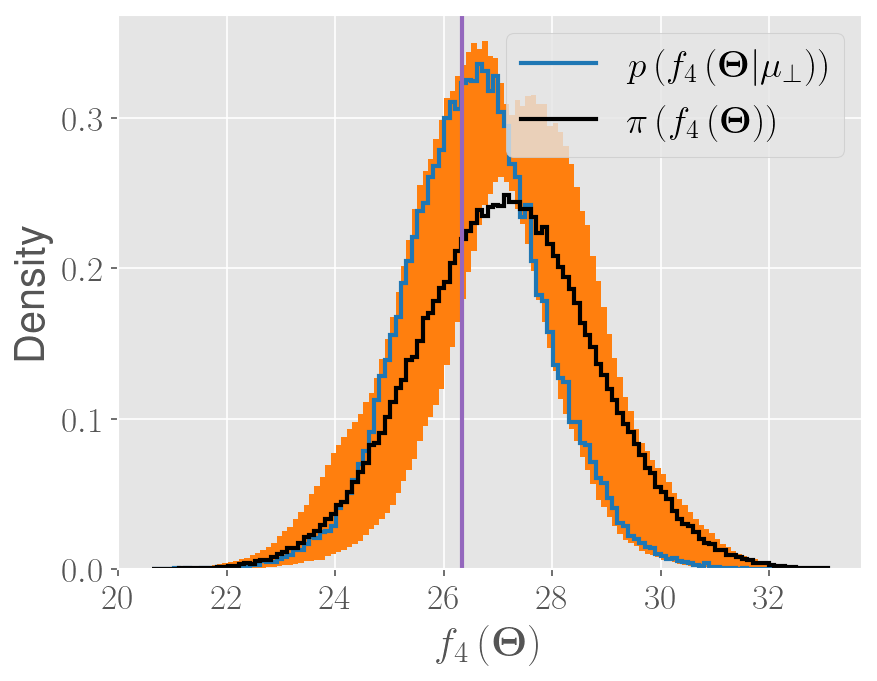}
  \caption{}
  \label{fig:example_2_case_2_conditional_posterior_output_nominal_and_errorbars_dim_4}
\end{subfigure}
\begin{subfigure}{.24\textwidth}
  \centering
  \includegraphics[width=\textwidth]{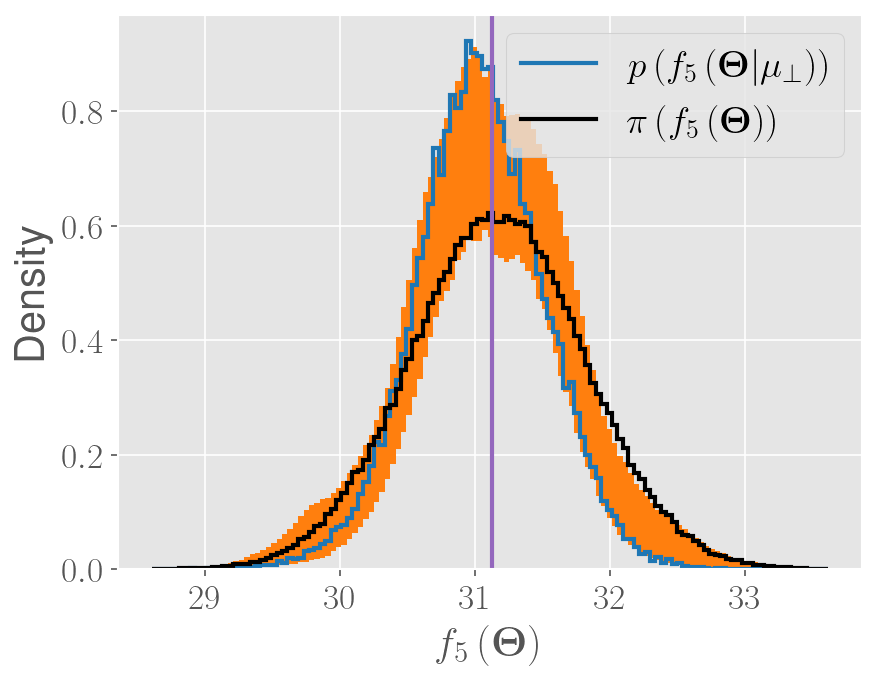}
  \caption{}
  \label{fig:example_2_case_2_conditional_posterior_output_nominal_and_errorbars_dim_5}
\end{subfigure}%
\begin{subfigure}{.24\textwidth}
  \centering
  \includegraphics[width=\textwidth]{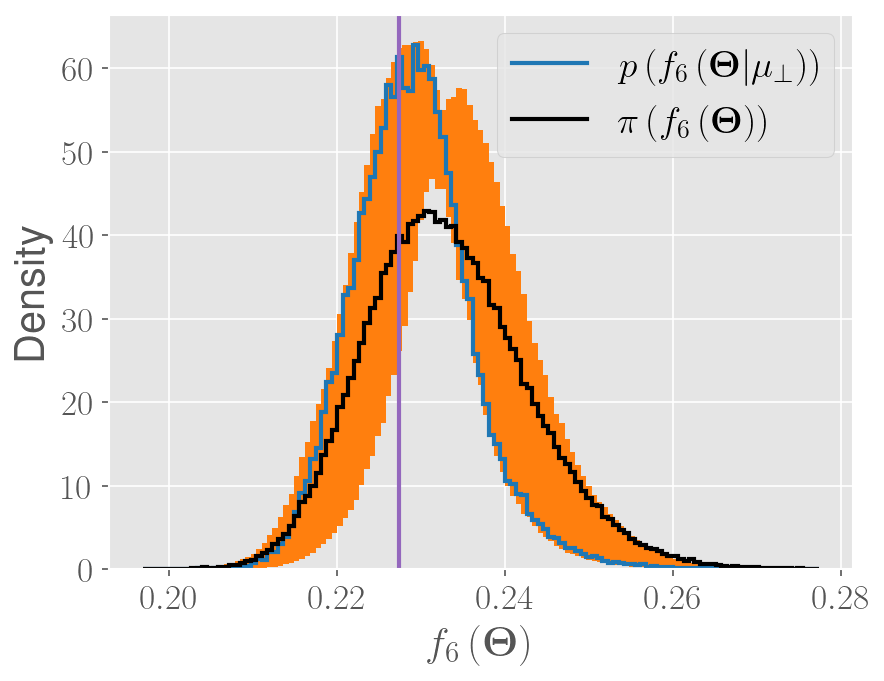}
  \caption{}
  \label{fig:example_2_case_2_conditional_posterior_output_nominal_and_errorbars_dim_6}
\end{subfigure}%
\begin{subfigure}{.24\textwidth}
  \centering
  \includegraphics[width=\textwidth]{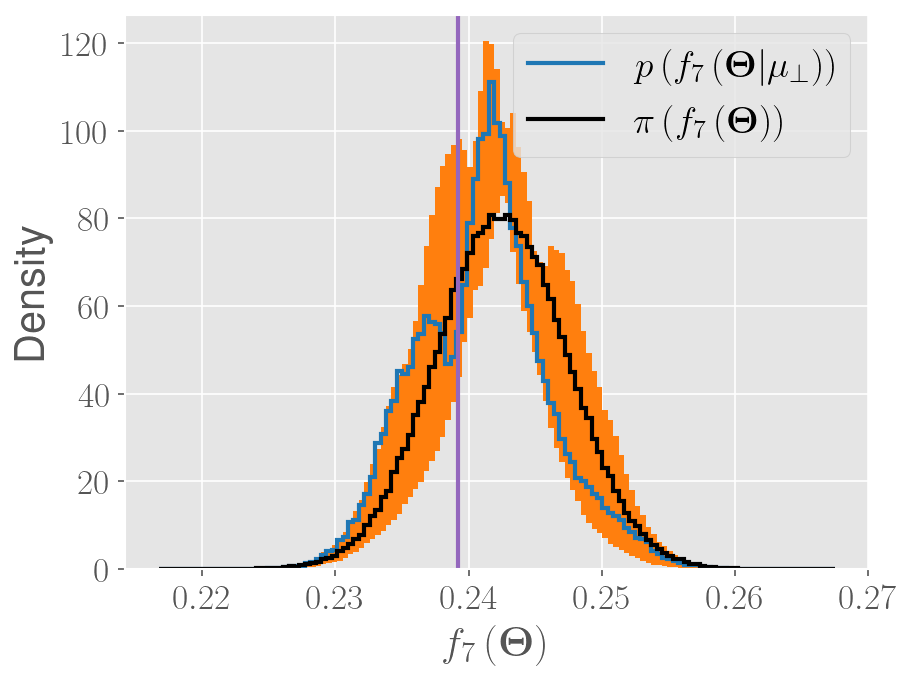}
  \caption{}
  \label{fig:example_2_case_2_conditional_posterior_output_nominal_and_errorbars_dim_7}
\end{subfigure}%
\begin{subfigure}{.24\textwidth}
  \centering
  \includegraphics[width=\textwidth]{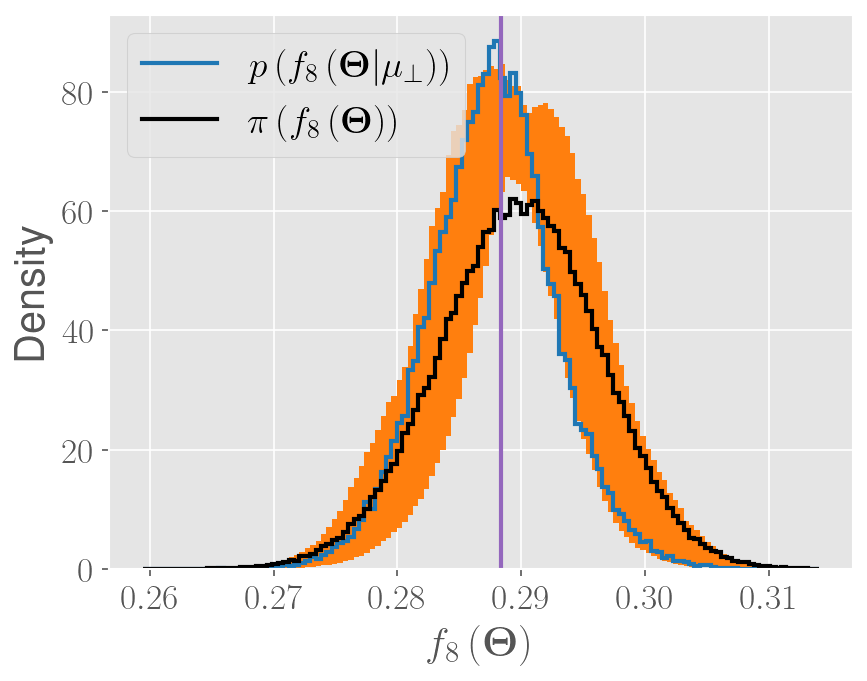}
  \caption{}
  \label{fig:example_2_case_2_conditional_posterior_output_nominal_and_errorbars_dim_8}
\end{subfigure}
\begin{subfigure}{.24\textwidth}
  \centering
  \includegraphics[width=\textwidth]{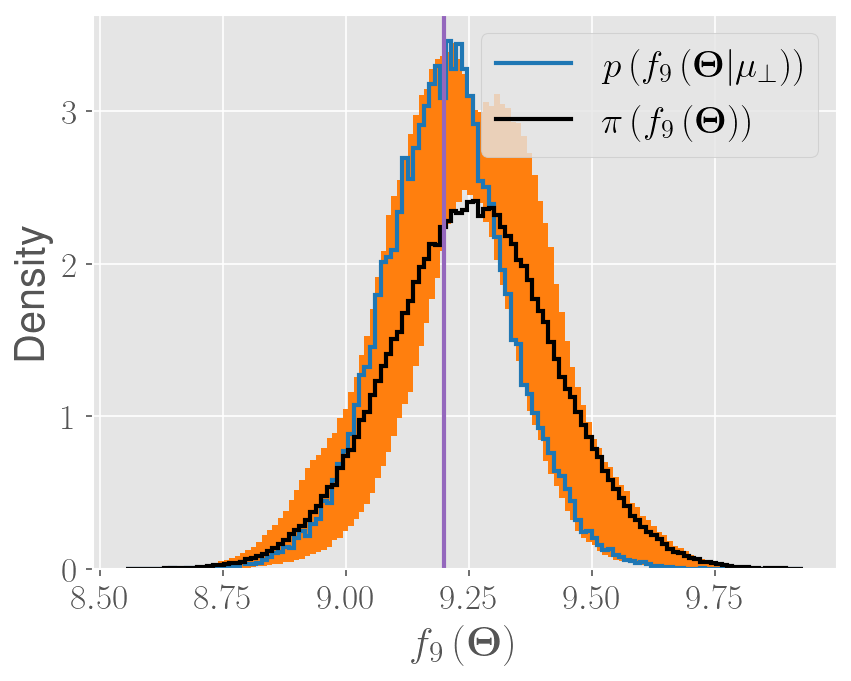}
  \caption{}
  \label{fig:example_2_case_2_conditional_posterior_output_nominal_and_errorbars_dim_9}
\end{subfigure}%
\begin{subfigure}{.24\textwidth}
  \centering
  \includegraphics[width=\textwidth]{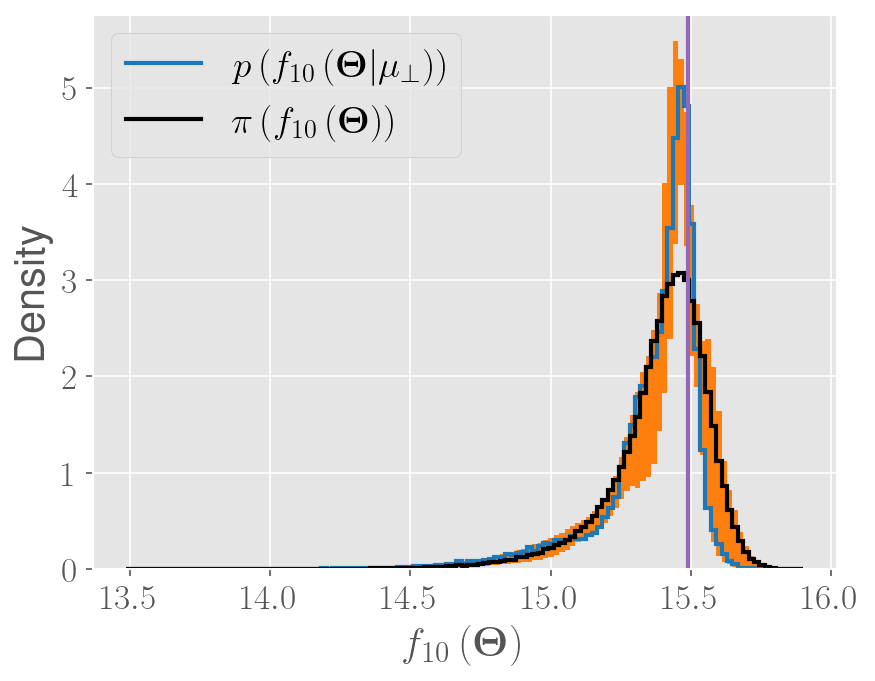}
  \caption{}
  \label{fig:example_2_case_2_conditional_posterior_output_nominal_and_errorbars_dim_10}
\end{subfigure}%
\begin{subfigure}{.24\textwidth}
  \centering
  \includegraphics[width=\textwidth]{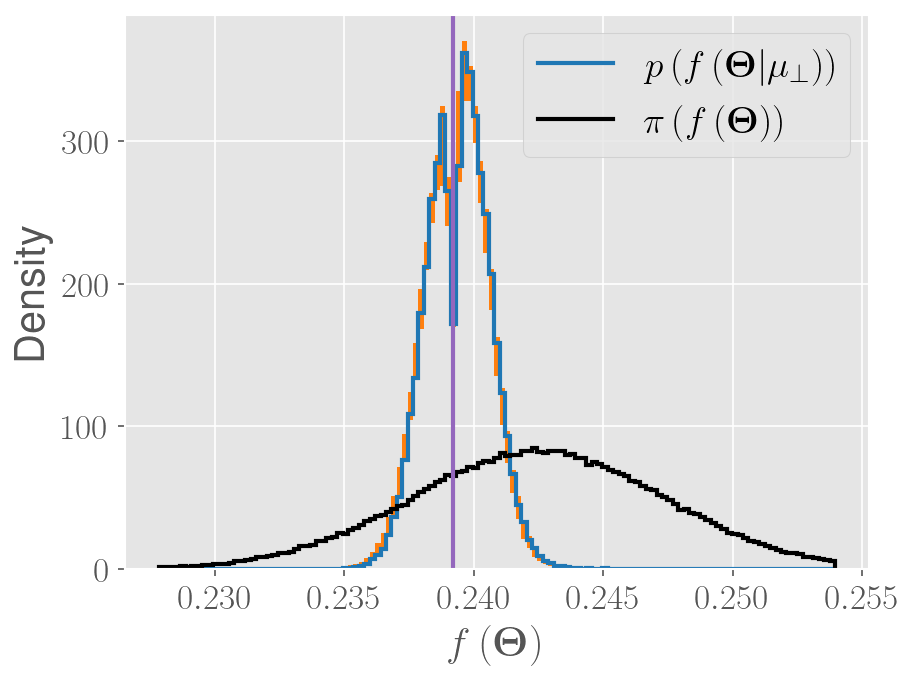}
  \caption{}
  \label{fig:example_2_case_1_conditional_posterior_output_nominal_and_errorbars}
\end{subfigure}
\caption{Plots of the density of $ f \left( \boldsymbol{\Theta} \right) $ corresponding to the conditional active posterior $ p \left( \boldsymbol{\Theta}_{\mu_{\perp}} \right) $ for the vehicle side impact test problem (see Section~\ref{section:example_2_vehicle_side_impact}). Figures (a) - (j) depict the marginal output densities of the 10 output components of $ f \left( \boldsymbol{\Theta} \right) $ (i.e., $ f_i \left( \boldsymbol{\Theta} \right) $, $ i = 1, \dots, 10 $) for Case 2, while figure (k) depicts the density of $ f \left( \boldsymbol{\Theta} \right) $ for Case 1. In each figure, the nominal calibrated posterior constructed using $ \mathfrak{C}_{\boldsymbol{\Theta}} $ is in blue, while the orange bands depict the $ 90 \% $ confidence interval of the posteriors constructed using the sample set replicates $ \mathfrak{C}_{\boldsymbol{\Theta}}^{(i)} $ is in orange. Additionally, the prior density is drawn in black, with the purple line indicating the observed value used for calibration.}
\label{fig:calibrated_conditional_output_posteriors_results_for_example_2}
\end{figure}

Supplementary figures for the vehicle side impact problem from Section~\ref{section:example_2_vehicle_side_impact} are included here. The LIS projection vectors not included in Figure~\ref{fig:active_subspace_results_for_example_2} (i.e., eigenvectors 4 - 11) are shown in Figure~\ref{fig:example_2_case_1_all_projection_vectors} for Case 1 and Figure~\ref{fig:example_2_case_2_all_projection_vectors} for Case 2. Figures~\ref{fig:calibrated_full_posteriors_results_for_example_2_case_1} and~\ref{fig:calibrated_conditional_posteriors_results_for_example_2_case_1} collect all one-dimensional marginals as well as the misfit probability density functions corresponding to the full posterior and the conditional active posterior for Case 1, respectively. Figures~\ref{fig:calibrated_full_posteriors_results_for_example_2_case_2} and~\ref{fig:calibrated_conditional_posteriors_results_for_example_2_case_2} do the same for Case 2. Finally, Figures~\ref{fig:calibrated_marginal_output_posteriors_results_for_example_2} and~\ref{fig:calibrated_conditional_output_posteriors_results_for_example_2} show the probability density of the output $ f \left( \boldsymbol{\Theta} \right) $ corresponding to the full posterior and the conditional active posterior of the parameters, respectively. For Case 2, the one-dimensional marginals of the output (i.e., corresponding to each individual response variable) are plotted. Notably, the one-dimensional marginal QoI posterior distributions are frequently bimodal, indicating that the joint posterior has a complex shape, with various combinations of discrepancies in the QoI components producing similar misfit values.

\bibliographystyle{unsrt}
\bibliography{references}

\end{document}